\documentclass[12pt,titlepage,twoside]{article}
\usepackage{float}
\usepackage{latexsym}
\usepackage{amsmath,mathrsfs,bm,amssymb,color}
\usepackage{graphicx}
\usepackage{bbm}
\usepackage{hyperref}

\usepackage{makeidx}
  \makeindex

\newtheorem{theorem}{Theorem}[section]
\newtheorem{proposition}[theorem]{Proposition}
\newtheorem{lemma}[theorem]{Lemma}
\newtheorem{corollary}[theorem]{Corollary}
\newtheorem{definition}[theorem]{Definition}
\newtheorem{example}[theorem]{Example}
\newtheorem{remark}[theorem]{Remark}

\newtheorem{assumption}[theorem]{Assumption}

\newcommand{\ko}[1]{\index{#1}}

\newcommand{\proof}{{\noindent \it Proof:\ }}
\newcommand{\nani}{H_{\rm f}^{\amass '}}
\newcommand{\navv}{\dv'_\k}

\newcommand{\ppmm}{\gamma_\pm}
\newcommand{\naka}{\eta}
\newcommand{\ho}{{\rm H}}
\newcommand{\vol}{|S_{d-1}|}
\newcommand{\eb}{e^{-\theta(x)}}
\newcommand{\MMM}[4]
{\left[ \!\!\!\begin{array}{cc}#1&#2\\
#3&#4\end{array}\!\!\!\right] }
\newcommand{\rieman}{\eta}
\newcommand{\sg}{\sqrt{|{\rm det} g|}}

\newcommand{\y}{{\sqrt\omega}}
\newcommand{\yy}{\frac{1}{\y}}
\newcommand{\kkt}{|k|^2}
\newcommand{\kkkt}{|k'|^2}
\newcommand{\kkkkt}{|k''|^2}
\newcommand{\tama}{Q_\Omega}

\renewcommand{\H}[1]{\sigma(#1)\cap [\is(#1),\is(#1)+\amass)\subset \sigma_{\rm disc}(#1)}
\newcommand{\qed}{\hfill $\Box$\par\medskip}
\newcommand{\BR}{{{\mathbb  R}^d}}

\newcommand{\BT}{{{\mathbb  R}^3}}
\newcommand{\RR}{{{\mathbb  R}}}

\renewcommand{\sp}{{\mathfrak{Sp}}}
\newcommand{\spp}{{\mathfrak{sp}}}
\newcommand{\CC}{{{\mathbb  C}}}
\newcommand{\NN}{{{\mathbb  N}}}
\newcommand{\Spec}{\sigma}

\newcommand{\sln}{(\Lambda N)^{1/4}}
\newcommand{\XX}{E^{\eff}_{[\Sigma, \Sigma+\delta)}}
\newcommand{\YY}{E^{\eff} _{[\Sigma+\delta,\infty)}}

\newcommand{\fdd}{\frac{d-2}{2}}
\newcommand{\KKK}{K}
\newcommand{\KKKK}{K^\perp}
\newcommand{\KKKKK}{\ms K}

\providecommand{\TTT}[1]{
{\rm (\textbf{#1})}}
\newcommand{\sumn}{\sum_{n=0}^\infty}
\newcommand{\JJJ}{C}
\newcommand{\TT}{{\ms T}}
\newcommand{\CA}{\CCA'}
\newcommand{\CCA}{\theta_\k}

\newcommand{\Jj}{{\rm J}}
\newcommand{\LM}{\lk \frac{\lambda}{\Lambda}\rk }
\renewcommand{\sl}{\sqrt{\Lambda}}
\newcommand{\sll}{\Lambda^{1/4}}
\newcommand{\fla}{f_\Lambda}
\newcommand{\gla}{h_\Lambda}
\newcommand{\liml}{\lim_{\Lambda\rightarrow\infty}}
\newcommand{\limln}{\lim_{\Lambda, N \rightarrow\infty}}
\newcommand{\limn}{\lim_{n\to \infty}}

\newcommand{\AAA}{{\rm A}}
\newcommand{\BBB}{{\rm B}}
\newcommand{\bdd}{b_\AAA^\ast}
\newcommand{\bbb}{b_\AAA}
\newcommand{\bsss}{b_\AAA^\#}
\newcommand{\LN}{\ell}
\newcommand{\eq}[1]{\begin{equation}\label{#1}}
\newcommand{\en}{\end{equation}}
\newcommand{\ba}{\begin{align*}}
\newcommand{\ea}{\end{align*}}

\newcommand{\eqq}[1]{\label{#1}}
\newcommand{\dd}[1]{\D^\ast _{#1}}
\newcommand{\da}[1]{N_{#1}}
\renewcommand{\aa}[1]{\D_{#1}}
\newcommand{\ddd}{{\cal D}_\infty}
\newcommand{\ui}[1]{e^{-\half \D_{#1}^\ast }}
\newcommand{\uii}[1] {\wick{e^{- N_{#1}}}}
\newcommand{\uiii}[1]{e^{-\half \D_{#1}}}

\newcommand{\eqn}{\begin{align*}}
\newcommand{\enn}{\end{align*}}
\newcommand{\bt}[1]{\begin{theorem}\label{#1}}
\newcommand{\et}{\end{theorem}}
\newcommand{\bl}[1]{\begin{lemma}\label{#1}}
\newcommand{\el}{\end{lemma}}
\newcommand{\bp}[1]{\begin{proposition}\label{#1}}
\newcommand{\ep}{\end{proposition}}
\newcommand{\bc}[1]{\begin{corollary}\label{#1}}
\newcommand{\ec}{\end{corollary}}
\newcommand{\bi}{\begin{itemize}}
\newcommand{\ei}{\end{itemize}}
\newcommand{\nn}{\Psi_{\varepsilon'}}

\newcommand{\IR}{\d \int\!\!\frac{\vp^2}{\omega^3}dk<\infty}
\newcommand{\SR}{\d \int\!\!\frac{\vp^2}{\omega^3}dk=\infty}

\newcommand{\LRT}{{L^2({\mathbb R}^3)}}
\renewcommand{\ll}{\LR}
\newcommand{\LR}{{L^2({\mathbb R}^d)}}

\newcommand{\mc}{M_0\cap M_{-1/2}}
\newcommand{\LRNd}{{L^2({\mathbb R}^{dN})}}

\newcommand{\dg}{d\Gamma}

\newcommand{\fff}{{\ms F}}
\newcommand{\ffff}{{{\ms F}_{\rm fin}}}
\newcommand{\fffb}{{\ms F}}
\newcommand{\hhh}{{\ms H}}
\newcommand{\hf}{{H_{\rm f}}}

\newcommand{\f}{^{-1}}
\newcommand{\ov}[1]{\overline{#1}}
\newcommand{\add}{a^{\ast}}
\newcommand{\ass}{a^{\sharp}}

\newcommand{\han}{{1\!/\!2}}
\newcommand{\ham}{\frac{1}{2m}}
\newcommand{\half}{\frac{1}{2}}
\newcommand{\lk}{\left(}
\newcommand{\rk}{\right)}
\newcommand{\lkk}{\left\{}
\newcommand{\rkk}{\right\}}
\newcommand{\is}{\inf\!\sigma}
\newcommand{\hi}{H_{\rm I}}
\newcommand{\kak}[1]{(\ref{#1})}
\newcommand{\hp}{H_{\rm p}}

\def\bbbone{{\mathchoice {\rm 1\mskip-4mu l} {\rm 1\mskip-4mu l}
{\rm 1\mskip-4.5mu l} {\rm 1\mskip-5mu l}}}
\def\one{\bbbone}

\newcommand{\pro}[1]{(#1_t)_{t\geq0}}
\newcommand{\ms}[1]{{\mathscr #1}}

\renewcommand{\d}{\displaystyle}
\newcommand{\non}{\nonumber}
\newcommand{\gr}{\varphi_{\rm g}}

\newcommand{\vp}{\hat\varphi}
\newcommand{\ground}{g}

\newcommand{\ovv}[1]{\ov{#1}}

\newcommand{\jmp}[5]{{#1}, {#2}, {\it J. Math. Phys.} {\bf #3} (#4), #5}
\newcommand{\jfa}[5]{{#1}, {#2}, {\it J. Funct. Anal.} {\bf #3} (#4),
#5}

\renewcommand{\k}{{\kappa }}

\newcommand{\eai}{e_a^i  }
\newcommand{\ems}{e_\mu^i  }

\newcommand{\enr}{e_\nu^j}
\newcommand{\ejn}{e_\nu^j }

\newcommand{\ejk}{e^j(k)}
\newcommand{\ej}{e^j}

\newcommand{\fin}{\ak }

\newcommand{\dmu}{d_{\mu\nu}}

\newcommand{\dbm}{d_{b\mu} }

\newcommand{\tmn}{T_{\mu\nu}}

\newcommand{\tab}{T_{ab}}
\newcommand{\btmn}{\overline{T}_{\mu\nu}}

\newcommand{\sumnn}{\sum_{n=1}^M}
\newcommand{\sumnnm}{\sum_{n=0}^M}

\newcommand{\limM}{\!\!\lim_{M\rightarrow\infty}}

\newcommand{\qqqqq}{\mmm}
\newcommand{\rrr}{\Kp^\nu(k,j)}

\newcommand{\ak}{\alpha}

\newcommand{\ddk}{dk}

\newcommand{\pij}{W_{+ij}}
\newcommand{\mij}{W_{-ij}}
\newcommand{\bmij}{\overline{W}_{-ij}}
\newcommand{\bpij}{\overline{W}_{+ij}}

\newcommand{\wwp}{{W_{+}}}

\newcommand{\jp}{{W_+^\ast}}
\newcommand{\jm}{{W_-^\ast}}

\newcommand{\ip}{{\overline{W}_+^\ast}}
\newcommand{\im}{{\overline{W}_-^\ast}}

\newcommand{\II}{\one}
\newcommand{\vm}{\vec{v}_\mu}
\newcommand{\vn}{\vec{v}_\nu}
\newcommand{\vf}{\vec{f}}

\newcommand{\wwm}{{W_{-}}}

\newcommand{\bwwp}{{\overline{W}_{+}}}

\newcommand{\bwwm}{{\overline{W}_{-}}}

\newcommand{\oooo}{\omega}
\newcommand{\JI}{Q}

\newcommand{\ooo}[1]{\omega^{#1}}

\newcommand{\T}[1]{\ov{#1^\ast}}

\newcommand{\Mm}{\frac{1}{m}\M}
\providecommand{\wick}[1]
{:\!\!{#1}\!\!:}

\newcommand{\CONS}{\sum_{\phi:{\rm CONS}}}
\newcommand{\nnn}{\sum_{\mu=1}^d}

\newcommand{\ta}{\widehat{A}}
\newcommand{\tp}{\widehat{\Pi}}

\newcommand{\sss}{\sum_{|l|\leq \aL}}
\newcommand{\aL}{2\pi L}

\newcommand{\ld}{D\times D}
\newcommand{\ka}{(s^2+A_0)^{-1}}

\renewcommand{\bm}{\M}
\newcommand{\tr}{{\rm tr}}

\newcommand{\ppp}{P}
\newcommand{\qqq}{Q}

\newcommand{\mm}{|\ak |}

\newcommand{\sh}{\frac{1}{\sqrt{2}}}

\newcommand{\nf}{{N}}

\newcommand{\hfn}{{H_{\rm f}^\amass}}

\newcommand{\alg}{\otimes_{\rm alg}}

\newcommand{\thn}{H^\amass}

\newcommand{\n}{\grn}

\newcommand{\amass}{\varepsilon}
\newcommand{\mren}{m_{\rm ren}}

\newcommand{\ph}{\hat{\varphi}}
\newcommand{\phk}{\hat{\varphi}(k)}
\newcommand{\rh}{{\rho}}
\newcommand{\eps}{\epsilon}

\newcommand{\re}{{\rho}_\epsilon}

\newcommand{\dpp}{D(-\Delta)\cap D(\hf)}

\newcommand{\ccc}{{C_0^\infty}(\BR)}

\newcommand{\dv}{\delta V}
\newcommand{\dvn}{\delta V}

\newcommand{\PH}{\chi_{\Gamma(l,a)}}
\newcommand{\LLL}{\sum_{|l|\leq  L} }
\newcommand{\Kp}{\varrho}
\newcommand{\limla}{\lim_{L\rightarrow \infty}\lim_{a\rightarrow \infty}}

\newcommand{\veff}{{V_{\rm eff}}}
\newcommand{\at}{\ak^2}
\newcommand{\ai}{A_\infty}

\newcommand{\A}{{A}}
\newcommand{\V}{P}

\newcommand{\q}{\ak }

\renewcommand{\S}{Schr\"odinger\ }
\newcommand{\pfh}{Pauli-Fierz\ }
\newcommand{\D}{\Delta}

\newcommand{\p}{{-i\nabla}}

\newcommand{\dip}{K}
\newcommand{\ds}{H_p}
\newcommand{\dsz}{H_0}

\newcommand{\two}{{\rm II}}
\newcommand{\three}{{\rm III}}
\newcommand{\four}{{\rm IV}}

\newcommand{\onu}{\omega}
\newcommand{\M}{\frac{\at}{m}}

\newcommand{\vv}[1]{\vec{v}_{\mu_#1}}

\newcommand{\mv}{(\max_\mu\|\nabla_\mu V\|_\infty)}
\newcommand{\lc}[2]{\left.{#1}\right\lceil_{#2}}

\newcommand{\on}{\omega_\epsilon}

\newcommand{\vpp}{\vec{p}}
\newcommand{\vq}{\vec{q}}

\newcommand{\N}{{C_N}}
\newcommand{\WWW}{{\mathfrak{h}}}

\newcommand{\UU}{\ms U}
\newcommand{\WWWWW}{W}

\newcommand{\eff}{{H_{{\rm eff}}}}
\newcommand{\beff}{\ov{H_{{\rm eff}}}}
\newcommand{\he}{H_{\eps}}
\newcommand{\hez}{H_{\eps}(0)}

\newcommand{\hep}{H_{\eps,p}}
\newcommand{\hren}{H_{\rm ren}}

\newcommand{\heff}{H_{\rm eff}}
\newcommand{\e}{\epsilon}
\newcommand{\eT}{\langle T\rangle_D}

\newcommand{\Ebb}{{\mathbb E}}

\newcommand{\HS}{{\rm I}_2}
\newcommand{\HSS}{\ov{{\rm I}_2}}
\newcommand{\ST}{\mat{S}{\ov T}{T}{\ov S}}
\newcommand{\STt}{\mat{S_t}{\ov T_t}{T_t}{\ov S_t}}
\newcommand{\WST}{\mat {W_+} {\ov{W_-}} {W_-} {\ov{W_+}}}
\newcommand{\WSTE}{\mat {W_+^\eps} {\ov{W_-^\eps}} {W_-^\eps} {\ov{W_+^\eps}}}
\newcommand{\XST}{\mat {X_+} {\ov{X_-}} {X_-} {\ov{X_+}}}
\newcommand{\Y}{E^{\beff}_{[|\Sigma|,\infty)}}

\newcommand{\mat}[4]{
\def\arraystretch{0.5}
\lk\!\!\begin{array}{cc}
#1&#2\\
#3&#4
\end{array}
\!\!\rk
}

\newcommand{\s}{\sigma}

\newcommand{\essinf}{\operatorname*{ess.inf}}

\newcommand{\cE}{\mathcal{E}}

\newcommand{\slim}{\mathop{\mbox{\rm s-lim}}}

\newcommand{\supp}{\mathop{\mathrm{supp}}}
\usepackage{graphicx,color}
\newcommand{\vep}{V_\epsilon}

\newcommand{\limcp}{\lim_{\substack{\epsilon,\delta\downarrow 0\\
 R\to\infty}}}

\newcommand{\cp}{{C_{\epsilon,\delta,R}}}
\newcommand{\limdt}{\lim_{t\downarrow 0}}

\newcommand{\ap}{T_\k}
\newcommand{\apm}{T_\k\f}

\newcommand{\KK}{{\mathcal K}}

\newcommand{\jjj}{\sum_{j=1}^{d-1}}
\newcommand{\X}{E^{\beff}_{[0,|\Sigma|)}}
\newcommand{\xx}{{\rm x}}
\newcommand{\yyy}{{\rm y}}

\newcommand{\mmm}{{\lk\frac{d-1}{d}\rk}}
\newcommand{\mass}{m_{\rm eff}}
\newcommand{\grn}{{\Psi_\amass}}
\newcommand{\hz}{-\frac{1}{2\mass}\Delta}

\newcommand{\be}{\begin{eqnarray*}}
\newcommand{\ee}{\end{eqnarray*}}
\newcommand{\la}{{\lambda}}
\newcommand{\Ew}[1]{{\mathbb E}\left[#1\right]}
\newcommand{\mup}{d{\mu}_{\rm p}}
\newcommand{\lkkk}{\left[}
\newcommand{\rkkk}{\right]}
\newcommand{\double}{{\ms W}}
\newcommand{\FFF}{{\mathcal F}}
\newcommand{\La}{\Lambda}
\newcommand{\EE}{\mathbb E}
\newcommand{\sqrtt}[1]
{\Big[\!#1\!\Big]^{\!\!\han}}
\newcommand{\eee}{e^{-(T/2^n)\lpb}}
\newcommand{\HP}{{{\ms H}_{\rm p}}}

\newcommand{\jj}{{\rm j}}
\newcommand{\JJ}{{\rm J}}
\newcommand{\domf}{{\rm Q}}

\newcommand{\ix}[1]
{\int\!\! \mup   {\mathbb E}^\xx
\!\!\left[
#1
\right]}
\newcommand{\wY}{\widetilde Y}

\newcommand{\ixp}[1]
{\int \mup   {\mathbb E}_{P^\xx }
\left[
#1
\right]}

\newcommand{\WTTI}{
\int_{-T}^T\! ds
\int_{-T}^T \!dt
\double_\infty( X_s,X_t,|s-t|)
}

\newcommand{\WTT}{
\int_{-T}^T\! ds
\int_{-T}^T \!dt
\double( X_s,X_t,|s-t|)
}

\newcommand{\WT}{
\int_{-T}^0 \!ds
\int_0^T \!dt \double( X_s,X_t,|s-t|)
}

\newcommand{\WTI}{
\int_{-T}^0 \!ds
\int_0^T \!dt \double_\infty( X_s,X_t,|s-t|)
}
\newcommand{\omegai}{\omega_\infty}

\newcommand{\WTZ}{
C_1 \int_{-T}^0\! ds
\int_0^T \! dt \double_\infty (X_s,X_t,C_2|s-t|)
}

\newcommand{\wtt}{
\int_{-T}^T \! ds
\int_{-T}^T \! dt \double}

\newcommand{\wt}{
\int_{-T}^0 \! ds
\int_0^T \! dt \double}

\newcommand{\wti}{
\int_{-T}^0 \! ds
\int_0^T \! dt \double_\infty}

\newcommand{\dom}{D}
\newcommand{\VV}{W}
\newcommand{\ab}[1]{\langle#1
\rangle}
\newcommand{\lp}{L_{\rm p}}
\newcommand{\lpb}{{L}_{\rm p}}

\makeatletter
\@addtoreset{equation}{section}
\makeatother

\title
{\sc Enhanced Binding in Quantum Field Theory}
\author{
\small Fumio Hiroshima\\
{\small\it Faculty of Mathematics, Kyushu University}    \\[-0.7ex]
{\small\it   Fukuoka,   Japan}
\\[0.3cm]
\small Itaru Sasaki\\
{\it \small Department of Engineering, Shinshu University} \\[-0.7ex]
{\it \small Matsumoto,   Japan }
 \\[0.3cm]
\small Herbert Spohn\\
{\small\it Mathematical Zentrum, Technische Universit\"at M\"unchen}    \\[-0.7ex]
{\small\it  M\"unchen,  Germany}
\\[0.3cm]
\small Akito Suzuki\\
{\it \small Department of Engineering, Shinshu University} \\[-0.7ex]
{\it \small Nagano,    Japan } \\[-0.7ex]
}

\begin{document}
\maketitle

\begin{abstract}
Enhanced binding in quantum field theory and related topics are  reviewed,
which suggests new possibility, beyond toy models, in the study of
the stability of
quantum field models.
This Lecture Note reviews
papers below:
\small{
\begin{description}
\item
[1.]
F. Hiroshima and H. Spohn,
Enhanced binding through coupling to quantum field,
{\it Ann. Henri Poincar\'e} {\bf 2}  (2001),
1159--1187.
\item
[2.]
F. Hiroshima and I. Sasaki,
Enhanced binding of an $N$ particle system interacting
with a scalar field I,
{\it Math. Z.} {\bf  259}  (2008),
657--680.
\item
[3.]
F. Hiroshima, H. Spohn and A. Suzuki,
 The no-binding regime of the Pauli-Fierz model,
 {\it J.  Math.  Phys.}  {\bf 52}  (2011),
 062104.
\item
[4.]
C. G\'erard, F. Hiroshima, A. Panati,
and A. Suzuki,
Absence of ground state of the Nelson model with variable coefficients,
{\it J.  Funct.  Anal.} {\bf 262}  (2012),
273--299.
\end{description}
}
This lecture note consists of three parts.
Fundamental facts on Boson Fock space are  introduced in Part I.
Ref. 1.and 3. are reviewed in Part II and, Ref.  2. and 4. in Part III.

In Part I  a symplectic structure of a Boson Fock space is studied and
 a projective unitary representation of an infinite dimensional symplectic group
through Bogoliubov transformations is constructed.

In Part II  the so-called Pauli-Fierz  model (PF model) with the dipole approximation in non-relativistic
quantum
electrodynamics is investigated.
This  model describes  a minimal interaction between a massless quantized radiation field and
a quantum mechanical particle (electron) governed by Schr\"odinger operator.
By applying the Bogoliubov transformation introduced in Part I we investigate the spectrum of
the PF model.
First  the translation invariant case is considered and  the dressed electron state
with a fixed momentum is studied.
Secondly
the absence of ground state is proven by extending the Birman-Schwinger principle.
Finally  the enhanced binding  of a ground state is discussed and
the transition from  unbinding  to binding is shown.

In Part III  the so-called $N$-body Nelson model is studied.
This model describes a linear interaction between a scalar field and $N$-body quantum mechanical particles.
First the enhanced binding is shown
by checking the so-called stability condition.
Secondly
the Nelson model with variable coefficients is  discussed,
which model can be derived when the Minkowskian space-time is replaced
by a static Riemannian manifold,
and the absence of ground state is proven, if
the variable mass decays to zero sufficiently fast.
The strategy is based on a path measure argument.
\end{abstract}

\newpage
\tableofcontents
\cleardoublepage

\setlength{\baselineskip}{16pt}
\part{Boson Fock space and symplectic structures}

\label{hukkatu}

\section{Boson Fock space}
 \subsection{Second quantization}
Let $\WWW$ be a separable Hilbert space
over the complex field $\CC$ with the scalar product $(\cdot,\cdot)_\WWW$.
Here the scalar product is linear in the second component  and antilinear in the first one.
We omit subscript $\WWW$ unless confusion may arise.
Consider the operation $\otimes _{\rm s}^n$ of
$n$-fold symmetric tensor product defined through the symmetrization operator
\eq{pa1}
\d S_n(f_1\otimes\cdots\otimes f_n) = \frac{1}{n!}\sum_{\pi\in \wp_n} f_{\pi(1)}\otimes \cdots \otimes
f_{\pi(n)}, \quad n\geq 1,
\en
where $f_1,...,f_n \in \WWW$ and $\wp_n$ denotes the permutation group of order $n$. Define
$
\otimes_{\rm s}^n \WWW =S_n(\otimes^n \WWW )$
with $\otimes _{\rm s}^0\WWW =\CC $. The space
\eq{N1}
\fffb(\WWW)=\bigoplus_{n=0}^\infty \otimes_{\rm s}^n \WWW ,
\en
 is
called the {boson Fock space}\index{boson Fock space} over $\WWW$. We simply denote $\fffb(\WWW)$ by $\fffb$.
The boson Fock space $\fffb$ can be identified
with the space of $\ell_2$-sequences $(\Psi^{(n)})_{n\geq 0}$ such that $\Psi^{(n)} \in \otimes_{\rm s}^n \WWW $
and
$
\|\Psi\|^2_\fffb =\sum_{n=0}^\infty\|\Psi^{(n)}\|_\fffb ^2<\infty$.
The boson Fock space $\fffb$ is a Hilbert space endowed with the scalar product
\eq{N2}
(\Psi, \Phi)_\fffb=\sum_{n=0}^\infty (\Psi^{(n)},\Phi^{(n)})_{\otimes_{\rm s}^n \WWW}.
\en
The element
$
\Omega = (1,0,0,...)\in\fffb$
is called {Fock vacuum}\index{Fock vacuum}.
In the description of the free quantum field the following operators acting in $\fffb$ are used. There are
two fundamental  operators, the {creation operator}\index{annihilation operator}
denoted by $\add(f)$, $f \in \WWW$, and the {annihilation operator}\index{creation operator} by $a(f)$, both acting on $\fffb$, defined by
\begin{align}
(\add(f)\Psi)^{(n)}=\lkk
\begin{array}{ll}
\sqrt n S_n(f\otimes \Psi^{(n-1)}),& n\geq1,\\
0,&n=0
\end{array}
\right.
\end{align}
with domain
$
D(\add(f))=\lkk
(\Psi^{(n)})_{n\geq0} \in\fffb\left| \sum_{n=1}^\infty n\|
S_n(f\otimes \Psi^{(n-1)})\|_{\fffb }^2<\infty\right.\rkk$,
and
$
a(f)=(\add(\bar f))^\ast$.
As the terminology suggests, the action of $\add(f)$ increases the number of bosons by one, while $a(f)$ decreases
it by one. Since one is the adjoint operator of the other, the relation $(\Phi, a(f)\Psi)_\fffb =(\add(\bar f)\Phi,\Psi)_\fffb $
holds. Furthermore, since both operators are closable by the dense definition of their adjoints, we will use and
denote their closed extensions by the same symbols. Let $D\subset \WWW$ be a dense subset. It is known that
\eq{dense}
\fffb=
\ov{
{\rm L.H.}
\{\add(f_1)\cdots\add(f_n)\Omega,\Omega|\, f_j\in D, j=1,..,n,n\geq1\}
},\en
where ${\rm L.H.}$ is a shorthand for the linear hull, and $\ov{\{\cdots\}}$ denotes the closure in $\fffb$.
The space
\eq{lll2}
\ffff=\left\{(\Psi^{(n)})_{n\geq 0} \in\fffb
\left| \; \Psi^{(m)}=0 \mbox{ for all } m \geq M \;\mbox{with some} \; M\right.\right\}
\en
is called {finite particle  subspace}.\index{finite particle subspace} The field operators $a,\add$ leave $\ffff$
invariant and satisfy the {canonical commutation relations}\index{canonical commutation relation}
\eq{lll3}
[a(f),\add(g)]=(\bar f,g)_\WWW\one,\quad
[a(f),a(g)]=0,\quad
[\add(f),\add(g)]=0
\en
on $\ffff$.
Given a bounded  operator $T$ on $\WWW$, the {second quantization}\index{second quantization} of $T$ is the
operator $\Gamma (T)$ on $\fffb$ defined by
\eq{gammat}
\Gamma(T)=\bigoplus_{n=0}^\infty
 \otimes ^n T.
\en
Here it is understood that $\otimes^0T=\one $. In most cases $\Gamma(T)$ is an unbounded operator resulting from the fact
that it is given by a countable direct sum. However, for a contraction operator $T$, the second quantization $\Gamma(T)$
is also a contraction on $\fffb$.
 The map $\Gamma$
 satisfies
 \eq{stst}
\Gamma(S)\Gamma(T)=\Gamma(ST),\quad \Gamma(S)^\ast=\Gamma(S^\ast),\quad \Gamma(\one_\WWW)=\one_\fffb.
\en
For a self-adjoint operator $h$ on $\WWW$ the structure relations
\kak{stst} imply in particular that
$
\{\Gamma (e^{it h})\}_{t\in\RR}
$
is a strongly continuous one-parameter unitary group on $\fffb$. Then by the Stone theorem
there exists a unique self-adjoint operator $d\Gamma(h)$ on $\fffb$ such that
\eq{gammas}
\Gamma(e^{it h})=e^{it\dg h},\quad t\in\RR.
\en
The operator $\dg (h)$ is called the {differential second quantization} \index{differential second quantization}
of $h$ or simply {second quantization} of $h$. Since
$
\d d\Gamma(h)=-i \frac{d}{dt} \Gamma(e^{ith})\lceil_{t=0}
$,
we have
\eq{N3}
d\Gamma(h)=0\oplus \left[\bigoplus_{n=1}^\infty \left(\ov{\sum_{j=1}^n\underbrace{
 \one \otimes\cdots\otimes
\stackrel{j}{h}\otimes\cdots\otimes \one}_{n} }\right)\right],
\en
where the overline denotes closure, and $j$ on top of $h$ indicates its position in the product. Thus the action of
$d\Gamma(h)$  is given by
\begin{align}
&d\Gamma(h)\Omega=0,\\
&
d\Gamma(h)\add(f_1)\cdots\add(f_n)
\Omega =\sum_{j=1}^n \add(f_1)\cdots \add(hf_j)\cdots\add(f_n)\Omega.
\end{align}
It can be also seen by \kak{N3} that
\begin{align*}
\Spec (d\Gamma(h))&=\ov
{\lkk
\left. \sum_{j=1}^n a_j\right|a_j\in \Spec (h),j=1,...,n,n\geq 1
\rkk
\cup\{0\}},
\\
\Spec_{\rm p}(d\Gamma(h))
&=
\lkk
\left.
\sum_{j=1}^n a_j\right|a_j\in \Spec _{\rm p}(h),j=1,...,n,n\geq 1\rkk\cup\{0\}.
\end{align*}
If $0\not\in \Spec_{\rm p}(h)$, the multiplicity of $0$ in $\Spec_{\rm p}(d\Gamma(h))$ is one.
A crucial operator in quantum field theory is the {boson number operator} \index{number operator} defined
by the second quantization of the identity operator on $\WWW$:
$
N=\dg (\one_\WWW)$.
Since
$N\Omega=0$ and
$
N\add(f_1)\cdots \add(f_n)\Omega=n\add(f_1)\cdots \add(f_n)\Omega$,
it follows that
$
\Spec(N)=\NN \cup\{0\}$.
We will use the following facts below.
\begin{proposition}
\label{A123}
Let $h$ be a nonnegative  self-adjoint operator, and $f\in D(h^{-\han})$, $\Psi\in D(d\Gamma(h)^\han)$. Then
$\Psi\in D(\ass(f))$ and
\begin{eqnarray}
\label{yy1}
\|a(f)\Psi\|&\leq& \|h^{-\han} f\|\|d\Gamma(h)^\han \Psi\|,\\
\label{yy2}
\|\add(f)\Psi\|&\leq& \|h^{-\han} f\| \| d\Gamma(h)^{\han}\Psi\|+\|f\| \|\Psi\|.
\end{eqnarray}
In particular, $D(d\Gamma(h)^\han)\subset D(\ass(f))$, whenever $f\in D(h^{-\han})$.
\end{proposition}
\noindent
To obtain the commutation relations between $\ass(f)$ and $d\Gamma(h)$,
suppose that $f\in D(h^{-\han})\cap D(h)$.
Then
\eq{con}
[d\Gamma(h), \add(f)]\Psi =\add(hf)\Psi, \quad [d\Gamma(h), a(f)]\Psi =-a(hf)\Psi,
\en
for $\Psi\in D(d\Gamma(h)^{3/2})\cap \ffff$.
By a limiting argument \kak{con} can be extended to $\Psi\in
D(d\Gamma(h)^{3/2})$, and it is seen that $\ass(f)$ maps $D(d\Gamma(h)^{3/2})$ into $D(d\Gamma(h))$.
In general
we can see that
$
\ass(f):D(d\Gamma(h)^{n+\han})\rightarrow D(d\Gamma(h)^n)
$
for all $n\geq 1$, when $f\in \bigcap_{n=1}^\infty D(h^{n/2}) $. In particular, $\ass(f)$
maps $\bigcap_{n=1}^\infty D(d\Gamma(h)^n)$ into itself.

Take now $\WWW = \LR$ and consider the boson Fock space $\fffb(\LR)$. In this case, for $n \in \NN$ the space
$\otimes_{\rm s}^n \LR$ can be identified with the set of symmetric functions on $L^2(\RR^{dn})$ through
\eq{N6-1}
\otimes_{\rm s}^n \LR\cong \{f\in L^2(\RR^{dn})| f(k_1, \ldots, k_n) = f(k_{\pi(1)}, \ldots, k_{\pi(n)}), \forall \pi\in\wp_n\}.
\en
The creation and annihilation operators \index{creation operator} \index{annihilation operator}
are realized as
\begin{eqnarray}
&&
\hspace{-2cm}
(a(f)\Psi)^{(n)}(k_1,...,k_n) = \sqrt{n+1}\int_\BR f(k) \Psi^{(n+1)}(k,k_1,...,k_n) dk,\ n\geq0,\\
&&
\hspace{-2cm}
(\add (f)\Psi )^{(n)}(k_1,...,k_n) = \lkk
\begin{array}{ll}
\frac{1}{\sqrt n} \sum_{j=1}^n f(k_j)
\Psi^{(n-1)}(k_1,...,\hat k_j,...,k_n),& n\geq1,\\
0,& n=0.
\end{array}
\right.
\end{eqnarray}
Here $\Psi\in \fffb $ is denoted as a pointwise defined function for convenience, however, all of these
expressions are to be understood in $L^2$-sense.
Let $\onu:\LR\rightarrow\LR$ be the multiplication operator called {dispersion relation}
\index{dispersion relation} given by
\eq{53}
\onu(k)=\sqrt{|k|^2+\nu^2},\quad k\in\BR,
\end{equation}
with $\nu\geq 0$. Here $\nu$ describes the {boson mass}. \index{boson mass} \index{second quantization}
The second quantization
of the dispersion relation is
\eq{freesec}
\lk d\Gamma(\onu) \Psi\rk ^{(n)}(k_1,...,k_n)=\lk \sum_{j=1}^n \onu(k_j)\rk \Psi^{(n)}(k_1,...,k_n).
\en
The self-adjoint operator $d\Gamma(\onu)$ is called the free field Hamiltonian\ko{free field Hamiltonian}
on
$\fffb(\LR)$ and we use the notation
\eq{freeham}
\hf=d\Gamma(\onu).
\en
The spectrum of the free field Hamiltonian is
$\Spec(\hf)=[0,\infty)$,
with component
$\Spec_{\rm p}(\hf)=\{0\}$,
which is of single multiplicity with $\hf\Omega=0$. Then formally we may write the free field Hamiltonian
as
\eq{formalhf}
\hf=\int \onu(k)\add(k) a(k) dk.
\en
Physically, this describes the total energy of the free field since $\add(k) a(k)$ gives the number of bosons
carrying momentum $k$, multiplied with the energy $\onu(k)$ of a single boson, and integrated over all momenta.
The commutation relations are
\eq{N11}
[\hf, a(f)]=-a(\onu f),\quad [\hf,\add(f)]=\add(\onu f).
\en
The relative bound of $\ass(f)$ with respect to the free field Hamiltonian $\hf$ can be seen from \kak{N71} and \kak{N72}.
If $f/\sqrt\onu\in \LR$, then
\begin{eqnarray}
\label{N71}
\|a(f)\Psi\|&\leq&\|f/\sqrt\onu\|
\|\hf^\han\Psi\|,\\
\label{N72}
\|\add(f)\Psi\|
&\leq& \|f/\sqrt\onu\|\|\hf^\han\Psi\|+\|f\|\|\Psi\|
\end{eqnarray}
hold.

\subsection{Segal fields}
\label{segal}
The creation and annihilation operators are not symmetric and do not commute. Roughly speaking, a creation operator
corresponds to $\frac{1}{\sqrt2}(x-\frac{d}{dx})$ and an annihilation operator to $ \frac{1}{\sqrt2}(x+\frac{d}{dx})$ in
$L^2(\RR)$. We can, however, construct symmetric and commutative operators by combining the two field operators and this leads
to {Segal fields}\index{Segal fields}.
The {Segal field} $\Phi(f)$ on the boson Fock space $\fffb(\WWW)$ is defined by
\eq{s90}
\Phi(f)=\frac{1}{\sqrt 2} (\add(\bar f)+a(f)),\quad f\in\WWW,
\end{equation}
and its {conjugate momentum} \index{conjugate momentum} by
\eq{s91}
\Pi(f)=\frac{i}{\sqrt 2}(\add(\bar f)-a(f)),\quad f\in\WWW.
\en
Here $\bar f$ denotes the complex conjugate of $f$.
By the above definition both $\Phi(f)$ and $\Pi(g)$ are symmetric, however, not linear in $f$ and $g$ over $\CC$.
Note that, in contrast, they are linear operators over $\RR$.
\index{canonical commutation relation}
It is straightforward to check that
$
[\Phi(f), \Pi(g)]=i{\rm Re}( f, g)_\WWW \one_\WWW$,
$
[\Phi(f), \Phi(g)]=i{\rm Im}( f, g)_\WWW \one_\WWW$
and $
[\Pi(f), \Pi(g)]=i{\rm Im}( f, g)_\WWW \one_\WWW$.
In particular, for real-valued $f$ and $g$ the canonical commutation relations \index{canonical commutation relation}
become
\eq{s94}
[\Phi(f), \Pi(g)]=i(f,g)_\WWW \one_\WWW,\quad [\Phi(f),\Phi(g)]=[\Pi(f),\Pi(g)]=0.
\en
Applying the inequalities \kak{N71} and \kak{N72} to $h=\one $, we see that $\ffff$ is the set of analytic vectors
 of $\Phi(f)$, i.e.,
$
\d \lim_{m\rightarrow \infty} \sum_{n=0}^m {\|\Phi(f)^n \Psi\|t^n}/{n!}<\infty
$
for $\Psi\in \ffff$ and $t\geq 0$. The following is a general result.
\bp{nelsonana}\TTT{Nelson's analytic vector theorem}\ko{Nelson's analytic vector theorem}
Let $K$ be a symmetric operator on a Hilbert space. Assume that there exists a dense subspace
${\cal D}\subset D(K)$ such that $\d \lim_{m\to\infty}
 \sum_{n=0}^m  \|K^n f\|t^n/n!<\infty$, for $f\in {\cal D}$ and
some $t>0$. Then $K$ is essentially self-adjoint on ${\cal D}$, and $\d e^{-tK}\Phi=\slim_{m\to \infty}\sum_{n=0}^m t^n K^n f /n!$ follows for $f\in {\cal D}$.
\ep
\noindent
By  Nelson's analytic vector theorem both $\Phi(f)$ and $\Pi(g)$ are essentially self-adjoint on $\ffff$. We keep
denoting the closures of $\Phi(f)\lceil_{\ffff}$ and $\Pi(g)\lceil_{\ffff}$ by the same symbols.

\subsection{Wick product}
Loosely speaking, the so-called Wick product\index{Wick product} $\wick{\ass(f_1)\cdots\ass(f_n)}$ is defined in a product of creation and
annihilation operators by moving the creation operators to the left and the annihilation operators to the right
without taking the commutation relations into account.
The {Wick product} $\wick{\prod_{i=1}^n\Phi(g_i)}$ is recursively defined
by the equalities
\begin{align*}
\wick{\Phi(f)}=\Phi(f),\quad
\wick{\Phi(f) \prod_{i=1}^n \Phi(f_i)}=\Phi(f)\wick{\prod_{i=1}^n \Phi(f_i)}-\half\sum_{j=1}^n (f,f_j) \wick{\prod_{i\not=j}\Phi(f_i)}.
\end{align*}
By the above definition we have
\eq{defwick2}
\wick{\Phi(f)^n}=\sum_{k=0}^{[n/2]}\frac{n!}{k!(n-2k)!} \Phi(f)^{n-2k} \lk -\frac{1}{4}\|f\|^2\rk^k.
\en
Note that $\wick{\Phi(f_1)\cdots\Phi(f_n)}\Omega= 2^{-n/2}\add(f_1)\cdots\add(f_n)\Omega$.
From this
\eq{N21}
\left(
\wick{\prod_{i=1}^n \Phi(g_i)}\Omega, \wick{\prod_{i=1}^m \Phi(f_i)}\Omega \right)=\delta_{nm}
2^{-n/2} \sum_{\pi\in\ms P_n} \prod_{i=1}^n (g_i,f_{\pi(i)})
\en
follows.
The Wick product of the exponential can be computed directly to yield
\eq{o61}
\wick{e^{\ak \Phi(f)}}\Omega =
e^{-(1/4)\ak ^2\|f\|^2}
e^{\ak \Phi(f)}\Omega.
\en
Hence for real-valued $f$ and $g$,
\eq{o63}
(\Omega, {e^{\ak \Phi(f)}}\Omega_{\rm b})=e^{(1/4)\ak ^2\|f\|^2},\quad \ak \in\CC.
\en
For example
$(\Omega, {e^{i\Phi(f)}}\Omega_{\rm b})=e^{-(1/4)\|f\|^2}$ and
$(\Omega, {e^{\Phi(f)}}\Omega_{\rm b})=e^{(1/4)\|f\|^2}$.

\cleardoublepage
\section{Symplectic structure}
\subsection{Infinite dimensional symplectic group}
\label{bogoliubov}
In this section we investigate an infinite dimensional symplectic group\index{symplectic group} and
its projective unitary representation on a Fock space.
Symplectic transformations leave canonical commutation relations invariant.
By
symplectic transform of the annihilation operators and the creation operators $\{\ass\}$
we can construct operators $\{b^\sharp\}$ satisfying the same canonical commutation relations.
However it is not necessarily unitarily equivalent with each others if the dimension of the configuration space is infinity.
We will see it in Proposition \ref{yuichan}. We will also give an application of symplectic group in Section 4 to study the spectrum of
some quadratic self-adjoint operator.
The general reference of this section is \cite{ara91, ber66,hi04,rui77,rui78,seg70,sha62}.

Let $\JJJ $
be a conjugation on $\WWW $, i.e.,
$\JJJ $ is an antilinear isometry on
$\WWW $ with $\JJJ ^2=\one$.
For $f\in\WWW$
and
$T\in B(\WWW )$
($B=B(\WWW )$
is the  set of bounded linear operators on
$\WWW $),
we define
$\bar{f}\in \WWW $
and $\overline{T}\in B(\WWW )$ by
$\bar{f}=\JJJ f$ and
$
\overline{T}=\JJJ T\JJJ$ .
Let
$\HS=\HS(\WWW)$  denote the set of Hilbert-Schmidt operators on $\WWW$.
We denote the norm (resp. Hilbert-Schmidt norm) of a bounded operator
$X$ on $\WWW$ by $\|X\|$ (resp. $\|X\|_2$).
For  $S,T\in B$ we define
\eq{ha1}
\AAA=\ST:\WWW\oplus\WWW\rightarrow\WWW\oplus\WWW
\en
by
\eq{ka}
\AAA (\phi\oplus \psi)=(S\phi+\ov T \psi)\oplus (T\phi+\ov S \psi).
\en
Let
\eq{jop}
\Jj =\mat \one 0 0 {-\one}.
\en
Then $\Jj (\phi\oplus \psi)=\phi\oplus(-\psi)$.
We define a symplectic group
\ko{symplectic group}$\sp$ by
\eq{ha2}
\sp=\lkk \left. \AAA=\ST \right|
\AAA \Jj \AAA^\ast=\AAA^\ast \Jj \AAA=\Jj \rkk
\en
and
the  subgroup $\sp_2\subset\sp$ by
\eq{ha3}
\sp_2=\lkk \left.\AAA=\ST\in\sp\right|T \in \HS \rkk.
\en
Here
$\AAA^\ast$ is the adjoint of $\AAA$, i.e., $$\AAA^\ast=
{\mat{S^\ast}{T^\ast}{\ov{T^\ast}}{\ov {S^\ast}}}.$$
Note also that
  the inverse  of  $\AAA\in\sp$ is given by
$$
\AAA\f=\Jj \AAA ^\ast \Jj =\mat{S^\ast}{-T^\ast}{-\ov {T^\ast}}{\ov {S^\ast}}.$$
We can see that
$\AAA$  induces  the following maps:
\begin{align}
\label{2266}
\lk\!
\begin{array}{ll}
a(f)\\
\add(f)
\end{array}
\!\rk
\mapsto
\lk\!\begin{array}{ll}
b_{\AAA}(f)\\
b_{\AAA}^\ast(f)
\end{array}
\!\rk=
\lk\!\begin{array}{ll}
\add(Tf)+a(Sf)\\
\add(\bar S f)+a(\bar T f)
\end{array}
\!\rk.
\end{align}
Crucial  fact   is that  the map leaves
both
canonical commutation relations
\ko{canonical commutation relation}
\begin{align}
[\bbb(f),\bdd(g)]=(\ovv f, g)\one,\quad
[\bbb(f),\bbb(g)]=0=
[\bdd(f),\bdd(g)],
\end{align}
and
 adjoint relation
$(\Psi, \bdd(f)\Phi)
 =(\bbb(\ovv f)\Psi,\Phi)$.
Furthermore $\ass$ can be represented in terms of $b^\sharp_A$:
\begin{align}
\label{koro1}
a(f)&=b_\AAA(S^\ast f)-b_\AAA^\ast (\ov{T^\ast} f),\\
\label{koro2}
\add (f)&=-b_\AAA(T^\ast f)+b_\AAA^\ast (\ov{S^\ast} f).
\end{align}
In particular
the Segal field and its conjugate are represented as
\begin{align}
&\phi(f)=\sh(b_\AAA^\ast(
\ov{S^\ast f-T^\ast \bar f})+b_\AAA(S^\ast f-T^\ast \bar f)),\\
&\pi(f)=\frac{i}{\sqrt 2}(b_\AAA^\ast(\ov{S^\ast f+T^\ast \bar f})-b_\AAA({S^\ast f+T^\ast \bar f})).
\end{align}

  We will see below that $b_\AAA^\sharp(f)$ and $\ass(f)$ are unitarily equivalent if and only if $T$ is Hilbert Schmidt operator and will construct the unitary operator implementing this unitary equivalence.
\begin{proposition}
\TTT{Necessary  and sufficient condition to the unitary equivalence}
\label{yuichan}
Let $\AAA=\ST\in\sp_2$ and define $b_\AAA^\sharp(f)$ by \kak{2266}.
Then
there exists a unitary operator $U$ such that
$U\f b_\AAA^\sharp (f) = \ass(f)$ if and only if  $T$ is the Hilbert Schmidt operator.
\end{proposition}
\proof
We give the proof of the necessary part only for the case of $\WWW=\LR$.
The proof of sufficient part is given in Proposition \ref{bog2}.

Set $\Omega_\AAA=U\Omega$. Then $b_\AAA(f)\Omega_\AAA=0$ for all $f\in\LR$.
Hence $(a(Sf)+\add(Tf))\Omega_\AAA=0$ for all $f\in\LR$. Let $P_n$ be the projection from $\fffb$ to the $n$-particle subspace.
Then we have $$a(Sf)P_{n+2}\Omega_\AAA=
P_{n+1}a(Sf)\Omega_\AAA=
-P_{n+1}\add(Tf)\Omega_\AAA=
-\add(Tf)P_n\Omega_\AAA.$$
When $P_m\Omega_\AAA=0$, $a(Sf)P_{m+2}\Omega_\AAA=0$ for all $f\in\LR$, and then
$a(f)P_{m+2}\Omega_\AAA=0$, since $S\f$ exists.  Hence
$P_m\Omega_\AAA=0$ implies that $P_{m+2}\Omega_\AAA=0$. Since
$b_\AAA(f)\Omega_\AAA=0$, $a(Sf)P_1\Omega_\AAA=0$ and then $P_n\Omega_\AAA=0$ for all odd number $n$. If furthermore $P_0\Omega _\AAA=0$, $P_m\Omega_\AAA=0$ for all even number $m$, and it implies that $\Omega_\AAA=0$. Since $\Omega_\AAA\not=0$,
$\k =P_0\Omega_\AAA\not\not=0$ follows.
Notice that $\Phi=P_2\Omega_\AAA$ is
a function belonging to  $L^2(\BR\times \BR)$ and
\eq{ha66}
a(Sf)P_2\Omega_\AAA=-\add(Tf)P_0\Omega_\AAA.
\en
We see that
$a(Sf)P_2\Omega_\AAA=\sqrt 2\int (Sf)(k')\Phi(k',k) dk'$ and
$-\add(Tf)P_0\Omega_\AAA(k)=\k  Tf(k)$.
Let $K_\Phi$ be the Hilbert Schmidt operator defined by $K_\Phi f(k)=\int f(k')\Phi(k',k) dk$.
We then conclude  that
$$(Tf)(k)=\frac{\sqrt 2}{\k }\int (Sf)(k')\Phi(k',k) dk'=K_\Phi Sf(k).$$
Since $S$ is bounded and $K_\Phi$ is Hilbert-Schmidt,
$T$ is Hilbert-Schmidt operator.
\qed

\subsection{Quadratic operators}
\ko{quadratic operator}
Let $K\in \HS $ and $S\in B$.
Then there exist two orthonormal systems $\{\psi_n\}$ and $\{\phi_n\}$ in $\WWW$ and a positive sequence $\lambda_1\geq\lambda_2\geq\cdots\geq 0$ such that $Kf=\sumn\lambda_n(\psi_n, f)\phi_n$ with $\sumn \lambda_n^2=\|K\|_2^2$, where $\|\cdot\|_2$ denotes the Hilbert-Schmidt norm.
\bl{quadratic}
{\rm \cite{rui77,rui78,ara90}}
Let  $\{e_n\}$ be an arbitrary  complete orthonormal system of $\WWW$.
Then
for $\Psi\in\ffff$,
sequences
\eq{seq1}
\lkk
\sumnn\lambda_n\add(\ovv\psi_n)\add(\phi_n)\Psi\rkk,
\lkk
 \sumnn\lambda_n a(\ovv\psi_n) a(\phi_n)\Psi\rkk,
 \lkk
  \sumnn \add(e_n)a(\ovv{S^\ast e_n})\Psi
  \rkk
  \en
  strongly converge as $M\to\infty$.
  \el
\proof
We check only the convergence of $\sumnn\lambda_n\add(\ovv\psi_n)\add(\phi_n)\Psi$. The others  are similar or rather simpler. We have
\begin{align*}
\left\| \sumnn\lambda_n\add(\ovv\psi_n)\add(\phi_n)\Psi
\right\|^2
=
\sum_{m,n}^M
\lambda_n\lambda _m(\Psi,
a(\ov{\phi_n})a(\psi_n)
\add(\ovv\psi_m)\add(\phi_m)\Psi)
\end{align*}
and
\begin{align}
a(\ovv\phi_n)a(\psi_n)
\add(\ovv\psi_m)\add(\phi_m)
&\label{sin1}
=\delta_{nm}a(\ov{\phi_n})\add(\phi_m)
+\delta_{nm}\add(\ov{\psi_m})a(\psi_n)\\
&\label{sin2}
+
(\ov{\psi_n},\phi_m)a(\ov{\phi_n})\add(\ov{\psi_m})+
(\phi_n,\ov{\psi_m})\add(\phi_m)a(\psi_n)\\
&\label{sin3}
+
\add(\ov{\psi_m})\add(\phi_m)a(\psi_n)a(\ov{\phi_n}).
\end{align}
We will estimate $\sum_{m,n}^M \lambda_m\lambda_n (\Phi, (\ast)\Phi)$ for
$(\ast)=
\kak{sin1}, \kak{sin2}, \kak{sin3}$ separately.
For \kak{sin1} we have
\begin{align*}
&\sum_{m,n}^M \lambda_m\lambda_n (\Phi, \kak{sin1}\Phi)
\\
&=\lambda_n^2 \|\add(\phi_n)\Phi\|^2+\sum_n \lambda_n^2 \|a(\psi_n)\Phi\|^2
\leq 2\sum_n\lambda_n^2\|(N+1)\Phi\|^2
=2\|K\|_2^2\|(N+\one)\Phi\|^2.
\end{align*}
For the first term of $\sum_{m,n}^M \lambda_m\lambda_n (\Phi, \kak{sin2}\Phi)$  we have
$$\sum_{m,n}^M \lambda_m\lambda_n (\ov{\psi_n},\phi_m)(\phi_n,\ov{\psi_m})
\|\Phi\|^2+
\sum_{m,n}^M \lambda_m\lambda_n(\ov{\psi_n},\phi_m)
(a(\psi_m)\Phi, a(\ov{\phi_n})\Phi).$$
We see that
\begin{align*}
&\sum_{m,n}^M \lambda_m\lambda_n (\ov{\psi_n},\phi_m)(\phi_n,\ov{\psi_m})
\|\Phi\|^2
=
\sum_{m,n}^M \lambda_m\lambda_n ((\phi_m, \ov{\psi_n})\phi_n,\ov{\psi_m})\|\Phi\|^2\\
&=
\sum_m \lambda_m (K
 \ov{\phi_m}, \ov{\psi_m})\|\Phi\|^2
\leq(\sum_n\lambda^2_n)^\han(\sum_m\|K\ov{\phi_m}\|^2)^\han\|\Phi\|^2
=\|K\|_2^2\|
\|\Phi\|^2.
\end{align*}
For the second term
let $\Phi=\add(f_1)\cdots\add(f_L)\Omega$.
We have
\begin{align*}
&\sum_{m,n}^M \lambda_m\lambda_n(\ov{\psi_n},\phi_m)
\add
({\ov\psi_m}) a(\ov{\phi_n})\add(f_1)\cdots\add(f_L)\Omega\\
&=
\sum_j \sum_{m,n}^M \lambda_m\lambda_n(\ov{\psi_n},\phi_m)
\add((\phi_n,f_j){\ov\psi_m}) \add(f_1)\cdots\widehat{\add(f_j)}\cdots\add(f_L)\Omega\\
&
=\sum_j \sum_{m,n}^M \lambda_m\lambda_n
\add((\phi_n,f_j)(\ov{\psi_n},\phi_m){\ov\psi_m}) \add(f_1)\cdots\widehat{\add(f_j)}\cdots\add(f_L)\Omega \\
&=
\sum_j \sum_{m}\lambda_m
\add(\lambda_m (K\ov{\phi_m}, f_j){\ov\psi_m}) \add(f_1)\cdots\widehat{\add(f_j)}\cdots\add(f_L)\Omega\\
&=\sum_j
\add(\ov{K^\ast \ov {K^\ast f_j}})
 \add(f_1)\cdots\widehat{\add(f_j)}
 \cdots\add(f_L)\Omega.
 \end{align*}
Then the second term converges.
The second term of
of $\sum_{m,n}^M \lambda_m\lambda_n (\Phi, \kak{sin2}\Phi)$
is similarly estimated.
Finally for $\sum_{m,n}^M \lambda_m\lambda_n (\Phi, \kak{sin3}\Phi)$,
we have
\begin{align*}
\sum_{m,n}^M \lambda_m\lambda_n (\Phi, \kak{sin3}\Phi)&=\sum_n \lambda_n a(\psi_n)a(\ov{\phi_n})\add(f_1)\cdots\add(f_L)\Omega\\
&=
\sum_j a(\sum_n\lambda_n({\phi_n},f_j){\psi_n})
 \add(f_1)\cdots\widehat{\add(f_j)}
 \cdots\add(f_L)\Omega\\
& =
\sum_j a(K^\ast f_j)
 \add(f_1)\cdots\widehat{\add(f_j)}
 \cdots\add(f_L)\Omega.
\end{align*}
Then $\sum_{m,n}^M \lambda_m\lambda_n (\Phi, \kak{sin3}\Phi)$ also converges.
\qed
We can define for $\Psi\in\ffff$
\begin{align}
\eqq{ss1}
&\dd K\Psi=
{\rm s}\!-\!
\limM\sumnn\lambda_n\add(\ovv\psi_n)\add(\phi_n)\Psi,\\
&
\eqq{ss222}
\aa K\Psi={\rm s}\!-\!\limM\sumnn\lambda_n a(\ovv\psi_n) a(\phi_n)\Psi,\\
&
\eqq{ss3}
\da S\Psi={\rm s}\!-\!\limM\sumnn \add(e_n)a(\ovv{S^\ast e_n})\Psi.
\end{align}
Let $\Psi=\add(f_1)\cdots\add(f_n)\Omega$.
Then it is seen that
\begin{align}
\label{1to}
 & \aa K \Psi= \sum_{i\not=j}
 (\bar f_i,
 (K+\T K) f_j)
\add(f_1)\cdots \widehat{\add(f_i)}\cdots \widehat{\add(f_j)}\cdots  \add (f_n)\Omega,\\
\label{2}
& \da S \Psi=\sum_{j=1}^n
\add(f_1)\cdots{\add}(Sf_j)\cdots  \add (f_n)\Omega.
\end{align}
We  note that
$
(\D_K)^\ast=\D_{K^\ast}^\ast $ and
$
(N_S)^\ast=N_{S^\ast}.
$
Set $K^T=\T K$.
It can be also checked that
on $\ffff$,
\begin{align}
\label{u1}
 & [\dd K, a(f)]=-\add((K+K^T) f),\\
 &
\label{u2}
[\aa{K}, \add(f)]=a((K+K^T) f),\\
 &
\label{u3}
[\da S, a(f)]=-a(S^T  f),\\
 &
\label{u4}
[\da S,\add(f)]=\add( S f).
\end{align}
From \kak{u1} and \kak{u2}  it follows that
\begin{align*}
\|\dd K\Omega\|^2&=
(\dd K\Omega, \dd K\Omega)=
\sum_n\lambda_n(\Omega, \dd K \add(\bar\psi_n)\add(\phi_n)\Omega)\\
&=\sum_n\lambda_n(\Omega, a((K+K^T)\bar\psi_n)\add(\phi_n)\Omega)
=\sum_n\lambda_n(\ov{(K+K^T)}\psi_n,\phi_n).
\end{align*}
Since $\tr(KT)=\sum_n\lambda_n(\psi, T\phi_n)$, we have
$\|\dd K\Omega\|^2
=
\tr(K(K+K^T))$.
Moreover
\begin{eqnarray*}
\left\|
\sum_{n=0}^N \frac{1}{n!}
\left(-\frac{1}{2}\dd K\right)
^n\Omega\right\|^2
=\sum_{n=0}^N a_n,
\end{eqnarray*}
where $a_{n}=
(2^n n!)^{-2}\|(\dd K)^n\Omega\|^2$.
We set
$\ddd=\bigcap_{k=1}^\infty D(N^k)$.
We introduce a subset
$\HSS (\WWW)\subset \HS(\WWW)$
by
\eq{hs}
\HSS  (\WWW )
=
\{K\in \HS (\WWW )
|K=K^T,\|K\|<1\}.
\en
\begin{proposition}
\label{2.2m}
Let
$K\in \HSS (\WWW )$.
Then (1) and (2) hold:
\bi
\item[(1)]
For all
$|z|\leq \|K\|^{-2}$ and $k\geq0,$
the limit
$\d \lim_{N\rightarrow\infty}
\sum_{n=0}^N n^ka_n z^n$
exists.
In particular
$\d \sum_{n=0}^\infty a_nz^n
=\det(\one -zK^\ast K)^{-\frac{1}{2}}$.
\item[(2)]
 For all $\Phi\in \ffff$,
the strong limit
\eq{tumetai}
\exp\lk -\frac{1}{2}\dd K\rk \Phi
=
 {\rm s}\!-\!\lim_{N\rightarrow\infty}\sum_{n=0}^N
\frac{1}{n!}
\lk
-\frac{1}{2}\dd K\rk ^n\Phi,
\en
exists and  belongs  to ${\cal D}_\infty$.
 \ei
\end{proposition}
\proof
Let $\d a_{n,N}=\left\|
\frac{1}{n!}\lk
-\half \sum_{m=1}^N \lambda_n\add(\ov{\phi_m})a(\psi_m)\rk ^n\Omega\right\|^2$. Then we can see that
$$\sum_{n=0}^\infty a_{n,N}\ak ^n=\frac{1}{\sqrt{\prod_{j=1}^N(1-\ak  \lambda_j^2)}}$$
for $|\ak |<1$.
By the limiting argument
we have
$$\sum_{n=0}^\infty a_{n}\ak ^n
=\frac{1}{\sqrt{\prod_{j=1}^\infty (1-\ak  \lambda_j^2)}}=
[{\rm det}(\one-\ak  K^\ast K)]^{-\han}.$$
In particular
$\sum_{n=k}^\infty n(n-1)\cdots (n-k+1) a_{n}\ak ^{n-k}<\infty$.
 Thus
$$\left\|
\sum_{n=k}^\infty
\sqrt{n(n-1)\cdots (n-k+1)}
\frac{1}{n!}
(-\half \Delta_K)^n \Omega
\right\|^2<\infty$$
and
$$\d \sum_{n=0}^\infty \frac{1}{n!}(-\half \Delta_K)^n \Omega\in {\cal D}_\infty.$$
Furthermore
$\d \sum_{n=0}^\infty \frac{1}{n!}\lk
-\half \Delta_K\rk ^n \Phi$ converges for $\Phi=\add(f_1)\cdots \add(f_L)\Omega$.
\qed
Suppose that $S\in B$ and $K\in\HS$.
Then for $\Psi\in\ffff$,
we can also see (rather easier  than \kak{tumetai})
that
\begin{align}
 & {\uii{S}}\Psi={\rm s}\!-\!\limM\sumnnm\frac{1}{n!}\wick{\lk-
  \da S\rk^n} \Psi
\\
 &
{\uiii{K}}\Psi={\rm s}\!-\!\limM\sumnnm\frac{1}{n!}\lk-\half \aa K\rk^n\Psi
\end{align}
exist, and
${\uii{S}}\Psi, \uiii{K}\Psi\in \ffff$.
By  \kak{u1}-\kak{u4} we can check
the following  commutation relations
  on $\ffff$:
\begin{align}
&
\label{tt1}
[\ui K, a(f)]=\half \add((K+K^T) f) \ui K, \\
&
\label{tt2}
[\uiii K, \add(f)]=- \half a((K+K^T) f)\uiii K,\\
&
\label{tt3}
[\uii S, a(f)]= a(S^T  f)\uii S, \\
&
\label{tt4}
[\uii S, \add(f)]=-\add(S f)\uii S.
\end{align}

\begin{corollary}
Let $K_1\in \HS$,
$K_2$ and $K_2^{-1}$ be in
$B$,
and $K_1K_2^{-1}
\in \HSS $. Then
\eq{wagon}
\{\add(K_1 f)+a(K_2 f)\}
\Omega(K_1K_2^{-1})=0, \quad  f\in\WWW,
\en
where $\Omega(K_1K_2^{-1})=
\exp\lk
-\half \dd{K_1K_2\f}\rk
\Omega$.
\end{corollary}
\proof
By \kak{tt1}, we can see that
$$
a(K_2f)
\exp\lk
-\half \dd{K_1K_2\f}\rk
\Omega
=-\add(K_1f)
\exp\lk
-\half \dd{K_1K_2\f}\rk
\Omega.
$$
The desired result follows.
\qed

\subsection{Bogoliubov   transformations}
In this section we construct a unitary operator implementing
the unitary equivalence between $\ass$ and $b_\AAA^\sharp$ when $A\in\sp_2$.
\index{Bogoliubov transformation}
\subsubsection{Homogeneous case}
Let $\AAA=\ST\in\sp$.
Then $\AAA$  induces  the following maps:
\begin{align}
\lk\!
\begin{array}{ll}
a(f)\\
\add(f)
\end{array}
\!\rk
\mapsto
\lk\!\begin{array}{ll}
b_{\AAA}(f)\\
b_{\AAA}^\ast(f)
\end{array}
\!\rk=
\lk\!\begin{array}{ll}
\add(Tf)+a(Sf)\\
\add(\bar S f)+a(\bar T f)
\end{array}
\!\rk.
\end{align}
Formally we may write as
$(
{\bbb (f)} \
{\bdd(f)})
=({a(f)}\  {\add(f)})\AAA$.
Since $\bbb(f)\lceil_{\ffff}$ (resp. $\bdd$)
 is closable, we denote its closed extension
 by the same symbol $\bbb(f)$ (resp $\bdd(f)$).
It is  seen in Proposition \ref{yuichan}
 that
there exists a unitary operator
$\UU_\AAA$ on $\fff$
such that
$
\UU_\AAA\f \bsss(f) \UU_\AAA=\ass(f)$
if
$\AAA\in\sp_2$.
The condition $\ST\in \sp$  is equivalent
to  the following algebraic relations:
\begin{eqnarray}
\eqq{s1}
& & S^\ast S-T^\ast T=\one,
\\
& &
\eqq{s2}\ov {S^\ast} T-\ov {T^\ast} S=0,
\\
& & \eqq{s30}SS^\ast-\ov {T {T^\ast}}=\one,
\\
& & \eqq{s40}TS^\ast-\ov{ S  T^\ast} =0.
\end{eqnarray}
Using these algebraic relations we can prove that
$S\f\in B$,
$\|TS\f\|<1$,
${(TS\f)^T}=TS\f$,
and ${(S\f \ov T)^T} =S\f \ov T$.
We set
\begin{align}
K_1=TS\f,\ \ \ K_2=\one-\ov{(S\f)^\ast},\ \ \
K_3=-S\f \ov T.
\end{align}
Let $\ST\in \sp_2$. Since $K_1\in \HS$, ${K_1^T}=K_1$  and $\|K_1\|<1$, i.e., $K_1\in\HSS $,
we can see that by Proposition \ref{2.2m},
\eq{for}
\UU_\AAA={\rm det}(\one -K_1^\ast K_1)^{1/4}
\ui{K_1}
{\uii{K_2}}
\uiii{K_3}
\en
is  well defined on $\ffff$ and
$\UU_\AAA\Psi\in {\cal D}_\infty$
for $\Psi\in\ffff$.
$\UU_\AAA$ is called the intertwining operator\ko{intertwining operator}associated with $\AAA\in \sp_2$.
\bp{bog2}
\TTT{Homogeneous case}
\index{Bogoliubov transformation!homogeneous case}Let $\AAA\in\sp_2$. Then
$\UU_\AAA$ can be uniquely extended to the  unitary operator on $\fff$ and
\eq{homog}
\UU_\AAA\f
b_\AAA^\sharp (f)\UU_\AAA= \ass(f)
  \en  holds for all $f\in\WWW$.
\ep
\proof
Let $\UU_1=\ui{K_1}$, $\UU_2=
{\uii{K_2}}$ and $\UU_3=
\uiii{K_3}$.
By commutation relations \kak{tt1}-\kak{tt4}
we see that
\begin{eqnarray}
\UU_1 \UU_2 \UU_3
\add(f)
&=&
\UU_1 \UU_2
\add(f) \UU_3 +
\UU_1 \UU_2 a(-K_3 f) \UU_3 \nonumber \\
&=&
\label{3.12}
\add((\one -K_2 )f)
\UU_1 \UU_2 \UU_3
+ \UU_1 \UU_2
a(-K_3 f) \UU_3.
\end{eqnarray}
Using
$\one -K_2=\ov{(S\f)^\ast}=\bar S-TS\f\bar T$ and $\add((\one -K_2)f)
=
\bdd (f)
+a(-\overline{T}f)
+\add(-TS^{-1}\overline{T}f),$
we  have
\begin{align*}
& \add((\one -K_{2})f)
\UU_{1}\UU_{2}\UU_{3}\\
&=
\bdd (f)
\UU_{1}\UU_{2}\UU_{3}
+ \UU_{1}\add(-TS^{-1}\overline{T}f)
\UU_{2}\UU_{3}
+ \UU_{1}
\{\add(K_{1}\overline{T}f)
 +a(-\overline{T}f)\}
\UU_{2}\UU_{3}.
\end{align*}
Hence  the right hand side above is identical with \eq{3.14}
\add((\one -K_{2})f)
\UU_{1}\UU_{2}\UU_{3}=
\bdd(f)\UU_1\UU_2\UU_3+
\UU_{1}\UU_{2}
\{a(-\overline{T}f)
+a(K_{2}^{T}\overline{T}f)\}\UU_{3}.
\en
Combining \kak{3.12} and \kak{3.14},
 we obtain
$$
\UU_{1}\UU_{2}\UU_{3}
\add(f)
=\bdd (f)
\UU_{1}\UU_{2}\UU_{3}
+\UU_{1}\UU_{2}
a(-\overline{T}f
+K_{2}^T \overline{T}f
-K_{3}f)\UU_{3}.
$$
Since
$-\overline{T}+K_{2}^T \overline{T}-K_{3}=0$,
we get
$
\UU_1\UU_2\UU_3\add(f)\Phi
=
\bdd (f)\UU_1\UU_2\UU_3\Phi$
for all $\Phi \in {\cal D}_\infty$ and $f\in\WWW $.
I.e.,
\eq{OY}
\UU_\AAA \ass(f)\Phi= \bsss(f) \UU_\AAA \Phi,\quad \Phi\in{\cal D}_\infty.
\en
From this,  and the canonical commutation relations it follows that
$$\|\UU_\AAA\add(f_1)\cdots \add(f_n)\Omega\|^2
=
\|\bdd(f_1)\cdots \bdd(f_n)\UU_\AAA\Omega\|^2
=
\|\add(f_1)\cdots \add(f_n)\Omega\|^2,
$$
where   we used that
$\|\ui{K_1}\Omega\|^2={\rm det}
(\one -K_1^\ast K_1)^{-\han}$
and
$\bbb(f) e^{-\half \D_{K_1}^\ast }\Omega=0$.
Then $\UU_\AAA$ is an isometry from $\ffff$ onto
the dense subspace:
\begin{align*}
{\ms E}= L.H.
\{\bdd(f_1)\cdots \bdd(f_n)\UU_\AAA\Omega,\UU_\AAA\Omega| f_j\in\WWW, j=1,...,n,n\geq 1\}.
\end{align*}
We notice that
$\bsss(f){\ms E}\subset {\ms E}$ and
 $\ass$ can be represented
in terms of  a  linear combination  of $\bsss$.
See \kak{koro1} and \kak{koro2}.
By this  we see that $\ass(f)$ also leaves
${\ms E}$ invariant:
$\ass(f) {\ms E}\subset {\ms E}$
for all $f\in\WWW$.
Let $\Psi\in{\ms E}$ and
$\Psi_N=\{\Psi^{(0)},\Psi^{(1)}, ...,\Psi^{(N)},0,0,...\}$.
Since $\Psi_N\in\ffff$, we see that
$\Psi_N$ is an analytic vector of $\phi(f)=\frac{1}{\sqrt 2}(\add(\bar f)+a(f))$,
which implies, together with $\ass(f)
{\ms E}\subset {\ms E}$,
 that
$e^{i\phi(f)}\Psi_N\in\ov{\ms E}$, and by a limiting argument
$e^{i\phi(f)}\Psi\in\ov{\ms E}$.
Thus
$e^{i\phi(f)}{\ms E}\subset \ov{\ms E}$ follows. By a limiting argument we have
$e^{i\phi(f)}\ov {\ms E}\subset \ov{\ms E}$.
Thus
$\ov{\ms E}=\fff$ by the irreducibility of $\{\phi(f)\}$.
Hence we conclude that
$\UU_\AAA$ can be uniquely extended to a unitary operator on $\fff$.
Then the  proposition  follows.
\qed

\subsubsection{Inhomogeneous case}
Let $\AAA=\ST\in \sp$ and $L\in\WWW$.
Then it induces the map
\begin{align}
\lk\!
\begin{array}{ll}
a(f)\\
\add(f)
\end{array}
\!\rk
\mapsto
\lk\!\begin{array}{ll}
b_{\AAA,L}(f)\\
b_{\AAA,L}^\ast(f)
\end{array}
\!\rk=
\lk\!\begin{array}{ll}
\add(Tf)+a(Sf)+(L,f)\\
\add(\bar S f)+a(\bar T f)+(\bar L,f)
\end{array}
\!\rk.
\end{align}
It is clear that $b_{\AAA,L}^\#$ satisfies canonical commutation relations:
$[b_{\AAA,L}(f),b_{\AAA,L}^\ast(g)]=(\bar f,g)$ and $[b_{\AAA,L}(f),b_{\AAA,L}(g)]=0=
[b_{\AAA,L}^\ast(f),b_{\AAA,L}^\ast(g)]$.
Moreover $\ass$ can be represented  in terms of a linear sum of $b_{\AAA,L}^\#$:
\begin{align}
\label{korokoro1}
a(f)&=b_{\AAA,L}(S^\ast f)-b_{\AAA,L}^\ast (\ov{T^\ast} f)+(-SL+\bar T \bar L,f),\\
\label{korokoro2}
\add (f)&=-b_{\AAA,L}(T^\ast f)+b_{\AAA,L}^\ast (\ov{S^\ast} f)+(-\bar S\bar L+TL,f).
\end{align}
We define the operator $\pi(L)
$ in $\fffb$
by
$\pi(L)=
i\{b_\AAA(L)
-b_\AAA^{\ast}(\overline{L})\}$, which
can be represented in terms of $\ass$ by
\eq{pi}
\pi(L)=
i\{a^\ast (TL-\overline{SL})-
a(\overline{TL}-{SL})\}.
\en
Let
$S_{\AAA,L}=
\exp(-i\pi(L))=\exp(b_\AAA(L)-
b_\AAA^\ast(\ov L))$ and we define
$\UU_{\AAA,L}$ by
\eq{uuu}
\UU_{\AAA,L}=S_{\AAA,L}\UU_\AAA.
\en
Here
$S_{\AAA,L}$ is called the displacement operator\index{displacement operator}
and $\UU_{\AAA.L}$ Bogoliubov transform\index{Bogoliubov transformation}.
 \bp{inhom}
\TTT{Inhomogeneous case}
\index{Bogoliubov transformation!inhomogeneous case}Let $\AAA\in \sp_2$ and $L\in\WWW$.
 Then we have
\eq{inhomo}
\UU_{\AAA,L}\f b_{\AAA,L}^{\sharp}(f)\UU_{\AAA,L}
=
\ass(f).
\en
\ep
\proof
Notice that
$
S_{\AAA,L}b_\AAA^\#(f)
S_{\AAA,L}^{-1}
=
b_{\AAA,L}^\sharp(f)$.
Then the proposition  follows  from
Proposition
\ref{bog2}.
\qed

Suppose that $\AAA\in \sp_2$ and
$L\in\WWW$. Let $\Phi=\UU_{\AAA,L}\Omega$.
For later use we compute
$(\add(f)\Omega, \Phi)$ and $(\add(f) \add(g)\Omega, \Phi)$.
\bl{exp}
Suppose that $\AAA=\ST\in \sp_2$ and $L\in\WWW$. Set  $\xi=TL-\ov S\ov L$ and $K=TS\f$.
Then
it follows that
\begin{align}
\label{ichi}
&\frac{(\add(f)\Omega, \Phi)}{(\Omega,\Phi)}
=(f,\xi)+(\bar K f, \bar \xi),\\
\label{ni}
&\frac{(\add(f) \add(g)\Omega, \Phi)}{(\Omega, \Phi)}\non \\
&=
(f,\xi)(g,\xi)+(g,\xi)(\bar K f,\bar \xi)+
(f,\xi)(\bar K g,\bar \xi)+(\bar K g, \bar \xi)(\bar K f,\bar \xi)-(f, K\bar g).
\end{align}
\el
\proof
We set ${\rm S}=S_{\AAA,L}=\exp(\add(\xi)-a(\bar \xi))$,$\UU=\UU_\AAA$ and $J=(\Omega,\Phi)$.
Notice that
${\rm S}\f a(f) {\rm S}=a(f)+(\bar f,\xi)$ and
${\rm S} a(f) {\rm S}\f =a(f)-(\bar f,\xi)$.
Then directly we have
\begin{align*}
(\add(f)\Omega, \Phi)&=
(\Omega, a(\bar f) {\rm S} \UU \Omega)\\
&=
(f,\xi) J +(\Omega, {\rm S} a(\bar f)\UU \Omega)\\
&=
(f,\xi) J +({\rm S}\f \Omega, -\add (K\bar f)\UU \Omega)\\
&=(f,\xi) J +(-a ({\bar K f}){\rm S}\f \Omega, \UU \Omega)\\
&=(f,\xi) J +{(\bar K f,\bar \xi)}J.
\end{align*}
Then
\kak{ichi} follows.
Next we see that
\begin{align*}
&(\add(f)\add(g)\Omega, \Phi )\\
&=
(\Omega, a(\bar g)a(\bar f) \Phi)\\
&=
({\rm S}\f \Omega, (a(\bar g)+(g,\xi))
(a(\bar f) +(f,\xi))
\UU \Omega)\\
&=({\rm S}\f \Omega, a(\bar g)a(\bar f)
\UU \Omega)
+(g,\xi)
({\rm S}\f \Omega, a(\bar f)
\UU \Omega)\\&
\hspace{3cm}+(f,\xi)
({\rm S}\f \Omega, a(\bar g)
\UU \Omega)+(g,\xi)(f,\xi)J\\
&=
(g,\xi)(f,\xi)J+
(g,\xi)(\bar K f,\bar \xi)J+
(f,\xi)(\bar K g,\bar \xi)J+
({\rm S}\f \Omega, a(\bar g)a(\bar f)
\UU \Omega)J.
\end{align*}
Moreover
we have
\begin{align*}
&({\rm S}\f \Omega, a(\bar g)a(\bar f)
\UU \Omega)\\
&=
({\rm S}\f \Omega, a(\bar g)(-\add(K\bar f))
\UU \Omega)\\
&=({\rm S}\f \Omega, -\add(K\bar f)a(\bar g)
\UU \Omega)-(f, K\bar g)J\\
&=({\rm S}\f \Omega, \add(K\bar f)\add(K\bar g)
\UU \Omega)-(f, K\bar g)J\\
&=(a (\bar K f) a ( \bar K g)
{\rm S}\f \Omega, \UU \Omega)-(f, K\bar g)J\\
&=({\rm S}\f (a (\bar K f)-(\ov{\bar K f},\xi))
(a (\bar K g)-(\ov{\bar K g},\xi))
  \Omega, \UU \Omega)-(f, K\bar g)J\\
&=(\bar K f,\bar \xi)
(\bar K g,\bar \xi) J-(f, K\bar g)J.
\end{align*}
Then \kak{ni} follows.
\qed
\bl{g-exp}
Suppose that $\AAA=\ST\in \sp_2$ and $L\in\WWW$.
Set
$\xi=TL-\ov S\ov L$ and $K=TS\f$.
We assume that $\bar \xi=\xi$ and $\bar f=f$. Then
\eq
{gamma}
((2\add(f)+\add(f)\add(f))\Omega, \Phi)=(2\gamma+\gamma^2-(f,Kf)) (\Omega,\Phi),
\en
where
$\gamma=(\xi, (\one+K)f)$.
In particular for all $p\in\RR$,
it follows that
\eq{gamma2}
\frac{
((p+a(\bar f)+\add(f))^2\Omega, \Phi)}{(\Omega, \Phi)}
=(p+\gamma)^2+(f,(\one-K)f).
\en
\el
\proof
By the assumptions we have
\begin{align*}
&(\add(f) \Omega, \Phi)=\gamma,\\
&
(\add(f) \add(f)\Omega, \Phi)=
(f,\xi)^2+2(\xi, f)(\xi, Kf)+(\xi, Kf)^2-(f,Kf)=\gamma^2-(f,Kf).
\end{align*}
 Then
\kak{gamma} follows.
Notice that
$$(p+a(\bar f)+\add(f))^2\Omega=
(p^2+2p\add(f)+\add(f)\add(f))
\Omega+\|f\|^2\Omega.$$
Then \kak{gamma2} follows from \kak{gamma}.
\qed

\subsection{One parameter symplectic groups and 2-cocycles}
In this section we review the pseudo unitary representation of symplectic groups $\sp_2$.
Let $\UU (\fff)$ be the set of unitary operators  on $\fff$. Then
we can define the map
\eq{SP2}
\UU_\cdot :\sp_2\to U(\fff).
\en
Since
$$\UU_{\AAA \BBB}\f\UU _\AAA\UU _\BBB\ass(f)=
\ass(f)\UU_{\AAA \BBB}\f\UU _\AAA\UU _\BBB$$
for all $f\in\WWW$,
it follows that
$\UU_{\AAA \BBB}\f\UU _\AAA\UU _\BBB=\omega(\AAA,\BBB)\one$
with $\omega(\AAA,\BBB)\in U(1)$.
Thus $\UU_\cdot $ gives a projective unitary representation\ko{projective unitary representation} of $\sp_2$, i.e.,
\eq{ui}
\UU_\AAA \UU_\BBB=\omega(\AAA,\BBB)\UU_{\AAA\BBB}
\en
where $\omega(\AAA,\BBB)\in U(1)$ is 2-cocycle\index{cocycle}.
Let
\begin{align}
\spp&=
\lkk \left. \AAA=\ST\right|
\AAA \Jj + \Jj \AAA^\ast=\AAA^\ast \Jj +\Jj \AAA=0
\rkk\\
&=
\lkk \left. \AAA=\ST\right|
S^\ast=-S, \ov{T^\ast}=T
\rkk.
\end{align}
If $\AAA\in\spp$,
then it can be shown
that
\eq{opsg}
e^{t\AAA}\in \sp,\quad
t\in\RR.
\en
\kak{opsg} is called the one-parameter symplectic group\ko{symplectic group!one-parameter}.
Set
\begin{align}
\spp_2&=
\lkk \left. \AAA=\ST\in\spp\right|
T\in\HS
\rkk.
\end{align}
If $\AAA\in\spp_2$,
then we can also show that
\eq{opsg1}
e^{t\AAA}\in \sp_2,\quad
t\in\RR.
\en
Let  $\UU_t=\UU_{e^{t\AAA}}$ be the intertwining operator associated with
$\AAA\in \spp_2$.
Thus  $\UU_t$ satisfies that
\eq{ts}
\UU_t \UU_s=e^{i\rho(t,s)}\UU_{t+s}
\en
with
the so called local exponent\index{local exponent}  $\rho(t,s)\in\RR$.
The local exponent $\rho(t,s)$ also satisfies relation:
\eq{ll1}
\rho(t,s)+\rho(t+s,r)=\rho(s,r)+\rho(t,s+r).
\en
We shall show the explicit form of the local exponent.
Let $\AAA=\ST\in \spp_2$. Then $\e^{tA}=\STt\in\sp_2$ induces the map
\begin{align}
\label{tsym}
\lk\!
\begin{array}{ll}
a(f)\\
\add(f)
\end{array}
\!\rk
\mapsto
\lk\!\begin{array}{ll}
b_{t}(f)\\
b_{t}^\ast(f)
\end{array}
\!\rk=
\lk\!\begin{array}{ll}
\add(T_tf)+a(S_tf)\\
\add(\bar S_t f)+a(\bar T_t f)
\end{array}
\!\rk.
\end{align}
The intertwining operator $\UU_t=\UU_{e^{tA}}$ implements
the map \kak{tsym}, i.e.,
\eq{uin}
\UU_t\f b^\sharp_t(f) \UU_t=\ass(f)
\en for all $f\in\WWW$.
Furthermore we see that
\begin{align}
&\frac{d}{dt} b_t(f)=a(Sf)+\add(Tf)\\
&\frac{d}{dt} b_t^\ast (f)=a(\bar Sf)+\add(\bar Tf).
\end{align}
Let $\AAA=\ST\in\spp_2$. We define \eq{ge}
\Delta(\AAA)=\frac{i}{2}(\dd T- \aa {\bar T})-i \da{\bar S}.
\en
Operator  $\Delta(A)$  is essentially self-adjoint on $\ffff$
and
we
 can also directly see commutation relations:
\begin{align}
&[i\Delta(A),a(f)]=a(Sf)+\add(Tf),\\
&[i\Delta(A),\add(f)]=a(\bar Tf)+\add(\bar S f).
\end{align}
By using commutation relations we have
\begin{align}
e^{it\Delta(A)} \ass(f) e^{-it\Delta(A)}=b_t^\sharp(f)
\end{align}
and hence
the equation $\UU_te^{it\Delta(A)}\ass(f)=\ass(f)\UU_te^{it\Delta(A)}$ is derived, i.e.,
$$[\UU_te^{it\Delta(A)},\ass(f)]=0.$$
Thus for $\AAA\in \spp_2$ there exists $\theta_\AAA(t)\in\RR$ such that
\begin{align}
\label{ccru}
\UU_t=e^{i\theta_\AAA(t)} e^{-it\Delta(A)}.
\end{align}
It is immediate from \kak{ccru} that
$[\UU_t,\Delta(A)]=0$.
\bl{diff}
Let $\AAA\in\spp_2$.
Then the function $\theta_\AAA(\cdot)$ is $C^1(\RR)$.
\el
\proof
By the definition of $\theta(t)$ we have
$$e^{i\theta_\AAA(t)}=
\frac{(\Omega, \UU_t\Omega)}
{(\Omega, e^{it\Delta(A)}\Omega)}.$$
We can check that
$(\Omega, \UU_t\Omega)$ and
$(\Omega, e^{it\Delta(A)}\Omega)$ are differentiable in $t$. Then the lemma follows.
\qed

\bp{symgr}
\TTT{Local exponent}
\ko{local exponent}
Let $\AAA=\ST\in \spp_2$ and we set $e^{tA}=\STt$.
Then
\bi
\item[(1)]
$
\d \theta(t)=
\int_0^t \tau_r
dr$ and $\tau_r=\half {\rm Im}{\rm tr}
(T^\ast T_r
S_r ^{-1} )$,
\item[(2)]
$\d \rho(t,s)=\int_0^t \tau_r dr
+\int_0^s \tau_r dr-\int_0^{s+t} \tau_r dr$.
\ei
\ep
\proof
Let $K_t=T_tS_t\f$.
We notice that $\RR\ni {\rm det}(\one-K_t^\ast K_t)^{1/4}=(\Omega,\UU_t\Omega)=
(\Omega, e^{i\theta_\AAA(t)}e^{-it\Delta(\AAA)}\Omega)$.
Thus we see that
\begin{align*}
\frac{d}{dt}
(\Omega, e^{i\theta_\AAA(t)}e^{-it\Delta(\AAA)}\Omega)
=i\theta_\AAA'(t)(\Omega, \UU_t\Omega)+
(\Omega, i\Delta(\A) \UU_t\Omega)\in\RR
\end{align*}
which implies that the imaginary part of the right hand side disappear and then  $$\theta_\AAA'(t)(\Omega,  \UU_t\Omega)=-{\rm Im}
(\Omega, i\Delta(\A) \UU_t\Omega)=
-{\rm Im} (\half \dd T \Omega, \dd {K_t}\Omega).
$$
From this
we have
$$\theta_\AAA'(t)=-{\rm Im}
\half \frac{
 (\dd T \Omega, \UU_t \Omega)}{(\Omega, \UU_t \Omega)}=-\half {\rm Im}(\dd T\Omega, e^{-\half\dd {K_t}}\Omega)=
 \frac{1}{4}{\rm Im}(\dd T \Omega, \dd {K_t}\Omega).$$
Since
$(\dd T \Omega, \dd {K_t}\Omega)=(\Omega, [\Delta_{T^\ast},\dd{K_t}]\Omega)=2\tr (T^\ast K_t)$, we complete (1).
The statement (2) follows from (1) immediately.
\qed
From this proposition
$2$-cocycle $e^{i\rho(t,s)}$ vanishes if $
{\rm Im}{\rm tr}
(T^\ast T_r
S_r ^{-1} )=0$.
We give a sufficient condition to vanish the $2$-cocycle.
\bc{coc}
Suppose $\AAA=\ST\in\spp_2$ and
$\bar S=S=-S^\ast$ and $\bar T=T=T^\ast$. Then $\rho(t,s)=0$, i.e.,
 $\UU_t=e^{it \Delta(\AAA)}$.
In particular $\UU_t$, $t\in\RR$,  is the one-parameter unitary group\ko{unitary group!one-parameter }.
\ec
\begin{example}
{\rm
Let $\WWW=L^2(\RR)$. Define the Hilbert-Schmidt operator $T$ by
$Tf(x)=\int K(x,y) f(y) dy$ with a real-valued function $K\in L^2(\RR\times \RR)$. Let $h$ be a real-valued function such that $h\in L^\infty(\RR)$ and $h(-x)=-h(x)$. Define the operator $S=h(d/dx)$.
Then $S$ and $T$ satisfy
the condition in Corollary \kak{coc}. }
\end{example}

\cleardoublepage
\part{The Pauli-Fierz model}

\section{The Pauli-Fierz Hamiltonian}
\subsection{Introduction}
\label{diagonal}
It was  well known that vacuum polarization divergences, self-energy divergences and another infinity plagued QED in 1930's.
When one attempted  to compute calculate the contribution of radiative effects to the scattering of electrons
by the Coulomb field of a nucleus, infrared divergences were  encountered.
In 1937 Bloch and Nordsieck \cite{bn37} showed that this infrared divergence arose from the illegitimate neglect of processes involving the simultaneous emission of many photons, i.e,
 an emission of photons of very low frequencies yields
  divergences of an electromagnetic mass,
a scattering cross section, etc.
In 1938 according to a certain model describing
an interaction between an electron and a quantized
radiation field
Pauli and Fierz\cite{pf38} recognized that the quantized radiation field
reacts back on the electron to produce an electromagnetic mass.
This  model is today the so-called
Pauli-Fierz model,  which is the main object in this paper\footnote{In this note we take the dipole approximation of the standard Pauli-Fierz Hamiltonian.}.
The concept of mass renormalization in QED has its origin in these researches of Pauli and Fierz.

\begin{figure}[t]
\centering
\includegraphics[width=150pt]{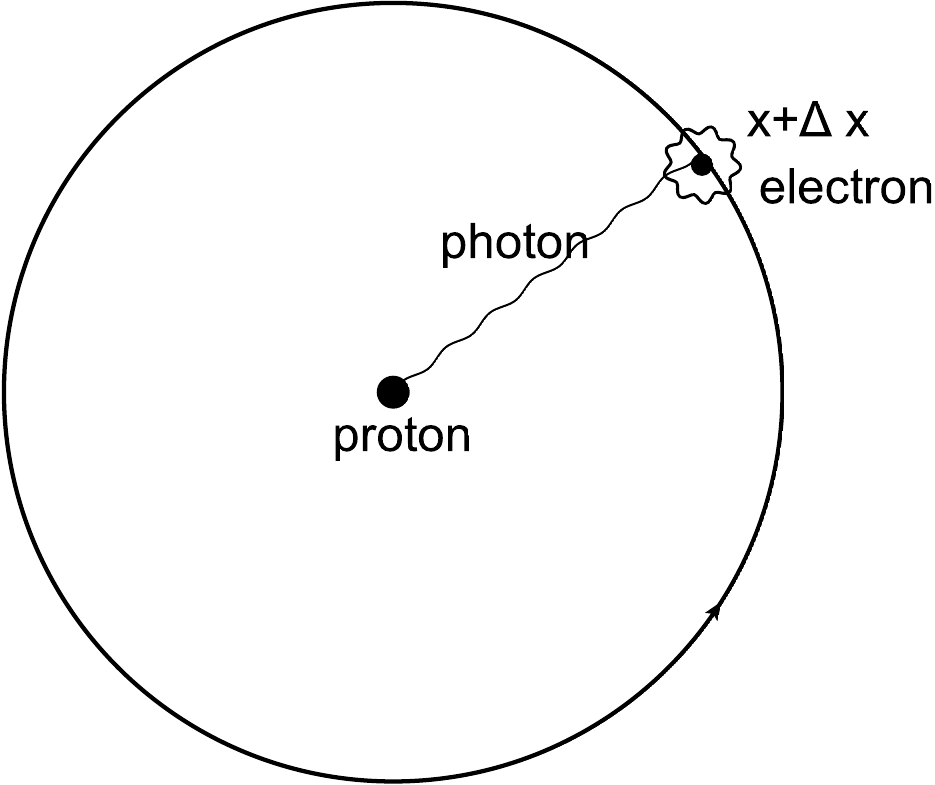}
\caption{Electron fluctuated by radiation in hydrogen atom}
\label{pichyd}
\end{figure}

Here we look at  a typical example of successes of the Pauli-Fierz
model.
Let $\alpha_\mu$, $\mu=1,2,3$,  and $\beta$ are $4\times 4$ Hermitian matrices obeying
the anticommutation relations
$\{\alpha_\mu,\alpha_\nu\}=2\delta_{\mu\nu}1$, $ \{\alpha_\mu,\beta\}=0$
and $\beta^2=1$.
The Dirac Hamiltonian\ko{Dirac Hamiltonian} of a hydrogen-like  atom\ko{hydrogen-like  atom} is given by
\eq{hyd}
D=\sum_{\mu=1}^3
\alpha_\mu
(-i\nabla_\mu)+
\beta m-\frac{Ze^2}{|x|},
\en
where
$Z$ is an atomic number,
$e$ the charge of an electron,
and $m$ the mass of an electron.
Then
$D$
has  eigenvalues
\eq{eigen}
E_{nj}=\frac{m}{
\sqrt{
1+{Z^2e^4}{\lk n-(j+\han)+\sqrt{(j+\han)^2-Z^2e^4} \rk^{-2}}}},
\en
where
$n=1,2,...,$ denotes the principal quantum number and
$j=l\pm\han$ the total angular-momentum  with
the angular-momentum $l=0,...,n-1$.
Eigenvalue $2S_{\han}$ corresponds to $n=2$,  $j=\han$, $l=0$,
and
$2P_{\han}$  to $n=2$,  $j=\han$, $l=1$.
Then the Dirac theory concludes that
two levels $2S_{\han}$ and $2P_{\han}$
in a hydrogen-like atom
sit at the same energy level, i.e.,
$$2S_{\han}=2P_{\han}.$$
In 1947,
by Lamb and Retherford\cite{lr47},
it was experimentally observed, however,
that
$$ 2S_{\han}>2P_{\han}.$$
This discrepancy is called the Lamb-shift\index{Lamb shift}.
Bethe\cite{bet47}
regarded the Lamb-shift
as an evidence  of
a radiation  reaction,
and
tentatively made a nonrelativistic calculation of
the difference of the two levels.
The  resulting value was in remarkable agreement with
the observation.
In 1948,   using the Pauli-Fierz model,
Welton\cite{wel48} gave an intuitive derivation
to the Lamb-shift.
He   argued  that a  position-fluctuation of an electron  through the
  radiation field will effectively modify the    external potential $V$ (Figure  \ref{pichyd}).
The fluctuation was thought of as a Gaussian random variable $\Delta x$,
then an  effective potential  is formally given by a mean value of
$V(x+\Delta x)$;
\begin{eqnarray}
V_{\rm eff}(x)&=&\langle V(x+\Delta x)\rangle_{\rm AVE}
=
\label{wel}
(2\pi C)^{-3/2}
\int_{\RR^3} V(y)e^{-|x-y|^2/(2C)}dy,
\end{eqnarray}
with a certain positive constant $C$.
Then an electron Hamiltonian   effectively turns out to be
governed by a  Hamiltonian with the external potential
$V_{\rm eff}$ instead of $V$.
Welton  gave an interpretation of the Lamb shift as the difference
between  the spectrum  of  the  original Hamiltonian and an  effective one.

In Part II  we study the  Pauli-Fierz Hamiltonian. The basic assumptions are
as follows.
\bi
\item[(1)] We take the dipole  approximation.
\item[(2)] We neglect spin.
\ei
The spectrum of this model is  studied by the series  of papers by A. Arai
\cite{ara81-a,ara81-b,ara81-c,ara83-a,ara83-b}.
Since we take the dipole approximation, the Pauli-Fierz Hamiltonian $H$ is reduced to simple.
Although it is not translation invariant, i.e., it does not commute with the total momentum,
it commutes with particle momentum.  Then $H$ without external potential can be decomposable with respect to the momentum of the particle
$H=\int^\oplus_\BR  \ds dp$ and we can diagonalize $\ds$ for each fiber $p\in\BR$ by applying
Bogoliubov transform studied in Section 2.
This is a key observation in this section.

\subsection{The Pauli-Fierz Hamiltonian  with the dipole approximation}
Let us assume that the electron moves in  dimension $d\geq 3$.
The physically reasonable dimension is $d=3$.
We
denote by $\fff$ the boson Fock space over the one particle space $L^2(\BR\times \{1,...,d-1\})$.
Here  a photon is regarded as a transversal wave in $d-1$ directions.
The Hilbert space $\hhh$ of the coupled system is then given by
\eq{ln1}
\hhh=\ll\otimes\fff.
\en
The annihilation operator $a(f,j)$ and the creation operator $\add(g,j)$
satisfies
canonical commutation relation:
\index{canonical commutation relation}
\eq{ccr2}
[a(f,j),\add(g,j')]=\delta_{jj'}(f,g)_\LR,\quad
[a(f,j),a(g,j')]=0=[\add(f,j),\add(g,j')]
\en
on the finite particle subspace $\ffff$ for $f,g\in\LR$ and $1\leq j,j'\leq d-1$.
The
$d$-dimensional polarization vectors\ko{polarization vector} are written
as
\eq{pol}
\ejk=(\ej_1(k),\ldots,\ej_d(k)),\quad
j=1,\ldots,d-1,
\en
which satisfy
$e^i(k)\cdot \ej(k)=\delta_{ij}$ and
 $\ej(k)\cdot k=0$ almost everywhere on $\BR$.
Let
\eq{hff}
\hf=d\Gamma(\omega)
\en
be the free  field Hamiltonian\ko{free field Hamiltonian!Pauli-Fierz Hamiltonian} with the dispersion relation
\ko{dispersion relation}
\eq{dispersion}
\omega(k)=|k|.
\en
Let\footnote{Throughout Part II in this lecture note  the summation over  repeated indices is understood. The Greek letters $\mu,\nu,...$ and $a,b$ run from $1$ to $d$, and $i,j,k$ from $1$ to $d-1$.}
\begin{align*}
&A_\mu(\eta)=\frac{1}{\sqrt 2}
\int \frac{e_\mu^j(k)}{\sqrt{\omega(k)}}
\lk{\hat\eta(-k)}
\add(k,j)
+
{\hat\eta(k)}
a(k,j)\rk dk,\\
&\Pi_\mu(\eta)=\frac{i}{\sqrt 2}\int \sqrt{\omega(k)}
e_\mu^j(k)
\lk{ \hat\eta(-k)}\add(k,j)
-
{\hat\eta(k)}
a(k,j)\rk dk.
\end{align*}
When $\eta$ is real, then $\ov{\hat \eta}(k)= {\hat \eta}(-k)$ and $A_\mu(\eta)$ and $\pi_\mu(\eta)$
are  symmetric.
  We have the following commutation relations on $\ffff$
  \begin{align}
&[A_\mu(\eta),\Pi _\nu(\rho)]=i\int \dmu(k)\hat \eta(-k)\hat \rho(k)\ddk=i (\dmu\hat{\bar \eta},\hat \rho), \\
&
[A_\mu(\eta), A_\nu(\rho)]=0,\\
&
[\Pi _\mu(\eta), \Pi _\nu(\rho)]=0,
\end{align}
where
\eq{tdf}
\dmu(k)=\delta_{\mu\nu}-\frac{k_\mu k_\nu}{|k|^2}
\en
denotes the transversal delta function\ko{transversal delta function},
and
\begin{align}
&
[\hf, A_\mu(\eta)]=-i \Pi_\mu(\eta),\\
&[\hf, \Pi _\mu(\eta)]=iA_\mu(-\Delta\eta).
\end{align}
The quantized radiation field\ko{radiation field}
 $A_\mu$ with a cutoff function $\vp$
is defined by
\eq{radfield}
A_\mu= \sh \int\frac{1}{\sqrt{\omega(k)}}e_\mu^j(k)\lkk\widetilde{\phk} a^\ast(k,j)+{\phk} a(k,j)\rkk\ddk,
\en
and the quantized electric field\ko{electric field}, as its  canonically conjugate,  by
\eq{pif}
\Pi_\mu=
i \sh  \int {\sqrt{\omega(k)}}
e_\mu^j(k)
\lkk \widetilde{\phk} \add(k,j)-{\phk} a(k,j)\rkk dk.
\en
Here
$\tilde f(k)=f(-k)$.
They satisfy that
\begin{align}
&[A_\mu,\Pi_\nu]=i\int \dmu (k) \vp(-k) \vp(k) dk,\\
&[A_\mu, A_\nu]=0,\\
&[\Pi_\mu,\Pi_\nu]=0.
\end{align}
We define the Pauli-Fierz Hamiltonian\index{Pauli-Fierz Hamiltonian}.
\begin{definition}
\TTT{Pauli-Fierz Hamiltonian!dipole approximation}\ko{Pauli-Fierz Hamiltonian}
The Pauli-Fierz Hamiltonian $H$ with the dipole approximation  is  defined
  by
\eq{pf}
H=\ham \lk -i\nabla \otimes \one -\ak \one\otimes A\rk^2+V\otimes \one +\one\otimes \hf,
\en
where $\ak\in\RR$ denotes the coupling constant and $V:\BR\to\RR$ is an external potential.
\end{definition}
In what follows we omit the tensor notation $\otimes$ for notational convenience.
Thus $H$ is simply written as
$$
\ham
(-i\nabla-\ak  A)^2+V+\hf.
$$

We first of all state the self-adjointness of $H$.
\begin{proposition}
\label{self}
\TTT{Self-adjointness}
{\rm \cite{ara81-a}}
Suppose that
$\ph/\omega,\sqrt{\omega}\ph\in \LR$ and $\vp(-k)=\ov{\vp(k)}$,
and that
 $V$ is relatively bounded with respect to
$-\ham  \Delta$
with a  relative bound strictly smaller than one\footnote{$D(V)\subset D(-\Delta)$ and $\|Vf\|\leq a\|-\ham\Delta f\|+b\| f\|$ with some $a<1$.}.
Then
$H$ is self-adjoint on $D(-\Delta )\cap D(\hf)$ and bounded
below for arbitrary $\fin\in\RR$.
\end{proposition}
\proof
Let $V=0$.
Let $L=-\Delta+\hf +\one$.
It can be seen that
$$|(HF, LG)-(LF,HG)|\leq C \|L^\han F\| \| L^\han G\|$$
with some constant $C$. Then $H$ is essentially self-adjoint on $D(-\Delta )\cap D(\hf)$
by the Nelson commutator theorem.
Furthermore
by the inequality
$\|H F\|\leq C\|L F\|$, the closedness of $H\lceil_{D(-\Delta )\cap D(\hf)}$ follows.
For nonzero $V$, by the diamagnetic inequality,
$$\|(\ham
(-i\nabla-\ak  A)^2+\hf-z)\f F\|\leq \|(-\ham\Delta+\hf-z)\f F\|,$$
we can see that $V$ is also relatively bounded
with respect to
$\ham
(-i\nabla-\ak  A)^2+\hf
$  with a relative bound strictly smaller than one.
Then the proposition  follows by the Kato-Rellich theorem.
\qed

Let
$$A_\mu(x)=\sh \int\frac{1}{\sqrt{\omega(k)}}
e_\mu^j(k)\lk \vp(-k) e^{-ikx}\add(k,j)+ \vp(k) e^{ikx} a(k,j)\rk dk$$
for each $x\in\BR$.
Under the identification $\hhh=\int^\oplus_\BR \fff dx$,
we define $$A_\mu =\int^\oplus_\BR A_\mu(x) dx.$$
Thus the Pauli-Fierz Hamiltonian without the dipole approximation\ko{Pauli-Fierz Hamiltonian} is defined by
\eq{nodip}
\ham(-i\nabla\otimes\one -\ak A)^2+V\otimes \one+\one\otimes\hf.
\en
The Pauli-Fierz Hamiltonian $H$ under consideration in this lecture note is defined by
\kak{nodip}
 with $A_\mu$ replaced by $\one\otimes A_\mu(0)$.

\subsection{Translation invariant Hamiltonian}
Suppose that $V=0$.
Let us define the operator $\dip$ in $\hhh$ by
\eq{dip}
\dip=\ham (\p -\ak  A)^2+ \hf.
\en
Since $\dip$ commutes with $\p_\mu$, the  Hilbert space $\hhh$ and
the  operator
$\dip $ are  decomposable with respect to the joint spectrum of $\p_\mu$, i.e.
\begin{align*}
&\hhh=\int^\oplus _\BR \fff dp,\\
&
\dip=\int_\BR^\oplus \ds dp
\end{align*}
where
\index{Pauli-Fierz Hamiltonian!translation invariant}\eq{fiber}
\ds=\ham\left(p-\ak{A}\right)^2+ \hf, \ \ \ p\in\BR.
\en
\bp{selfdip}
\TTT{Self-adjointness}
{\rm \cite{ara83-a}}
Suppose that
$\ph/\omega,\sqrt{\omega}\ph\in \LR$ and $\vp(-k)=\ov{\vp(k)}$.
Then
$\ds$ is self-adjoint on $ D(\hf)$ and bounded
below for arbitrary $\fin\in\RR$.
\end{proposition}
\proof
Let $L=\hf +\one$.
It can be seen that
$$|(\ds F, LG)-(LF,\ds G)|\leq C \|L^\han F\| \| L^\han G\|$$
with some constant $C$. Then $\ds$ is essentially self-adjoint on
$D(\hf)$
by the Nelson commutator theorem.
Furthermore
by the inequality
$\|\ds  F\|\leq C\|L F\|$, the closedness of $\ds \lceil_{ D(\hf)}$ follows.
Then the proposition  follows.
 \qed

The quadruple
$${\ms D}=\{\add,a,\hf,\Omega\}$$ satisfies the algebraic relations:
  $[a(f,j),\add(g,j')]=\delta_{jj'}(\bar f, g)$,
  $[\hf , a(g)]=-a(\omega g)$, $[\hf, \add(g)]=\add(\omega g)$ and $\hf \Omega=a(f)\Omega =0$.
In this section we construct operators $B_p$ and $B_p^\ast $ and a vector $\Omega_p$ such that
the quadruple
$${\ms D}_p=\{B_p,B_p^\ast , \ds-E_p,\Omega_p\}$$
satisfies the same algebraic relations as those of ${\ms D}$ for each $p\in\BR$, where $E_p$ denotes the
ground  state energy of $\ds$, i.e., $E_p=\is(\ds)$.
  We need in addition some technical assumptions on $\vp$.
\begin{assumption}
\label{1v1}
We suppose (1),(2), (3) or
(1), (2'), (3):
\bi
\item[(1)]
$\y \vp, \vp/\omega\in \ll$,
$\ov{\ph(k)}=\ph(-k)$ and
$\vp$ is rotation invariant, i.e.  $\ph(k)=\ph(|k|)$,
\item[(2)]
$\vp(k)\not=0$ for $k\not=0$,
and  $\rho (s)=|\vp(\sqrt{s})|^2s^{\fdd }\in L^\epsilon([0,\infty),ds)$ for some
$1<\epsilon$,
 and
there exists $0<C<1$ such that
$|\rho (s+h)-\rho (s)|\leq K|h|^C$ for all $s$ and $0\leq h\leq 1$,
\item[(2')]
$\vp(k)\not=0$ for $\lambda \leq |k|\leq \Lambda$, and $\vp(k)=0$ for $|k|>\Lambda$ and $|k|< \lambda$ with some $\Lambda>0$ and $\lambda>0$,
\item[(3)]
$\|\vp\ooo{(d-3)/2}\|_\infty<\infty$ and
$\|\vp\ooo{(d-1) /2}\|_\infty<\infty$.
\ei
\end{assumption}
Assumption (1) is used for self-adjointness of $H$, (2) or (2') for the definition of
 a function $D_+$ and $Q$ in Section \ref{bobo},  and (3) for
 an operator $T_{\mu\nu}$ in Lemma \ref{DD}.
The main theorem in Section 3 is as follows. \bt{mainp-ari}
\TTT{Diagonalization of $\ds$}
\index{diagonalization!translation invariant}Suppose Assumption \ref{1v1}.
\begin{description}
\item[(1)]
Let $p=0$. Then there
 exists a unitary operator $\UU_0:D(\hf) \to D(\hf)$ such that
\eq{ln111}
\UU_0 \f \dsz\UU_0=\hf+\ground .
\en
\item[(2)]
Suppose, in addition,  $\IR$.
Then for all $p\in\BR$, there exists a unitary operator $\UU_p:D(\hf)\to D(\hf)$ such that
\eq{ln11}
\UU_p \f \ds\UU_p=\frac{1}{2\mass}p^2+\hf +
\ground .
\en
\end{description}
Here
the effective mass\index{effective mass} $\mass$ is given by
\begin{align}
\mass=m+\at\mmm \|\vp/{\omega}\|^2,
\end{align}
and the additional constant  $\ground$ by
\begin{align}
\ground =\frac{d}{2\pi}\int_{-\infty}^\infty
\frac{\at \mmm  \left\|\frac{t\vp}{t^2+\omega^2}\right\|^2}
{m+\at  \mmm\left\|\frac{\vp}{\sqrt{t^2+\omega^2}}\right\|^2}dt.
\end{align}
\et
The condition  $\IR$ is called the
infrared regular condition\index{infrared regular condition}, on the other hand $\SR$ infrared singular condition\index{infrared singular condition}. The term $\at\mmm \|\vp/{\omega}\|^2$
in $\mass$ is called self-energy.
\ko{self-energy}
The self energy $\at\mmm  \int \frac{|\vp(k)|^2}{|k|^2} dk$ has no singularity at the origin $k=0$, since $d\geq3$.

We furthermore define the unitary operator $\UU$ on $\hhh=\int^\oplus_\BR \fff dp$ by
\eq{unit}
\UU=\int ^\oplus_\BR \UU_p e^{i\frac{\pi}{2}N}dp,
\en
where $N$ denotes the number operator\ko{number operator} in $\fff$.
\bt{mainp-nasi}
\TTT{Diagonalization of $H$}
\index{diagonalization}Suppose Assumption \ref{1v1} and
that  $V$ is
relatively bounded with respect to $-\ham  \Delta$
with a relative bound strictly smaller than one.
Assume furthermore  $\IR$.
Then, for each $\ak\in\RR$,
$\UU$ maps $D(-\Delta)\cap D(\hf)$ onto itself and
\eq{ln12}
\UU\f H\UU=\eff+ \hf+\dv+\ground ,
\en
where $\eff$ denotes the effective Hamiltonian  given by
\begin{align*}
\eff=-\frac{1}{2\mass}\Delta+V,
\end{align*}
and
$\dv$ is
the perturbation given by
\eq{dv2}
\dv=T\f VT-V,
\en
with
\begin{align}
&\label{matsu1}
T=\exp\left(-i(\p)\cdot K\right),\\
&
\label{matsu2}K_\mu=\sh   \int
\frac{e_\mu^j(k)}{\sqrt{\omega(k)}}
\lk
\frac{\ak \vp(k)}{\ov{\mass(k)}\omega(k)}
 \add  (k,j)+
\frac{\ak \vp(k)}{{\mass(k)}\omega(k)}
 a(k,j)\rk \ddk.
\end{align}
Here  the function $\mass(k)$  is given by \kak{rmass} below.
\et
We shall give proofs of Theorems \ref{mainp-nasi} and \ref{mainp-ari} in Section \ref{rikisi}.

Formally
$$T\f V T(x)=e^{\nabla\cdot K}Ve^{-\nabla\cdot K}(x)=V(x+K).$$
Thus
$$V(x+K)=\sum_{n=0}^\infty \frac{1}{n!}(\nabla\cdot K)^n V(x).$$
Thus the discrepancy between $V$ and $\veff$ is given by
$\sum_{n=1}^\infty \frac{1}{n!}(\nabla\cdot K)^n V(x)$. In particular
$$(\Omega,
\sum_{n=1}^\infty \frac{1}{n!}(\nabla\cdot K)^n V(x)\Omega)
=
(\Omega,
\sum_{n=2}^\infty \frac{1}{n!}(\nabla\cdot K)^n V(x)\Omega)$$
and
$(\Omega,
\sum_{n=2}^\infty \frac{1}{n!}(\nabla\cdot K)^n V(x)\Omega)\sim
\half (\Omega,
(\nabla\cdot K)^2 V(x)\Omega)$.
Approximately the radiative effect changes
$V$ to
$V+\half (\Omega,
(\nabla\cdot K)^2 V(x)\Omega)$.
This
 gives an interpretation of the Lamb shift\ko{Lamb shift}. See \cite{bet47,wel48}.

\subsection{Bogoliubov transformation}
\index{Bogoliubov transformation}\label{bobo}
\subsubsection{Algebraic relations}
In order to prove
Theorems \ref{mainp-nasi} and \ref{mainp-ari} we prepare several lemmas.

Let us consider the time evolution of $A_\mu(f)$ by the Hamiltonian $\ds$.
Let
$$A_\mu(f,t)=e^{it\ds}A _\mu(f) e^{-it\ds}$$ for $p\in\BR$, and
 set
$A _\mu(f,t)=\int A_\mu(x,t) f(x)dx$.
 Formally
 we have
 \begin{align}
\label{arai}
( \frac{\partial^2}{\partial t^2}-\Delta)
A_\mu(x,t)=\frac{\ak}{m}
(p_\nu-\ak A_\nu(x,t))\rho_{\nu\mu}(x),
\end{align}
where
$\rho_{\nu\mu}(x)=(2\pi)^{-d/2}\int d_{\nu\mu}(k) \vp(k) e^{ikx}dk$.
We shall operator-theoretically
 solve \kak{arai} in what follows.
 Let us define
\eq{Dz}
D(z)=
m-\ak ^2  \mmm
\int_\BR \frac{|\vp(k)|^2}{z-\omega(k)^2 }
\ddk,\quad z\in \CC\setminus[0,\infty).
\en
\bl{DD}
Suppose Assumption \ref{1v1}.
\bi
\item[(1)]
The function $D(z)$ is analytic
 and has no zero points in $\CC\setminus[0,\infty)$.
\item[(2)]
The function $\d D_\pm(s)=\lim_{\epsilon\downarrow 0}D(s\pm i\epsilon)$ exists for all
$s\in [0,\infty)$, $D_\pm(0)=\mass$ and
$\d \lim_{s\to\infty}D_\pm(s)=m$.
\item[(3)]
It follows that
$\d D_+(s)-D_-(s)=\pi i \ak ^2 \mmm \vol |\vp(\sqrt{s})|^2s^{\frac{d-2}{2}}$,
\ei
where $\vol =2\pi^{\frac{d+1}{2}}/
\Gamma(\frac{d+1}{2})$
the volume of the
$d-1$-dimensional unit sphere $S_{d-1}$.
\el
\proof
(1) is fundamental.
We directly see that
\begin{align}
\label{taku}
D_\pm(s)
=
m-\frac{\ak ^2}{2}
 \mmm \vol  \left(
H\rho(s)\mp \pi i \rho(s)\rk,
\end{align}
where
$\rho(s)=|\vp(\sqrt s)|^2s^\fdd$ and
$H\rho$ denotes the Hilbert transform\index{Hilbert transform} of $\rho$, i.e.,
$$H\rho(s)=\lim_{\epsilon\downarrow 0}\int _{|s-x|>\epsilon}\frac{\rho(x)}{s-x}dx.$$
Namely\footnote{\cite[(4.1)]{hs01} is incorrect. 
$\mp 2\pi i|\vp(\sqrt s)|^2s^{(d-2)/2}$ is changed to
$\mp \pi i|\vp(\sqrt s)|^2s^{(d-2)/2}$. 
}
\begin{align}
D_\pm(s)
=
m-\frac{\ak ^2}{2}
 \mmm \vol  \left(
\lim_{\epsilon\downarrow 0}\int_{|s-x|>\epsilon}
\!\!\!\!\!\!
\!\!\!\!\!\!
\frac{|\vp(\sqrt{x})|^2|x|^{\fdd }}{s-x}dx
\mp  \pi i|\vp(\sqrt{s})|^2 s^{\fdd }\right).
\end{align}
Then (2) and (3) follow from this.
\qed
\bl{2v2-}
\bi
\item[(1)]
Suppose
(1),(2) and (3) of Assumption \ref{1v1} .
Then
there exists $\epsilon>0$ such that $|D_\pm(s)|>\epsilon$ for $s\in [0,\infty)$.
\item[(2)]
Suppose
(1),(2') and (3) of Assumption \ref{1v1} .
Then
there exists $\epsilon>0$ such that $|D_\pm(s)|>\epsilon$ for $s\in [\lambda^2,\Lambda^2]$ and
$D_\pm(s)$ has  at most one zero point
on  open interval  $(\Lambda^2,\infty)$.
\ei
\el
\proof
By  Assumption  \ref{1v1}  (2), the imaginary part of $D_\pm(s)$ does not vanish,
and
Assumption \ref{1v1} (2) implies that
the real part of $D_\pm$ is also Lipshitz continuous with the same order $C$ as that of $\rho(s)=|\vp(\sqrt s)|^2s^{\fdd}$,
since the real part is the Hilbert transformation of $\rho$.
Then
the real part of $D_\pm(s)$ goes to $m>0$ as $s\to\infty$ \cite[p.145,5.15]{tit37}.
In particular,
there exists $\epsilon>0$ such that
$\sup_{s\in [0,\infty)}|D_\pm(s)|>\epsilon$.

Next Assumption \ref{1v1} (2') implies
that the imaginary part of $D_+(s)\not=0$ for $s\in[\lambda^2,\Lambda^2]$.
The real part of $D_+(s)$ is
$
m-\frac{\at}{2}\mmm\vol \int_{\lambda^2}^{\Lambda^2}\frac{|\vp(\sqrt x)|^2x^{\fdd}}{s-x}dx$.
Then
it  is
monotonously increasing   on $[0,\lambda^2)\cup (\Lambda^2,\infty)
$
with $D_+(0)=m+\at \mmm \|\vp/\omega\|^2$ and $\lim_{s\to\infty}D_+(s)=m$.
Then the lemma follows.
\qed
Define
\eq{defG}
Gf(k)=\lim_{\epsilon\downarrow 0}
\int_\BR\frac{f(k')}{(\omega(k)^2 -
\omega(k')^2 +i\epsilon)(\omega(k)\omega(k'))^{\fdd }}\ddk'.
\en
It is seen that
\begin{align}
\label{gk}
Gf(k)
=\frac{1}{2|k|^{\fdd }}
\left(
H  F(|k|^2)
-\pi i[f]
(|k|)|k|^{\fdd }\right),
\end{align}
where
$HF$ denotes the Hilbert transform of $F(x)=[f](\sqrt x)x^{\frac{d-2}{4}}$ and $[f](r)=\int_{S_{d-1}} f(r,v)dv$ and  $v$
is the volume element on $S_{d-1}$.
Define the running effective mass\index{effective mass!running}
  by
\eq{rmass}
m_{\rm eff}(k)=D_+(\omega(k)^2)
\en
and
we set
\eq{tmn}
\tmn f=\delta_{\mu\nu}f+\ak  \JI
 \ooo{\fdd } G \ooo{\fdd }\dmu  \vp f,
 \en
where
\eq{Qq}
\JI (k)= \frac{\ak  \vp(k)}
{m_{\rm eff}(k)}.
\en
The operator $\tmn$ is then given by
$$\tmn f(k)=\delta_{\mu\nu}f(k)+
\at\int
\frac{\dmu(k')\vp(k)\vp(k')f(k')}
{D_+(\kkt )
(
\kkt -\kkkt +i0)}dk'.
$$
\begin{remark}
We give a comment on the definition of $\JI$.
Under (2) of Assumption \ref{1v1}, $\mass(k)\not=0$ for all $k\in\BR$, then $\JI$ is well defined.
On the other hand
under (2') of Assumption \ref{1v1},
$\mass(k)$ probably has one   zero point
 in $(\Lambda^2,\infty)$
Then $\JI$ is  well defined, since
the  support of the numerator is
${\rm supp} \vp=[\lambda,\Lambda]$.
Notice also that
 $\mass(k)\not=0$ at least for
 $\lambda-\delta \leq |k|\leq \Lambda+\delta$ with some $\delta>0$.
 \end{remark}

\begin{example}
\TTT{Running effective mass for sharp cutoff}
\label{goodexample}
\ko{effective mass!running!sharp cutoff}
{\rm
Let us compute an effective mass $\mass(k)$ with sharp cutoff\ko{effective mass!sharp cutoff}.
Let
\eq{UVsharp}
\vp(k)=\one_{[\lambda,\Lambda]}(|k|)
\en
be the indicator function on $\lambda\leq |k|\leq \Lambda$.
Suppose $d=3$.
Then
\begin{align*}
H\rho(s)=
\lim_{\eps\downarrow 0}
\int_{|s-x|>\eps}\frac{\one_{[\lambda,\Lambda]}(\sqrt x)\sqrt x}{s-x}dx =
\lim_{\eps\downarrow 0}
\int_{|s-x|>\eps}\frac{\one_{[\lambda^2,\Lambda^2]}(x) \sqrt x}{s-x} dx.
\end{align*}
Let $s\in [0,\lambda)$. Then we see that
\begin{align*}
H\rho(s)
&=\int_{\lambda^2}^{\Lambda^2} \frac{\sqrt x}{s-x}dx
=
\int_{\lambda}^{\Lambda} \frac{2t^2}{s-t^2} dt
=
-2(\Lambda-\lambda)-
2s
\int_{\lambda}^{\Lambda} \frac{1}{t^2-s} dt\\
&=
-2(\Lambda-\lambda)-\sqrt s
\int_{\lambda}^{\Lambda} \frac{1}{t-\sqrt s}-\frac{1}{t+\sqrt s} dt\\
&=
-2(\Lambda-\lambda)+\sqrt s
\log\lk\frac {(\lambda-\sqrt s)(\sqrt s+\Lambda)}{(\sqrt s+\lambda)(\Lambda-\sqrt s)}\rk.
\end{align*}
Let $s\in (\Lambda,\infty)$.
Then in a similar computation we have
\begin{align*}
H\rho(s)=-2(\Lambda-\lambda)+\sqrt s
\log\lk\frac {(\sqrt s-\lambda)(\sqrt s+\Lambda)}{(\sqrt s+\lambda)(\sqrt s-\Lambda)}\rk.
\end{align*}
Finally let $s\in [\lambda,\Lambda]$.
\begin{align*}
&\int_{|s-x|>\eps}\frac{\one_{[\lambda,\Lambda]}(\sqrt x)\sqrt x}{s-x}dx
=
\lk \int_{\lambda^2}^{s-\eps}+\int_{s+\eps}^{\Lambda^2} \rk \frac{\sqrt x}{s-x}dx\\
&=
\lk \int_{\lambda}^{\sqrt{s-\eps}}+\int_{\sqrt{s+\eps}}^{\Lambda} \rk \frac{2t^2}{s-t^2}dt\\
&
=
2\lk \int_{\lambda}^{\sqrt{s-\eps}}+\int_{\sqrt{s+\eps}}^{\Lambda} \rk (-1+\frac{s}{s-t^2}) dt\\
&\to -2(\Lambda-\lambda)+
\lim_{\eps\downarrow 0}
2s
\lk  \int_{\lambda}^{\sqrt{s-\eps}}+\int_{\sqrt{s+\eps}}^{\Lambda} \rk \frac{1}{s-t^2}ds
\end{align*}
as $\eps\to0$.
Directly we have
\begin{align*}
&2s
\lk  \int_{\lambda}^{\sqrt{s-\eps}}+\int_{\sqrt{s+\eps}}^{\Lambda} \rk \frac{1}{s-t^2} ds\\
&=\sqrt s
\lk
\log\lk\frac{(\sqrt s +\sqrt{s-\eps})(\sqrt s-\lambda)}{(\sqrt s-\sqrt {s-\eps})(\sqrt s+\lambda)}\rk-
\log\lk\frac{(\sqrt s +\sqrt{s+\eps})(\Lambda -\sqrt s)}{(\sqrt {s+\eps}-\sqrt s)(\sqrt s+\Lambda)}\rk
\rk\\
&=\sqrt s
\log\lk
\frac{(\sqrt s-\lambda)(\sqrt s+\Lambda)}{(\sqrt s+\lambda)(\Lambda-\sqrt s)}
\frac{\sqrt s+\sqrt{s-\eps})(\sqrt {s+\eps}-\sqrt s)}{\sqrt s-\sqrt{s-\eps})(\sqrt{s+\eps}+\sqrt s)}\rk
\\
&\to
\sqrt s
\log\lk
\frac{(\sqrt s-\lambda)(\sqrt s+\Lambda)}{(\sqrt s+\lambda)(\Lambda-\sqrt s)}
\rk
\end{align*}
as $\eps\to 0$. Thus we obtain that
\begin{align}
\label{goodex}
H\rho
(s)=
-2(\Lambda-\lambda)+
\sqrt s\log
\left|
\frac{(\sqrt s+\Lambda)(\sqrt s-\lambda)}
{(\sqrt s+\lambda)(\sqrt s-\Lambda)}
\right|.
\end{align}
Since $\mmm \vol=8\pi/3$ for $d=3$,
the running effective mass with sharp cutoff
$\kak{UVsharp}$
\ko{UV cutoff!sharp}
is given by
\begin{align}
\label{tobira}
\mass(k)=
m+\frac{8\pi\at }{3}
 (\Lambda-\lambda)
 -\frac{4\pi \at}{3}
\!\!\lk\!\!
|k|\log
\!\left|
\!\frac{(|k|+\Lambda)(|k|-\lambda)}
{(|k|+\lambda)(|k|-\Lambda)}
\!\right|
-i\pi
\one_{[\lambda,\Lambda]}(|k|)\sqrt{|k|}\rk.
\end{align}
}\end{example}
\bl{gbound}
The operator $G$ is a bounded and antisymmetric operator on $\LR$, i.e., $G^\ast=-G$.
\el
\proof
Let $f\in\LR$.
The imaginary part of  $Gf$ is $-\frac{\pi}{2}[f](|k|)$ and
$\|[f]\|\leq {\vol}\|f\|$.
The real part  of  $Gf$ is given by
the Hilbert transform;
$HF$ with $F(x)=[f](\sqrt x)x^{\frac{d-2}{4}}$.
The Hilbert transformation is bounded operator on $\LR$ with $\|Hf\|=\pi\|f\|$
and then
we have
\begin{align*}
\|HF\|^2
&=\frac{1}{4}\int \frac{|HF(|k|^2)|^2}{|k|^{d-2}}dk
=\half \int _0^\infty |HF(s)|^2 ds \vol
=\half\vol \pi^2\|F\|^2\\
&=\half\vol \pi^2\int[f](\sqrt s)^2 s^\fdd ds
=
\vol \pi^2 \int [f](r)^2 r^{d-2}dr\\
&\leq \vol^2\pi^2\int f(r,v)^2 r^{d-2}dr dv
= \vol^2\pi^2 \|f\|^2.
\end{align*}
Then the lemma follows.
\qed
The operator $\tmn$,  functions  $\JI $ and  $\vp$ satisfy some  algebraic relations.
We list up  them in the lemma below, where
we assume that $m$ is not only positive but also negative. Precisely we assume that
\eq{negative}
m>-\mmm \at \|\vp/\omega\|^2, \quad m\not=0
\en
  for mathematical generality.
When $0>m>-\mmm \at \|\vp/\omega\|^2$,
 we see that
$D(s)\in\RR$ for $s\leq 0$, $\d\lim_{s\to-\infty}D(s)=m<0$
and
$D(0)=
m+\mmm \at \|\vp/\omega\|^2>0$.
Then $D(z)$ has a unique zero point of order one in $(-\infty,0)$.
We denote the zero point by $-E^2$ $(E>0)$, and $\gamma$ is defined only in the case of $m<0$ by
\eq{gammadef}
\gamma=D'(-E^2)^{-\han},
\en
where we can directly see that
$$\d D'(-E^2)=\at \mmm \int\frac{|\vp(k)|^2}{E^2+\omega(k)^2}dk.$$
Let $\tmn^\ast=(\tmn)^\ast$, i.e.,
$$\tmn^\ast f=\delta_{\mu\nu} f-\ak\dmu\vp\omega^\fdd G\omega^\fdd \ov \JI  f.$$
Note that $G^\ast=-G$.
\bl{algebra}
\TTT{Algebraic relations}
{\rm \cite{ara83-a,ara83-b}}
Suppose Assumption \ref{1v1} and \kak{negative}.
Let  $\theta(m)=\lkk\begin{array}{ll}
1,&m<0\\
0,&m>0.\end{array}
\right.$
Set
\begin{align*}
&\rho_{\mu \nu}(k)=
\delta_{\mu\nu}-\dmu(k)=
k_\mu k_\nu/|k|^2,\quad
\naka=\frac{1}{\mmm \vol },\quad
\ppmm=D_+/D_-,\\
&\d F_{\mu\nu}=
\dmu\frac{\ak  \vp }{E^2+\omega^2},\quad
[f](|k|)=\int_{S_{d-1}}f(|k|,v)dv.
\end{align*}
Then the
operator $T_{\mu\nu}$ has the following properties:
\begin{enumerate}
\item 
 $\|\omega^{n/2}T_{\mu\nu}f\|\leq C\|\omega^{n/2}f\|$ for $n=-1,0,1$.

\item 
$
\widetilde{T_{\mu\nu}f}=T_{\mu\nu}\widetilde{f}$,
$T_{\mu\nu}=T_{\nu\mu}$
and
if $ f$ is rotation invariant,
then so is $\tmn f$.

\item 
$\d T^{\ast}_{\mu\nu}d_{\nu a}T_{ab }
f=
d_{\mu b}f-
\theta(m)
\gamma^2
(F_{ab }, f)F_{a \mu}$.
In particular
\eq{tu3}
e_\mu ^r T^{\ast}_{\mu\nu}d_{\nu a}
T_{ab }e_b^s f=\delta_{rs }f
-\theta(m) \gamma^2
(e_b^sF_{ab},  f)e_\mu ^rF_{a\mu}.\en

\item 
$\d T_{\mu\nu} d_{\nu a} T_{ab}^\ast f =d_{\mu b} f +\ak (\rho_{\mu b}\vp \hat G \ov \JI - \JI  \hat G \vp\rho_{\mu b})f$. In particular
\eq{tu1}
\d e_{\mu}^{r}T_{\mu\nu}d_{\nu a}
T^{\ast}_{ab }e_{b}^{s}
f=
\delta_{rs}f.\en

\item 
$\d \overline{T}_{\mu\nu}f=
\ppmm  T_{\mu\nu}f+(1-\ppmm )
\left(\delta_{\mu\nu}f-
\naka
[d_{\mu\nu}f]\right)$.

 \item 
$\d \ov {T_{\mu\nu}} d_{\nu a} T_{ab}^\ast f =d_{\mu b} f -(1-\ppmm)\naka[d_{\mu b}f]+\ak (\rho_{\mu b}\vp\hat G \ov \JI -\ov \JI  \hat G \vp \rho_{\mu b})f$. In particular
\eq{tu2}
e_\mu^r \ov {T_{\mu\nu}} d_{\nu a} T_{ab}^\ast \tilde {e_b^s}=
e_\mu ^r d_{\mu b}\tilde{e_b ^s}-(1-\ppmm )\naka e_\mu ^r[d_{\mu b} \tilde {e_b^s}f].
\en

\item 
$
\d T^\ast_{\mu\nu}d_{\nu a}hT_{ab }=
\overline{T^\ast_{\mu\nu}}
d_{\nu a}h
\overline{T_{ab }}$
for
rotation invariant function $h$.

\item 
$[\omega^2 , T_{\mu\nu}]
 f=
\ak   ( d_{\mu\nu}\vp , f)\JI $ and
$[  \omega^2 , T_{\mu\nu}^\ast]
 f=
-\ak   (\JI , f) d_{\mu\nu}\vp$.

\item 
$\d \ak  T_{\mu\nu}\vp =\delta_{\mu\nu}m \JI $. In particular
$\ak  e_\mu^r T_{\mu\nu}d_{\nu a }\vp=\delta_{\mu a}m \JI $.

\item 
$\d e_{\mu}^{r}
T_{\mu\nu}T^{\ast}_{\nu a }e_{a}^{s}
f=\delta_{rs}f$.

\item 
$\d \left(
d_{\nu a}\frac{\JI }{{\omega}},
\frac{1}{{\omega}}T_{\mu\nu}f \right)=
\frac{\ak }{\mass}
\left(
d_{\mu a }
\frac{\vp}{\omega },
\frac{f}{{\omega}}
\right)
-
\theta(m)
\frac{\gamma^2}
{ E^2}\left(
F_{\mu a},
f\right)$.

\item 
$\d
\left(d_{\nu a}\frac{\JI }{\sqrt{\omega^n}}, h\frac{1}{\omega}
 T_{\mu\nu}f\right)
=
\left(d_{\nu a}\frac{\overline{\JI }}{\sqrt{\omega^n}},
h\frac{1}{\omega} \overline{T}_{\mu\nu}f\right)$
for
  rotation invariant function $h$ with $n=0,1,2$.

\item 
$
\d e_{\mu}^rT_{\mu\nu}F_{\nu a}=0\quad  (m<0).
$

\item 
$\d \ak  (\vp ,F_{\mu\nu})
=-m\delta_{\mu\nu}\quad  (m<0).$

\item 
$\d T_{\mu\nu}^{\ast}d_{\nu a}\JI
=\theta(m)
\gamma^2F_{\mu a}$.

\item 
$\d (F_{\mu a} , F_{a \nu})
=\delta_{\mu\nu}\frac{1}{\gamma^2}\quad
 (m<0).$
 \end{enumerate}
\end{lemma}
Statements (3)-(7) are used in Lemma \ref{symp} to show some symplectic structure, (8) and (9) in Lemma \ref{ccc} to show  some commutation relations,
(10) in the proof of Theorem \ref{mainp-nasi},  (11) and (12) in Lemma \ref{displace} and the proof of Theorem \ref{mainp-nasi}, and (13)-(16) in Section \ref{negativemass}.

{\it Proof of Lemma \ref{algebra}}:

Note that
for  rotation invariant functions $f$ and $g$,
$$(d_{\mu\nu} f, g)=\delta_{\mu\nu}\mmm (f,g)$$ and
the identity
\eq{ddd}
\frac{\at}{2}\mmm  \vol  \frac  {\vp (\sqrt s)^2s^\fdd}{|D_+(s)|^2}=\frac {1}{2\pi i}\lk\frac{1}{D_-(s)}-\frac{1}{D_+(s)}\rk
\en
holds\footnote{When $\vp (k)=0$ for $|k|>\Lambda$ or $|k|< \lambda$
((2') of Assumption \ref{1v1}),
\kak{ddd} is valid for $s\in[\lambda^2,\Lambda^2]$.}
 by $D_+(s)-D_-(s)=\pi i \at \mmm  \vol \vp (\sqrt s)^2 s^\fdd$.
We set $\hat G=\omega^\fdd G\omega^\fdd$
 for the notational convenience.
\begin{enumerate}
\item 
By  (3) of Assumption \ref{1v1} we can see that
\begin{align*}
&
\|\y T_{\mu\nu}f\|
\leq \|\y \delta_{\mu\nu}f\|+|\ak|
\|\frac{1}{D_+(\omega^2(\cdot))}\|_\infty
\|\omega^{\frac{d-1}{2}}\vp  \|_\infty \|\omega^{\frac{d-3}{2}}\vp  \|_\infty \|G\|_2
\|\y  f\|,\\
&\|\yy T_{\mu\nu}f\|\leq
\|\yy \delta_{\mu\nu}f\|+|\ak|
\|\frac{1}{D_+(\omega^2(\cdot))}\|_\infty
\|\omega^{\frac{d-1}{2}}\vp  \|_\infty \|\omega^{\frac{d-3}{2}}\vp  \|_\infty \|G\|_2
\|\yy f\|,\\
&\|T_{\mu\nu}f\|\leq
\|\delta_{\mu\nu}f\|+|\ak|
\|\frac{1}{D_+(\omega^2(\cdot))}\|_\infty
\|\omega^{\fdd}\vp  \|_\infty \|\omega^{\fdd}\vp  \|_\infty \|G\|_2
\|f\|.
\end{align*}
Then 1.  follows\footnote{When $\vp (k)=0$ for $|k|>\Lambda$ or $|k|< \lambda$
((2') of Assumption \ref{1v1}),
it is understood that $\|\frac{1}{D_+(\omega^2(\cdot))}\|_\infty=
\sup_{\lambda\leq|k|\leq\Lambda}
|\frac{1}{D_+(\omega^2(k))}|$.
}.
\item 
This follows from the definition of $\tmn$.

\item 
We have
\begin{align*}
&(T^{\ast}_{\mu\nu}d_{\nu a}T_{ab }f, g)\\
&=(d_{\nu a}T_{ab }f, T_{\mu\nu} g)
\\
&=(d_{\mu b}f,g)+
\ak (f, d_{\nu b }\JI \hat G \dmu \vp  g)+
\ak (d_{a\mu}\JI \hat G d_{ab}\vp  f, g)\\
&\hspace{5cm}+\at\mmm (\JI \hat G d_{ab}\vp  f, \JI \hat G d_{a \mu}\vp  g)\\
&=
{\rm I}+\two +\three +\four .
\end{align*}
We compute $\four $ as
\begin{align*}
\four &=
\limdt \at \int\frac{\mmm  |\JI (k)|^2 F(k',k'')}{(\kkt-\kkkt -it)(\kkt-\kkkkt +it)}
 dk dk' dk''\\
 &=
 \limdt \frac{\at}{2}
 \int\lk
 \int_0^\infty
 \frac{ \mmm \vp ^2(\sqrt s)s^\fdd \vol F(k',k'')}{(s-\kkkt -it)(s-\kkkkt +it)|D_+(s)|^2}
 ds\rk
  dk' dk''\\
&=
\limdt
\frac{1}{2\pi i}
\int
\!\!
\lk
 \int_0^\infty
\!\!\!
\frac{F(k',k'')}
{
(s-\kkkt -it)(s-\kkkkt +it)}
\lk\frac{1}{D_-(s)}-\frac{1}{D_+(s)}\rk
ds
\rk
 dk'dk'',
 \end{align*}
where $F(k',k'')=d_{ab}(k')d_{a\mu}(k'')\vp (k')\vp (k'')\bar f(k') g(k'')$.
\begin{figure}[t]
\centering
\includegraphics[width=200pt]{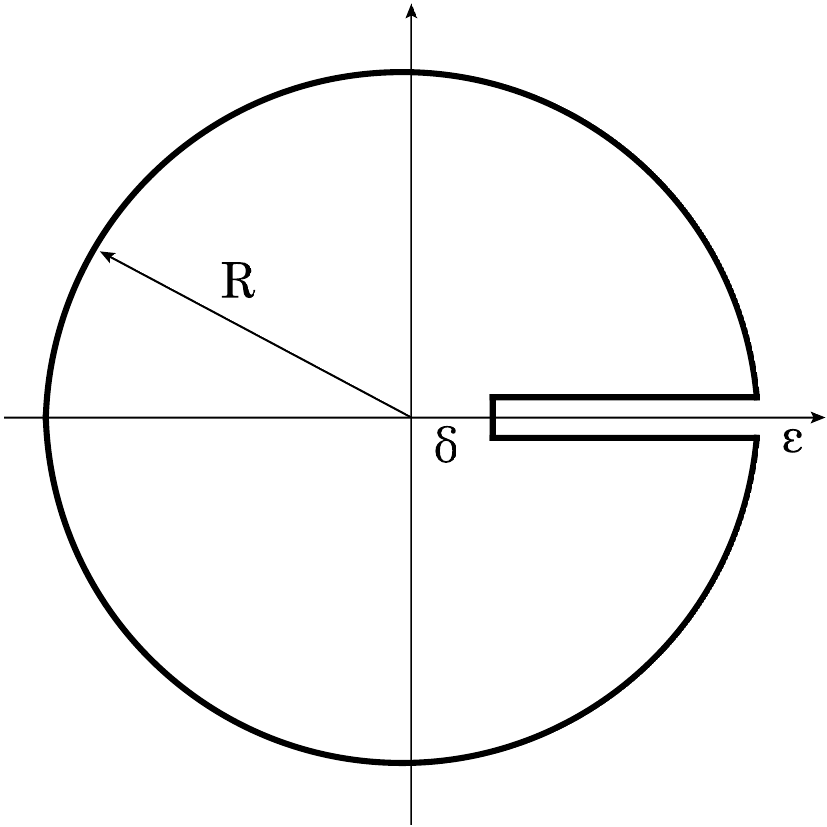}
\caption{Cut plane $\cp$}
\label{piccut}
\end{figure}
By a contour integral\ko{contour integral} on the cut plane\ko{cut plane} $\cp$ (Figure  \ref{piccut}),
we have
\begin{align*}
\four &=
\limdt
\limcp
\frac{1}{2\pi i}
 \int\lk\int_\cp
 \frac{-F(k',k'')}
{
(z-\kkkt -it)(z-\kkkt +it)D(z)}dz\rk
 dk'dk''\\
&=
\limdt
\int \frac{-\at F(k',k'')}{D(\kkkkt -it)(\kkkkt -\kkkt -2it)}dk'dk''\\
&\quad +
\limdt
\int \frac{-\at F(k',k'')}{D(\kkkt +it)(\kkkt -\kkkkt +2it)}dk'dk''\\
&\quad +
\limdt
\int \frac{-\at F(k',k'')\gamma^2\theta(m)}
{(E^2+\kkkt +it)
(E^2+\kkkkt -it)}dk'dk''\\
&=-\at\lk
f,d_{ab}
\frac{\vp }{D_+}\hat Gd_{a\mu}
 \vp  g\rk
-\at\lk d_{a\mu}\frac{\vp }{D_+}
\hat G d_{ab}\vp  f,g\rk
-\theta(m)\gamma^2(f,F_{ab})(F_{a\mu},g).
\end{align*}
Then
$\four =-\two -\three -\theta(m)\gamma^2
(f,F_{ab})(F_{a\mu},g)$. Hence  the desired result is obtained.

\item 
We see that
\begin{align*}
&(
T_{\mu\nu}d_{\nu a}
T^{\ast}_{ab}
f,g)\\
&=
(d_{\mu b}f,g)
-\ak(d_{\mu b}\vp  \hat G \ov \JI   f, g)
-\ak(f, d_{\mu b}\vp   \hat G \ov \JI  g)
+\at (d_{\mu b}\vp  \hat G \ov \JI   f, \vp  \hat G\ov \JI   g)\\
&={\rm I}+\two +\three +\four .
\end{align*}
We have
\begin{align*}
\four &=
\limdt \at \delta_{\mu b}
 \int \frac{\mmm  \vp (k)^2 H(k',k'')}{
(\kkt -\kkkt -it)
(\kkt -\kkkkt +it)}dk dk' dk'',
\end{align*}
where $H(k',k'')=\JI (k')\ov \JI (k'')\bar f(k') g(k'')$.
We can see that
\begin{align*}
\four &=\limdt \frac{\at}{2} \delta_{\mu b}
 \int
 \lk
  \int_0^\infty
 \frac{\mmm  \vp (\sqrt s)^2 s^\fdd \vol H(k',k'')}{
(s-\kkkt -it)
(s-\kkkkt +it)}ds\rk
 dk' dk''\\
&=\limdt \frac{\at}{2} \delta_{\mu b}
\int \lk
 \int_0^\infty
\frac{|D_+(s)|^2
\mmm  \vp (\sqrt s)^2 s^\fdd \vol H(k',k'')}{
(s-\kkkt -it)
(s-\kkkkt +it)|D_+(s)|^2}ds\rk
 dk' dk''\\
&=\limdt \frac{1}{2\pi i}
\delta_{\mu b} \int \lk
 \int_0^\infty
 \frac{|D_+(s)|^2
H(k',k'')}{
(s-\kkkt -it)
(s-\kkkkt +it)}
\right.\\
 &\hspace{5cm}\times\left.
\lk\frac{1}{D_-(s)}-\frac{1}{D_+(s)}\rk
ds\rk
 dk' dk''\\
&=\limdt \frac{1}{2\pi i}
\delta_{\mu b} \int \lk
 \int_0^\infty
 \frac{(D_+(s)-D_-(s))
 H(k',k'')}{
(s-\kkkt -it)
(s-\kkkkt +it)}
 ds\rk
  dk' dk''.
 \end{align*}
 It can be computed by a contour integral on the cut plane $\cp$ as
 \begin{align*}
 \four &=
\limdt \limcp
\frac{1}{2\pi i}
 \delta_{\mu b}
 \int \lk
 \int_\cp \frac{D(z)
 H(k',k'')}{
(z-\kkkt -it)
(z-\kkkkt +it)} dz \rk
 dk' dk''\\
&=\limdt
\delta_{\mu b}\int\frac{D(\kkkt +it)-D(\kkkkt -it)}{\kkkt -\kkkkt +2it}H(k',k'') dk' dk''\\
&= \ak \delta_{\mu b}(f, \vp \hat G \ov \JI   g)+\ak  \delta_{\mu b}(\vp  \hat G \ov \JI   f,g).
\end{align*}
Then
  the desired result is obtained.

\item 
We have
$\d \overline{T}_{\mu\nu}f(k)
=\delta_{\mu\nu}f(k)+\limdt\ak \ov {\JI (k)}\int \frac{\dmu(k')\vp (k')f(k')}
{\kkt -\kkkt -it}dk'$, and
\begin{align*}
& \int \frac{\dmu(k')\vp (k')f(k')}
{\kkt -\kkkt -it}dk'\\
&=
  \int \frac{\dmu(k')\vp (k')f(k')}
{\kkt -\kkkt +it}dk'
+2i
\int\frac{t \dmu(k')\vp (k')f(k')}
{(\kkt -\kkkt )^2+t^2}
dk'.
\end{align*}
Since $\pi\f \int\frac{t}{x^2+t^2}f(x)dx\to f(0)$ as $t\to0$, we see that
\begin{align*}
2i
\int\frac{t \dmu(k')\vp (k')f(k')}
{(\kkt -\kkkt )^2+t^2}
dk'
&=
i \int_0^\infty
 \frac{t[\dmu f](\sqrt s)\vp (\sqrt s)s^\fdd }{(\kkt -s)^2+t^2} ds\\
&
\to \pi i [\dmu f](\omega)\vp (\omega)\omega^{d-2}.
\end{align*}
Then
$$\ov T_{\mu \nu }f=\delta_{\mu\nu} f +\ak  \ov \JI \lk
\hat G \dmu \vp  f+\pi i \omega^{d-2}[\dmu f](\omega)\vp \rk $$
Notice that $\ov \JI =\ppmm \JI $. Hence
\begin{align*}
\ov \tmn f&=
\delta_{\mu\nu}f+\ak \ppmm \JI \lk
\hat G \dmu \vp  f+\pi i \omega^{d-2}\vp [\dmu f](\omega)
\rk
\\
&=
\ppmm  \tmn f
+\delta_{\mu\nu}(1-\ppmm ) f
+\pi i
\ak
\ppmm
\JI \vp [\dmu f](\omega)
\omega^{d-2}.
\end{align*}
Since
$D_+(s)-D_-(s)=\pi i \at\mmm  \vp ^2(\sqrt s)s^\fdd  \vol $,
we see that
\begin{align*}
\pi i \ak \omega^{d-2} \ppmm  \JI \vp
=
\pi i \ak \omega^{d-2} \ak \frac{\vp ^2}{D_-(\omega^2)}
=
\naka  \frac{D_+(\omega^2)-D_-(\omega^2)}{D_-(\omega^2)}
=\naka  (\ppmm-1).
\end{align*}
Then the desired result is obtained.

\item 
 Note that $[\dmu]=\delta_{\mu\nu}/\naka $.
\begin{align*}
\ov{T_{\mu\nu}}d_{\nu a}T_{ab}^\ast f=\ppmm {T_{\mu\nu}}d_{\nu a}T_{ab}^\ast f
+(1-\ppmm)(\delta_{\mu \nu} d_{\nu a}T_{ab}^\ast f-\naka[ d_{\mu \nu} d_{\nu a}T_{ab}^\ast f])
\end{align*}
Notice that
$$\delta_{\mu \nu} d_{\nu a}T_{ab}^\ast f-\naka[ d_{\mu \nu} d_{\nu a}T_{ab}^\ast f]=d_{\mu b}f-\naka[d_{\mu b}f]+\ak \rho_{\mu b}\vp \hat G \ov \JI  f.$$
Thus
\begin{align*}
&\ov{T_{\mu\nu}}d_{\nu a}T_{ab}^\ast f\\
&=
d_{\mu b}f -(1-\ppmm)\naka[d_{\mu b}f]+\ak(1-\ppmm)\rho_{\mu b}\vp  \hat G \ov \JI  f+\ak \ppmm (\rho_{\mu b}\vp  \hat G \ov \JI -\JI \hat G \vp \rho_{\mu b})\\
&=
d_{\mu b}f -(1-\ppmm)\naka[d_{\mu b}f]+\ak\rho_{\mu b}\vp  \hat G \ov \JI  f
-\ak \ov \JI \hat G \vp \rho_{\mu b}f.
\end{align*}
Since $e_\mu ^r \rho_{\mu b}=0$, we in particular obtain  \kak{tu2}.

\item 
We see that
\begin{align*}
(\ov{T^\ast_{\mu\nu}} d_{\nu a} h \ov{T_{ab}}f,g)
=&
(d_{\nu a} h \ov{T_{ab}}f,\ov{T_{\mu\nu}}g)\\
=&
(\ppmm d_{\nu a} h {T_{ab}}f, \ppmm {T_{\mu\nu}}g)\\
&+(d_{\nu a}h(1-\ppmm)(\delta_{ab}f-\naka[d_{ab}f]), \ppmm T_{\mu\nu}g)\\
&+(d_{\nu a}h\ppmm T_{ab} f, (1-\ppmm)(\delta_{\mu\nu}g-\naka[ d_{\mu\nu}g]))\\
&+(d_{\nu a}h(1-\ppmm)(\delta_{ab}f-\naka[d_{ab}f]),
 (1-\ppmm)(\delta_{\mu\nu}g-\naka[ d_{\mu\nu}g]))\\
 =&{\rm I}+\two +\three +\four .
\end{align*}
We have
${\rm I}=(d_{\nu a} h {T_{ab}}f, {T_{\mu\nu}} g)$.
We will show that $\two +\three +\four =0$.
Since $h$ and $\ppmm$ are rotation invariant, we have
\begin{align*}
\two &=\int \bar h (1-\ov{\ppmm}) \ppmm([d_{\mu b}\bar f g]-\naka  [d_{\mu a} g][d_{ab}\bar f])dk\\
\three &=\int \bar h (1-{\ppmm}) \ov{\ppmm}([d_{\mu b}\bar f g]-\naka  [d_{\mu \nu} g][d_{\nu b}\bar f])dk\\
\four
&=
\int \bar h |1-{\ppmm}|^2 \\
&\times ([d_{\mu b}\bar f g]-\naka  [d_{\mu a} g][d_{ab}\bar f]
-\naka  [d_{\mu \nu} g][d_{\nu b}\bar f]
+\naka^2[d_{\nu a}][d_{\mu\nu}g][d_{ab}\bar f]
)dk.
\end {align*}
Notice that
\begin{align*}
&(1-\ov \ppmm)\ppmm =\ppmm -1,\\
&
(1-\ppmm)\ov \ppmm=\ov  \ppmm-1,\\
&
|1-\ppmm|^2+\ppmm-1+\ov \ppmm-1=0,\\
&-\naka  [d_{\mu \nu} g][d_{\nu b}\bar f]
+\naka^2[d_{\nu a}][d_{\mu\nu}g][d_{ab}\bar f]=
-\naka  [d_{\mu \nu} g][d_{\nu b}\bar f]
+\naka\delta_{\nu a} [d_{\mu\nu}g][d_{ab}\bar f]=0.
\end{align*}
Then $\two +\three +\four =0$ follows.

\item 
We see that
$$[  \omega^2 , T_{\mu\nu}]
 f=\limdt \ak \int \frac{(\kkt -\kkkt )\JI (k)\dmu(k')\vp (k')f(k')}{\kkt -\kkkt +it}dk'
 =\ak (\dmu\vp ,f)\JI .$$

\item 
We see that
$$\d \ak  T_{\mu\nu}\vp  = \ak \delta_{\mu\nu}\vp +\at  \frac{\vp }{D_+}
\hat G \dmu \vp ^2=
\ak \delta_{\mu\nu}
\vp \lk
1+\frac{\at \mmm  \hat G\vp ^2}{m-\at \mmm  \hat G\vp ^2}\rk=
m\delta_{\mu\nu}\JI .$$
Here we used
that  $D_+=m-\at \mmm \hat G \vp ^2$.
In particular
it follows that
$$\ak  e_\mu^r T_{\mu\nu} d_{\nu a}\vp =\ak  e_\mu^r T_{\mu a}\vp -e_\mu^r\rho_{\mu a}\vp =
\ak  e_\mu^r T_{\mu a}\vp =me_a^r \JI .$$

\item 
This is shown in the same way as $4.$

\item 
We have
$$\d \left(
d_{\nu a}\frac{\JI }{{\omega}},
\frac{1}{{\omega}}T_{\mu\nu}f \right)=
\lk d_{\mu a}\frac{\JI }{{\omega}},
\frac{f}{{\omega}}\rk+
\lk d_{\nu a}\frac{\JI }{{\omega}}
,
\frac{\ak \JI }{\omega}
\hat G  \dmu \vp  f\rk=
{\rm I}+\two .$$
By \kak{ddd}
 we have
\begin{align*}
\two &=\lim_{t\downarrow 0}\ak
\int\frac{|\JI (k)|^2\dmu(k')d_{\nu a}(k)\vp (k')f(k')}{(\kkt -\kkkt +it)\kkt }dkdk'\\
&=
\lim_{t\downarrow 0}\frac{\ak ^3}{2}
 \int\lk
  \int_0^\infty
\frac{\mmm  \vp ^2(\sqrt s)s^\fdd d_{\mu a}(k')\vp (k')f(k')\vol }
 {(s-\kkkt +it)s|D_+(s)|^2}ds\rk dk'\\
&=
\limdt\frac{\ak}{2\pi i}
 \int\lk
  \int_0^\infty
\frac{1}{(s-\kkkt +it)s}\lk\frac{1}{D_-(s)}-\frac{1}{D_+(s)}\rk F(k')ds\rk dk',
 \end{align*}
 where
 $F(k')=d_{\nu a}(k')\vp (k')f(k')$.
By a contour integral on the cut
 plane $\cp$, we have
\begin{align*}
&\frac{1}{2\pi i}
 \int\frac{1}{(s-\kkkt +it)s}\lk\frac{1}{D_-(s)}-\frac{1}{D_+(s)}\rk ds \\
 &=-\frac{1}{2\pi i}\limcp \int _\cp
 \frac{1}{(z-\kkkt +it)zD(z)}dz\\
 &=-\frac {1}{(\kkkt -it)D(\kkkt -it)}-\frac{\theta(m)\gamma^2}{(E^2+\kkkt +it)E^2}+\frac{1}{\mass(\kkkt -it)}.
\end{align*}
Then
\begin{align*}
&\two \\
&=\limdt
\int\lkk
\frac {\ak F(k')}{\mass(\kkkt -it)}
-
\frac{\ak F(k')}{(\kkkt -it)D(\kkkt -it)}
-\frac{\ak \theta(m)\gamma^2 F(k')}{(E^2+\kkkt +it)E^2}\rkk dk'\\
&=\frac{\ak}{\mass} \lk{d_{\mu a}}
\frac{\vp }{{\omega}},
\frac{f}{\omega}\rk
-\lk d_{\mu a}\frac{\JI }{{\omega}},
\frac{f}{\omega}\rk-
\theta(m)
\frac{\gamma^2}{E^2}\lk F_{\mu a}, f\rk.
\end{align*}
Hence we have
$$ {\rm I}+\two =
\frac{\ak }{\mass}
\left(
d_{\mu a }
\frac{\vp }{{\omega}},
\frac{f}{{\omega}}
\right)
-
\theta(m)
\frac{\gamma^2}
{ E^2}
\left(
F_{\mu a},
f\right).$$

\item 
Note that
$\int h(k) \dmu (k) f(k)dk=
\naka \int h(k) d_{\mu a}(k) [d_{a\nu}f](|k|)dk$ for rotation invariant function $h$.
From 5. it follows that
\begin{align*}
\lk
d_{\nu a}\frac{\ov \JI }{\sqrt{\omega^n}}, h \frac{1}{\omega} \ov{\tmn} f
\rk
&=
\lk
\ppmm d_{\nu a}\frac{\JI }{\sqrt{\omega^n}},
\ppmm \frac{1}{\omega} h{\tmn} f
\rk\\
&+
\lk
\ppmm d_{\nu a}\frac{\JI }{\sqrt{\omega^n}}, h
\frac{1}{\omega} (1-\ppmm) (\delta_{\mu\nu}f-\naka [d_{\mu \nu}f])
\rk.
\end{align*}
Since $\ppmm d_{\nu a}\frac{\JI }{\sqrt{\omega^n}} h\yy (1-\ppmm) $ is rotation invariant,
the second term vanishes. Then the claim is proven.

\item 
 We have
$$
 (e_{\mu}^rT_{\mu\nu}F_{\nu a},g)=(F_{\mu a},e_\mu^r g)-(F_{\nu a },\ak d_{\mu\nu}\vp  \hat G \ov \JI  e_\mu^r g)={\rm I}-\two .$$
 Then
\begin{align*}
\two =\at
\int\frac{\mmm \vp (k)^2 \ov \JI (k') e_a^r(k') g(k')}{(\kkt +E^2)(\kkt -\kkkt +it)}
 dk dk'.
\end{align*}
We see that
\begin{align*}
&
\int\frac{\vp (k)^2}
{(\kkt +E^2)(\kkt -\kkkt +it)}
 dk\\
 &=
\int \lk \frac{\vp (k)^2}{\kkt -\kkkt +it}-\frac{\vp (k)^2}{\kkt +E^2}\rk
\frac{1}{E^2+\kkkt -it}dk.
\end{align*}
By the definitions of $D_-$ and
$-E^2$, we have
\begin{align}
&
\limdt \int  \frac{\at \mmm  \vp (k)^2}
{\kkt -\kkkt +it}dk
=D_-(\kkkt )-m,\\
&\label{toshi}
\int
\frac{\at \mmm \vp (k)^2}{\kkt +E^2}dk=
D_-(-E^2)-m=-m.
\end{align}
Thus we get
$$\two =
\int \frac{\vp (k')e_a^r(k')g(k')}
{E^2+\kkkt }dk'=
(F_{ba},e_b^r g)$$
and ${\rm I}-\two=0$ follows.

\item 
We see that
$\d \ak  (\vp  ,F_{\mu\nu})
=\at \delta_{\mu\nu}\mmm  \int\frac{\vp (k)^2}{E^2+\kkt }dk=-m\delta_{\mu\nu}
$
by \kak{toshi}.

\item 
We see that
$$ (T_{\mu\nu}^{\ast}d_{\nu a}\JI , f)=(d_{\nu a}\JI ,\delta_{\mu\nu}f)+(d_{\nu a}\JI ,\ak \JI \hat G \dmu \vp  f)={\rm I}+\two.$$
We have
\begin{align*}
\two &=\limdt\ak\int\frac{|\JI (k)|^2\dmu(k')d_{\nu a}(k)\vp (k')f(k')}{\kkt -\kkkt +it}dk dk'\\
&=\limdt
\frac{\ak ^3}{2} \int\lk
 \int_0^\infty
\frac{\mmm \vp (\sqrt s)^2 s^\fdd d_{\mu a}(k') \vp (k')f(k')\vol }{(s-\kkkt +it)|D_+(s)|^2}ds\rk
dk'\\
&=\limdt
\frac{\ak}{2\pi i}\int
 \lk
  \int_0^\infty
\frac{1}{s-\kkkt +it}
\lk \frac{1}{D_-(s)}-\frac{1}{D_+(s)}\rk F(k') ds \rk
dk',
\end{align*}
where $F(k')=d_{\mu a}(k')\vp (k') f(k')$. By a contour integration on the cut plane $\cp$, we compute as
\begin{align*}
&
\frac{1}{2\pi i}
 \int_0^\infty
 \frac{1}{s-\kkkt +it}
\lk \frac{1}{D_-(s)}-\frac{1}{D_+(s)}\rk  ds\\
&=
-\limcp
\frac{1}{2\pi i}\int_\cp  \frac{1}{(z-\kkkt +it)D(z)}dz\\
&=
-\frac{1}{D(\kkkt -it)}+\frac{\gamma^2}{E^2+\kkkt +it}.
\end{align*}
Then we have
\begin{align*}
\two &=\limdt
\ak \int \lk
-\frac{1}{D(\kkkt -it)}+\frac{\gamma^2}{E^2+\kkkt +it}
\rk F(k') dk'\\
&=
-(d_{\mu a}\JI ,f)+
\gamma^2\lk F_{\mu a},f\rk.
\end{align*}
Hence  ${\rm I}+\two =\gamma^2(F_{\mu a },f)$ follows.

\item 
We have
\begin{align*}
 (F_{\mu a} , F_{a \nu})
&=\delta_{\mu\nu}\at \mmm  \int \frac{\vp (k)^2}{(\kkt +E^2)^2}dk\\
&=
\delta_{\mu\nu}\mmm  \frac{\at}{2}
 \int_0^\infty
 \frac{\vp (\sqrt s)^2s^\fdd}{(s +E^2)^2}ds
\\
&=\frac{1}{2\pi i}
\delta_{\mu\nu}
 \int_0^\infty
 \frac{D_+(s)-D_-(s)}{(s+E^2)^2}ds.
\end{align*}
Hence by a contour integral on the cut plane $\cp$, we have
\begin{align*}
 (F_{\mu a} , F_{a \nu})
=\limcp \frac{1}{2\pi i}\delta_{\mu\nu}\int_ \cp \frac{D(z)}{(z+E^2)^2}dz
=\delta _{\mu\nu}D'(-E^2).
\end{align*}
 \end{enumerate}
Then the proof is complete.
\qed

\subsubsection{Intertwining operator}
Now we introduce the class of functions.
Let
\eq{mclass}
M_n=\{f|\omega^n \hat f\in\LR\}.
\en
Let  $\ta_\mu (f)=A_\mu (\hat f)$ and $\tp_\mu(g)=\Pi_\mu(\hat g)$, i.e.,
\begin{align*}
&\ta_\mu(f)=\frac{1}{\sqrt 2} \int\frac{1}{\sqrt{\omega(k)}}e_\mu^j (k)\lk \add(k,j)f(k)+a(k,j)f(-k)\rk dk,\\
&\tp_\mu(f)=\frac{i}{\sqrt 2} \int{\sqrt{\omega(k)}}e_\mu^j (k)\lk \add(k,j)f(k)-a(k,j)f(-k)\rk dk.
\end{align*}
Then
$[\ta_\mu (f), \tp_\nu(g)]=i(\dmu \bar f, \tilde g)$
holds, and
$$
[A_\mu,\tp_\nu(g)]=
i(d_{\mu\nu}\hat{\bar\varphi},\hat{\hat g})=
i(d_{\mu\nu}\tilde {\bar\vp},\tilde g)=
i(d_{\mu\nu}\bar\vp,g)=
i(d_{\mu\nu}\vp,g),$$
 where we used $\bar\vp=\vp$.
Then we  define
\begin{align}
&B_p(f,j)=\frac{1}{\sqrt{2}}\left\{
\ta_\mu\left(
\tmn^\ast\sqrt{\omega} \ejn f\right)+
i\tp_\mu\left(
\tmn^\ast\frac{1}{\sqrt{\omega}}\enr f\right)
-   \left(
p\cdot e^j\frac{\JI }{\omega}, \frac{f}{\y}\right)\right\},\\
&
B_p^\ast (f,j)=\frac{1}{\sqrt{2}}\left\{
\ta_\mu\left(
\btmn^\ast\sqrt{\omega} \TT {\ejn f}\right)-
i\tp_\mu
\left(
\btmn^\ast\frac{1}{\sqrt{\omega}} \TT {\ejn f}\right)-
\left(
p\cdot e^j\frac{\ov \JI }{\omega},
\frac{f}{\y}\right)\right\}
\end{align}
for $f\in M_0\cap M_{-1/2}$, where $\TT f(k)=\widetilde f(k)=f(-k)$.
\begin{remark}
Note that condition $f\in M_{-1/2}$ is not needed
for the definition of $B_p^\#$ for  $p=0$.
\end{remark}
We have
\begin{align}
&B_p(f,j)=
a(\pij f,i)+\add({\mij  f}, i)-
\left(
p\cdot e^j\frac{\JI }{\sqrt 2\omega},
\frac{ f}{\y}\right), \\
&
B_p^\ast (f,j)=
a ({\bmij f}, i)+\add (\bpij f, i)-
\left(
p\cdot e^j\frac{\ov \JI }{\sqrt 2\omega},
\frac{ f}{\y}\right)
\end{align}
for $f\in M_0\cap M_{-1/2}$, where
\begin{align}
&\pij =\half \ems
\lk
\frac{1}{\sqrt{\oooo}}
\tmn^\ast
\sqrt{\oooo}
+
\sqrt{\oooo}
\tmn^\ast
\frac{1}{\sqrt{\oooo}}
\rk
\TT \enr, \\
&
\mij =\half
\ems
\lk
\frac{1}{\sqrt{\oooo}}
\tmn^\ast
\sqrt{\oooo}-
\sqrt{\oooo}
\tmn^\ast
\frac{1}{\sqrt{\oooo}}
\rk
{\enr }.
\end{align}
Let
$W_\pm=
\lk
W_{\pm ij}
\rk_{1\leq i,j\leq d-1}:\oplus^{d-1}\LR\to\oplus^{d-1}\LR$ and
$$W=\mat {W_+} {\ov{W_-}} {W_-} {\ov{W_+}}:
\bigoplus^2 (\oplus^{d-1}\LR)
\to
\bigoplus^2
(\oplus^{d-1}\LR).
$$
We set
\begin{align}
&b_W(f,j)=
a(\pij f,i)+\add({\mij  f}, i),\\
&
b_W^\ast (f,j)=
a ({\bmij f}, i)+\add (\bpij f, i).
\end{align}
for $f\in M_0$.
\bl{symp}
\TTT{Symplectic structure}
\index{symplectic structure}
Suppose Assumption \ref{1v1} and that $m>-\mmm \at \|\vp/\omega\|^2$,  $m\not =0$.
Then
\begin{align}
\label{w1}
&\jp\wwp-\jm\wwm=\one,\\
\label{w3}
&\ip \wwm-\im\wwp=0,\\
&\label{w2}
\wwp\jp-\bwwm\im=\one+\theta(m)Z_+,\\
&\label{w4}
\wwm\jp-\bwwp\im=\theta(m)Z_-,
\end{align}
where
\begin{align*}
&\theta(m)=\lkk\begin{array}{ll} 1&m<0,\\
0&m>0,
\end{array}
\right.\\
&Z_{\pm,ij}f =
\mp\half\gamma^2\lk\sqrt\omega F_\mu^i\lk\frac{F_\mu^j}{\sqrt\omega},f\rk
\pm\frac{1}{\sqrt\omega}
F_\mu^i(\sqrt\omega F_\mu^j,f)\rk,\\
&F_\mu^j=\frac{\ak\vp}{E^2+\omega^2}e_\mu^j,\\
&
\gamma=D'(-E^2)^{-\han}=\lk
\mmm \int\frac{\at \vp(k)^2}{E^2+\omega(k)^2}dk\rk^{-\han}
.
\end{align*}
In particular in the case of $m>0$,
\begin{align}
\label{w111}
&\jp\wwp-\jm\wwm=\one, \\
\label{w110}
&\ip \wwm-\im\wwp=0,\\
&\label{w222}
\wwp\jp-\bwwm\im=\one, \\
\label{w220}
&\wwm\jp-\bwwp\im=0,
\end{align}
holds, i.e., $W\in \sp$.
\el
\proof
These relations are proven by making use of algebraic relations stated in Lemma \ref{algebra}.
By \kak{tu1} in Lemma \ref{algebra} we see that
\begin{align*}
&W_{+jk}^\ast W_{+kj'}-W^\ast_{-jk} W_{-kj'} \\
&=
\half \lk
\y e_\nu^j T_{\mu\nu} d_{\mu a} T_{ab}^\ast e_b^{j'}\yy+
 \yy e_\nu^j T_{\mu\nu} d_{\mu a} T_{ab}^\ast e_b^{j'}\y\rk\\
 &=\delta_{jj'}\one.
 \end{align*}
Then
\kak{w1} follows.
By \kak{tu2} of Lemma \ref{algebra} we see that
$$e_\mu ^r\y \ov{T_{\mu\nu}}d_{\nu a} T_{ab}^\ast \yy \tilde {e_b^s}
=
e_\mu ^r\yy \ov{T_{\mu\nu}}d_{\nu a} T_{ab}^\ast \y \tilde {e_b^s}.
$$
Then
\begin{align*}
&\ov{W_{+jk}^\ast} W_{-kj'}^\ast -
\ov {W^\ast_{+jk}} {W_{-kj'}}\\
&=
\half \lk
\y e_\nu^j \ov{T_{\mu\nu}}
 d_{\mu a} {\omega}
 T_{ab}^\ast  \yy \TT e_b^{j'}-
 \y e_\nu^j T_{\mu\nu}^\ast  d_{\mu a} T_{ab} \TT e_b^{j'}\yy\rk =0.
 \end{align*}
 Then
\kak{w1} follows.
By (7) of Lemma \ref{algebra} we have $$W_{+jk} W_{+kj'}^\ast -\ov {W^\ast_{-jk}} \ov{W_{-kj'}^\ast}=
\half \lk
\y e_\nu^j T_{\mu\nu}^\ast d_{\mu a} T_{ab} e_b^{j'}\yy+
 \yy e_\nu^j T_{\mu\nu}^\ast  d_{\mu a} T_{ab} e_b^{j'}\y\rk.
$$
By \kak{tu3} yields that
$$W_{+jk} W_{+kj'}^\ast -\ov {W^\ast_{-jk}} \ov{W_{-kj'}^\ast}=\delta_{jj'}\one+\theta(m) Z_{+jj'}.$$
Then
\kak{w1} follows.
By (7) of Lemma \ref{algebra} we have
\begin{align*}
&W_{-jk} W_{-kj'}^\ast
-\ov {W^\ast_{+jk}}
\ov{W_{-kj'}^\ast}\\
&=
\half \lk
e_\mu^ j\yy T_{\mu\nu}^\ast
d_{\nu a} T_{ab}\y \TT e_b^{j'}-
 e_\mu^ j\y T_{\mu\nu}^\ast d_{\nu a} T_{ab}\yy \TT e_b^{j'}\rk.
 \end{align*}
 Then from \kak{tu3} it follows that
$$W_{-jk} W_{-kj'}^\ast -\ov {W^\ast_{+jk}} \ov{W_{-kj'}^\ast}=\theta(m)Z_{-ij}.
$$
Then
\kak{w1} follows.
\qed

\bl{hilbert}
Suppose Assumption \ref{1v1} and $m>0$. Then $W\in\sp_2$, i.e., $W_-\in \HS$.
\el
\proof
It is enough to show that $W_{-ij}$ is a Hilbert-Schmidt operator  on $\LR$ for each $i,j$.
By the definition of $W_-$, we can see
that $W_{-ij}$ is the integral operator with the integral kernel:
$$ W_{-ij}(k,k')=\frac{\at}{2}
\frac{\vp(k)\vp(k')e_\mu^i(k)e_\mu^j(k')}
{\sqrt{|k|}
\sqrt{|k'|}
(|k|+|k'|)
D_-(|k'|^2)}.
$$
Since
$$|W_{-ij}(k,k')|\leq C(\at/2)(\vp(k)/|k|)(\vp(k')/|k'|)$$ with some constant $C$ by the assumption $\vp/\omega\in\LR$,
$W_{-ij}(\cdot,\cdot)\in L^2(\BR\times\BR)$.
Then $W_{-ij}$ is a Hilbert-Schmidt operator.
\qed
Let $m>0$.
By the general result obtained in Section \ref{bogoliubov}
we can see that  canonical commutation relations   hold:
\begin{align*}
&[B_p (f,j), B_p^\ast(g,j')]=\delta_{jj'}(\bar{f}, g), \\
&[B_p^\ast(f,j), B_p^\ast(g,j')]=0,\\
&[B_p(f,j), B_p(g,j')]=0,
\end{align*}
and the  adjoint relation
$(F, B_p(f,j)G)=(B^\ast _p(\bar f,j)F,G)$
is satisfied.
Furthermore
for $W=\WST$ we have
\begin{align}
a(f,j)&=b_W(W_{+ij}^\ast f,i)-b_W^\ast(\ov{W_{-ij}^\ast}f,i),\\
\add (f,j)&=-b_W(W_{-ij}^\ast f,i)+b_W^\ast(\ov{W_{+ij}^\ast}f,i)
\end{align}
for $f\in\LR$.
\bl{inttwin}
\TTT{Intertwining operator}
\index{intertwining operator}
Let  $m>0$.
Then
 the intertwining operator
$\UU_W $  associated with $W=\WST\in \sp_2$ is given by
\begin{align}
\UU_W
=C
\exp\lk
-\half \dd{
W_-W_+\f}
\rk
\wick{\exp\lk
-N_{\one-\ov{(W_+\f)^\ast}}
\rk}
\exp\lk
-\half \Delta_{-W_+\f \ov {W_-}}\rk
\end{align}
and the normalizing constant $C$  by $C={\rm det}(\one-(W_-W_+^-)^\ast (W_-W_+^-))^{1/4}$.
\el
\proof
This follows from Proposition \ref{bog2}.
\qed
\subsubsection{Displacement operator}
\index{displacement operator}We will construct the displacement operator associated with $
W=\mat {W_+} {\ov{W_-}} {W_-} {\ov{W_+}}$
and
the vector
\eq{nost}
L=-p_\mu \lk
\begin{array}{c}
 e^1_\mu \frac { \JI }{
\sqrt{\omega^3}}
\\
\vdots\\
\vdots\\
 e_\mu^{d-1}\frac { \JI }
{\sqrt{\omega^3}}
\end{array}
\rk\in \oplus^{d-1}\LR.
\en
\bl{displace}\TTT{Displacement operator}
\index{displacement operator}Suppose that $\IR$.
Then the displacement operator
associated with $W=\mat {W_+} {\ov{W_-}} {W_-} {\ov{W_+}}
$ and $L$ is given by
\eq{sppp}
S_p=\exp(-i\Pi_p)
\en
with
the generator
\eq{pipi}
\Pi_p=\frac {i}{\sqrt 2}
\frac{\ak}{\mass}\lkk
\add(\frac{ p \cdot e^j\vp}{ \omega^{3/2}},j)-
a(\frac{ p \cdot e^j\vp}{ \omega^{3/2}},j)\rkk.
\en
\el
\proof
The generator of the displacement operator is given by
$$\Pi_p=-\frac{i}{\sqrt 2}(
b_W(\frac{p\cdot e^j \JI }{\omega^{3/2}},j)-
b_W^\ast (\frac{p\cdot e^j \ov \JI }{\omega^{3/2}},j)).$$
We compute the right hand side above.
Then
$\Pi_p=\frac {i}{\sqrt 2}
(\add(\bar \xi_j,j)-a(\xi_j,j))$ with
$$\d \xi_i=W_{+ij}
\frac{p\cdot e_j\JI }{\omega^{3/2}}
-\ov{W_{-ij}}
\frac{p\cdot e_j\ov \JI }{\omega^{3/2}}.
$$
By (7),  (9) and (12) of Lemma \ref{algebra}, we can see that
$\ov{T_{\mu\nu}^\ast} d_{\nu a}\frac{\ov \JI }{\omega}=
{T_{\mu\nu}^\ast} d_{\nu a} \frac{\JI }{\omega}$ and
$\ov{T_{\mu\nu}^\ast} d_{\nu a}
\frac{\ov \JI }{\omega^2}=
{T_{\mu\nu}^\ast} d_{\nu a} \frac{\JI }{\omega^2}$,
and we have
$$\xi_j=e_\mu^j \y T_{\mu\nu}^\ast d_{\nu a} p_a\frac{\JI }{\omega^2}=
\frac{\ak}{\mass}p\cdot e^j \frac{\vp}{\omega^{3/2}}$$ by (11) of Lemma \ref{algebra}
under the condition $m>0$.
\qed
\begin{definition}
\TTT{Bogoliubov transformation}
\index{Bogoliubov transformation}
Let $W=\WST$ and $p\in\BR$.
Suppose
$\IR$.
 Then
 we define the unitary operator $\UU_p$ by
\eq{up}
\UU_p=S_p\UU_W .
\en
\end{definition}
\begin{remark}
In the case of $p=0$, we do not need
to assume that $\IR$ in the definition of $\UU_p$ in \kak{up}.
\end{remark}

\bl{mainp}
Suppose Assumption \ref{1v1}.
(1) Let $p=0$. Then
$\UU_0$ maps $D(\hf)$ onto itself and
\begin{align}
\label{u}
\UU_0^{-1} B_0^\sharp (f,j) \UU_0=\ass (f,j).
\end{align}
(2) In addition to  Assumption \ref{1v1},
suppose that $\IR$.
Then for  all  $p\in\BR$,
$\UU_p$ maps $D(\hf)$ onto itself and
\begin{align}
\label{uu}
&\UU_p^{-1} B_p^\sharp (f,j) \UU_p=\ass(f,j).
\end{align}
\el
\proof
This follows from the general results of Theorem \ref{inhom}.
\qed

\subsection{Diagonalization and time evolution of radiation fields}
\label{rikisi}
\subsubsection{Diagonalization}
In this section we diagonalize $\ds$ by a unitary operator.
\begin{lemma}
\label{ccc}
Suppose Assumption \ref{1v1}.
Let $f\in M_1\cap M_{1/2}\cap M_0\cap M_{-1/2}$.
Then
for all $p\in\BR$,
\begin{align}
&\label{mari1}
[\ds, B_p(f,j)]=- B_p(\omega f,j),\\
&
\label{mari2}
[\ds, B_p^\ast(f,j)]= B_p^\ast(\omega f,j).
\end{align}
\end{lemma}
\proof
By the algebraic relations in Lemma \ref{algebra} we will check commutation relations,
$[A_\mu, B^\sharp (f,j)]$ and
$[\hf, B^\sharp (f,j)]$.
By (9) of Lemma \ref{algebra} we see that
\begin{align*}
&[A_\mu, B_p(f,j)]\\
&=
-\frac{1}{\sqrt 2}
(d_{\mu a}\vp, T_{ab}^\ast \yy e_b^j f)=
-\frac{1}{\sqrt 2}
(e_b^j T_{ab}d_{\mu a}\vp, \yy  f)=
-\frac{1}{\sqrt 2}
\frac{m}{\ak}
(e_\mu^j \JI ,  \yy  f).
\end{align*}
By taking the adjoint we also obtain
$$[A_\mu, B_p^\ast(f,j)]=\frac{1}{\sqrt 2}\frac{m}{\ak}(e_\mu^j\ov \JI , \yy f).$$
Next we see that
by the definition of $B_p(f,j)$,
\begin{align*}
[\hf, B_p(f,j)]&
=
\frac{1}{\sqrt 2}
[\hf, \ta_\mu(T_{\mu\nu}^\ast \y e_\nu^j f)+i\tp_\mu(T_{\mu\nu}^\ast \yy e_\nu^j f)]\\
&=
\frac{1}{\sqrt 2}
\lkk
 -i\tp_ \mu(T_{\mu\nu}^\ast \y e_\nu^j f)- \ta _\mu(\omega^2T_{\mu\nu}^\ast \yy e_\nu^j f)\rkk.
 \end{align*}
 By (8) of Lemma \ref{algebra} we have
\begin{align*}
[\hf, B_p(f,j)]=
\frac{-1}{\sqrt 2}
\lkk
 i\tp_ \mu(T_{\mu\nu}^\ast \yy e_\nu^j \omega f)
 + \ta _\mu(T_{\mu\nu}^\ast \y  e_\nu^j \omega f)\rkk+\frac{\ak}{\sqrt 2}(e_\mu ^j \JI , \yy f)A_\mu.
 \end{align*}
 Together with them we can see that
\begin{align*}
&
[\ds, B_p(f,j)]\\
&=
\frac{1}{m}
(p_\mu-\ak  A_\mu)
(-\ak)[A_\mu, B_p(f,j)]
+[\hf, B_p(f,j)] \\
&=
-\frac{1}{\sqrt 2}
\lkk
 i\tp_ \mu
 (T_{\mu\nu}^\ast \yy e_\nu^j \omega f)
 +
 \ta _\mu
 (T_{\mu\nu}^\ast \y  e_\nu^j \omega f)
 \rkk
+ \frac{1}{\sqrt 2}
( \frac{p \cdot e ^j \JI }{\omega},
\frac{\omega   f}{\y})\\
&=-B_p(\omega f,j).
\end{align*}
Then \kak{mari1} follows.
\kak{mari2} is similarly proven.
\qed
\bl{timeevolution}
Suppose Assumption \ref{1v1}.
Let $f\in M_0\cap M_{-1/2}$.
Then
for all $p\in\BR$,
it follows that
\begin{align}
&e^{it\ds} B_p^\ast (f,j)e^{-it\ds}=B_p^\ast(e^{it\omega}f,j),\\
&
e^{it\ds} B_p(f,j)e^{-it\ds}=B_p(e^{-it\omega}f,j).
\end{align}
\el
\proof
Fix $j$ and $f$.
Let $A=B_p^\ast (\bar f,j)+B_p(f,j)$ and $\Pi=i(B_p^\ast (\bar f,j)-B_p(f,j))$. Then
$A$ and $\Pi$ are essentially self-adjoint. We denote the self-adjoint extensions by the same symbols.
Let $A_t=B_p^\ast (e^{it\omega}\bar f,j)+B_p(e^{-it\omega}f,j)$ and
$\bar A_t=e^{it\ds}A e^{-it\ds}$.
Then for $\Phi\in\ffff$
 we can see that
$\frac{d }{dt }A_t\Phi=[\ds, A]\Phi$ by Lemma \ref{ccc},
and $\frac{d }{dt }\bar A_t\Phi=[\ds, A]\Phi$.
Thus
the function
$F(t)=(\Phi, (A_t-\bar A_t)\Psi)$  $\Phi,\Psi\in\ffff$, satisfies that
$\frac{d}{dt}F(t)=0$, and hence $F(t)=F(0)=0$ for all $t$.
 Thus
$A_t=\bar A_t$ on $\ffff$.
By a limiting argument $A_t=\bar A_t$ follows.
Similarly we can see that
$e^{it\ds}\Pi e^{-it\ds}i(B_p^\ast (\bar e^{it\omega} f,j)-B_p(e^{it\omega} f,j))$.
Thus the lemma follows.
\qed

Set  $\Omega_p=\UU_p\Omega$.

\bl{ev}
Suppose Assumption \ref{1v1}.
Then (1) and (2) follow.
\bi
\item[(1)]
 $\Omega_0\in \s_{\rm p}(\dsz)$.
\item[(2)]
 Suppose that $\IR$.
 Then $\Omega_p\in \s_{\rm p}(\ds)$
 for all $p\in\BR$.
\ei
\el
\proof
We prove (2).
Statement (1) is similarly proven.
Since
$$B_p(f,j) \Phi =\UU_pa(f,j) \UU_p\f \Phi,$$
$B_p(f,j)\Phi=0$ for all $f\in \mc$ implies that $\Phi=a\Omega_p$ with some $a\in\CC$.
By Lemma \ref{timeevolution} we see that
$$B_p(f,j) e^{-it\ds}\Omega_p=
e^{-it\ds}B_p(e^{-it\omega} f,j)
\Omega_p=0$$ for all $f\in \mc$. Thus
$e^{-it\ds}\Omega_p=a_t(p)\Omega_p$ with some $a_t\in\CC$.
By  the unitary properties of $e^{it\ds}$
we see that $a_t(p)$ can be represented $a_t(p)=e^{it E_p}$ with some
$E_p\in\RR$.
Thus $\ds\Omega_p=E_p\Omega_p$ follows.
\qed

{\it Proof of Theorem \ref{mainp-ari}}

\proof
Let $\ms M={\rm L.H}\{\prod_{i=1}^n  B_p^\ast (f_i,j_i)\Omega_p|f_i\in\mc, 1\leq j_i\leq
 d-1, n\geq0\}$.
Since $\ass$ leaves invariant,
$\ms M$ is dense in $\fff$.
Let $\Phi\in\ms M$. Then we have
\begin{align*}
e^{it\ds}\prod_{i=1}^n B_p^\ast (f_i,j_i)\Omega_p
&=
\prod_{i=1}^n B_p^\ast (e^{it\omega} f_i,j_i) e^{itE_p} \Omega_p
=
\UU_p\prod_{i=1}^n a^\ast (e^{it\omega} f_i,j_i) e^{itE_p} \Omega\\
&=
\UU_pe^{it(\hf+E_p)} \prod_{i=1}^n a^\ast ( f_i,j_i)  \Omega
=
\UU_pe^{it(\hf+E_p)}\UU_p\f  \Phi.
\end{align*}
Hence the theorem follows on $\ms M$. By a limiting argument the theorem is proven.
\qed

We define the unitary operator on $\hhh\cong \int_\BR^\oplus \fff dx$ by
\eq{def}
\UU=
\int_\BR^\oplus \UU_p
 e^{i\frac{\pi}{2}\nf} dp.
\en

{\it Proof of Theorem \ref{mainp-nasi}:}

\proof Let $V\in L^\infty(\BR)$.
By Theorem  \ref{mainp-ari}, we have
\eq{sttt}
\UU\f H \UU=\eff+\hf+\UU \f V \UU-V+\ground
\en
on a core of the right-hand side above,
e.g., $\ccc\alg [\ffff\cap D(\hf)]$.
Since $H$ is self-adjoint on $D(-\Delta)\cap D(\hf)$,
a limiting argument tells us
that $\UU$ maps $D(-\Delta )\cap D(\hf)$ onto itself and \kak{sttt} is
valid on
$\dpp$.
We see that
$$
\UU \f e^{-ikx}\UU =
e^{-ikx}e^{-i\frac{\pi}{2}\nf}
\UU_W  ^{-1}
\exp\left(-i \frac{\ak}{\mass}k
\cdot \Pi\right)
\UU_W  e^{i\frac{\pi}{2}\nf},
$$
where
$\Pi_\mu=
\frac {i}{\sqrt 2}
\lkk
\add(\frac{  e_\mu^j\vp}{ \omega^{3/2}},j)-
a(\frac{ e_\mu^j\vp}{ \omega^{3/2}},j)\rkk$.
We have
\begin{align*}
&\UU_W  \f \Pi_\mu
\UU_W  \\
&=
\frac{i}{\sqrt{2}}
\UU_W  \f
\lkk
\add(e_\mu^j\frac{\vp}{\omega^{3/2}},j)-
a(e_\mu^j\frac{\vp}{\omega^{3/2}},j)
\rkk
\UU_W  \\
&
=\frac{i}{\sqrt{2}}
\UU_W \f\\
&
 \lkk
b^\ast_W\lk
 \ip_{ij}e_\mu^j
 \frac{\vp}{\omega^{3/2}}
 +
 \im_{ij}e_\mu^j
 \frac{\vp}{\omega^{3/2}}, i\rk
 -
b_W\lk
\jm_{ij}e_\mu^j \frac{\vp}{\omega^{3/2}}
+\jp_{ij} e_\mu^j \frac{\vp}{\omega^{3/2}}, i\rk
\rkk\UU_W \\
&
=\frac{i}{\sqrt{2}}
 \lkk
 \add\lk
 \ip_{ij}e_\mu^j
 \frac{\vp}{\omega^{3/2}}
 +
 \im_{ij}e_\mu^j
 \frac{\vp}{\omega^{3/2}}, i\rk
 -
a\lk
\jm_{ij}e_\mu^j \frac{\vp}{\omega^{3/2}}
+\jp_{ij} e_\mu^j \frac{\vp}{\omega^{3/2}}, i\rk
\rkk\\
&=
\frac{i}{\sqrt{2}}
 \lkk
\add \lk
\ov{\ooo{1/2}\eai\tab\dbm
\frac{\vp}{\omega^{2}}},i\rk
-
a\lk
{\ooo{1/2}\eai\tab\dbm
\frac{\vp}{\omega^{2}}},i
\rk
\rkk.
\end{align*}
By (10) and (11) of Lemma \ref{algebra}
we see that
$$\ooo{1/2}e_a^i T_{ab}d_{b\mu}\frac{\vp}{\omega^2}
=
\frac{\mass}{\ak}
\ooo{1/2}e_a^i T_{ab}T_{b\nu}^\ast
d_{\nu \mu}\frac{\JI }{\omega^2}
=
\frac{\mass}{\ak}
\ooo{1/2}e_a^i T_{ab}T_{b\nu}^\ast e_\nu^j e_\mu^j \frac{\JI }{\omega^2}
=
\frac{\mass}{\ak}
e_\mu^i
\frac{\JI }{\omega^{3/2}}.$$
Since
\eq{hihi}
\d e^{-i\frac{\pi}{2}\nf}i \lkk a^\ast(\bar g,j)-a(g,j)\rkk
e^{i\frac{\pi}{2}\nf}=a^\ast(\bar g,j)+a( g,j),
\en
we have
$$e^{-i\frac{\pi}{2}\nf}
\UU_W  \f \Pi_\mu
\UU_W
e^{i\frac{\pi}{2}\nf}=\frac{\mass}{\ak}
\sh\lkk
\add(e_\mu^j \frac{\ov \JI }{\omega^{3/2}},j)+
a(e_\mu^j \frac{\JI }{\omega^{3/2}},j)\rkk
=\frac{\mass}{\ak} K_\mu.
$$
Hence
$$
\UU\f   e^{-ikx} \UU =e^{-ikx}e^{-i  k\cdot K}=T\f e^{-ikx} T.$$
Let $\rh\in\ccc$ be such that $\rh(x)\geq 0$,
${\rm supp}\rh\subset\{x\in\BR\|x|\leq 1\}$ and
$\int_\BR\rh(x)dx =1$.
Define $\re(x)=\rh(x/\epsilon)/\epsilon^d$,
$\epsilon>0$,
and
$\vep=\re\ast V$.
We see that
$$\UU\f \rh\UU =(2\pi)^{-d/2}\int_\BR\check{\rh}(k)\UU\f e^{-ikx}\UU \ddk
=\int_\BR\check{\rh}(k)T\f e^{-ikx}T \ddk
=T\f \rh T.$$
Thus
$\UU\f \vep \UU\Psi=T\f \vep T\Psi$.
By a limiting argument, we obtain \kak{ln12} on $\dpp$.
\qed

\subsubsection{Time evolution of quantized radiation field}
Now we can construct  the solution
to formal equation:
$$(\frac{\partial^2}{\partial t^2}-\Delta)A_\mu(x,t)=\frac{\ak}{m}(p_\nu-A_\nu(x,t))\rho_{\nu\mu}(x).$$
 \bt{asymptotic field}
\TTT{Time evolution of $A$} {\rm \cite{ara83-a,ara83-b}}
\index{time evolution of radiation field}Suppose Assumption \ref{1v1} and that $f$ is real-valued and $f\in M_0\cap M_{-1/2}$.
Then
 for all $p\in\BR$,
\begin{align*}
&e^{it\ds}A_\mu(f) e^{-it\ds}\\
&=
\frac{1}{\sqrt 2}
\lkk B_p^\ast(e^{it\omega}
e_\nu^j\yy \ov{T_{\nu\mu}}\hat f,j)
+B_p(e^{-it\omega}
e_\nu^j \yy T_{\nu\mu}\tilde {\hat f},j)\rkk
+\frac{\ak}{\mass}p_\nu(d_{\mu\nu}\frac{\vp}{\omega},\frac{\hat f}{\omega}).
\end{align*}
\et
\proof
 By the symplectic structure, $W\in \sp$,
 $\ass$ can be represented\footnote{\cite[(4.7) and (4.8)]{hs01} is incorrect. 
 $-\frac{\alpha}{\mass}(\frac{p\cdot e_j\vp}{\sqrt 3\omega^{3/2}},\bar f)$ is changed to 
 $+\frac{\alpha}{\mass}(\frac{p\cdot e_j\vp}{\sqrt 3\omega^{3/2}},\bar f)$
and
 $-\frac{\alpha}{\mass}(\frac{p\cdot e_j\vp}{\sqrt 3\omega^{3/2}}, f)$ 
 to 
 $+\frac{\alpha}{\mass}(\frac{p\cdot e_j\vp}{\sqrt 3\omega^{3/2}}, f)$.
}
  in terms of $B_p^\sharp$:
 \begin{align}
\label{a1-1}
&
a(f,j)=-B_p^\ast(\ov{W_{-ij}^\ast}  {f}, i)+B_p(\jp_{ij} {f}, i)+
\frac{\ak}{\mass}
\left(\frac{p\cdot \ej \vp}{\sqrt{2}\omega }, \frac{f}{\y}\right),\\
\label{a2-2}
&
\add(f,j)=
B_p^\ast(\ip_{ij} f,i)-
B_p({W_{-ij}^\ast} f,i)+
\frac{\ak}{\mass} \left(
\frac{p\cdot \ej \vp}{\sqrt{2}\omega}, \frac{f}{\omega}\right).
\end{align}
Here we used \kak{korokoro1}  and  \kak{korokoro2} under $S=W_+$, $T=W_-$ and $L=-p\cdot e^jQ/\omega^{3/2}$ and that
$$-W_{+ij}(-p\cdot e^j\frac{\JI}{\omega^{3/2}})+
\ov{W_{-ij}}(-p\cdot e^j\frac{\bar {\JI}}{\omega^{3/2}})=
e_\mu^i
\y T_{\mu\nu}^\ast \yy e_\nu^j p\cdot e^j
\frac{\JI}{\omega^{3/2}}=
\frac{\ak}{\mass}p\cdot e^i \frac{\vp}{\omega^{3/2}}.$$
The quantized radiation field $A_\mu$ can be represented in terms of $B_p^\#$.
Inserting \kak{a1-1} and \kak{a2-2} into $A_\mu$ we can see that
\begin{align*}
&A_\mu(f)\\
&=\frac {1}{\sqrt 2}
\lkk
B^\ast _p(\ov{W_{+ij}^\ast}\tilde R_\mu^j-\ov{W_{-ij}^\ast} R_\mu^j,i)
+
B _p({W_{+ij}^\ast} R_\mu^j-{W_{-ij}^\ast \tilde R_\mu^j},i)\rkk
+\frac{\ak}{\mass}p_\nu(d_{\mu\nu}\frac{\vp}{\omega}, \frac{f}{\omega}),
\end{align*}
where $R_\mu^j=e_\mu^j\hat f/\y$. Explicitly
we can compute as
\begin{align*}
{W_{+ij}^\ast} R_\mu^j-{W_{-ij}^\ast \tilde R_\mu^j}=e_\nu ^i \yy T_{\nu\mu}\tilde {\hat f},\quad
\ov{W_{+ij}^\ast}\tilde R_\mu^j-\ov{W_{-ij}^\ast} R_\mu^j=e_\nu ^i \yy \ov{T_{\nu\mu}}
{\hat f}.
\end{align*}
Then
\begin{align*}
A_\mu =
\frac{1}{\sqrt 2}
\lkk B_p^\ast(
e_\nu^j\yy \ov{T_{\nu\mu}}\hat f,j)
+B_p(
e_\nu^j \yy T_{\nu\mu}\tilde {\hat f},j)\rkk
+\frac{\ak}{\mass}p_\nu(d_{\mu\nu}\frac{\vp}{\omega},\frac{\hat f}{\omega})
\end{align*}
follows and the  theorem is obtained
 from
Lemma \ref{timeevolution}.
\qed

From Theorem \ref{asymptotic field} we immediately see the corollary below.
 \bc{asymptotic field2}
\TTT{Time evolution of $A$}
\index{time evolution of radiation field}
Let $V=0$.
Suppose Assumption \ref{1v1} and that $f$ is real-valued and $f\in M_0\cap M_{-1/2}$.
Then
\begin{align*}
&e^{it H}A_\mu(f) e^{-it H}\\
&=
\frac{1}{\sqrt 2}
\lkk B_p^\ast(e^{it\omega}
e_\nu^j\yy \ov{T_{\nu\mu}}\hat f,j)
+B_p(e^{-it\omega}
e_\nu^j \yy T_{\nu\mu}\tilde {\hat f},j)\rkk
+\frac{\ak}{\mass}(-i\nabla_\mu)(d_{\mu\nu}\frac{\vp}{\omega},\frac{\hat f}{\omega}).
\end{align*}
\ec
\proof
Since $e^{it H}=\int^\oplus_\BR e^{it\ds}dp$,
$$e^{itH}A_\mu(f) e^{-itH}
=\int _\BR^\oplus
e^{it\ds} A_\mu(f) e^{-it\ds} dp.$$
Then the corollary follows from Theorem \ref{asymptotic field}.
\qed

\subsection{Dressed electron states}
\label{DESS}
Let us study the relationship between
the infrared regular/singular condition and the ground state of $\ds$.
 \begin{definition}
 \TTT{Dressed electron state}
\ko{dressed electron state}
The ground state of $\ds$ is called the dressed  electron state (DES).
\end{definition}
\bl{cu}
Suppose Assumption \ref{1v1}.
Let $\Phi$ be   an eigenvector of $\ds$. Then
 $B_p(f,j)\Phi=0$, $j=1,..,d-1$,  for all $f\in \mc$.
\el
\proof
Let $\ds\Phi=E\Phi$.
Then $$B_p(f,j)\Phi=
e^{it\ds}e^{-it\ds}B_p(f,j)e^{it\ds}e^{-itE}\Phi=
e^{it\ds}B_p(e^{it\omega }f,j)e^{-itE}\Phi.$$
Let $\ms M=LH\{\prod_{i=1}^n b^\ast_W(f_i,j_i)\UU_W \Omega |
f_i\in\mc,1\leq i\leq n, n\geq 0\}$,
where $\UU_W $ denotes the intertwining operator associated with $W=\WST$.
Since $\ass$ can be represented in terms of  $B_p^\sharp$ and
$$B_p(f,j)\UU_W \Omega=-(p\cdot e^j\frac{\JI}{\omega},\frac{f}{\y})\UU_W \Omega,$$
operator $\ass$ leaves $\ms M$ invariant, and  $\ms M$ is dense in $\fff$.
Let $\Psi=
\prod_{i=1}^n  B_p(f_i,j_i) \UU_W \Omega$.
We see that
\begin{align*}
B_p(e^{it\omega }f,j)\Psi&=
\sum_{i=1}^n
(e^{-it\omega}\bar f, f_i)B_p^\ast(f_1,j_1)\cdots
\widehat{B_p^\ast(f_i,j_i)}
\cdots B_p^\ast(f_n,j_n)\UU_W \Omega\\
&-(p\cdot e^j\frac{\JI}{\omega},e^{-it\omega} \frac{f}{\y})
\Psi\to 0
\end{align*}
as $t\to\infty$ by the Riemann-Lebesgue lemma.
By a limiting argument we see that
$e^{it\ds}B_p(e^{it\omega }f,j)e^{-itE}\Phi\to 0$ as $t\to\infty$. Hence we conclude that $B_p(f,j)\Phi=0$.
\qed

\bt{DES}
\TTT{Existence and absence of DES}
\index{dressed electron state}
{\rm \cite{ara83-a,ara83-b}
}Suppose Assumption \ref{1v1}.
\bi
\item[(1)]
Let $p=0$. Then $\dsz$ has a dressed electron state  and it is unique.
\item[(2)]
Suppose in addition that  $\IR$. Then $\ds$ has a dressed electron state for all $p\in\RR$, and it is unique.
\item[(3)]
Suppose  $\SR$ and $p\not=0$. Then  $\s_{\rm p}(\ds)=~\emptyset$. In particular $\ds$ has no ground state.
\ei
\et
\proof
In the case of (1) and (2) we have
the unitary equivalence $\UU_p\f \ds \UU_p=\hf+E_p$.
Since  the Fock vacuum $\Omega$ is the ground state of $\hf+E_p$,
$\UU_p\Omega$ is the dressed electron state of $\ds$.
Next we shall show (3).
Let $\Phi$ be any  bound state of $\ds$.
Then $B_p(f,j)\Phi=0$ for all $f\in \mc$ by Lemma \ref{cu}.
Then
$$0=(F, B_p(f,j)\Phi)=(F, b_W(f,j)\Phi)-(p\cdot e^j\frac{\JI}{\omega},\frac{f}{\y})(F, \Phi)$$
for $F\in\fff$ such that $F\in D(N)$ and $(F,\Phi)\not=0$. Hence
$$|(p\cdot e^j\frac{\JI}{\omega},\frac{f}{\y})|\leq
C\|f\|$$
holds   with some constant $C$, since $\|\ass(f)F\|\leq \|f\| \|(N+\one)^\han F\|$.
Hence the  functional $f\mapsto (p\cdot e^j\frac{\JI}{\omega},\frac{f}{\y})$ can be extended on  $M_0(=\LR)$.
Note that
$$\|p\cdot e^jQ/\omega^{3/2}\|^2=p^2\int \frac{|\vp(k)|^2}{\omega(k)^3}\frac{1}{D_+(\omega(k)^2)}dk=\infty.$$
The Riesz lemma  yields
that there exists $g\in\LR$ such that
$(g, f)=(p\cdot e^j\frac{\JI}{\omega},\frac{f}{\y})$. It is however contradiction, since
$p\cdot e^j\frac{\JI}{\omega^{3/2}}\not\in\LR$.
\qed
See Figure \ref{picdes}.
\begin{figure}[t]
\begin{center}
\arrayrulewidth=1pt
\def\arraystretch{2.0}
\begin{tabular}{|c|c|c|}
\hline
\       & $\IR$ & $\SR$ \\
\hline
\hline
$p=0$  & exist & exist \\
\hline
$p\not= 0$& exist & not exist\\
\hline
 \end{tabular}
\end{center}
\caption{DES of $\ds$}
\label{picdes}
\end{figure}%

\subsection{Ground state energy}
\ko{ground state energy}
\subsubsection{Holomorphic  property}
 The ground state energy $E_p$ can be represented by $W_\pm$ and $\vp$.
\bl{g-expectation}
Suppose Assumption \ref{1v1} and $\IR$.
Then\footnote{\cite[Lemma 5.12]{hir93} is incorrect. It should be changed to 
\kak{ichi1}.
}
\eq{ichi1}
E_p=\frac{1}{2m}(p+\gamma(p))^2+\ground ,
\en
where
\begin{align*}
&\gamma(p)_\mu
=-\frac{\at}{2\mass}
p_\nu\lk
\frac{ e_\nu^j \vp}{\sqrt{\omega^3}},
(\one+W_-W_+\f)_{ij}
\frac{e_\mu^i \vp}{\sqrt\omega}
\rk,\\
&\ground
=\frac{1}{4m}
\lk
\frac{e_\mu^j \vp}{\y},
(\one-W_-W_+\f)_{ij}
\frac{e_\mu^i \vp}{\y}
\rk
\end{align*}
\el
\proof
We notice that
$E_p=(\ds\Omega, \UU_p\Omega)/(\Omega, \UU_p\Omega)$,
$$\ds\Omega=\frac{1}{2m}
\sum_{\mu=1}^d \lk
(p_\mu+\add(f_{j,\mu},j)+a(f_{j,\mu},j))^2
\Omega, \UU_p\Omega
\rk,$$ and
$\UU_p=\exp(\add(\xi_j,j)-a(\xi_j,j)) \UU_W $,
where
$\d \xi_j
=\frac{1}{\sqrt 2}
\frac{\ak}{\mass}p\cdot e^j
\frac{\vp}{\sqrt{\omega^3}}$ and
$\d f_{j,\mu}=-\frac{\ak}{\sqrt 2}e_\mu^j\frac{\vp}{\y}$. Then the lemma follows from
Lemma \ref{g-exp}.
\qed
We will show that $E_p$ is
holomorphic function of $\ak$ on some neighborhood of the real line.
In what follows in this section we suppose (1),(2) and (3) of Assumption~ \ref{1v1}.
Under (1), (2') and (3) of  Assumption \ref{1v1} a similar procedure is also shown.
Set $\hat G=\ooo{\fdd}G\ooo{\fdd}$.
Let
\eq{horo}
\ho(s)=\lim_{\epsilon\downarrow 0}
\mmm \int \frac{\vp(k)^2}{s+i\epsilon-\omega(k)^2}dk.
\en
Then  $D_+(s)=m-\at \ho(s)$
and
\eq{dor1}
T_{\mu\nu}f=\delta_{\mu\nu }f+
\lk
\frac{1}{\frac{m}{\at}-\ho(\omega^2)}\rk
\vp \hat G
\vp  d_{\mu\nu} f.
\en
For $\zeta\in\CC$ we define
$\tmn(\zeta)$ by $\tmn$ in \kak{dor1} with $m/\at$ replaced  by $\zeta$, and $W_{\pm ij}(\zeta)$ by $W_{\pm ij}$ with $\tmn$ replaced by $\tmn(\zeta)$.
We see that
\bi
\item[(1)] ${\rm Im} \ho(s) \not=0$ for $s\not=0$,
\item[(2)] $\d \ho(0)=-\mmm\|\vp/\omega\|^2<0$,
\item[(3)] $\d \lim_{s\to \infty}\ho(s)=0$.
\ei

\begin{figure}[t]
\centering
\includegraphics[width=300pt]{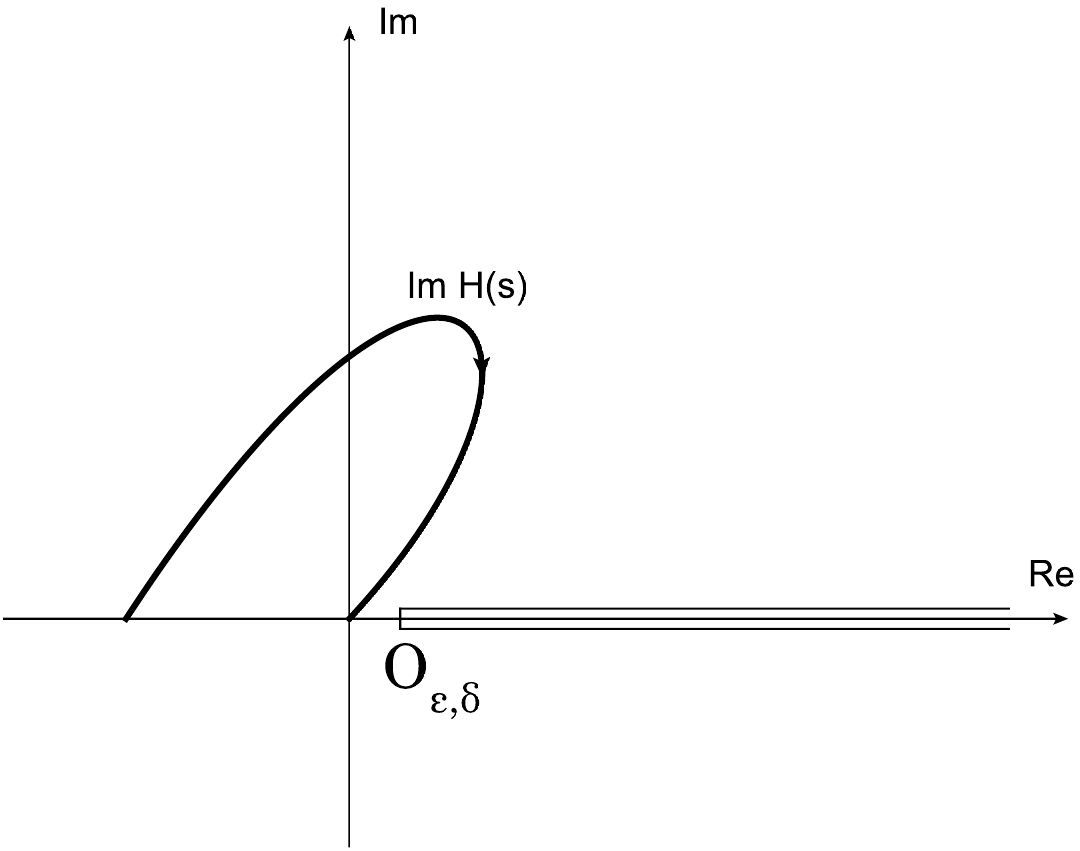}
\caption{${\rm Im} H(s)$ and
${\ms O_{\epsilon,\delta}}$}
\label{pichol}
\end{figure}
Let the image of $\ho(s)$, $s\geq 0$,  be denoted by $\ho$.
From (1)-(3) we can see that
for each given $\epsilon>0$,
there exists $\delta>0$ such that
 ${\ms O}_{\epsilon,\delta}=\{x+iy\in\CC|x>\epsilon, |y|<\delta\}$
 satisfies
 ${\rm dist}({\ms O}_{\epsilon,\delta}, \ho)>0$.
See Figure \ref{pichol}.
We have
$$W_{\pm ij}(\zeta)=
\delta_{ij}\one+e_\mu^i Y^\pm_{\mu\nu}(\zeta)e_\nu^j,$$
where
$Y^\pm_{\mu\nu}(\zeta)=
Y^\pm (\zeta)d_{\mu\nu}$ and
\begin{align*}
Y^+(\zeta)&=\frac{1}{\zeta-\ho(\omega^2)} \vp \lk
\yy \hat G  \y
+
\y \hat G  \yy\rk
\vp\TT,\\
Y^-(\zeta)&=\frac{1}{\zeta-\ho(\omega^2)} \vp \lk
\yy \hat G  \y
-
\y \hat G  \yy\rk\vp.
\end{align*}
Let $\zeta_0\in {\ms O}_\epsilon$. Then
we expand  $Y^\pm(\zeta)$ around $\zeta_0$ as
\begin{align*}
Y^+ (\zeta)&= \sum_{n=0}^\infty
\lkk
\frac{(-1)^n}{(\zeta_0-\ho(\omega^2))^{n+1}}
\vp
\lk
\yy \hat G \y+
\y \hat G \yy
\rk\vp
\TT\rkk
(\zeta-\zeta_0)^n
,\\
Y^- (\zeta)&= \sum_{n=0}^\infty
\lkk
\frac{(-1)^n}{(\zeta_0-\ho(\omega^2))^{n+1}}
\vp
\lk
\yy \hat G \y-
\y \hat G \yy
\rk\vp
\rkk
(\zeta-\zeta_0)^n.
\end{align*}
Thus $Y^\pm(\zeta)$ is analytic on $
{\ms O}_{\epsilon,\delta}$, and then
so is $W_\pm(\zeta)$.
Furthermore
$W_\pm(\zeta)\in\sp$ for $\zeta\in\RR$.
Thus we see that
\begin{align}
\label{z1}
&\jp(\zeta)\wwp(\zeta)-\jm(\zeta)\wwm(\zeta)=\one+\Delta_1(\zeta),\\
\label{z22}
&\ip (\zeta) \wwm(\zeta)-\im(\zeta)\wwp(\zeta)=\Delta_2(\zeta),
\end{align}
where
$\Delta_j(\zeta)$, $j=1,2$, are
  bounded self-adjoint operators with
\eq{374}
\|\Delta_j(\zeta)\|<1
\en
for
 $\zeta$ with sufficiently small imaginary part.
Thus for each $R>0$ we can
define
the open set
${\ms O}_{\epsilon,\delta',R}
=\{x+iy\in\CC|\epsilon<x<R,|y|<\delta'\}$ such that
${\ms O}_{\epsilon,\delta',R}\subset
{\ms O}_{\epsilon,\delta}$
and for all $\zeta\in
{\ms O}_{\epsilon,\delta',R}$,
\kak{z1}, \kak{z22} and \kak{374} hold.
 Then
 $W_+\f(\zeta)$ exists and is also holomorphic  on
${\ms O}_{\epsilon,\delta', R}$.
Define
$$E_p(\zeta)=\frac{1}{2m}(p+\gamma(p,\zeta))^2+\ground (\zeta),$$
where $\gamma(p,\zeta)$ and $g(\zeta)$ are defined with
$W_\pm$ replaced by $W_\pm(\zeta)$ for $\zeta\in {\ms O}_{\epsilon,\delta', R}$.

\bl{analytic}
Suppose Assumption \ref{1v1} and $\IR$.
Then for each $\epsilon>0$ and $R>0$, there exits $\delta'$ such that
$E_p(\zeta)$ is holomorphic on
${\ms O}_{\epsilon,\delta', R}$.
\el
\proof
Since
\begin{align*}
&\gamma(p,\zeta)_\mu
=-\frac{\at}{2\mass}
p_\nu\lk
\frac{ e_\nu^j \vp}{\sqrt{\omega^3}},
(\one+W_-(\zeta)W_+(\zeta)\f)_{ij}
\frac{e_\mu^i \vp}{\sqrt\omega}
\rk,\\
&g(\zeta)
=\frac{1}{4m}
\lk
\frac{e_\mu^j \vp}{\y},
(\one-W_-(\zeta)W_+(\zeta)\f)_{ij}
\frac{e_\mu^i \vp}{\y}
\rk,
\end{align*}
the theorem follows from the holomorphic properties of $W_-(\zeta)$ and $W_+(\zeta)\f$.
\qed

\bl{diagonalization1}
Suppose  Assumption \ref{1v1} and $\IR$.
Then there exists $m_\ast>0$ such that $\d E_p=\frac{1}{2m_\ast}p^2+\ground $.
\el
\proof
Notice that
\begin{align*}
\gamma (p)_\mu&=-\frac{\at}{2\mass}p_\nu(\frac{e_\nu^j\vp}{\sqrt{\omega^3}},
(\one+W_-W_+\f)_{ij}\frac{e_\mu^i \vp}{\sqrt\omega})\\
&=
-\frac{\at}{\mass}p_\mu \mmm \|\vp/\omega\|^2-
\frac{\at}{2\mass}
p_\nu
(\frac{e_\nu^j\vp}{\sqrt{\omega^3}},
e_a^i Y_{ab}^-e_b^k
(W_+\f)_{kj}\frac{e_\mu^i \vp}{\sqrt\omega}).
\end{align*}
Let $\at$ be sufficiently small.
We notice that $W_+=\one+{\rm Y}$, where ${\rm Y}_{ij}=e_\mu^i Y^+_{\mu\nu}e_\nu^j$.
The second term is computed as
\begin{align*}
(\frac{e_\nu^j\vp}{\sqrt{\omega^3}},
e_a^i Y_{ab}^-e_b^k
(W_+\f)_{kj}\frac{e_\mu^i \vp}{\sqrt\omega})
&=(\frac{e_\nu^j\vp}{\sqrt{\omega^3}},
e_a^i Y_{ab}^-e_b^k
((\one + {\rm Y})\f )_{kj}\frac{e_\mu^i \vp}{\sqrt\omega})\\
&=\sum_{n=0}^\infty
(-1)^n
( \frac{e_\nu^j\vp}{\sqrt{\omega^3}},
e_a^i Y_{ab}^-e_b^k
({\rm Y}^{n}) _{kj}\frac{e_\mu^i \vp}{\sqrt\omega}).
\end{align*}
Thus
we directly see that
\begin{align*}
&(\frac{e_\nu^j\vp}{\sqrt{\omega^3}},
e_a^i Y_{ab}^-e_b^k
(W_+\f)_{kj}\frac{e_\mu^i \vp}{\sqrt\omega})
\\
&=\sum_{n=0}^\infty
(-1)^n
(e_\nu^i
\frac{
\vp}{\sqrt{\omega^3}},
e_a^i Y_{ab}^-e_b^k
e_{\mu_1}^k Y^+_{\mu_1\nu_1} e_{\nu_1}^{k_1}
e_{\mu_2}^{k_1} Y^+_{\mu_2\nu_2} e_{\nu_2}^{k_2}
\cdots
e_{\mu_n}^{k_{n-1}} Y^+_{\mu_n\nu_n} e_{\nu_n}^{k_n}
\frac{e_\mu^i \vp}{\sqrt\omega})\\
&=\sum_{n=0}^\infty
(-1)^n
(
\frac{
\vp}{\sqrt{\omega^3}},
d_{a\nu}Y^- d_{ab} d_{b\mu_1}Y^+d_{\mu_1\nu_1}d_{\nu_1\mu_2} Y^+d_{\mu_2\nu_2}d_{\nu_2\mu_3}Y^+\cdots
Y^+ d_{\mu_n\nu_n}d_{\nu_n\mu}
\frac{ \vp}{\sqrt\omega})\\
&=\delta_{\mu\nu}
\sum_{n=0}^\infty
(-1)^n
\mmm^{n+2}
(
\frac{
\vp}{\sqrt{\omega^3}},
Y^- Y^{+n}
\frac{\vp}{\sqrt\omega}).\end{align*}
We set the right hand side by
$\delta_{\mu\nu}M$.
Then
we have
$$E_p=\frac{1}{2m}\lk
p+\gamma(p)\rk^2+\ground =
\frac{p^2}{2m}\lk
1-\frac{\at}{\mass}\mmm
\|\vp/\omega\|^2-
\frac{\at}{2\mass}M\rk^2+\ground .$$
Hence the corollary follows for sufficiently small $\at$. Since $E_p$ is holomorphic on ${\ms O}_{\epsilon,\delta,R}$ for arbitrary $\epsilon>0$ and $R>0$, the corollary follows for all $\ak \in\RR$.
\qed
In Lemma \ref{diagonalization1} we suppose $\IR$. This condition is, however, removed
in the next section.

\subsubsection{Explicit form of
effective mass and ground state energy}
\ko{effective mass}
\ko{ground state energy}
In the present section we show that
 $E_p$ is of the form $\frac{1}{2m^\ast}p^2 +\ground$.
The main theorem in this section is as follows.
\bt{explicit ground state}
\TTT{Explicit form of $E_p$}
Suppose Assumption \ref{1v1}.
Then
$
\d E_p=\frac{1}{2\mass}p^2+\ground $ for all $\ak\in\RR$, where $\ground$ is given by
$$\d \ground =\frac{d}{2\pi}\int_{-\infty}^\infty
\frac{\at \mmm  \left\|\frac{t\vp}{t^2+\omega^2}\right\|^2}
{m+\at  \mmm\left\|\frac{\vp}{\sqrt{t^2+\omega^2}}\right\|^2}dt.$$
\et
Note that we do not assume $\IR$ in Theorem \ref{explicit ground state}.
Throughout this section
we assume that $\at$ is sufficiently small unless otherwise stated.

Since a momentum lattice approximated
$\ds $ can be identified
with a harmonic oscillator in $L^2(\RR^D)$ for some $D$,
$E_p$ can be  obtained through calculating
the ground state energy of the harmonic oscillator.

First
$\omega$ is
replaced by
$\on(k)=\omega(k)+\epsilon$ for $ \epsilon>0$.
For $l=(l^1,\cdots, l^d)\in \BR$, let
$|l|=\max_j|l^j|$.
For the time being we suppose
$l\in(2\pi {\mathbb Z}/a)^d$,
$ |l|\leq 2\pi  L$
with some $a$ and $L$; $l$ is a lattice point with the width
$2\pi/a$ of
the $d$-dimensional rectangle centered at
the origin with the width $4\pi L$.
The lattice points are named
$l_1,l_2,\cdots,l_{\LN }$, where $\LN=(2[aL]+1)^d$ denotes the number of  lattice points and
$[z]$ denotes the integer part of $z\in\RR$.
For $l\in(2\pi {\mathbb Z}/a)^d$,
 we define the rectangle:
$\Gamma(l)=\left[l^1,l^1+\frac{2\pi}{a}\rk\times\cdots \times \left[l^d,l^d+\frac{2\pi}{a}\rk$.
Let
\begin{align}
&\qqq_{jl}=\sh
\frac{1}{\sqrt{\on (l)}}
\lk\frac {a}{2\pi}\rk^{d/2}
\left\{\add(\chi_{\Gamma(l)},j)+
a(\chi_{\Gamma(l)},j)\right\},\\
&
\ppp_{jl}=
\frac{i}{\sqrt{2}}
\sqrt{\on (l)}
\lk\frac {a}{2\pi}\rk^{d/2}
\left\{\add(\chi_{\Gamma(l)},j)
-a(\chi_{\Gamma(l)},j)\right\}.
\end{align}
Then the Weyl relations hold,
\begin{equation}
\label{vn}
\exp\lk {it\ppp_{jl}}\rk
\exp\lk {is\qqq_{j'l'}}\rk
=
\exp\lk {its\delta_{l_1l_1'}...\delta_{l_dl'_d}
\delta_{jj'}}\rk
\exp\lk {is\qqq_{j'l'}}\rk
\exp\lk {it\ppp_{jl}}\rk, \ \ \ t,s\in\RR.
\end{equation}
Let $D=(d-1)\LN $.
We define the $\ld$-diagonal  matrix by\footnote{$A_0$ given in \cite[p.1176 in Appendix]{hs01} is incorrect. Matrix elements $\omega_\epsilon(l_j)$ are  changed to
$\omega_\epsilon(l_j)^2$.}
$$A_0=
\lk
\begin{array}{cccc}
\on(l_1)^2\II&\ &\ & \ \\
\ & \on(l_2)^2\II&\ &\ \\
\ &\ &\ddots & \ \\
\ &\ &\  & \on(l_{\LN })^2\II
\end{array}\rk,
$$
where
$\II$ denotes the $(d-1)\times(d-1)$-identity matrix.
Since $\epsilon>0$, $A_0$ is a strictly positive matrix.
We denote by $(f,g)_D$ the $D$-dimensional scalar product.
Let
$v_{jl}^\mu=\vp(l)\ej_\mu(l)$,
and
$$\vm=
\lk
\begin{array}{c}
v_{1l_1}^\mu\\
\vdots\\
v_{d-1l_1}^\mu\\
\vdots\\
v_{1l_{\LN }}^\mu\\
\vdots\\
v_{d-1l_{\LN }}^\mu
\end{array}
\rk
\in \RR^D,\ \ \ \mu=1,...,d.$$
For linear operator $T$, let
$\eT= ( \vm ,T \vm )_D$.
Suppose that $T:\BR\rightarrow\RR$ is a  rotation invariant function.
Let $T_{\rm diag}$ be the $\ld$-diagonal matrix with diagonal elements
$T(l)$:
$$T_{\rm diag}=
\lk
\begin{array}{cccc}
T(l_1)\II&\ &\ & \ \\
\ & T(l_2)\II&\ &\ \\
\ &\ &\ddots & \ \\
\ &\ &\  & T(l_{\LN })\II
\end{array}\rk.
$$
Then
$
(\vm,T_{\rm diag}\vn)_D=\delta_{\mu\nu}
\mmm
\sss T(l)|\ph(l)|^2$.
 Let
$
\ppp=(\ppp_{jl})_{1\leq j\leq d-1,|l|\leq \aL}$ and
$\qqq=(\qqq_{jl})_{1\leq j\leq d-1,|l|\leq \aL}$.
Then the momentum lattice approximated $\ds$ is written as
$$H_{L,a}^\epsilon(p)=
\ham
\lk
p-\ak (\vec v, \qqq)_D\rk^2+
\half \lk
(\ppp, \ppp)_D+(\qqq, A_0\qqq)_D\rk-{\rm tr}\sqrt{A_0},$$
where $p\in\BR$ and $(\vec v, \qqq)_D=
((\vec v_1, \qqq)_D,...,(\vec v_d, \qqq)_D)$.
\bl{91}
Suppose that  $\epsilon>0$.
Let $\vpp$ and $\vq$ be the momentum operator and its canonical position operator in
$L^2(\RR^D)$, respectively.
Then there exist\footnote{Possibly $M=\infty$.}
 $M\leq \infty$, a  $\ld$  nonnegative symmetric matrix $A$
and $\vf \in \RR^D$ such that
\eq{har}
H_{L,a}^\epsilon(p)\cong
\bigoplus^M\lk
\half (\vpp,\vpp)_D+\half (\vq, A\vq)_D+\frac{1}{2m}p^2-\half(\vf,A\vf)_D
-\half  {\rm tr}\sqrt{A_0}\rk.
\en
\el
\proof
Define the  $\ld $-matrix by
$\V=\nnn |\vm \rangle\langle \vm |$.
Set
$A=A_0+\M \V$.
Note that $A$ is a strictly positive symmetric matrix,
 since $A_0$ is strictly positive and $\V$ is nonnegative.
In particular,  $(A+a)^{-1}$ exists for $a\geq 0$.
Let
$\vf=\vf(p)= \frac{\q}{m}A^{-1}p_\mu \vm \in \RR^D$.
Then we have
$$H_{L,a}^\epsilon(p)=
\half (\ppp,\ppp)_D+\half ((\qqq-\vf), A(\qqq-\vf))_D+
\frac{1}{2m}p^2-\half(\vf,A\vf)_D
-\half  {\rm tr}\sqrt{A_0}.$$
By (\ref{vn}) and the von Neumann uniqueness theorem\ko{von Neumann uniqueness theorem},
there exists $M\leq \infty$ and a unitary operator
$U :\fff\rightarrow \bigoplus^M L^2(\RR^D)$ implementing
that
\begin{align*}
&U  \ppp_{jl} U \f= \bigoplus^M (-i\nabla_{ x_{jl}}),\\
&U  \qqq_{jl} U \f =\bigoplus^M x_{jl}.
\end{align*}
Then
$H_{L,a}^\epsilon(p)$ is unitarily equivalent with the direct sum of the harmonic oscillator\ko{harmonic oscillator}:
$$
\bigoplus^M \lk
\half (\vpp,\vpp)_D+\half ((\vq-\vf), A(\vq-\vf))_D+\frac{1}{2m}p^2-\half(\vf,A\vf)_D
-\half  {\rm tr}\sqrt{A_0}\rk$$
in $\bigoplus^M
L^2(\RR^D)$.
By the shift $\vq\rightarrow \vq+\vf$ implemented
by a  unitary operator,
we obtain
\kak{har}.
\qed
\bl{92}
Suppose $\epsilon>0$.
Then
\eq{har2}
\is(H_{L,a}^\epsilon(p))=\frac{1}{2m}p^2-\half (\vf,A\vf)_D+
\half {\rm tr}(\sqrt{A}-\sqrt{A_0}).
\en
\el
\proof
Generally for the harmonic oscillator
$H_T=\half(\vpp,\vpp)_D+\half(\vq,T\vq)_D$
with a symmetric nonnegative matrix $T$,
$\is (H_T)=\half\tr{\sqrt{T}}$.
Hence
$$
\is\left(\half(\vpp,\vpp)_D+\half (\vq, A \vq)_D\right)
=
\half
\tr\sqrt{A}.$$
Thus
the ground state energy of $H_{L,a}^\epsilon(p)$ is given by \kak{har2} by Lemma \ref{91}.
\qed

We  calculate $(\vf,A\vf)_D$ and ${\rm tr}(\sqrt{A}-\sqrt{A_0})$ as follows.
We note that
\begin{align}
&\label{a11}
(\vm, A_0\f \vn)=\delta_{\mu\nu}\qqqqq\sss \frac{|\ph(l)|^2}{\on (l)^2},\\
&\label{a22}
(\vm, (s^2+A_0)\f \vn)=\delta_{\mu\nu}\qqqqq \sss \frac{|\ph(l)|^2}{s^2+\on (l)^2},\\
&
\label{a33}
(\vm, (s^2+A_0)\f A_0\vn)=\delta_{\mu\nu}\qqqqq \sss \frac{\on (l)^2|\ph(l)|^2}
{s^2+\on (l)^2}.
\end{align}
Furthermore
$
\d A\f={\rm s}\!-\!\lim_{N\rightarrow \infty} \sum_{n=1}^N (-\M A_0\f P)^{n-1}A_0\f$.
\bl{93}
Suppose $\epsilon>0$.
 Then
\eq{har3}
\frac{1}{2m}p^2-\half(\vf,A\vf)_D=\frac{p^2}{2m}\frac{1}{1+\M \theta},
\en
where
$\d \theta=
\qqqqq\sss\frac{|\vp(l)|^2}{\on (l)^2}$.
\el
\proof
By \kak{a11} we have
\begin{align*}
&(\vf,A\vf)_D\\
&=\Mm p_\mu p_\nu(\vm, A\f \vn)_D\\
&= \Mm
p_\mu p_\nu\sum_{n=1}^\infty
(-\M)^{n-1}
( \vm, (A_0^{-1}\V)^{n-1}A_0^{-1} \vn)_D\\
&
=
\Mm
\sum_{n=1}^\infty
p_\mu p_\nu
(-\M)^{n-1}
(\vm, A_0\f\vv 1)_D
(\vv 1, A_0\f\vv 2)_D
\cdots
(\vv{{n-1}}, A_0\f\vn)_D\\
&
=
\Mm
\sum_{n=1}^\infty
p_\mu p_\nu
\delta_{\mu\mu_1}\delta_{\mu_1\mu_2}\cdots\delta_{\mu_{n-1}\nu}
(-\M)^{n-1}\theta^n\\
&
=
\Mm
 \sum_{n=1}^\infty \left(-\M\right)^{n-1}p^2
\theta^n\\
&=
\frac{\M \theta}{1 +\M \theta}\frac{p^2}{m}.
\end{align*}
Hence \kak{har3} follows.
\qed
\bl{94}
Suppose $\epsilon>0$.
Then
\eq{har4}
\half\tr\lk \sqrt{A}-\sqrt{A_0}\rk
=\frac{d-1 }{2\pi}\int_{-\infty}^\infty
\frac{\M  s^2}{1+\M\xi}
\sss
\frac{|\vp(l)|^2}{(s^2+\on (l)^2)^2}ds,
\en
where
$\d \xi=\qqqqq \sss \frac{|\vp(l)|^2}{s^2+\on (l)^2}$.
\el
\proof
We see that
$$\tr\sqrt{A}-\tr\sqrt{A_0}=\frac{1}{\pi}\int_{-\infty}^\infty\tr\left\{A(s^2+A)^{-1}-A_0\ka\right\}ds.$$
Let $\ai=
\sum_{n=1}^\infty\left\{
-\M \V\ka\right\}^n.$
We have
$$A(s^2+A)^{-1}-A_0\ka =\M \V\ka+A\ka \ai.$$
It follows that
$$
\tr \M \V\ka =
\M \nnn \CONS (\phi,\vm)_D(\vm,\ka\phi)_D,$$
where
$\CONS$ means to sum up all the  vectors $\phi_n$ in a complete
orthonormal system (CONS).
Take a CONS such that
$\d \lkk\phi_1=\frac{\vm}{\|\vm\|},\phi_2,\phi_3,\cdots\rkk$.
Then we have by \kak{a22}
\eq{11}
\tr \M \V\ka=\M\e{\ka}=d\M \xi.
\en
We see that
$$ A\ka\ai=
 A_0 \ka\ai
+
\M \V\ka
\ai .$$
It follows that
\begin{align*}
&\tr A_0 \ka\ai\\
&=
\sum_{n=1}^\infty
(-\M)^n\CONS \lk\ka A_0\phi,(P\ka)^n\phi\rk_D\\
&
=
\sum_{n=1}^\infty
(-\M)^n\CONS
(\ka A_0\phi,\vv 1)_D\times \\
&\hspace{4cm}\times  (\ka \vv 1,\vv 2)_D\cdots
(\ka  \vv n,\phi)_D.
\end{align*}
Take a CONS such that
$\d \lkk \phi_1=
\frac{\ka  A_0 \vv n}{\|\ka A_0 \vv n\|},\phi_2,\phi_3,\cdots,\rkk$.
Since $|\ak|$ is sufficiently small,
from  \kak{a33} it follows that
\begin{align}
\tr A_0 \ka\ai
&=
\sum_{n=1}^\infty
\left(-\M \right)^n
(\ka \vv n,\ka A_0\vv 1)_D\times\non \\
&\quad \quad \times(\ka \vv 1,\vv 2)_D
\cdots (\ka \vv{{n-1}},\vv n)_D\non \\
&
=
\sum_{n=1}^\infty
\left(-\M \right)^n
\delta_{\mu_1\mu_2}\cdots\delta_{\mu_{n-1}\mu_n}
\xi^{n-1}\e{(s^2+A_0)^{-2}A_0}\non \\
&
=-\M \frac{\e{(s^2+A_0)^{-2}A_0}}{1+\M\xi}\non\\
\label{22}
&=\frac{-d\M}{1+\M\xi}\sss
\frac{\on (l)^2|\ph(l)|^2}{(s^2+\on (l)^2)^2},
\end{align}
and
\begin{align*}
&
\tr\bm \V\ka\ai\\
&=\sum_{n=1}^\infty(-\M)^n\M\CONS\lk \phi,P\ka\lk P\ka\rk^n\phi\rk_D\\
&
=
\sum_{n=1}^\infty
(-\M)^n\M \CONS(\phi,\vv 1)_D(\vv 1,\ka\vv 2)_D
\cdots (\vv{{n+1}},\ka\phi)_D.
\end{align*}
Take a CONS such that
$\lkk\phi_1=\frac{\vv 1}{\|\vv 1\|},\phi_2,\phi_3,\cdots,\rkk.$
Then
we see that
\begin{align}
\tr\bm \V\ka\ai
=
-
\sum_{n=1}^\infty(-\M)^{n+1}
\delta_{\mu_1\mu_2}\cdots\delta_{\mu_n\mu_{n+1}}\delta_{\mu_{n+1}\mu_1}
\xi^{n+1}
=
d \frac{-\M^2\xi^2}{1+\bm\xi}.
\end{align}
Hence we have
\begin{align*}
&\tr\left\{A(s^2+A)^{-1}-A_0\ka\right\}\\
&
=
\frac{\e{(s^2+A_0)^{-2}A_0}}{1+\bm\xi}
\left(-\bm\right)-d\frac{\left(\bm \xi\right)^2}{1+\bm \xi}+d\bm\xi\\
&=\bm\frac{d\mmm}{1+\bm\xi}
\sss
\left\{\frac{|\vp(l)|^2}{s^2+\on (l)^2}-
\frac{\on (l)^2|\vp(l)|^2}{(s^2+\on (l)^2)^2}\right\}\\
&
=\frac{(d-1) \M  s^2}{1+\M\xi}
\sss \frac{|\vp(l)|^2}{(s^2+\on (l)^2)^2}.
\end{align*}
Thus the lemma follows.
\qed
\bl{95}
Suppose the same assumptions as in Lemma \ref{91}.
Then
$$\is(H_{L,a}^\epsilon(p))=\frac{p^2}{2(m+ \at   \theta)}+
\frac{d-1}{2\pi}
\int_{-\infty}^\infty  \frac{\at s^2}{m+\at\xi}\sss
\frac{|\vp(l)|^2}{(s^2+\on (l)^2)^2}ds.$$
\el
\proof
It follows from Lemmas \ref{93} and \ref{94}.
\qed

{\it The proof of Theorem \ref{explicit ground state}}

\proof
Suppose that $\IR$ and $\at$ is sufficiently small.
We set
\begin{align*}
&
\mass(a,L,\epsilon)=m+\at \theta,\\
&
g(a,L,\epsilon)=
\frac{d-1}{2\pi}
\int_{-\infty}^\infty  \frac{\at s^2}{m+\at\xi}\sss
\frac{|\vp(l)|^2}{(s^2+\on (l)^2)^2}ds.
\end{align*}
Note that
$\mass(a,L,\epsilon)\rightarrow \mass$
and $g(a,L,\epsilon)\rightarrow g$
as $a\rightarrow\infty, L\rightarrow\infty,\epsilon\rightarrow 0$.
Taking $a\rightarrow\infty$ and then $L\rightarrow\infty$, we see that
$H_{L,a}^\epsilon(p)\rightarrow \ds+\epsilon \nf$
uniformly in the resolvent
sense, which yields that
$\is(H_{L,a}^\epsilon(p))\rightarrow\is(\ds+\epsilon \nf)$.
Hence
\begin{align*}
\is(\ds+\epsilon\nf)=
\lim_{L\rightarrow\infty}\lim_{a\rightarrow \infty}\lk
\frac{p^2}{2\mass(a,L,\epsilon)}+\ground (a,L,\epsilon)\rk
=
\frac{p^2}{2\mass(\epsilon)}+\ground (\epsilon),
\end{align*}
where $\mass(\epsilon)$ and $g(\epsilon)$ are  defined by $\mass$ and $g$ with
$\omega$ replaced by $\on$.
Since
$
\ds+\epsilon \nf\rightarrow \ds
$
strongly on $D(\hf)$ as $\epsilon\rightarrow 0$,
$
\ds+\epsilon \nf\rightarrow \ds
$
 holds in the strong resolvent sense. Then
it follows
\eq{444}
\limsup_{\epsilon\rightarrow 0}\is(\ds+\epsilon \nf)\leq \is(\ds).
\en
Furthermore, since $\nf\geq 0$,
we have
\eq{44}
\liminf_{\epsilon\rightarrow 0}\is(\ds+\epsilon \nf)\geq \is(\ds).
\en
Combining \kak{444} and \kak{44} we have
$\is(\ds +\epsilon \nf)\rightarrow \is (\ds )=E_p$
as $\epsilon\rightarrow 0$.
Then
$$E_p=\lim_{\epsilon\rightarrow 0}
\lk \frac{p^2}{2\mass(\epsilon)}+\ground (\epsilon)\rk =\frac{p^2}{2\mass}+\ground .$$
Hence the theorem  follows
 for sufficiently small $\at$.
 The theorem  is valid however for all $\ak$ since $E_p$ is holomorphic on ${\ms O}_{\epsilon,\delta',R}$ for
 arbitrary $\epsilon>0$ and $R>0$ by Lemma \ref{diagonalization1}.
Next we do not assume $\IR$.
Let $\vp_n$ be a sequence such that
$\vp_n\omega^l\to \vp/\omega^l$ for $l=0,-1/2,-1$ and $\vp_n/\omega^{3/2}\in\LR$.
Then the ground state energy of $\ds$ with cutoff function $\vp_n$, $\ds(n)$,  is given by
\eq{kob}
\frac{1}{2(m+\at \mmm\|\vp_n/\omega\|^2)}p^2+
\frac{d}{2\pi}\int_{-\infty}^\infty
\frac{\at \mmm  \left\|\frac{t\vp_n}{t^2+\omega^2}\right\|^2}
{m+\at  \mmm\left\|\frac{\vp_n}
{\sqrt{t^2+\omega^2}}\right\|^2}dt.
\en
We can see that $\ds(n)\to\ds$  as $n\to\infty$ in the uniform resolvent sense. Then
$\is (\ds(n))\to \is(\ds)$ and \kak{kob}
converges to $E_p$ as $n\to\infty$. Hence the theorem follows.

\subsubsection{Ultraviolet cutoffs}
\index{UV cutoff}
In this subsection
we assume that $d=3$, $\ak=1$, the ultraviolet cutoff is given by
the sharp cutoff\ko{UV cutoff!sharp} \eq{sharp2}
\vp(k)=(2\pi)^{-3/2}\one_{[\lambda,\Lambda]}(k).
\en
See Example \ref{goodexample}.
Set $\ground=\ground(\Lambda)$ for the emphasis of the dependence of the ultraviolet cutoff parameter $\Lambda$.
We estimate the asymptotic behavior of $\ground(\Lambda)$
as $\Lambda\rightarrow \infty$.
In the case of $V=0$, from Theorem \ref{mainp-nasi} it follows that
$\ground(\Lambda)=\is(H)$.
It is seen that $\ground(\Lambda)$ is monotonously increasing in $\Lambda$.
Since
$$\|\ph/\sqrt{t^2+\omega^2}\|^2
=
4\pi\lkk(\Lambda-\lambda)+t
\lk \arctan\frac{\lambda}{t}-\arctan\frac{\Lambda}{t}\rk \rkk,$$
and
$$\|\ph/(t^2+\omega^2)\|^2=
4\pi\lkk\frac{1}{2t}
\lk
\arctan\frac{\Lambda}{t}-\arctan\frac{\lambda}{t}
\rk +\half
\lk
\frac{\lambda}{t^2+\lambda^2}-\frac{\Lambda}{t^2+\Lambda^2}\rk
\rkk,$$
changing  variable $t$ to  $r=\Lambda/t$,
we have the explicit form of ground state energy of $H$ with
ultraviolet cutoff \kak{sharp2} and $V=0$.
\bp{expgr}
\TTT{Ground state energy}
\ko{ground state energy!sharp cutoff}
Suppose that $V=0$,   $d=3$, \kak{sharp2}, and $\ak=1$.
Then
\eq{gro}
\ground(\Lambda)=4\Lambda^2
\int_0 ^\infty
\frac{
(\arctan r-\frac{r}{1+r^2})
-
\lk \arctan r\LM-\frac{r\LM}{1+r^2\LM^2}\rk
}{mr+\frac{8\pi}{3}\Lambda
\lkk (r-\arctan r)-(r\LM-\arctan r\LM)\rkk} \frac{dr}{r^2}.
\en
\ep
\bt{ezawa22}
\TTT{Asymptotic behavior of \ko{ground state energy!asymptotics}
$\ground(\Lambda)$}
Assume that
$m>8\pi\lambda /3$.
Then
$$
\frac{8}{3}\lk\frac{3}{8\pi}\frac{1}{m}\rk^\han\frac{\pi}{2}
\leq
\lim_{\Lambda\rightarrow\infty}
\frac{\ground(\Lambda)}{\Lambda^{3/2}}
\leq \frac{8}{3}\lk\frac{9}{8\pi}\frac{1}{m}\rk^\han\frac{\pi}{2}.
$$
\et
\proof
We decompose $\frac{\ground(\Lambda)}{4\Lambda}$ as $ \frac{\ground(\Lambda)}{4\Lambda} =
\int_0^{1/\sll}+\int_{1/\sll}^\infty=I_1(\Lambda)+I_2(\Lambda)$.
It is enough to show that
\eq{mu1}
\frac{2}{3}\lk\frac{3}{8\pi}
\frac{1}{m}\rk^\han\frac{\pi}{2}
\leq
\lim_{\Lambda\rightarrow\infty}\frac{ I_1(\Lambda)}{\sl}
\leq
\frac{2}{3}\lk\frac{9}{8\pi}\frac{1}{m}\rk^\han\frac{\pi}{2},
\en
and that
$
 \lim_{\Lambda\rightarrow\infty}\frac{I_2(\Lambda)}{\sl}=0$.
Note that
$\arctan x=\frac{x}{1+x^2}+\frac{2}{3}\frac{x^3}{(1+x^2)^2}+\frac{5}{4}\frac{3}{2}\frac{x^5}{(1+x^2)^3}+\cdots$. We define functions $f$ and $h$ by
\begin{align*}
&\arctan r-\frac{r}{1+r^2}=\frac{2}{3}\frac{r^3}{(1+r^2)^2}+f(r),\\
&
r-\arctan r=\frac{r^3}{1+r^2}- h(r).
\end{align*}
It is satisfied that $f(r)\geq 0$,  $h(r)\geq 0$,
$\lim_{r\rightarrow 0} \frac{f(r)}{r^3}=0$,
$
\lim_{r\rightarrow 0} \frac{h(r)}{r^3}=\frac{2}{3}
$
and
$$h(r)=
\frac{2}{3}\frac{r^3}{(1+r^2)^2}+f(r)=\arctan r-\frac{r}{1+r^2}.$$
Let us set
\begin{align*}
&\fla(r)= \arctan r\LM-\frac{r\LM}{1+r^2\LM^2}>0,\\
&
\gla(r)= r\LM-\arctan r\LM >0.
\end{align*}
Then $I_1(\Lambda)$ is written as
$$I_1(\Lambda)
=\int_0^{1/\sll}\frac{\frac{2}{3}+\frac{(1+r^2)^2}{r^3}(f(r)-\fla(r))}{
\frac{m}{\Lambda}(1+r^2)+\frac{8\pi}{3}r^2-
\frac{8\pi}{3}
\frac{1+r^2}{r}
(h(r)+\gla(r))}\frac{dr}{1+r^2}.$$
It follows that for $0\leq r\leq 1/\sll$,
\eq{rh}
\frac{8\pi}{3}
\frac{1+r^2}{r}
(h(r)+\gla(r))=
\frac{8\pi}{3}
r^2 (1+r^2)
\frac{h(r)+\gla(r)}{r^3}
\leq
\frac{8\pi}{3}
\lk 1+\frac{1}{\sl}\rk
r^2
\theta(\Lambda),\en
where
$
\theta(\Lambda) =\sup_{0\leq r \leq 1/\sll}  \frac{h(r)+\gla(r)}{r^3}.
$
We set the right hand side of \kak{rh}
by $r^2 \delta(\Lambda)$.
Note that
\begin{align}
&
\lim_{r\downarrow 0} \frac{h(r)+\gla(r)}{r^3} =\frac{2}{3}+\frac{1}{3}\LM^3,\quad
\liml \delta(\Lambda)= \frac{8\pi}{3} \frac{2}{3}.
\end{align}
Moreover we have
$$
-\sup_{0\leq r \leq (1/\sll)} \frac{(1+r^2)^2}{r^3}\fla(r)\leq
\frac{(1+r^2)^2}{r^3}(f(r)-\fla(r))\leq
\sup_{0\leq r \leq (1/\sll)} \frac{(1+r^2)^2}{r^3}
f(r).$$
Set
$\epsilon(\Lambda)=\max\lkk\sup_{0\leq r \leq (1/\sll)} \frac{(1+r^2)^2}{r^3}
f(r),
\sup_{0\leq r \leq (1/\sll)} \frac{(1+r^2)^2}{r^3}\fla(r) \rkk$.
It is trivial to see that
$
\liml\epsilon (\Lambda)=0$.
Hence we have
$$
\frac{\frac{2}{3}-\epsilon(\Lambda)}{1+1/\sl}
\int_0^{1/\sll}
\!\!\!\!\!\!
\!\!\!\!\!\!
\frac{dr}{\frac{m}{\Lambda}+
\lk \frac{m}{\Lambda}+\frac{8\pi}{3}\rk r^2}
\leq I_1(\Lambda)
\leq
\frac{\frac{2}{3}+\epsilon(\Lambda)}{1-1/\sl}
\int_0^{1/\sll}
\!\!\!\!\!\!
\!\!\!\!\!\!
\frac{dr}{\frac{m}{\Lambda}+
\lk \frac{m}{\Lambda}+\frac{8\pi}{3}   -\delta(\Lambda)  \rk r^2}.$$
Then  a direct calculation yields that
\begin{align*}
&\liml \frac{1}{\sl}
\int_0^{1/\sll}\frac{1}{\frac{m}{\Lambda}+
\lk \frac{m}{\Lambda}+\frac{8\pi}{3}  - \delta(\Lambda) \rk r^2} dr\\
&=\liml
\frac{1}{\sqrt{m (\frac{m}{\Lambda}+\frac{8\pi}{3}- \delta(\Lambda) )}}
\arctan\sqrt{\frac{\frac{m}{\Lambda}+\frac{8\pi}{3}- \delta(\Lambda)}
{m/\sl}}\\
&=
\lk\frac{9}{8\pi}\frac{1}{m}\rk^\han\frac{\pi}{2}.
\end{align*}
Similarly  we have
$$
\liml
\frac{1}{\sl}\int_0^{1/\sll}\frac{1}{\frac{m}{\Lambda}+
\lk \frac{m}{\Lambda}+\frac{8\pi}{3}\rk r^2} dr=
\lk\frac{3}{8\pi}\frac{1}{m}\rk^\han\frac{\pi}{2}.$$
Thus
$$
\frac{2}{3}\lk\frac{3}{8\pi}\frac{1}{m}\rk^\han\frac{\pi}{2}\leq
\liml\frac{1}{\sl}I_1(\Lambda)
\leq
\frac{2}{3}\lk\frac{9}{8\pi}\frac{1}{m}\rk^\han\frac{\pi}{2}.$$
Hence
\kak{mu1} follows.
Next we show
that $ \d \lim_{\Lambda\rightarrow\infty}\frac{I_2(\Lambda)}{\sl}=0$.
Since
$$
(\arctan r-\frac{r}{1+r^2})
-
\lk \arctan r\LM-
\frac{r\LM}{1+r^2\LM^2}\rk
\leq \frac{2}{3}
\frac{r^3}{(1+r^2)^2}+
\frac{5}{4}
\frac{3}{2}
\frac{r^5}{(1+r^2)^2}$$
and by the assumption
$
m>8\pi \lambda/3,
$
\begin{align*}
&\frac{m}{\Lambda} r+\frac{8\pi}{3}
\lkk (r-\arctan r)-\lk r\LM-\arctan r\LM \rk \rkk\\
&
>
\frac{m}{\Lambda} r-\frac{\lambda}{\Lambda} \frac{8\pi}{3}r +\frac{8\pi}{3}
\lkk \frac{r^3}{1+r^2}-
\frac{2}{3}\frac{r^3}{1+r^2}+
\arctan (r\LM)\rkk
>\frac{8\pi}{9}\frac{r^3}{1+r^2}.
\end{align*}
 Then
\begin{align*}
\liml
\frac{1}{\sl}I_2(\Lambda)
& \leq
\liml \frac{1}{\sl}\int_{1/\sll}^\infty
\frac{\frac{2}{3}
\frac{r^3}{(1+r^2)^2}+
\frac{5}{4}
\frac{3}{2}
\frac{r^5}{(1+r^2)^2}}{
\frac{8\pi}{9}\frac{r^3}{1+r^2}
}\frac{dr}{r^2}\\
&
=
\liml \frac{9}{8\pi}
\frac{1}{\sl}\int_{1/\sll}^\infty
\frac{1}{1+r^2}\lk \frac{2}{3}+\frac{15}{8}r^2\rk
\frac{dr}{r^2}=0.
\end{align*}
Thus  $\lim_{\Lambda\rightarrow\infty}\frac{I_2(\Lambda)}{\sl}=0$
 follows, and then the theorem is proven.
\qed

\subsubsection{Many particle system}
We next consider an $N$ particle  system.
We assume simply that each particle has mass $m$
and there is no external potential.
The Hamiltonian, $H$,  is defined
as  a self-adjoint operator acting on
$L^2(\RR^{3N})\otimes\fff$,
and is given by
\eq{bein}
H_N=\sum_{j=1}^N\frac{1}{2m}(-i\nabla_j
-\ak A_j)^2+\hf,
\en
where
$$A_{j,\mu}=\sh \int\frac{e_\mu^{j'}(k)}{\sqrt{\omega(k)}}
\lkk
\vp_j(k)
a^\ast(k,j')+ \vp_j(k)a(k,j')
\rkk dk.$$
Let
$\is (H_N)=\ground (\Lambda, N)$.
We consider the two cases:
\bi
\item[(1)]
$\vp_j=\vp,\ \ \ j=1,...,N,$
\item[(2)]
${\rm supp}\vp_j\cap {\rm supp} \vp_{i}\cap \{0\}=\emptyset$,
$i\not=j$.
\ei
We will see below that the asymptotic behavior of $g(\Lambda, N)$
as $N\rightarrow \infty$ depends on ultraviolet cutoffs.
In the case of (2) we intuitively expect that
$g(\Lambda, N)\approx N$, since $N$  particles
may have no interaction through quantized radiation fields.
\bp{hiro}
\TTT{Ground state energy for many particle system}
\ko{ground state energy!many particle}In the case of (1),
$$g(\Lambda, N)=
\frac{N}{\pi}\int_{-\infty}^\infty
\frac{
\|t\vp/(t^2+\omega^2)\|^2}{m+\frac{2}{3}N\|\vp/
\sqrt{t^2+\omega^2}\|^2}dt,$$
in the case of (2),
$$g(\Lambda, N)=
\sum_{j=1}^N \frac{1}{\pi}\int_{-\infty}^\infty
\frac{
\|t\vp_j/(t^2+\omega^2)\|^2}
{m+\frac{2}{3}\|\vp_j/\sqrt{t^2+\omega^2}\|^2}dt.
$$
\ep
\proof
This is proven in a similar manner to
Theorem \ref{explicit ground state}.
\qed
In the case of (1) the following theorem holds.
\bt{ezawa23}
\TTT{Asymptotic behavior of $g(\Lambda)$}
\ko{ground state energy!asymptotics for many particles}
We assume case (1) and
$
m> 8\pi \lambda/3$.
Then
$$
\frac{8}{3}\lk\frac{3}{8\pi}\frac{1}{m}\rk^\han\frac{\pi}{2}
\leq
\limln \frac{g(\Lambda, N)}{\sqrt N \Lambda^{3/2}}
\leq
\frac{8}{3}\lk\frac{9}{8\pi}\frac{1}{m}\rk^\han\frac{\pi}{2}.
$$
\et
\proof
By Proposition \ref{hiro} we have
$$\frac{g(\Lambda, N)}{4\Lambda}=
\int_0 ^\infty
\frac{
(\arctan r-\frac{r}{1+r^2})
-
\lk \arctan r\LM-\frac{r\LM}{1+r^2\LM^2}\rk
}{\frac{m}{N \Lambda} r+\frac{8\pi}{3}
\lkk (r-\arctan r)-(r\LM-\arctan r\LM)\rkk} \frac{dr}{r^2}.$$
Then in the similar way as the proof of Theorem \ref{ezawa23}
we decompose it such as
$$\frac{g(\Lambda, N)}{4\Lambda}=\int_0^{1/\sln}+\int_{1/\sln}^\infty=I_1(\Lambda,N)+I_2(\Lambda, N),$$
and it can be seen that
$$
\frac{2}{3}\lk\frac{3}{8\pi}\frac{1}{m}\rk^\han\frac{\pi}{2}\leq
\limln \frac{I_1(\Lambda, N)}{\sqrt{N\Lambda}}\leq
\frac{2}{3}\lk\frac{9}{8\pi}\frac{1}{m}\rk^\han\frac{\pi}{2},
$$
and that
$
\limln \frac{I_2(\Lambda, N)}{\sqrt{N\Lambda}}=0$.
Then the theorem follows.
\qed

\subsection{Self-energy term}
\index{self-energy term}
\subsubsection{Diagonalization and DES}
In this section we investigate the $A^2$-dependence of the  ground state energy and DES.
Let us define
\begin{align}
\he=\hp -\frac{\ak}{m} (-i\nabla)\cdot A+\eps \frac{\at}{2m}A^2+\hf+V,
\end{align}
where $0\leq \eps\leq 1$ denotes a parameter. In the case of $\eps=1$,
$\he$ describes $H$
  and in the case of $\eps=0$, $H$
   without self-energy term $A^2$.
So the parameter $\eps$ interpolates between them.
Neglecting external potential $V$, we  first study the translation invariant Hamiltonian defined by
\begin{align}
\hep=\frac{1}{2m}p^2-\ak \frac{p}{m}\cdot A+
 \eps\frac{\at}{2m}A^2+\hf,\quad p\in\BR.
\end{align}
We can also diagonalize $\hep$ in a similar manner to $\ds$.
Let
\eq{de}
D^\eps(z)=m-\eps \frac{\at}{2}\mmm\int\frac{\vp(k)^2}{z-\omega(k)^2}dk.
\en
Then we define $D_+^\eps(s)$ and  $\JI_\eps$ by $D_+(s)$ and $Q$, respectively,  with $D(z)$ replaced by $D^\eps(z)$. We also define $T_{\mu\nu}^\eps f=\delta_{\mu\nu}f+\eps\ak \JI_\eps \omega^\fdd G \omega^\fdd d_{\mu\nu}\vp f$.
Furthermore
the ground state energy $E_{\eps,p}$
of $\hep$ can be explicitly computed
 in the same manner  as that of $H$.
 The net result is as follows:
\eq{eep}
E_{\eps,p}=\frac{1}{2m_\eps}p^2+\ground_\eps,
\en
where
\begin{align*}
\frac{1}{m_\eps}&=
\frac{1}{m}-
\frac{\at\mmm\|\vp/\omega\|^2}
{m+\eps \at \mmm\|\vp/\omega\|^2}\frac{1}{m},\\
\ground_\eps
&=\frac{d}{2\pi}\int_{-\infty}^\infty
\frac{\eps \at \mmm  \left\|\frac{t\vp}{t^2+\omega^2}\right\|^2}
{m+\eps \at  \mmm\left\|\frac{\vp}{\sqrt{t^2+\omega^2}}\right\|^2}dt.
\end{align*}
By replacing $T_{\mu\nu}$ with $T_{\mu\nu}^\eps$, we define
$W^\eps=\WSTE\in \sp_2$ and  the intertwining operator $\UU(W^\eps)$.
On the other hand the displacement operator is given by
$e^{-i\Pi_\eps}$ under the assumption $\IR$,
where
\eq{displacement V}
\Pi_\eps=\frac{i}{\sqrt 2}\frac{\ak}{\mass^\eps}\lkk
\add(\frac{p\cdot e^j\vp}{\omega^{3/2}},j)-
a(\frac{p\cdot e^j\vp}{\omega^{3/2}},j)\rkk,\quad p\in\BR,
\en
with $\mass^\eps=m+\eps\at\mmm\|\vp/\omega\|^2$.
Then we define $\UU_{\eps,p}$ by
$\UU_{\eps,p}=e^{-i\Pi_\eps}\UU_{W^\eps}$ for $p\in\BR$.

\bt{diagonalization V}
\TTT{Diagonalization of $\hep$}
\index{diagonalization!translation invariant!$\eps$ self-energy term}
Suppose Assumption \ref{1v1}.
\begin{description}
\item[(1)] Let $p=0$.
Then
\eq{ln111ep}
\UU_{\eps,0} \f
\lk
 \eps\frac{\at}{2m}A^2+\hf
 \rk\UU_{\eps,0}=
\ground_\eps+\hf.
\en
\item[(2)]
Suppose  $\IR$.
Then
\eq{ln11ep}
\UU_{\eps,p} \f \hep \UU_{\eps,p}=\frac{1}{2m_\eps}p^2+
\ground_\eps+\hf.
\en
\end{description}
\et
\proof
The proof is similar to Theorem \ref{mainp-ari}.
\qed

We define the unitary operator $\UU_\eps$ on $\hhh$ by
\eq{unit2}
\UU_\eps=\int ^\oplus_\BR \UU_{\eps,p} e^{i\frac{\pi}{2}N}dp.
\en

\bt{diagonalization VR}
\TTT{Diagonalization of $\he$}
\index{diagonalization!$\eps$ self-energy term}
Suppose Assumption \ref{1v1} and
that  $V$ is infinitesimally small
 with respect to $- \Delta$.
Assume furthermore that $\IR$.
Then $\he$ is self-adjoint on $D(-\Delta)\cap D(\hf)$,
and  for each $\ak\in\RR$,
$\UU_\eps$ maps $D(-\Delta)\cap D(\hf)$ onto itself and
\eq{ln122}
{\UU_\eps}\f H\UU_\eps=-\frac{1}{2\mass^\eps}\Delta
+ \hf+T_\eps\f V T_\eps+\ground_\eps,
\en
where
\begin{align}
&\label{matsu11}
T_\eps=\exp\left(-i(\p)\cdot K^\eps\right),\\
&
\label{matsu22}
K_\mu^\eps=\sh   \int
\frac{e_\mu^j(k)}{\sqrt{\omega(k)}}
\lk
\frac{\ov{\JI_\eps(k)}}{\omega(k)}
 \add  (k,j)+
\frac{{\JI_\eps(k)}}{\omega(k)}
 a(k,j)\rk \ddk.
\end{align}
\et
\proof
We set  $\he$ with $V=0$ by $\hez$.
In
a similar way to Proposition \ref{self}
 we can show that $\hez$ is self-adjoint on $D(-\Delta)\cap D(\hf)$.
 By the closed graph theorem there exists $C>0$ such that
$\|(-\Delta+\hf)F\|\leq C(\|\hez\Phi\|+\|\Phi\|)$.
Hence
$V$ is infinitesimally small with respect to $\hez$ and
$\he$ is self-adjoint on $D(-\Delta)\cap D(\hf)$ by the Kato-Rellich theorem.
Statement \kak{ln122} is proven in a similar manner to Theorem \ref{mainp-nasi}.
\qed

\bc{ground state2}
\TTT{Existence and absence of DES of $\hep$}
\index{dressed electron state}\\
Suppose Assumption \ref{1v1}.
Then (1)-(3) follow:
\bi
\item[(1)]
The Hamiltonian $\hep$ for $p=0$  has a dressed electron state and it is unique.
\item[(2)]
Suppose   $\IR$. Then $\hep$ has a dressed electron state for all $p\in\BR$ and it is unique.
\item[(3)]
Suppose   $\SR$. Then $\hep$ with $p\not=0$ has no  bound state.
\ei
\ec
\proof
The proof is similar to that of Theorem \ref{DESS}.
\qed
\bc{abs}
Let $0\leq \eps<1$.
Suppose Assumption \ref{1v1} and
that  $V$ is nonnegative and
infinitesimally small
 with respect to $-  \Delta$.
Assume also that $\IR$.
Let
$$\d \ak_\ast=\frac{m}{1-\eps}\frac{1}{\mmm\|\vp/\omega\|^2}.$$
Then
$H_\eps$ is bounded from below for
$\at <\ak_\ast$ and unbounded from below for $\at >\ak_\ast$. In particular
$\he $ has no ground state for $\at >\ak_\ast$.
\ec
\proof
Let $\hez$ be $\he$ with $V=0$.
 When $\at < \ak_\ast$ (resp.
$\at > \ak_\ast$),
$\frac{1}{m_\eps}>0$ (resp.$\frac{1}{m_\eps}<0$) follows.
Then
$\hez$ is bounded from below (resp. unbounded from below).
Since $V\leq 0$,
$\he$ is also unbounded from  below for
$\at > \ak_\ast$.
Contrary  to this,
$\he$ is bounded from below for
$\at < \ak_\ast$, since $T\f VT $ is infinitesimally small
 with respect to $-\frac{1}{2m_\eps}\Delta+\hf$.
 \qed

\subsubsection{No self-enery term}
We consider the special case:  $\eps=0$.
Thus
Hamiltonians
are
\begin{align}
H_0&=-\ham\Delta -\frac{\ak}{m}(-i\nabla)\cdot A+\hf +V,\\
H_{0,p}&=\ham p^2 -\frac{\ak}{m}p \cdot A+\hf.
\end{align}
Then
$D^\eps(z)=m$,
$\JI_\eps=\ak\vp/m$, $T_{\mu\nu}^\eps=\delta_{\mu\nu}$,
$\ground=0$, $W_+^\eps=\one$ and $W_-^\eps=0$. Thus
the intertwining operator is the identity and
$\UU_{0,p}=e^{-i\Pi}$, where
\eq{sakebi}
\Pi=\frac{i}{\sqrt 2}\frac{\ak}{m}\lkk
\add(\frac{p\cdot e^j\vp}{\omega^{3/2}},j)-
a(\frac{p\cdot e^j\vp}{\omega^{3/2}},j)\rkk.
\en
Suppose $\IR$.
Then
\eq{tatunammi}
\UU_{0,p}\f H_{0,p} \UU_{0,p}=\frac{1}{2m_0}p^2+\hf. \en
Let $\UU_0=\int \UU_{0,p}e^{i\frac{pi}{2}N}dp$. Then
\eq{daka}
\UU_0\f (-\ham\Delta -\frac{\ak}{m}(-i\nabla)\cdot A+\hf +V
)
\UU_0=-\frac{1}{2m_0}\Delta+\hf + T\f  VT,
\en
where
\eq{miji}
\frac{1}{m_0}=\frac{1}{m}-\frac{\at \mmm \|\vp/\omega\|^2}{m}\en
and $T=\exp(-i(-i\nabla) \cdot K)$ with
$$
K_\mu=\sh \frac{\ak}{m}  \int
\frac{e_\mu^j(k)}{\sqrt{\omega(k)^3}}
\lk
{\vp(k)}
 \add  (k,j)+
{\vp(k)}
 a(k,j)\rk \ddk.
$$
Thus formally we have
$T\f V T= V(x+\frac{\ak}{m} Z)$, where
$$Z_\mu =\sh   \int
\frac{e_\mu^j(k)}{\sqrt{\omega(k)^3}}
\lk
{\vp(k)}
 \add  (k,j)+
{\vp(k)}
 a(k,j)\rk \ddk.
$$
Let $V$ be infinitesimally small with respect to $-\Delta$.
By \kak{miji}
 we see that
$H_0$ is unbounded from below for
$\at >(\mmm \|\vp/\omega\|^2)\f$ and bounded from below for $\at < (\mmm \|\vp/\omega\|^2)\f$.

\subsection{Scaling limits}
\ko{scaling limit}
In this section we investigate scaling limits of Hamiltonian $H$ and derive effective Hamiltonians.
The general references in this section are  \cite{ara90,dav77,dav79,hir93,hir97,hir98,hir99,hir02}.
 \subsubsection{Weak  coupling limit}
\label{pfweak}
It is shown  that one of the useful tool to derive effective objects is scaling limits.
We introduce the scaling by
$
\ass\to \k  \ass$.
 Then the scaled Hamiltonian is of the form
\eq{ss2}
H(\k )=\frac{1}{2m}(\p-\ak\k  A)^2+V+\k ^2\hf.
\en
We consider the asymptotic behavior of $H(\k )$ as $\k \to \infty$.
The scaling \kak{ss2} is equivalent to the substitution
\eq{sub2}
\omega\to\k ^2\omega,\quad
\vp\to\k ^2\vp.
\en
The operator $\tmn$ leaves invariant under this scaling, and hence  $W_\pm$ and the intertwining operator $\UU_W $ also leave invariant under  \kak{sub2}.
 On the other hand the displacement operator $S_p$ is scaled as
 \eq{ss322}
 S_p\to
\exp\lk
\frac{1}{\k }
\frac {1}{\sqrt 2}
\frac{\ak}{\mass}\lkk
\add(\frac{ p \cdot e^j\vp}{ \omega^{3/2}},j)-
a(\frac{ p \cdot e^j\vp}{ \omega^{3/2}},j)\rkk\rk
\en
and then
we have
\eq{ss4}
\slim_{\k \to\infty}S_p=\one.
\en
Let
$\UU=\int^\oplus  S_p\UU_W e^{i\frac{\pi}{2}N}dp$.
We recall that $\ground=\frac{d}{2\pi}\int_{-\infty}^\infty
\frac{\at \mmm  \left\|\frac{t\vp}{t^2+\omega^2}\right\|^2}
{m+\at  \mmm\left\|\frac{\vp}{\sqrt{t^2+\omega^2}}\right\|^2}dt$. Then
$g$ is scaled as $g\to\k ^2 g$.
If we make the corresponding substitution in $\UU$ and $\dv$, we denote them  by
$\UU_\k$ and $\dv_\k $, respectively.
\bl{scalinglimit}
It follows that
$\d\slim_{\k \to\infty}
\UU_\k=\UU_W $.
\el
\proof
The lemma follows from \kak{ss4} and the invariance of $\UU_W $ under  \kak{sub2}.
\qed
In order to consider the asymptotic behavior of $H(\k)$ as $\k\to\infty$,
we introduce the  energy renormalization $\k^2\ground$.
\bt{wcl}
\TTT{Weak coupling limit}
\index{weak coupling limit}
{\rm \cite{hir02}}
Let $V$ be relatively bounded with respect to $-\ham\Delta$ with a relative bound strictly smaller than one.
Suppose Assumption \ref{1v1} and  $\IR$.
Then
for
$z\in\CC\setminus\RR$,
\eq{ss5}
\slim_{\k \to\infty}(H(\k )- \k ^2 g-z)\f=(\heff-z)\f\otimes P_W,
\en
where $P_W$ denotes the projection to the one dimensional subspace
$\{a\UU_W \Omega |a\in\CC\}$.
\et
\proof
By the unitary transformation $\UU_\k$
we have
\eq{ss7}
 (H(\k )-\k^2 g-z)\f  =\UU_\k(\heff+\k^2 \hf +\dv_\k-z)\f \UU_\k\f.
 \en
We have already seen that $\slim_{\k \to\infty}\UU_\k =\UU_W $.
We can directly see that
\bi
\item[(1)] $D(\dv_\k )\supset D(\heff)$ and $\dv_\k (\heff+\lambda)\f$ is bounded in $\hhh$ for large $\lambda>0$ with $\lim_{\lambda\to\infty}\|\dv_\k (\heff+\lambda)\f\|=0$,
\item[(2)] $\dv_\k (\heff+\lambda)\f$ is strongly continuous in $\k $,
\item[(3)] $\slim_{\k \to\infty}\dv_\k (\heff+\lambda)\f=0$.
\ei
In
 the abstract formula in \cite{ara90},
it  has been  established that
(1)-(3) above imply that
\eq{ss6}
\slim_{\k \to\infty}(\heff+\k^2 \hf+\dv_\k -z)\f
=(\heff-z)\f\otimes P_\Omega,
\en
where $P_\Omega$ denotes the projection to $\{a\Omega|a\in\CC\}$.
Hence we can see that
\begin{align*}
\slim_{\k \to\infty}
(H(\k )-\k ^2 g-z)\f
&=
\slim_{\k \to\infty}
\UU_\k
(\heff+\k^2 \hf+\dv_\k -z)\f
\UU_\k \f\\
&=
\UU_W
\lk
(
\heff-z)\f\otimes P_\Omega
\rk
\UU_W \f\\
&=
(
\heff-z)\f\otimes
(\UU_W  P_\Omega
\UU_W \f)\\
&=
(
\heff-z)\f\otimes
P_W.
\end{align*}
Then the theorem is proven.
\qed

We notice that the projection $P_W$ denotes the projection to the one dimensional subspace spanned by the unique ground state of operator
$H_{p=0}=\frac{1}{2m}A^2+\hf$.

\subsubsection{Strong coupling limit}
In the previous section we study the  weak coupling limit which is
given by the asymptotic behavior of the scaled Hamiltonian:
$$-\ham \Delta-\k \frac{\ak}{m}(-i\nabla)\cdot A+\k^2( \frac{\at}{2m}A^2+\hf) +V.$$
We introduce another scaling.
Let
\eq{scl}
\he(\k)=-\ham \Delta-\k^2 \frac{\ak}{m}(-i\nabla)\cdot A+\k^2(\eps \frac{\at}{2m}A^2+\hf) +V.
\en
The scaling \kak{scl} may reflect  the  interaction $(-i\nabla)\cdot A$ between the particle and the field rather than the weak coupling limit. As will be seen below under \kak{sub2} the displacement operator does not disappear, and an effective potential appears instead of effective mass.
The scaling in \kak{scl}
corresponds to the substitution:
\eq{sca}
\omega\to\k^2\omega,\quad
\vp\to\k^{3}\vp,\quad
\eps\to\eps/\k^2.\en
We investigate the scaling limit of $\he(\k)$ as $\k\to\infty$.
Instead of energy renormalizations, in this scaling limit we need a mass renormalization.
Under the scaling \kak{sca},
$\ground$ and $\UU$ are  invariant.
Let us define $\mren$ by
\eq{massre}
\frac{1}{\mren}=\frac{1}{m}+
\frac{\at\mmm\|\vp/\omega\|^2}
{m+\eps \at \mmm\|\vp/\omega\|^2}\frac{1}{m}.
\en
Under the scaling \kak{sca}, $\mren$ is scaled as
$$\frac{1}{\mren}\to \frac{1}{\mren(\k)}=\frac{1}{m}+
\k^2 \frac{\at\mmm\|\vp/\omega\|^2}
{m+\eps \at \mmm\|\vp/\omega\|^2}\frac{1}{m}.
$$
We define
the renormalized Hamiltonian
by
\eq{hren}
\hren=-\frac{1}{2\mren(\k)}\Delta-\k^2
\frac{\ak}{m}(-i\nabla)\cdot A+
\k^2\lk
\frac{\at}{2m}A^2+\hf
\rk+V.
\en
\bt{strong coupling limit}
\TTT{Strong coupling limit}
\index{strong coupling limit}
{\rm \cite{ara90,hir93,hir97,hir02}}Let $V\in L_{\rm loc}^1(\BR)$  be relatively bounded with respect to $-\ham\Delta$ with a relative bound strictly smaller than one.
Suppose Assumption \ref{1v1} and $\IR$.
Then for $z\in\CC\setminus\RR$,
\eq{scl1}
\slim_{\k\to\infty}(\hren-z)\f=
e^{-i(-i\nabla)\cdot \Pi}
\lk
(-\ham\Delta+V_{\rm eff}+g-z)\f\otimes P_W
\rk
e^{i(-i\nabla)\cdot \Pi},
\en
where
\begin{align}
&\Pi_\mu =\frac{i}{\sqrt2} \frac{\ak}{m}\lkk
\add
(\frac{e_\mu^j \vp}{\omega^{3/2}})-
a(\frac{e_\mu^j \vp}{\omega^{3/2}})
\rkk,\\
&P_W=\UU_W  P_\Omega \UU_W \f,\\
&V_{\rm eff}(x)=V\ast P_C(x),\\
&P_C(x)=(2\pi C)^{-d/2}
e^{-|x|^2/(2C)},\\
&C=\half\mmm\|Q_\eps/\omega^{3/2}\|^2.
\end{align}
\et
 \proof
By the unitary transformation $\UU$
we have
\eq{ss71}
 (\hren(\k )-z)\f =
\UU(
-\ham \Delta +\k^2 \hf +T\f V T +g-z)\f\UU\f.
\en
By the abstract formula  \cite{ara90} again,
it  has been  established that
$$
\slim_{\k\to\infty}(
-\ham \Delta +\k^2 \hf +T\f V T +g-z)\f =
(-\ham \Delta+(\Omega, T\f V T\Omega)_\fff+g-z)\f \otimes P_\Omega.$$
Hence we can see that
\begin{align*}
&\slim_{\k \to\infty}
(\hren(\k)-z)\f\\
&=
\UU
\lkk
(-\ham \Delta+(\Omega, T\f V T\Omega)_\fff+g-z)\f \otimes P_\Omega\rkk
\UU\f \\
&=
e^{-i(-i\nabla)\cdot \Pi}
\lk
(-\ham\Delta+V_{\rm eff}+g-z)\f\otimes P_W
\rk
e^{i(-i\nabla)\cdot \Pi}.
\end{align*}
Then the theorem is proven.
\qed
Potential $V_{\rm eff}$
is called the effective potential
\ko{effective potential}.
\begin{figure}[t]
\begin{center}
\arrayrulewidth=1pt
\def\arraystretch{2.0}
\begin{tabular}{|c|c|c|}
\hline
\       & Scaling  and renormalization& Effective Hamiltonian \\
\hline
WCL
&
$\ham(-i\nabla-\ak \k A)^2+V+\k^2 \hf-\k^2 g$ &
$-\frac{1}{2\mass}\Delta+V$
 \\
\hline
SCL
&
$-\frac{1}{2\mren(\k)} \Delta -\k^2\frac{\ak}{m}(-i\nabla) A+\k^2(\frac{\at}{2m}A^2+\hf)+V
$
&
$-\frac{1}{2m}\Delta+V\ast P_C+g$\\
\hline
\end{tabular}
\end{center}
\caption{Scaling limits}
\label{picsca}
\end{figure}%
In the case of $\epsilon=1$
the effective potential $\veff$ and $C$ are
 given by
\begin{align}
V_{\rm eff}(x)&=V\ast P_C,\\
C&=\at \half\mmm\int \frac{ |\vp(k)|^2}
{\mass(k) \omega(k)^3}dk,
\end{align}
and in the case of $\eps=0$,
\begin{align}
V_{\rm eff}(x)&=V\ast P_C,\\
C&=\at \half\mmm\int \frac{ |\vp(k)|^2}
{m^2 \omega(k)^3}dk.
\end{align}
The strong coupling limit gives a mathematical interpretation of
the Lamb shift\ko{Lamb shift} derived by \kak{wel}.
Namely the difference of the spectrum of $-\ham \Delta+V$ and $-\ham\Delta+\veff$
approximately  gives an interpretation of the Lamb shift.
This was done in \cite{wel48} and see  historical review \cite[p.306,(7.4.20)]{sch94}.
See also \cite{ara90, ara11}.

\subsection{Negative mass}
\label{negativemass}
We are interested  in  investigating  the Hamiltonian with negative mass $m<0$ from mathematical point of view.
This is of course an unphysical assumption.
In this section we suppose that $-\sum_{\mu=1}^3 \at \|\lambda /\omega\|^2<m<0$.
In this case $W=\WST \not\in \sp$.
Then we define a new operators.
 Let us define
$X_\pm$ by
\eq{xxx}
X_{\pm ij} f=W_{\pm ij}f+\half \yy e_\mu^i F_{\mu\nu}(\y e_\nu^j \JI ,f).
\en
\bl{symp2}
\TTT{Symplectic structure}
\index{symplectic structure!negative mass}
Suppose Assumption \ref{1v1}.
Then
it follows that
\begin{align}
\label{x1}
&X_+^\ast X_+-X_-^\ast X_-=\one,\\
\label{x2}
&\ov{X_+^\ast}X_--\ov{X_-^\ast}X_+=0,\\
&\label{x3}
X_+X_+^\ast-\ov{X_-}\ov{X_-^\ast}=\one,\\
&\label{x4}
X_-X_+^\ast-\ov{X_+}\ov{X_-^\ast}=0.
\end{align}
I.e., $X=\XST\in \sp$.
\el
\proof
Let $\xi=(\xi_{ij})_{1\leq i,j\leq d-1}$ and $\xi_{ij}=\half \yy e_\mu^i F_{\mu\nu}(\y e_\nu^j \JI ,\cdot)$. Then
we have $(\xi^\ast)_{ij}=\half \y e_\mu ^i \JI (\yy e_\nu^j F_{\mu\nu},\cdot)$.
Thus $X_\pm=W_\pm+\xi$.
We have
\begin{align*}
{\rm LHS}\kak{x1}=
\one +\xi^\ast (W_+-W_-)+(W_+^\ast-W_-^\ast)\xi
\end{align*}
By (13) of Lemma \ref{algebra} we see that
\begin{align*}
\xi^\ast (W_+-W_-)f&=
\half \y e_\mu^i \JI (\yy e_\nu^k F_{\mu\nu}, (W_+-W_-)_{kj}f)\\
&=
\half \y e_\mu^i \JI
(\yy e_\nu^k F_{\mu\nu}, e_a^k \y T_{ab}^\ast\yy e_b^j f)\\
&=
\half \y e_\mu^i \JI (e_b^j T_{ab} F_{a\mu },\yy f)=0.
\end{align*}
We also see that
\begin{align*}
(W_+^\ast-W_-^\ast) \xi f=\half (\y e_\nu^j \JI ,f)
(W_+^\ast-W_-^\ast)_{ik} \yy e_\mu^k F_{\mu\nu}=0.
\end{align*}
Then \kak{x1} follows. Identity \kak{x2} is similarly proven.
We have \begin{align*}
{\rm LHS}\kak{x3}=\one+Z_++(W_+\xi^\ast-\bar W_-\bar \xi^\ast)+(\xi W_+^\ast-\bar \xi\bar W_-^\ast)+(\xi\xi^\ast-\bar\xi\bar \xi^\ast).
\end{align*}
We see that
$\xi\xi^\ast-\bar\xi\bar \xi^\ast=0$.
By (12) and (15) of Lemma \ref{algebra} we have
\begin{align*}
&\bar W_-\bar \xi^\ast f\\
&=
\frac{1}{4}
\ov{e_\mu^i \yy T_{\mu\nu}^\ast \omega d_{\nu a} \JI (\yy e_b^j F_{ab},\bar f)}
-\frac{1}{4}
\ov{e_\mu^i \y  T_{\mu\nu}^\ast d_{\nu a} \JI (\yy e_b^j F_{ab}, \bar f)}\\
&=
\frac{1}{4}
{e_\mu^i \yy \bar T_{\mu\nu}^\ast \omega d_{\nu a} \bar \JI (\yy e_b^j F_{ab}, f)}
-\frac{1}{4}
{e_\mu^i \y  \gamma^2 F_{\mu a}
(\yy e_b^j F_{ab}, f)}\\
&=\frac{1}{4}
{e_\mu^i \yy  T_{\mu\nu}^\ast \omega d_{\nu a}  \JI (\yy e_b^j F_{ab}, f)}
-\frac{1}{4}
{e_\mu^i \y  \gamma^2 F_{\mu a}
(\yy e_b^j F_{ab}, f)}.
\end{align*}
On the other hand
\begin{align*}
W_+\xi^\ast f
=\frac{1}{4}
{e_\mu^i \yy  T_{\mu\nu}^\ast \omega d_{\nu a}  \JI (\yy e_b^j F_{ab}, f)}
+\frac{1}{4}
{e_\mu^i \y  \gamma^2 F_{\mu a}
(\yy e_b^j F_{ab}, f)}.
\end{align*}
Hence
$$W_+\xi^\ast-\bar W_-\bar \xi^\ast=\half e_\mu^i\y\gamma^2 F_{\mu a}(\yy e_b^j F_{ab},\cdot)$$ follows.
In a similar manner we have
\begin{align*}
\bar \xi \bar W_-^\ast f&=
-\frac{1}{4}\yy e_\mu^i F_{\mu\nu}(\gamma^2 F_{\nu b},\y e_b^j f)+
\frac{1}{4}\yy e_\mu^i F_{\mu\nu} (d_{\nu a}\y \JI , \y  T_{ab}\yy e_b^jf)\\
\xi W_+^\ast f&=
\frac{1}{4}\yy e_\mu^i F_{\mu\nu}(\gamma^2 F_{\nu b},\y e_b^j f)+
\frac{1}{4}\yy e_\mu^i F_{\mu\nu} (d_{\nu a}\y \JI , \y  T_{ab}\yy e_b^jf).
\end{align*}
Then
$$\xi W_+^\ast -\bar \xi \bar W_-^\ast=\half \yy e_\mu^i \gamma^2 F_{\mu\nu}(\y e_b^j F_{\nu b}, \cdot).$$
Together with them
we have
$$(W_+\xi^\ast-\bar W_-\bar \xi^\ast)+(\xi W_+^\ast-\bar \xi\bar W_-^\ast)+(\xi\xi^\ast-\bar\xi\bar \xi^\ast)=-Z_+.$$
 Then \kak{x3} follows.
Finally we  prove \kak{x4}.
We have
\begin{align*}
{\rm LHS}\kak{x4}=Z_++(W_-\xi^\ast-\bar W_+\bar \xi^\ast)+(\xi W_+-\bar \xi \bar W_-^\ast)+(\xi\xi^\ast -\bar \xi \bar \xi^\ast).
\end{align*}
By (12) and (15) of Lemma \ref{algebra}
we have
\begin{align*}
\bar W_+\bar \xi^\ast f
&=\frac{1}{4}e_\mu^i \yy \bar T_{\mu\nu}^\ast \omega
 d_{\nu a}
 \bar \JI (\yy e_b^j F_{ab},f)+
\frac{1}{4}e_\mu^i \y \bar T_{\mu\nu}^\ast d_{\nu a}\bar \JI  (\yy e_b^j F_{ab},f)\\
&=\frac{1}{4}e_\mu^i \yy  T_{\mu\nu}^\ast \omega
 d_{\nu a}
  \JI (\yy e_b^j F_{ab},f)+
\frac{1}{4}e_\mu^i \y  T_{\mu\nu}^\ast d_{\nu a} \JI  (\yy e_b^j F_{ab},f)\\
&=\frac{1}{4}e_\mu^i \yy  T_{\mu\nu}^\ast \omega
 d_{\nu a}
  \JI (\yy e_b^j F_{ab},f)+
\frac{1}{4}e_\mu^i \y
\gamma^2 F_{\mu a}
  (\yy e_b^j F_{ab},f)\\
W_-\xi^\ast f&=
  \frac{1}{4}e_\mu^i \yy  T_{\mu\nu}^\ast \omega
 d_{\nu a}
  \JI (\yy e_b^j F_{ab},f)-
\frac{1}{4}e_\mu^i \y
\gamma^2 F_{\mu a}
  (\yy e_b^j F_{ab},f).
\end{align*}
Hence
$$W_-\xi^\ast-\bar W_+\bar \xi^\ast=
-\half \gamma^2 e_\mu^i \y F_{\mu a}(\yy e_b^j F_{ab},f).$$
Similarly
\begin{align*}
\bar \xi \bar W_-^\ast f&
=-\frac{1}{4}
\ov{
\yy e_\mu^i F_{\mu\nu}(\y e_\nu^k \JI , e_a^k(\yy T_{ab}\y -\y T_{ab}\yy)e_b^j \bar f)}\\
&=
-\frac{1}{4}
\yy e_\mu^i F_{\mu\nu}(\bar T_{ab}^\ast d_{\nu a} \bar \JI , \y e_b^j f)
+
\frac{1}{4}
\yy e_\mu^i F_{\mu\nu}
(d_{\nu a}\bar \JI ,\omega \bar T_{ab}\yy e_b^j f)\\
&=
-\frac{1}{4}
\yy e_\mu^i F_{\mu\nu}( T_{ab}^\ast d_{\nu a}  \JI , \y e_b^j f)
+
\frac{1}{4}
\yy e_\mu^i F_{\mu\nu}
(d_{\nu a} \JI ,\omega  T_{ab}\yy e_b^j f)\\
&=
-\frac{1}{4}
\yy e_\mu^i F_{\mu\nu}( \gamma^2 F_{b\nu}, \y e_b^j f)
+
\frac{1}{4}
\yy e_\mu^i F_{\mu\nu}
(d_{\nu a} \JI ,\omega  T_{ab}\yy e_b^j f)\\
\xi W_+ f &=
\frac{1}{4}
\yy e_\mu^i F_{\mu\nu}( \gamma^2 F_{b\nu}, \y e_b^j f)
+
\frac{1}{4}
\yy e_\mu^i F_{\mu\nu}
(d_{\nu a} \JI ,\omega  T_{ab}\yy e_b^j f).
\end{align*}
Hence
$$\xi W_+-\bar \xi \bar W_-^\ast=\half
\yy e_\mu^i \gamma^2 F_{\mu\nu}(\y e_b^j F_{b\nu}, \cdot)$$ and then
$$(W_-\xi^\ast-\bar W_+\bar \xi^\ast)+(\xi W_++\bar \xi \bar W_-^\ast)+(\xi\xi^\ast -\bar \xi \bar \xi^\ast)=-Z_-.$$
Hence \kak{x4} follows.
\qed

Although $\WST\not\in \sp$, it is  shown that $B_p^\sharp (f,j)$ still satisfies
canonical commutation relations and adjoint relation.
We notice that  $\ass$ can not be realized however in terms of $B_p^\sharp$.
For $p\in\BR$, we define
\begin{align*}
&C_\mu (p)=-\frac{1}{E}p_\mu+E\ta_a(F_{a\mu})-\tp_a(F_{a\mu}),\\
&D_\mu (p)=-\frac{1}{E}p_\mu+E\ta_a(F_{a\mu})+\tp_a(F_{a\mu}).
\end{align*}
Both $C_\mu(p)$ and $D_\mu(p)$ are essentially self-adjoint.
\bl{ccrnegative}
Suppose Assumption \ref{1v1}.
Then  it follows that
\begin{align}
\label{sad1}
&[C_\mu(p), D_\nu(p)]=2i \frac {E}{\gamma^2}\delta_{\mu\nu},\\
\label{sad2}
&[C_\mu(p), C_\nu(p)]=0,\\
\label{sad3}
&[D_\mu(p), D_\nu(p)]=0,\\
\label{sad4}
&[B_p(f,j), D_\nu(p)]=[B_p(f,j), C_\nu(p)]=0,\\
\label{sad5}
&[B^\ast_p(f,j), D_\nu(p)]=[B^\ast_p(f,j), C_\nu(p)]=0.
\end{align}
 \el
\proof
We can directly see that
\begin{align*}
[C_\mu(p), D_\nu(p)]
&=2E[\tp_a(F_{a\mu}), \ta_b(F_{b\nu})]
=2Ei(d_{ab}\ov{F_{a\mu}}, \tilde{F_{b\nu}})\\
&=2Ei (F_{b\mu},F_{b\nu})=
2i \frac{E}{\gamma^2}\delta_{\mu\nu}
\end{align*}
by (16) of Lemma \ref{algebra}.
Both \kak{sad2} and \kak{sad3} also follow from (16) of
Lemma \ref{algebra}.
Both \kak{sad4} and \kak{sad5} follow from (13) of Lemma \ref{algebra}.
\qed

\bl{hcom}
Suppose Assumption \ref{1v1}.
Then it follows that
\begin{align}
\label{matu1}
&[H_0, C_\mu(p)]=iE C_\mu(p),\\
\label{matu2}
&[H_0, D_\mu(p)]=-iE D_\mu(p).
\end{align}
\el
\proof
We have
$[\hf, C_\mu(p)]=-iE\tp_a(F_{a\mu})-i\hat A_a(\omega^2 F_{a\mu})$.
Note that
$$\omega^2 F_{a\mu}=\ak \vp  d_{a\mu}-E^2 F_{a\mu}.$$
Then
$$[\hf, C_\mu(p)]=-iE\tp_a(F_{a\mu})-i\ak A_\mu+iE^2 \hat A_a( F_{a\mu}).$$
On the other hand we can see that
$$[A_\nu, C_\mu(p)]=-i(\vp , F_{\mu\nu})=i\frac{m}{\ak}\delta_{\mu\nu}$$
by (14) of Lemma \ref{algebra}.
Together with them we have
\begin{align*}
[H_0, C_\mu(p)]
&=\frac{1}{m}(p_\nu-\ak A_\nu)(-\ak)
\frac{mi}{\ak}
\delta_{\mu\nu}-iE\tp_a(F_{a\mu})-i\ak A_\mu +iE^2 \hat A_a(F_{a\mu})\\
&=
-ip_\mu -iE\tp_a(F_{a\mu})
+iE^2\hat A_a(F_{a\mu})
=iE C_\mu(p).
\end{align*}
Then \kak{matu1} follows.
\kak{matu2} is similarly proven.
\qed
\bl{time}
Suppose Assumption \ref{1v1}.
Then it
 follows that
\begin{align*}
&e^{itH_0} C_\mu(p) e^{-itH_0}=e^{-tE}C_\mu(p),\\
&
e^{itH_0} D_\mu(p) e^{-itH_0}=e^{tE}D_\mu(p).
\end{align*}
\el
\proof
Set $C_t=e^{tcE}C_\mu(p)$ and $\bar C_t=e^{itH_0} C_\mu(p) e^{-itH_0}$.
Then $$\frac{d}{dt} C_t=\frac{d}{dt} \bar C_t=i[H_0,C_\mu(p)]$$ and $C_0=\bar C_0$.
Then $C_t=\bar C_t$ follows.
The equality $e^{itH_0} D_\mu(p) e^{-itH_0}=e^{tE}D_\mu(p)$ is similarly proven.
\qed
\bt{time2}
\TTT{Time evolution of $A$}
\index{time evolution of radiation field!negative mass}
 Suppose Assumption \ref{1v1}.
 Then for all $p\in\BR$ and
real-valued $f$ such that $f\in M_0\cap M_{-1/2}\cap M_{-1}$,
\begin{align}
e^{itH_0}A_\mu e^{-itH_0}&=
\frac{1}{\sqrt 2}
\lkk B_p^\ast(e^{it\omega}
e_\nu^j\yy \ov{T_{\nu\mu}}\hat f)
+B_p(e^{-it\omega}
e_\nu^j \yy T_{\nu\mu}\tilde {\hat f})\rkk\non \\
&\label{dann5}
+\frac{\ak}{\mass}p_\nu(d_{\mu\nu}\frac{\vp }{\omega},\frac{\hat f}{\omega})+
\frac {\gamma^2}{2E}(F_{\mu\nu},f)(e^{-tE}C_\nu(p)+e^{tE}D_\nu(p)).
\end{align}
\et
\proof
We shall show that
\begin{align}
A_\mu=&
\frac{1}{\sqrt 2}
\lkk B_p^\ast(
e_\nu^j\yy \ov{T_{\nu\mu}}\hat f)
+B_p(
e_\nu^j \yy T_{\nu\mu}\tilde {\hat f})\rkk\non \\
&
+\frac{\ak}{\mass}p_\nu(d_{\mu\nu}\frac{\vp }{\omega},\frac{\hat f}{\omega})+
\frac {\gamma^2}{2E}(F_{\mu\nu},f)(C_\nu(p)+D_\nu(p))\non \\
=&\frac{1}{\sqrt 2}
\lkk B_p^\ast(
e_\nu^j\yy \ov{T_{\nu\mu}}\hat f)
+B_p(
e_\nu^j \yy T_{\nu\mu}\tilde {\hat f})\rkk\non \\
&
\label{dann1}
+\frac{\ak}{\mass}p_\nu
(d_{\mu\nu}\frac{\vp }{\omega},\frac{\hat f}{\omega})
+\lk
-\frac{\gamma^2}{E^2}p_\nu+\gamma^2 \hat A_a(F_{a\nu})\rk (F_{\mu\nu},\hat f)
.
\end{align}
We see that
\begin{align*}
&B_p^\ast(
e_\nu^j\yy \ov{T_{\nu\mu}}\hat f)\\
&=
a(\bar W_{-ij} e_\nu^j\yy \bar T_{\mu\nu}\hat f,i)
+
\add(\bar W_{+ij}e_\nu^j \yy\bar T_{\mu\nu}\hat f,i)
-(p\cdot e_\nu^j\frac{\bar \JI}{\sqrt 2\omega},\frac{1}{\omega}
\bar T_{\mu\nu} \hat f),\\
&
B_p(
e_\nu^j\yy \ov{T_{\nu\mu}}\hat f)\\
&=
a(W_{+ij} e_\nu^j\yy  T_{\mu\nu}\tilde{\hat f},i)+
\add( W_{-ij}e_\nu^j \yy T_{\mu\nu}\tilde{\hat f},i)
-
(p\cdot e_\nu^j\frac{ \JI}{\sqrt 2\omega},\frac{1}{\omega}
 T_{\mu\nu} \tilde{\hat f}).
\end{align*}
By using (3) and (7) of Lemma \ref{algebra},
we compute  the sum of test function of the creation operators  as
\eq{dann2}
\bar W_{+ij}e_\nu^j \yy\bar T_{\mu\nu}\hat f
+W_{-ij}e_\nu^j \yy T_{\mu\nu}\tilde{\hat f}=
\yy e_\mu^i\tilde{\hat f}-\gamma^2(F_{\mu\nu}, \tilde{\hat f})
\frac {e_a^i F_{\nu a}}{\sqrt\omega}
\en
and that of  the annihilation  operators as
\eq{dann3}
\bar W_{-ij}e_\nu^j \yy\bar T_{\mu\nu}\hat f
+W_{+ij}e_\nu^j \yy T_{\mu\nu}\tilde {\hat f}=
\yy e_\mu^i{\hat f}-\gamma^2(F_{\mu\nu}, {\hat f})
\frac {e_a^i F_{\nu a}}{\sqrt\omega}.
\en
By (11) and (12) of Lemma \ref{algebra} we have
\begin{align}
&-(p\cdot e_\nu^j\frac{\bar \JI}{\sqrt 2\omega},\frac{1}{\omega}
\bar T_{\mu\nu} \hat f)
-
(p\cdot e_\nu^j\frac{ \JI}{\sqrt 2\omega},\frac{1}{\omega}
 T_{\mu\nu} \tilde{\hat f})\non \\
 &\label{dann4}
 =-\sqrt 2
 p_a\lkk
\frac{\ak}{\mass}(d_{a\nu}\frac{\vp }{\omega},
\frac{\hat f}{\omega})
-\frac{\gamma^2}{E^2}(F_{a\nu},\hat f)
 \rkk.
\end{align}
From \kak{dann2}-\kak{dann4}, \kak{dann1} follows.
\kak{dann5} is derived from
Lemmas \ref{timeevolution} and \ref{time}.
\qed

\cleardoublepage
\section{Binding and non-binding}
\subsection{Enhanced binding}
\index{enhanced binding}
Non-perturbative analysis of perturbation of eigenvalues embedded in the continuous spectrum has
been developed in the last decade and has been applied to the mathematical rigorous analysis of Hamiltonians  in quantum field theory.  Among other things, stability and instability of quantum mechanical  particle coupled to quantum fields
have  been  investigated from mathematical point of view.

This section  is the review of \cite{hs01,hir03,hss11}
and we also revise  small errors found in \cite{hs01,hir03,hss11}.

Atoms consist of charged particles and they are necessarily
coupled
to the  quantized radiation field.
In the lowest approximation,
 this interaction can be ignored and one
is led to a \S operator of the form
\eq{hpp}
\d \hp(m)=-\frac{1}{2m}\Delta  +V
\en
 for the particles only.
Under suitable conditions on $V$
the \S operator has a state of the lowest energy,
the ground state of
the atom. There has been renewed interest within mathematical physics
to understand whether this  ground state persists when
the coupling to the radiation field is included.
We will investigate here a related, but distinct problem.

In the non-relativistic approximation, the coupling to the radiation field is
described by the \pfh Hamiltonian discussed in the previous section:
\eq{1-0}
H=\frac{1}{2m}\left(\p-\ak   \A\right)^2+V+ \hf
\en
acting on the Hilbert space $\hhh$.
In essence,  $V$ is short ranged
and sufficiently shallow.
The problem of the existence of the
ground state for $H$ is usually regarded as
a stability property. One assumes that $H$ has a ground state for $\ak=0$,
which amounts to the existence of a ground state for $\hp(m)$
and
proves that $H$ has also a ground state for $\ak\not=0$.
It is then necessarily unique, since $e^{-tH}$ has a
positivity improving kernel in a suitable function space.

In contrast we  assume  that
$H$
has no  ground state for $\ak=0$.
In fact, this will be the case for a sufficiently shallow $V$ in the space dimension three.
We expect the interaction with the quantized radiation field to enhance binding.
The  non-binding potential should become binding at a sufficiently strong
coupling strength.
The enhanced binding is studied in e.g., \cite{ak03,blv05,bv04,ceh04,cvv03,hvv03,hs08,hs12,hs01,hss11}.

\begin{figure}[t]
\centering
\includegraphics[width=200pt]{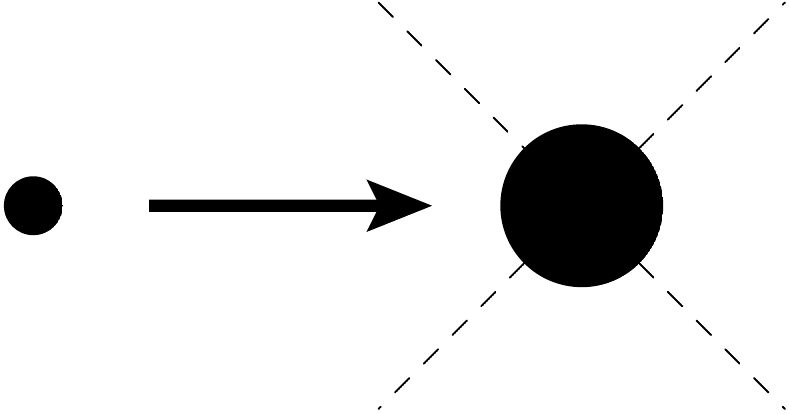}
\caption{Enhanced binding by effective mass}
\label{piceff}
\end{figure}
The physical reasoning behind such a result is simple.
As the particle binds photons it acquires
an effective mass $\mass=m+\at \mmm\|\vp/\omega\|^2$
which is increasing in $\mm$ (Figure \ref{piceff}).
Roughly speaking,
 $H$ may be replaced by
\eq{heff}
\d \eff=-\frac{1}{2\mass}\Delta+V,
\en
which binds for sufficiently strong
$\ak $. Indeed we can see that $\heff$ can be derived through the weak coupling limit of $H$ in
Section \ref{pfweak}.

\begin{figure}[t]
\centering
\includegraphics[width=200pt]{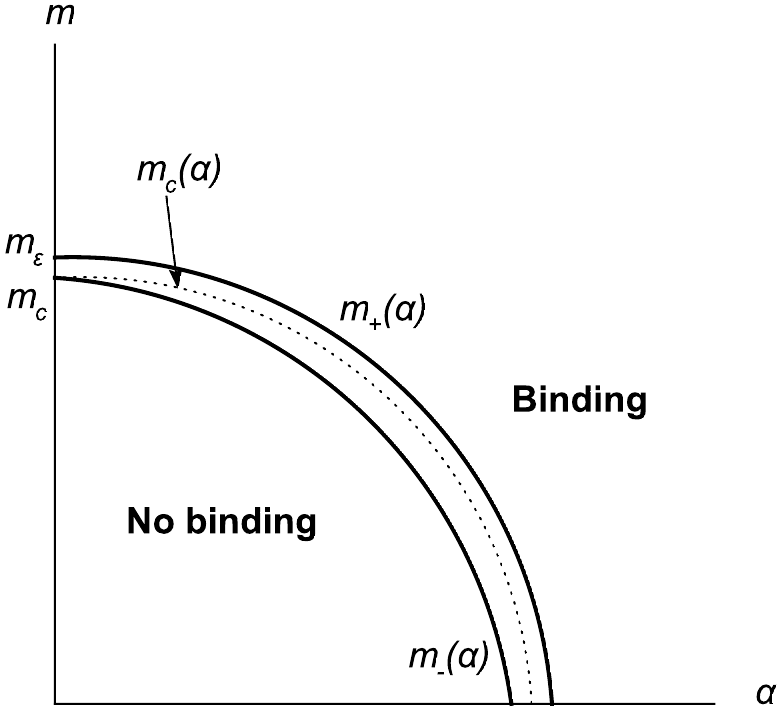}
\caption{Transition from unbinding to binding}
\label{pictra}
\end{figure}

Next let us consider
a transition from unbinding to binding as the mass $m$ is increased (Figure \ref{pictra}).
More precisely,
there is some critical mass, $m_{\mathrm{c}}$,
such that
$\hp(m)$ has no ground state for
$0 < m < m_{\mathrm{c}}$ and a unique ground state for $m_{\mathrm{c}} < m$.
In fact, the critical mass is given by
\eq{ha11}
\frac{1}{2m_{\mathrm{c}}}=
\left\|
|V|^\han \lk
-\Delta\rk\f |V|^\han
\right\|.
\en
On a heuristic level, through the dressing by photons
the particle becomes effectively more heavy,
which means that there is critical mass $m_\mathrm{c}(\ak )$ for the existence of a ground state.
 $m_\mathrm{c}(\ak )$
is expected to be decreasing as a function of $\ak $
with $m_{\mathrm{c}}(0) = m_{\mathrm{c}}$.
In particular, for fixed $m < m_{\mathrm{c}}$,  there should be an unbinding-binding transition as the coupling
$\ak $ is increased.
In case $m>m_{\mathrm{c}}$
more general techniques are available  and the existence of a unique ground state
for the full Hamiltonian $H$ is proven in   \cite{bfs99, gll01}.
The heuristic picture also asserts that the full Hamiltonian has a regime of couplings with no ground state.

\subsection{Absence of ground state}
The unbinding for the Schr\"odinger operator $\hp(m)=-\frac{1}{2\mass}\Delta+V$
is proven by the Birman-Schwinger principle\index{Birman-Schwinger principle}.
Formally one has
 $$ \hp(m) =
 \frac{1}{2m}
 (-\Delta)^\han
 \big(\one +
 2m (-\Delta)^{-\han}
 V
 (-\Delta)^{-\han}
  \big)
 (-\Delta)^\han.
 $$
 If $m$ is sufficiently small, then
 $
 2m(-\Delta)^{-\han}
 V
 (-\Delta)^{-\han}
 $
 is a strict contraction.
 Hence
the  operator $\one +
 2m (-\Delta)^{-\han}
 V
 (-\Delta)^{-\han}
 $ has a bounded inverse
 and $\hp(m)$ has no eigenvalue in $(-\infty,0]$.
More precisely the Birman-Schwinger principle
\index{Birman-Schwinger principle}
states that
\eq{a6}
{\rm dim}\one_{[\frac{1}{2m},\infty)
}
(V^\han (-\Delta)\f V^\han)
\geq
{\rm dim}
\one_{(-\infty,0]}
(\hp(m)).
\en
For small $m$ the left hand side equals $0$ and thus
$\hp(m)$ has no eigenvalues in $(-\infty,0]$.
Our approach will be to generalize
\kak{a6} to the Pauli-Fierz model with the dipole approximation.
We already see that
 $H$ can be transformed
  by $\UU$ and
one arrives at
$
\UU\f H\UU =
H_0(\ak) + W +\ground
$
as the sum of the free Hamiltonian
\eq{aa7}
 H_0(\ak) = -\frac{1}{2\mass}\Delta+\hf,
\en
involving the effective mass of the dressed particle,
the transformed interaction
$
W = T \f V  T$,
and the global energy shift $\ground $.
Effective mass  $\mass$ is an increasing function of $\ak $.

Let $h_0 = - \frac{1}{2} \Delta$.
We assume that $V \in L^1_{\rm loc}(\RR^d)$ and $V$ is relatively form-bounded
with respect to $h_0$ with relative bound $a<1$, i.e.,
$D(|V|^{1/2}) \supset D(h_0^{1/2})$ and
\begin{equation}
\label{1336}
| |V|^{1/2} \varphi \|^2 \leq a \| h_0^{1/2}\varphi \|^2 + b \| \varphi\|^2,
	\quad \varphi \in D(h_0^{1/2}),
\end{equation}
with some $b>0$.
Under  \ref{1336}
the operators
$
 R_E = \left(h_0 -E\right)^{-1/2} |V|^{1/2}$ for $
  E < 0$
are densely defined.
From \eqref{1336} it follows that
$R_E^* = |V|^{1/2}(h_0-E)^{-1/2}$ is bounded and thus
$R_E$ is closable.
We denote its closure by
the same symbol.
Let
\eq{s9}
 K_E = R_E^* R_E.
 \en
 Then $K_E$ ($E<0)$ is a bounded, positive self-adjoint operator
and it holds
\[ K_E f = |V|^{1/2}\left( h_0 -E \right)^{-1}|V|^{1/2} f, \quad f \in C_0^\infty(\RR^d). \]
Now let us consider the case $E=0$.
Let
$R_0 = h_0^{-1/2} |V|^{1/2}$.
The self-adjoint operator $h_0^{-1/2}$ has the integral kernel
$
\d  h_0^{-1/2}(x,y) =  \frac{a_d}{|x-y|^{d-1}}$
 for $ d\geq3$,
where $a_d = \sqrt{2}\pi^{(d-1)/2}/\Gamma(\frac{d-1}{2})$.
It holds that
\[ \left|(h_0^{-1/2}g, |V|^{1/2} f) \right| \leq a_d \|g\|_2 \| |V|^{1/2}f\|_{2d/(d+2)} \]
for $f, g \in C_0^\infty(\RR^3)$
by the Hardy-Littlewood-Sobolev inequality.
Since $f\in C_0^\infty(\RR^3)$ and $V \in L^1_{\rm loc}(\RR^3)$, one concludes
$\||V|^{1/2}f\|_{2d/(d+2)}
< \infty$.
Thus
$|V|^{1/2} f \in D(h_0^{-1/2})$ and
$R_0$ is densely defined.
Since $V$ is relatively form-bounded with respect to $h_0$,
$R_0^*$
is also densely defined, and
 $R_0$ is closable.
We denote the closure by the same symbol. We  define
\eq{ichinichi}
K_0 = R_0^* R_0.
\en

Next let us introduce assumptions on the external potential $V$.
\begin{assumption}
\label{cc}
$V$ satisfies
that (1) $V\leq 0$ and
(2) $R_0$ is compact.
\end{assumption}

\begin{lemma}
\label{lem1339}
Suppose Assumption \ref{cc}.
Then
\begin{itemize}
\item[(1)] $R_E$, $R_E^*$ and $K_E$ ($E \leq 0$) are compact.
\item[(2)] $\|K_E\|$ is continuous and     monotonously increasing in $E \leq 0$ and it holds that
$
\d  \lim_{E \to -\infty}\|K_E\| = 0$ and
 $ \lim_{E \uparrow 0}\|K_E\| = \|K_0\|$.
 \end{itemize}
\end{lemma}
\proof
Under (2) of Assumption \ref{cc},
$R_0^*$ and $K_0$ are compact.
Since  
\begin{equation}
\label{1240}
(f, K_E f)
 \leq (f, K_0 f), \quad f \in C_0^\infty(\RR^d),
\end{equation}
extends to $f \in L^2(\RR^3)$, $K_E$, $R_E$ and $R_E^*$ are also compact.
Thus (1) is proven.

We will prove (2).
It is clear from \kak{1240} that $K_E$ is monotonously increasing in $E$.
Since $R_0$ is bounded, \eqref{1240} holds on $L^2(\RR^d)$ and
\eq{2035}
K_E = R_0^* \left((h_0-E)\f h_0\right)
R_0
\en
 for $ E \leq 0$.
From this  one concludes that
$
\| K_E - K_{E^\prime} \|
  \leq  \|K_0\| \frac{|E-E^\prime|}{|E^\prime|}
$
for $E, E^\prime < 0$.
Hence $\|K_E\|$ is continuous in $E<0$.
We have to prove the left continuity at $E=0$.
Since $\|K_E\| \leq \|K_0\|$ ($E < 0$),
one has $ \lim\sup_{E \uparrow 0}\|K_E\| \leq \|K_0\|$.
By \eqref{2035} we see that
$K_0 = \mbox{s-}\lim_{E \uparrow 0}K_E$ and
\begin{align*}
\| K_0 f \| = \lim_{E \uparrow 0}\| K_E f \|
	\leq \left(\liminf_{E \uparrow 0}\|K_E\|\right) \|f\|, \quad f \in L^2(\RR^d).
\end{align*}
Hence we have $\|K_0\| \leq \liminf_{E \uparrow 0}\|K_E\|$
and $\lim_{E\uparrow 0}\|K_E\| = \|K_0\|$.
It remains to prove that $\lim_{E \to -\infty}\|K_E\|=0$.
Since $R_0^*$ is compact, for any $\epsilon>0$,
there exists a finite rank operator $T_\epsilon = \sum_{k=1}^n (\varphi_k, \cdot) \psi_k$
such that $n =n(\epsilon)<~\infty$, $\varphi_k, \psi_k \in L^2(\RR^d)$
and $\|R_0^* - T_\epsilon\| < \epsilon$.
Then it holds that $\| K_E \| \leq  \left(\epsilon + \|T_\epsilon h_0(h_0-E)^{-1}\|\right) \|R_0\|$.
For any $f \in L^2(\RR^d)$, we have
\[ \|T_\epsilon h_0(h_0-E)^{-1} f\|
\leq  \left( \sum_{k=1}^n \| h_0 (h_0-E)^{-1}\varphi_k \| \|\psi_k\| \right) \|f\| \]
and
 $\lim_{E \to - \infty} \|T_\epsilon h_0(h_0-E)^{-1}\| = 0$,
which completes  (2).
\qed

Let
$
\hp(m)$ be in \kak{hpp}.
By (2) of Lemma \ref{lem1339}, we have
$$\lim_{E \to -\infty}\| |V|^{1/2}(h_0-E)^{-1/2} \| = 0.$$
Therefore $V$ is infinitesimally form bounded with respect to $h_0$
and $\hp(m)$ is the self-adjoint operator associated with the quadratic form
$$f, g\mapsto
\frac{1}{m}(h_0^{1/2}f,h_0^{1/2}g) + (|V|^{1/2}f, |V|^{1/2}g)$$ for
 $f, g \in D(h_0^{1/2})$.
Note that the domain $D(\hp(m))$ is independent of $m$.

Under (2) of Assumption \ref{cc}, the essential spectrum of $\hp(m)$ coincides with that of $-\frac{1}{2m} \Delta$,
hence $\s_{\rm ess}(\hp(m))=[0,\infty)$.
Next we will estimate the spectrum of $\hp(m)$ contained in $(-\infty, 0]$. Let $\one_{({\cal O})}(T)$, ${\cal O}\subset \RR$,  be the spectral resolution of self-adjoint operator $T$ and set
$
 N_{\cal O}(T) = \dim {\rm Ran} \one_{\cal O}(T)$.
 The Birman-Schwinger principle\index{Birman-Schwinger principle} states that
\eq{s3}
\begin{array}
{ll}
\d (E<0)
&
N_{(-\infty,\frac{E}{m}]}
\lk
\hp(m)\rk=
N_{[\frac{1}{m},\infty)}
(K_{E}),\\
& \\
\d (E=0)&
N_{(-\infty,0]}
\lk
\hp(m)
\rk
\leq
N_{[\frac{1}{m},\infty)}(K_0).
\end{array}
\en
Now let us define the constant $m_{\mathrm{c}}$ by the inverse of the operator norm of $K_0$,
\eq{mzero}
 m_{\mathrm{c}} = \|K_0\|^{-1}.
 \en
\begin{lemma}\label{schr}
Suppose Assumption \ref{cc}.
\begin{itemize}
\item[(1)] If $m < m_{\mathrm{c}}$,
then
$N_{(-\infty,0]}(\hp(m))=0$.
\item[(2)] If $m > m_{\mathrm{c}}$,
then
$N_{(-\infty,0]}(\hp(m))\geq 1$.
\end{itemize}
\end{lemma}
\proof
It is immediate to see (1)
by the Birman-Schwinger principle \kak{s3}.
Suppose
 $m > m_{\mathrm{c}}$.
 Then, using the continuity and monotonicity of $ E \to \|K_E\|$, see Lemma \ref{lem1339},
there exists $\epsilon > 0$ such that
$m_{\mathrm{c}}<  \| K_{-\epsilon}\|\f \leq m$.
Since $K_{-\epsilon}$ is positive and compact,
$\| K_{-\epsilon}\| \in \s_{\rm p}( K_{-\epsilon})$ follows and hence
$N_{[\frac{1}{m}, \infty)}
(K_{-\epsilon}) \geq 1$.
Therefore (2) follows again from
the Birman-Schwinger principle.
\qed
By Lemma \ref{schr}, the critical mass at zero coupling is $m_{\mathrm{c}}(0)=m_{\mathrm{c}}$.
In the case $m>m_{\mathrm{c}}$,
by the proof of Lemma \ref{schr} one concludes that the bottom of the spectrum of $\hp(m)$
is strictly negative.
For $\epsilon>0$ we set
$
 m_\epsilon = \|K_{-\epsilon}\|^{-1}$.
 \bc{schr1}
Suppose Assumption \ref{cc} and $m>m_\epsilon$.
Then
\eq{oto}
\inf \s\left(\hp(m) \right) \leq \frac{-\epsilon}{m}.
\en
\ec
\proof
The Birman-Schwinger principle states that
$1\leq
N_{(-\infty,-\frac{\epsilon}{m}]}\left(\hp(m)
 \right)$, since $ 1/m < \|K_{-\epsilon}\|$,
 which implies
the corollary.
\qed

We extend the Birman-Schwinger type estimate to the Pauli-Fierz Hamiltonian.
\bl{koko}
Suppose Assumption \ref{cc}.
If $m < m_{\mathrm{c}}$,
then the zero coupling Hamiltonian
$
\hp(m)+\hf
$
has
no ground state.
\el
\proof
Since the Fock vacuum $\Omega$ is the ground state of $\hf$,
$\hp(m)+\hf
$ has a ground state
if and only if $\hp(m)$ has a ground state. But
$\hp(m)$ has no ground state by Lemma \ref{schr}.
Therefore
$\hp(m)+ \hf
$ has no ground state.
\qed
From now on we discuss
$\UU \f H\UU$ with $\ak \not=0$.
We have
\eq{sp4}
\UU \f H\UU=H_0(\ak)+W+\ground ,
\en
where
$\d H_0(\ak)=\hz+ \hf$ and
$
W=T\f V  T$ and $T$ is given in \kak{matsu1}.
\bt{absence}
\TTT{Absence of ground state}
\index{absence of ground state!Pauli-Fierz Hamiltonian}
{\rm \cite{hss11}}
Suppose Assumptions \ref{1v1} and \ref{cc}.
If $\mass < m_{\mathrm{c}}$,
then
$H$ has no ground state.
\et
\proof
Since $g$ is a constant,
we prove the absence of ground state of $H_0(\ak)+W$.
Since $V$ is negative, so is $W$.
Hence
$\inf \s(H_0(\ak)+W) \leq \inf \s(H_0(\ak)) = 0$.
Then it suffices to show that $H_0(\ak)+W$ has no  eigenvalues in $(-\infty,0]$.
Let $E \in (-\infty,0]$ and set
\eq{s4}
 \KK _E = |W|^{1/2}(H_0(\ak) - E)^{-1}|W|^{1/2},
 \en
 where $|W|^{1/2}$ is defined by the functional calculus.
We shall prove now that
if $H_0(\ak)+W$ has eigenvalue $E\in (-\infty,0]$, then $\KK _E$ has
eigenvalue $1$.
  Suppose that $(H_0(\ak)+W-E)\varphi = 0$ and $\varphi \not=0$, then
$\KK _E |W|^{1/2}\varphi = |W|^{1/2}\varphi$ holds.
Moreover if $|W|^{1/2}\varphi = 0$, then $W\varphi=0$ and hence $(H_0(\ak)-E) \varphi = 0$, but
$H_0(\ak)$ has no eigenvalue by Lemma \ref{koko}.
Then $|W|^{1/2}\varphi \not = 0$ is concluded
and  $\KK_E$ has eigenvalue $1$.
Then it is sufficient to see $\|\KK _E\| < 1$ to show that
$H_0(\ak)+W$ has no eigenvalues in $(-\infty,0]$.
Notice that
$\hz $ and $T$ commute, and
\[   \left\|
\lk-\Delta
\rk^\han(H_0(\ak)-E)^{-1}\lk-\Delta\rk^{1/2}
\right\| 
\leq 2\mass. \]
Then we have
\begin{align*}
 \|\KK _E\|
\leq
\left\|
|V|^\han \lk\hz\rk^{-\han}\right\|^2
=
\mass\|K_0\|=\frac{\mass}{m_\mathrm{c}}<1
\end{align*}
and the proof is complete.
\qed

Now we give examples of potentials $V$ satisfying Assumption \ref{cc}.
The self-adjoint operator $h_0^{-1}$ has the integral kernel
\[ h_0^{-1}(x,y) = \frac{b_d}{|x-y|^{d-2}}, \quad d\geq 3,  \]
with $b_d = 2{\Gamma(\frac{d}{2}-1)/\pi^{\frac{d}{2}-2}}$.
It holds that
\begin{equation}
\label{210}
(f, K_0f)
= \int dx \int dy \ov{f(x)}K_0(x,y)f(y),
\end{equation}
where
\eq{s12}
K_0(x,y) =
	b_d \frac{|V(x)|^{1/2}|V(y)|^{1/2}}{|x-y|^{d-2}},\quad d\geq3,
\en
is the integral kernel of operator $K_0$.
We recall that the Rollnik class\ko{Rollnik class} $\mathscr{R}$ of potentials is defined by
\[ \mathscr{R} = \left\{ V \Big| \int_{\BR} dx \int_{\BR} dy \frac{|V(x)V(y)|}{|x-y|^{2}} < \infty \right\}. \]
Let $d=3$.
By the Hardy-Littlewood-Sobolev inequality,
   $\mathscr{R} \supset L^p(\RR^3) \cap L^r(\RR^3)$ with $1/p + 1/r = 4/3$. In particular, $L^{3/2}(\RR^3)\subset \mathscr{R}$.

\begin{example}{\bf  ($d=3$ and Rollnik class)}
{\rm
Let $d=3$. Suppose that $V$ is negative and $V \in \mathscr{R}$.
Then
$K_0\in
L^2(\RR^3 \times \RR^3)$.
Hence $K_0$ is Hilbert-Schmidt
and  Assumption \ref{cc} is satisfied.
}
\end{example}

The example  can be extended to dimensions $d\geq 3$.
\begin{example} {\bf ($d\geq 3$ and $V\in L^{d/2}(\BR)$)}
{\rm
Let $L_w^p(\BR)$ be the set of Lebesgue measurable function $u$ such that
$\sup_{\beta>0}
\beta\left|
\{x\in\BR\|u(x)>\beta\}\right|_L^{1/p}<\infty$,
where $|E|_L$ denotes the Lebesgue measure of $E\subset\BR$.
Let $g\in L^p(\BR)$ and $u\in L_w^p(\BR)$ for $2<p<\infty$.
Define the operator $B_{u,g}$ by
$$B_{u,g}h=(2\pi)^{-d/2}\int e^{ikx} u(k)g(x)h(x)dx.$$
It is shown in \cite[Theorem, p.97]{cwi77} that
$B_{u,g}$ is a compact operator on  $\LR$.
It is known that $u(k)=2|k|^{-1}\in L_w^d(\BR)$ for $d\geq 3$.
Let $F$ denote Fourier transform on $\LR$, and suppose that $V\in L^{d/2}(\BR)$.
Then
$B_{u, |V|^{1/2}}$ is compact on $\LR$ and then
$R_0^\ast=F B_{u, V^{1/2}} F\f$ is  compact. Thus $R_0$ is also compact.
}
\end{example}

Assume that $V \in L^{d/2}(\RR^d)$.
Let us now see the critical mass of zero coupling $m_\mathrm{c}=m_0$.
By the Hardy-Littlewood-Sobolev inequality,
we have
\eq{lieb0}
|(f, K_0 f)|
\leq
D_V
\|f\|_2^2,
\en
where
$
\d
D_V=
\sqrt{2} \pi
\frac{\Gamma(\frac{d}{2}-1)}{\Gamma(\frac{d}{2}+1)}
\lk
\frac{\Gamma(d)}{\Gamma(\frac{d}{2})}\rk^{2/d}\|V\|_{d/2}^2$,
 \kak{lieb0}
is proved by Lieb
\cite{lie83}.
Then $
\|K_0 \|\leq D_V$.
By this bound  we have
$
m_\mathrm{c} \geq D_V\f$.
In particular
in the case of $d=3$,
\eq{mz}
m_\mathrm{c} \geq \frac{3}
{
\sqrt 2\pi^{2/3}{4^{5/3}}
}
\|V\|_{3/2}^{-2}.
\en

\subsection{Existence  of ground state}
In this section we investigate the existence of ground state of $H$ for sufficiently large $|\ak|$.
Let us define the Pauli-Fierz Hamiltonian
with scaled external potential $V_\k (x) = V(x/\k )/\k ^2$ by
\eq{ca}
\frac{1}{2m}(-i\nabla- \ak   A)^2+V_\k +\hf.
\en
We also define
$H(\k) $ by
$H$ with $a^\sharp$ replaced by
$\k  a^\sharp$.
Then
\eq{ki1}
H(\k)  = \frac{1}{2m}(-i\nabla-\k  \ak  A)^2 + V + \k ^2 H_{\rm f}.
 \en
We can see the unitary equivalence:
$$ \k^{-2}H(\k)\cong
\frac{1}{2m}(-i\nabla- \ak   A)^2+V_\k +\hf.
$$
Then $H(\k)$ has a ground state
if and only if
\kak{ca} has a ground state.
We furthermore introduce assumptions
on
the external potential $V$ and ultraviolet cutoff $\vp$.
Recall that $Q(k)=\ak \vp(k)/\mass(k)$.
\begin{assumption}
\label{vs}
The external potential $V$ and
the ultraviolet cutoff $\vp$ satisfy:
\bi
\item[(1)]
 $V\in C^1(\BR)$ and ${\nabla V}\in L^\infty(\BR)$,
\item[(2)] $\vp/\omega^{5/2}\in\LR$,
    \item[(3)]
$\sup_{\alpha}\|Q/\omega^{n/2}\|  <\infty$, $n=3,4,5$\footnote{\cite[Theorem 4.14]{hs01} is incorrect. The effective mass $\mass $ in \cite[Theorem 4.14]{hs01} should be changed to $\mass(k)$, and we need assumption (3) to show the enhanced binding.}.
\ei
\end{assumption}
\begin{example}
\label{goodexample2}
{\rm
We give an example of ultraviolet cutoff
satisfying both of  Assumption \ref{1v1} and  Assumption \ref{vs} (2) and (3).
Suppose that $d=3$ and
 $\vp(k)=\one_{[\lambda,\Lambda]}(|k|)$
 is
the sharp cutoff function\ko{UV cutoff!sharp}.
 See Example \ref{goodexample}.
Then
$$|\mass(k)|\geq \at \frac{4\pi^2}{3}
\one_{[\lambda,\Lambda]}(\omega(k))\sqrt{\omega(k)}.$$
We have
$$\|Q/\omega^{n/2}\|^2\leq
\frac{1}{\at}\lk\frac{3}{4\pi^2}\rk^2
\int_{\lambda\leq|k|\leq\Lambda}
\frac{1}{\omega(k)^{n+1}}dk.$$
In particular it follows that
$\lim_{\ak\to\infty}
\|Q/\omega^{n/2}\|=0$.
}
\end{example}

We drop $g$ for instance.
We reset
\eq{ss8}
H(\k)=\heff+\k ^2\hf +\dv_\k .
\en
In Theorem  \ref{wcl} we show that $H(\k)$  converges to $\heff$ as $\k\to\infty$ in some sense.
It suggests that $H(\k)$ with sufficiently large $\k$ has a ground state if $\heff$ does.
Let $m<m_\mathrm{c}$ and $\epsilon>0$.
We define
\begin{eqnarray}
\label{a10}
\ak _\epsilon  &=& \lk
\mmm\|\hat{\varphi}/\omega\|^2\rk
^{-1/2} \sqrt{m_\epsilon -m},\quad \epsilon>0,\\
\label{a20}
\ak _0&=& \lk
 \mmm\|\hat{\varphi}/\omega\|^2
 \rk
 ^{-1/2} \sqrt{m_\mathrm{c} - m},
\end{eqnarray}
where
we
recall that $m_\epsilon = \|K_{-\epsilon}\|^{-1}$ for $\epsilon\geq0$.
Note that
\begin{itemize}
\item[(1)] $|\ak|  < \ak _0$ if and only if $\mass < m_\mathrm{c}$;
\item[(2)] $|\ak | > \ak _\epsilon $ if and only if $\mass > m_{\rm \epsilon}$.
\end{itemize}
Note that $\ak _0 < \ak _\epsilon $ because of $m_\epsilon > m_{\rm c}$.
Since $\lim_{\epsilon \downarrow 0}m_\epsilon =m_\mathrm{c}$,
   it holds that
$  \lim_{\epsilon \downarrow 0}\ak _\epsilon  = \ak _0$.
We note that for $|\ak|>\ak_\eps$, $\heff$ has a ground state with negative ground state energy.

\subsubsection{Massive case}
 We introduce an artificial mass of photon, $\amass>0$,
 and
define
$$\thn(\k)=\eff+\dvn_\k+\k^2 \hfn,$$
where
 $$\hfn=\hf +\eps N=\int (\omega(k)+\amass)\add(k,j)a(k,j) dk.$$
  Using
a  momentum lattice approximation\ko{momentum lattice approximation} we will prove that
$\thn(\k)$
has
a ground state.
Let $\Gamma(l,a)$, $l=(l_1,\cdots,l_d)\in{\mathbb Z}^d$, $a>0$,
be the momentum  lattice with spacing $1/a$, i.e.,
$\Gamma(l,a)=
[\frac{l_1}{a}, \frac{(l_1+1)}{a})
\times
\cdots\times [\frac{l_d}{a}, \frac{(l_d+1)}{a})$
and
$$\PH(k) =\lkk\begin{array}{ll}0,&k\not\in \Gamma(l,a),\\
a^{d/2} ,&k\in\Gamma(l,a).
\end{array}\right.$$
For $L>0$ we define the momentum-lattice-approximated Hamiltonian
by
\eq{lattice}
\thn_{a,L}(\k)=\eff+\k^2 \nani +
\navv,
\en
where $\nani$ and $\navv$ are momentum-lattice-approximated operators
given by
\begin{align*}
\nani &=H_{{\rm f},a,L}^\amass=\int
\lk
\LLL
\PH(k) (\omega(l)+\amass)\rk
 \add(k,j)a(k,j)\ddk,\\
\navv &=
\dv_{\k, a,L}
=V(\cdot+K_{a,L}/\k)-V
 \end{align*}
and $K_{a,L}=(K_{a,L,1}, \cdots, K_{a,L,d})$ is the column of the field operator defined by
$$K_{a,L,\mu}=\frac{1}{\sqrt 2}
 \int \LLL\PH(k)(\Kp_\mu(l,j)\add(k,j)+
\ov{\Kp_\mu(l,j)} a(k,j)) \ddk.$$
Here we set
$\Kp_\mu (k,j)=e_\mu^j(k)\JI  (k)/\omega(k)^{3/2}$.
 We can show that
\begin{align}
&\|K_{a,L,\mu} \Psi\|
\leq C
\lk
\left\|
\LLL \frac{\PH \JI (l)}{\y \omega(l)^{3/2}}\right\|
+
\left\|
\LLL \frac{\PH \JI (l)}{\omega(l)^{3/2}}
\right\|\rk
(\|(\nani )^\han\Psi\|+\|\Psi\|)
\label{aef2}
\end{align}
with some constant $C$.
Here we used the bound
$$c_1\|(\nani )^\han\Psi\|\leq \|(\hfn)^\han  \Psi\|\leq c_2\|(\nani )^\han\Psi\|$$
with some constants $c_1$ and $c_2$.
\bl{l7}
It follows that
$\d \lim_{L\rightarrow \infty}
\lim_{a\rightarrow \infty}
\thn_{a,L}(\k)=\thn(\k)$
in the uniform resolvent sense.
\el
\proof
It can be seen
 that
there exists a constant $c_{a,L}$ such that
$$
\|(\nani -\hfn)\Psi\|\leq c_{a,L}\|\hfn\Psi\|$$
and
$\d \limla c_{a,L}=0$.
Moreover
\begin{align*}
\|(\dvn'_\k-\dvn_\k)\Psi\|&=
\|(V(\cdot+K_{a,L}/\k)-
V(\cdot+K/\k))\Psi\|\\
&\leq
\frac{1}{\k}
{\|\nabla_\mu V\|_\infty}
\|(K_\mu-K_{a,L,\mu})\Psi\|.
\end{align*}
Since
$
\|(K_\mu-K_{a,L,\mu})\Psi\|
\leq
c_{a,L}' (\|(\hfn)^\han\Psi\|+\|\Psi\|)$,
where
$$
c_{a,L}' =
 C
\lk \left\|\yy
\lkk
\frac{
\JI}{\omega^{3/2}}-
\LLL\frac{ \PH \JI (l)}
{\omega(l)^{3/2}}
\rkk \right\|
+\left\|\frac{\JI}{\omega^{3/2}}-\LLL
\frac{
\PH \JI(l)}{
\omega(l)^{3/2}}
\right
\|\rk
$$
with some constant $C$,
  and
$c_{a,L}'$ satisfies that
$\d \limla c_{a,L}'=0$,
we have
\begin{align*}
&\|(\thn(\k)-z)\f\Psi-(\thn_{a,L}(\k)-z)\f\Psi\|\\
&
\leq \|(\thn_{a,L}(\k)-z)\f\|
\|(\thn_{a,L}(\k)- \thn(\k))(\thn(\k)-z)\f  \Psi\|\\
&
\leq \frac{\mv  c_{a,L}'}{|{\rm Im} z|}
\lk
 \|(\hfn)^\han (\thn(\k)-z)\f \Psi\|
 +\| (\thn(\k)-z)\f \Psi\|\rk\\
&\hspace{3cm}
 +\frac{c_{a,L}}{|{\rm Im} z|}
 \|(\hfn)^\han(\thn(\k)-z)\f \Psi\|.
 \end{align*}
Since
$\|\hfn (\thn(\k)-z)\f \Psi\|\leq C'\|\Psi\|$
with some constant $C'$, we have
$$\|(\thn(\k)-z)\f\Psi-(\thn_{a,L}(\k)-z)\f\Psi\|\leq c''_{a,L}\|\Psi\|$$
with $c''_{a,L}$ such that
$\lim_{L\rightarrow \infty}
\lim_{a\rightarrow \infty} c''_{a,L}=0$.
Hence the lemma follows.
\qed
Let $f\in\LR$. We identify
$$\ell^2({\mathbb Z}^d)\ni \{f(l)\}_{l\in{\mathbb Z}^d}\cong
 a^{d/2}\sum_{l\in{\mathbb Z}^d}f(l) \PH(\cdot)\in \LR.$$
By this identification we regard $\ell^2=\ell^2({\mathbb Z}^d)$ as
the subspace of $\LR$.
Let
\begin{align}
&\hhh_a=\LR\otimes \fff(\ell^2\otimes\CC^{d-1}),\\
&
\KKKKK=
\oplus_{n=1}^\infty \fff^{(n)}
(\ell^{2\perp}\otimes\CC^{d-1}).
\end{align}
Then  the following fundamental identification follows:
\begin{align*}
\hhh&=
\LR\otimes \fff(L^2(\BR\times\{1,...,d-1\}))\\
&\cong
\LR\otimes \fff(\LR\otimes\CC^{d-1})\\
&\cong
\LR\otimes\fff([\ell^2\oplus \ell^{2\perp}]\otimes\CC^{d-1})\\
&\cong
\LR\otimes[
\fff(\ell^2\otimes\CC^{d-1})
\otimes
\fff( \ell^{2\perp}\otimes\CC^{d-1})]
\\
&\cong \hhh_a\otimes \fff( \ell^{2\perp}\otimes\CC^{d-1})\\
&=\hhh_a
\otimes
(\KKKKK\oplus \CC)
\\
&
\cong
\lk
\hhh_a \otimes
\KKKKK
\rk
\oplus
\hhh_a.
\end{align*}
We have
$$\hhh\cong \lk \hhh_a\otimes
\KKKKK
 \rk \oplus \hhh_a.$$
In particular
we can see that
\eq{l3}
\hhh_a^\perp\cong \hhh_a\otimes
\KKKKK
\en
and
that
$\thn_{a,L}(\k)$ is reduced by $\hhh_a$.
We set
\begin{align*}
\KKK&= \left.\thn_{a,L}(\k)\right\lceil_{\hhh_a},\\
\KKKK&= \left.  \thn_{a,L}(\k)\right\lceil_{\hhh_a^\perp}.
\end{align*}
Then
$$\thn_{a,L}(\k)=\KKKK\oplus \KKK.$$
We can immediately see the lemma below:
\bl{reduce2}
Under the identification \kak{l3},  we have
$$\KKKK
\cong
\KKK
\otimes \one +\one \otimes
\lc{\k^2 \nani }
{\KKKKK}
.$$
In particular
$\is(\KKKK)
\geq
\is(\KKK)+\amass$.
\el
In what follows we estimate the spectrum of $\KKK$.
\bl{42k}
Let $\Psi\in D(-\Delta )\cap D(\hf^{\han})$.
Then
\bi
\item[(1)] $\Psi\in D(\dvn_\k)$ and
\eq{bb1}
\|\dvn_\k\Psi\|\leq \CCA
(\|(\hfn)^{1/2}\Psi\|+\|\Psi\|),
\en
where
$\CCA=\frac{1}{\k}
{C\|\nabla V\|_\infty}
(\|\JI /\omega^2\|+\|\JI /\omega^{3/2}\|)$
with some constant $C$,
\item[(2)]
 $\Psi\in D(\navv)$ and
\begin{equation}
\label{b1}
\|\navv  \Psi\|\leq \CA
(\|(\nani )^{1/2}\Psi\|+\|\Psi\|),
\end{equation}
where
\begin{align*}
\CA= \theta_{\k,a,L}=\frac{1}{\k}
{C' \|\nabla V\|_\infty}
\lk
\left \|\frac{1}{\sqrt\omega}
\LLL\frac{\PH \JI (l)}{\omega(l)^{3/2}}
\right\|
+
\left \|
\LLL\frac{\PH \JI (l)}{\omega(l)^{3/2}}
\right\|\rk
\end{align*}
with some constant $C'$.
\ei
\end{lemma}
\proof
We have
$\|\dvn_\k\Psi\|\leq
\frac{1}{\k}
{\|\nabla_\mu V\|_\infty}
 \|K_{\mu,a,L} \Psi\|$.
Then \kak{bb1} follows.
\kak{b1} is similarly proven.
\qed

\begin{figure}[t]
\centering
\includegraphics[width=200pt]{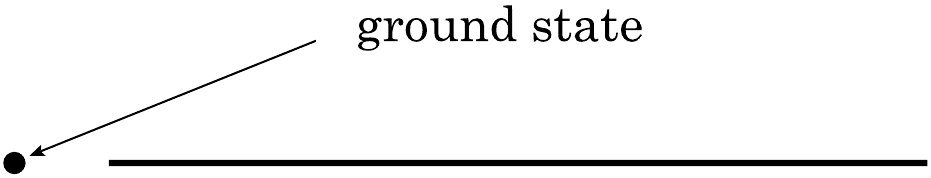}
\caption{Spectrum of massive Hamiltonian}
\label{picmassive}
\end{figure}

\bl{42L}
It follows that
\begin{align*}
\is(\thn(\k))&\leq \is(\eff) + \frac{3\CCA}{2},\\
\is(\thn_{a,L}(\k)\lceil_{\hhh_a})&\leq \is(\eff) + \frac{3\CA}{2}.
\end{align*}
\el
\proof
We write
$A\leq B$, if $D(B)\subset D(A)$ and
$(\psi, A\psi)\leq(\psi, B\psi)$ for
$\psi\in D(B)$.
We have
\begin{align*}
|(\Psi, \dvn_\k \Psi)|
\leq
\CCA
\lkk \|\Psi\| ( \|\hfn^\han\Psi\|+\|\Psi\| )\rkk
\leq
(\Psi,\CCA(\frac{3}{2}+\half \hfn)\Psi).
\end{align*}
Thus
we can get
the bound
\eq{23make}
-\CCA \left(\frac{1}{2}\hfn+\frac{3}{2}\right)
\leq \dvn_\k
\leq
\CCA \left(
\frac{1 }{2}\hfn+\frac{3}{2}\right).
\en
Hence for $f\in\ccc$,
$$\is (\thn(\k))\leq (f\otimes \Omega,
\thn(\k) f\otimes \Omega)\leq
(f,(\eff+\frac{3}{2}\CCA) f).$$
In particular, since $\ccc$ is a core of $\eff$,
 we have
$$
\is(\thn(\k))\leq \is(\eff) + \frac{3\CCA}{2}.
$$
Similarly we have
\eq{25make}
-\CA \left(\frac{1}{2}\nani +\frac{3}{2}\right)
\leq \dvn'_\k
\leq
\CA \left(
\frac{1 }{2}\nani  +\frac{3}{2}\right)
\en
and hence
$$
\is(\thn_{a,L}(\k))\leq \is(\eff) + \frac{3\CA}{2}.
$$
Then the lemma follows.
\qed
We set
$\Sigma=\is(\eff)$ and $\beff=\eff-\Sigma$.
Suppose $|\ak|>\ak_\eps$.
Since $\mass > m_{\epsilon} > m_{\epsilon/2}$,
\eq{yoshida}
\Sigma\leq \is(\hp(m_\epsilon))
\leq  -\frac{\epsilon}{2m_{\epsilon}}
\en
by Corollary \ref{schr1}.
In particular
\eq{sigma}
|\Sigma|>0.
\en
For a self-adjoint operator $M$,
the spectral projection of $M$
on a Borel set ${\cal B}\subset \RR$ is denoted by
$E^M_{\cal B}$.
\bl{gr}
Suppose $|\ak|>\ak_\eps$.
Let $a$,  $L$ and $\k$  be sufficiently large such that
$\min \{|\Sigma|/3, 2\k^2 \}>\CA$.
Then for $\amass$ such that $|\Sigma|>3\CA+\amass$,
we have
\begin{align*}
&\KKK-\is(\KKK)
-\amass
\geq
\X\otimes\lk(\k^2-\frac{\CA}{2})
 \nani -3\CA-\amass\rk.
\end{align*}
\end{lemma}
\proof
We directly see by Lemma \ref{42L} that
\begin{align*}
&\KKK-\is(\KKK)-\amass\\
&= \eff+\navv +\k^2 \nani   -\is(H)-\amass\\
&
\geq \eff+\navv +\k^2 \nani  -\frac{3}{2}\CA -\Sigma -\amass\\
&
\geq \eff+(\k^2-\frac{\theta}{2})
\nani  -\frac{3}{2}\CA-\frac{3}{2}\CA-\Sigma -\amass\\
&
= \beff +(\k^2-\frac{\CA}{2})  \nani  -3\CA -\amass\\
&
\geq |\Sigma|\Y\otimes\one
 -\CA'(\X+\Y)\otimes \one
  +  (\k^2-\frac{\CA}{2})(\X+\Y)\otimes  \nani ,
  \end{align*}
where $\CA'=3\CA +\amass$.
Then
\begin{align*}
&\KKK-\is(\KKK)-\amass\\
&\geq (|\Sigma|-\CA') \Y \otimes 1
+
(\k^2-\frac{\CA}{2})\Y \otimes  \nani\\
&\hspace{3cm}+\X\otimes \lk
(\k^2-\frac{\CA}{2}) \nani -\CA'\rk.
\end{align*}
Since
$|\Sigma|-\CA'=|\Sigma|-3\CA-\amass>0$
and
$\k^2-\frac{\CA}{2}>0$ by the assumption,
we have
$$
\KKK-\is(\KKK)-\amass \geq
\X\otimes \lk  (\k^2-\frac{\CA}{2}) \nani -\CA'\rk.$$
Thus the lemma follows.
\qed
Set
$T=\KKK-\is(\KKK)
-\amass$ as an operator in $\hhh_a$.
Define
$\hhh_a(+)=E^T_{[0,\infty)}\hhh_a$ and
$\hhh_a(-)=E^T_{[-\amass,0)}\hhh_a$.
\bl{l5}
Suppose $|\ak|>\ak_\eps$
and  that
$\min \{|\Sigma|/3, 2\k^2\}>\CA$.
Then
for $\amass$ such that
$|\Sigma| > 3 \CA + \amass$,
$\lc{T}{\hhh_a(-)} $ has a purely discrete spectrum, i.e.,
$$
\s(\KKK)\cap [\is(\KKK), \is(\KKK)+\amass)
\subset \s_{\rm disc}(\KKK).
$$
\el
\proof
Let $\{\phi_n\}_n$ be a complete orthonormal system  of $\hhh_a(-)$
and $\{\psi_m\}_m$ that of
 $\hhh_a(+)$.
We see that by Lemma \ref{gr},
$$0\geq \tr \lc{T}{\hhh_a(-)}=  \sum_n(\phi_n, T\phi_n)
\geq\sum_n(\phi_n, T'\phi_n),$$
where
$T'=\X\otimes\lk(\k^2-\frac{\CA}{2}) \nani -3\CA-\amass\rk$.
Set
$T'_-=T'E^{T'}_{(-\infty, 0)}$.
Then
$$0\geq \tr \lc{T}{\hhh_a(-)}\geq \sum_n(\phi_n, T'_-\phi_n)
\geq \sum_n(\phi_n, T'_-\phi_n)+
\sum_m(\psi_m, T'_-\psi_m)
=
\tr T_-'.$$
Hence we obtain that
\begin{align*}
\left|
\tr \lc{T}{\hhh_a(-)}
\right|
\leq
\left |\tr T_-'
\right|
=\tr \X\times
\left|
\tr
\lk (\k^2-\frac{\CA}{2})
\lc {\nani }{\hhh_a}
-3\CA-\amass
\rk_-
\right|,
\end{align*}
where
$\lk \cdots \rk_-$ denotes the negative  part of
 $\lk \cdots \rk$.
Since
$\sigma(\lc{\nani }{\hhh_a})=
\sigma_{\rm disc}(\lc{\nani }{\hhh_a})$ and
$\left|\tr\X\right|<\infty$,
it follows that
$\left|
\tr \lc{T}{\hhh_a(-)}\right|<\infty
$.
Thus the lemma follows.
\qed
\bl{l6}
Suppose that
$\min \{|\Sigma|/3, 2\k^2\}>\CA$.
Then
for $\amass$ such that
$|\Sigma| > 3 \CA + \amass$,
it follows that
$$\H{\thn_{a,L}(\k)}.$$
\el
\proof
We have by Lemmas \ref{reduce2} and \ref{l5},
\begin{align*}
&\s(\thn_{a,L}(\k))=
\s(\KKKK)
\cup
\s(\KKK),\\
&
\s(\KKKK)
\subset[\is(\KKK)+\amass,\infty),\\
&\s(\KKK)
\cap [\inf\s(\KKK),
\inf\s(\KKK)
+\amass)
\subset \s_{\rm disc}(\KKK).
\end{align*}
Notice that $\is(\KKK)=\is(\thn_{a,L}(\k))$. Then the lemma follows.
\qed
Now we can show the existence of ground state of massive Hamiltonian $H^\amass(\k)$.
\bl{l8}
Suppose $|\ak|>\ak_\eps$
and that
$\min\{|\Sigma|/3,2\k^2\}>\CCA$.
Then
for $\amass$ such that
$|\Sigma| > 3 \CCA + \amass$,
$$\H{\thn(\k)}.$$
In particular $\thn(\k)$ has a ground state.
\el
\proof
Note that
$\d \lim_{L\rightarrow \infty}\lim_{a\rightarrow \infty}\CA=\CCA$.
Then by Lemmas \ref{l7} and \ref{l6}, the lemma follows.
\qed
See Figure \ref{picmassive} for the spectrum of massive Pauli-Fierz Hamiltonian.

\subsubsection{Massless case}
 A ground state of  $\thn$ is denoted by $\grn$.
\begin{lemma}
\label{N}
Suppose $|\ak|>\ak_\eps$,
 Assumptions \ref{1v1} and  \ref{vs},
and that $\min\{|\Sigma|/3,2\}>\theta$.
Then
for $\amass$ such that
$|\Sigma| > 3 \CCA + \amass$,
\eq{nen}
\frac{\|\nf^\han\grn\|}{\|\grn\|}
\leq
\frac{1}{\k^2}
{C \|\JI /\omega^{5/2}\| \mv }
\en
with some constant $C$.
\end{lemma}
\proof
We set
$E=\is(\thn(\k))$. Since
$$[\thn(\k), a(k,j)]=-(\omega(k)+\amass)a(k,j) +[\dvn_\k,a(k,j)],$$
we have
\begin{align*}
\thn(\k)  a(k,j)\grn=&-(\omega(k)+\amass) a(k,j)\grn
+
Ea(k,j)\grn+[\dvn_\k, a(k,j)]\grn.
\end{align*}
Hence we derive that
\eq{pth}
(\thn(\k)-E+\omega(k)+\amass)a(k,j) \grn=[\dvn_\k, a(k,j)]\grn
\en
and
\begin{align*}
\left[\dvn_\k, a(k,j)\right]
=\left[V(\cdot+\frac{1}{\k}K),a(k,j)\right]
=\apm\left[V,\ap a(k,j)\apm\right]\ap.
\end{align*}
Since
$$\ap a(k,j)\apm=a(k,j)-\frac{1}{\k}\frac{i}{\sqrt 2 } (-i\nabla_\nu)  \rrr,$$
it follows that
\begin{align*}
\left[\dvn_\k,a(k,j)\right]&=
\apm\left[V,
-\frac{i}{\sqrt 2\k } (-i\nabla_\nu)  \rrr\right]\ap\\
&=\frac{1}{\k}
\apm\lk   \frac{1}{\sqrt 2} (\nabla_\nu V) \rrr  \rk\ap.
\end{align*}
Thus we obtain the pull-through formula:
\begin{align}
\label{id}
a(k,j)\grn=
\frac{1}{\k}
(\thn(\k)-E+\omega(k)+\amass)\f
\apm\lk \frac{1}{\sqrt 2} (\nabla_\nu V) \rrr  \rk\ap
\grn.
\end{align}
Using identity \kak{id} we see that
\begin{align*}
&\|N^\han \grn\|^2
=\jjj\int\|a(k,j)\grn\|^2 dk \\
&
=\frac{1}{2\k^2}
 \jjj\int \left\|
(\thn(\k)-E+\omega(k)+\amass)\f
 \apm
  (\nabla_\nu V)\rrr
  \ap
\grn
\right\|^2 dk
\end{align*}
We then estimate as
\begin{align*}
\frac{\|N^\han \grn\|^2}{\|\grn\|^2}
&
\leq
\frac{1}{2\k^2}
\jjj\int \lk\frac{1}{\omega(k)}
\|\nabla_\nu  V\|_\infty\rk^2
\left|
  \Kp^\nu (k,j)
  \right |^2
dk\\
 &
 \leq
\frac{1}{\k^2}
C {\mv }^2
 \|\JI/\omega^{5/2}\|^2.
 \end{align*}
 Hence the lemma follows.
\qed

\bl{Q}
 Suppose $|\ak|>\ak_\eps$,
 Assumptions \ref{1v1} and \ref{vs}.  Let $P_\Omega$ be the projection onto
$\{\ak \Omega\ |\ \ak \in\CC\}$
and
$\tama=E_{[\Sigma+\delta, \infty)}^{\eff}\otimes P_\Omega$
with some $\delta>0$ such that
$\delta>\frac{3}{2}\CCA$.
Suppose that
$\min\{|\Sigma|/3,2\k^2\}>\CCA$.
Then
for $\amass$ such that
$|\Sigma| > 3 \CCA  + \amass$,
it follows that
\eq{q1}
\frac{\|\tama\grn\|}{\|\grn\|}
\leq
\sqrt{\frac{\CCA}{\delta -\frac{3}{2}\CCA}}.
\en
\el
\proof
Since $(\grn, \tama(\thn(\k)-\is(\thn(\k)))\grn)=0$, we have
$$(\grn, \tama (\eff-\is(\thn(\k)))
\grn)=-(\grn, \tama \dvn_\k  \grn).$$
The left-hand side above is estimated as
\begin{align*}
(\n, \tama (\eff-\is(\thn(\k)))\grn)
\geq
 (\Sigma+\delta -\is(\thn(\k))) (\grn, \tama \grn).
 \end{align*}
 Note that
\begin{align*}
 \Sigma+\delta -\is(\thn(\k))\geq
\Sigma+\delta-\Sigma- \frac{3}{2}\CCA
=\delta -\frac{3}{2}\CCA>0.
\end{align*}
Then
$$
(\n, \tama     (\eff -\is(\thn(\k))+ g)  \n)\geq
(\delta -\frac{3}{2}\CCA)\|\tama \n\|^2>0.$$
Moreover
\begin{align*}
|(\n, \tama \dvn_\k\n)|&
=|(\dvn_\k \tama \n, \n)|
\leq
\|\dvn_\k \tama \n\|\|\n\|\\
&\leq
\CCA\lk \|\hf^\han \tama \n\|+\|\tama \n\|\rk\|\n\|\\
&
=\CCA\|\tama \n\|\|\n\|\leq \CCA\|\n\|^2.
\end{align*}
Hence we have
$$0< (  \delta -\frac{3}{2}\CCA)\|\tama \n\|^2\leq \CCA\|\n\|^2.$$
The lemma follows.
\qed
We normalize $\grn$, i.e., $\|\grn\|=1$.
 Take a subsequence $\amass'$ such that
$\nn$ weakly converges to a vector $\gr$ as $\amass'\rightarrow \infty$.
\bp{ahi}
{\rm \cite[Lemma 4.9]{ah97}}
Let $S_n$ and $S$ be self-adjoint operators on a Hilbert space $\WWW$, which have a common core $D$ such that $S_n\to S$ on $D$ strongly as $n\to \infty$.
Let $\psi_n$ be a normalized eigenvector of $S_n$ such that
$S_n=E_n\psi_n$,  $E=\limn E_n$ and the weak limit $\psi=w-\limn \psi_n\not=0$ exist.
Then
$S\psi=E\psi$. In particular if $E_n$ is the ground state energy,
then $E$ is the ground state energy of $S$ and $\psi$ is a ground state of $S$.
\ep
\proof
Since $S_n$ converges to $S$ in the strong resolvent sense by the assumption,
we can see that
$\limn (\phi, (S_n-z)\f \psi_n)=(\phi, (S_n-z)\f \psi)$ for any $\phi\in\WWW$.
This implies that $(S_n-z)\f \psi=(E-z)\f\psi$ and then $S\psi=E\psi$.
 \qed

Now we are in the position to state the main theorem in Section 4.
\begin{theorem}
\label{maintheorem}
\TTT{Enhanced binding}
\index{enhanced binding}
{\rm \cite{hs01}}
Suppose
Assumptions~\ref{1v1} and
  \ref{vs}. Then
for any $\epsilon>0$,
there exists $\k _\epsilon$ such that
for all $\k  >\k _\epsilon$,
$H(\k)$ has a unique ground state
for all
$\ak $ such that $|\ak|  >\ak _\epsilon $.
\end{theorem}
\proof
Let $E_\amass=\is(H^\amass(\k))$ and $E=\is(H(\k))$.
Since $H^\amass(\k)\to H(\k)$ as $\amass\to0$ in the strong resolvent sense,
$\limsup _{\amass\to 0}E_\amass\leq E$ follows.
On the other  hand  we notice that $E_\amass\geq E+\amass(\grn, N\grn)$. Since
$\amass(\grn, N\grn)\to 0$ as $\amass\to0$,
$\liminf _{\amass\to 0}E_\amass\geq  E$ follows.
Thus $\lim_{\amass\to 0}E_\amass=E$.
By Proposition \ref{ahi}
it is enough to prove $\gr\not=0.$
Note that
$N+P_\Omega\geq \one $.
Hence
\begin{align*}
\one \otimes N+(\XX+\YY)\otimes P_\Omega
=\one \otimes N+\XX\otimes P_\Omega+\tama \geq \one,
\end{align*}
and
\eq{st}
\XX\otimes P_\Omega\geq \one -\one \otimes N-\tama .
\en
Suppose that
$\min\{|\Sigma|/3,2\k^2\}>\CCA$.
and  $\delta>\frac{3}{2}\CCA$.
Then
for $\amass'$ such that
$|\Sigma| > 3 \CCA + \amass'$,
we have by \kak{st}, Lemmas \ref{N} and \ref{Q},
\begin{align*}
&(\nn, \XX\otimes P_\Omega\nn)\\
&
\geq 1-(\nn, N\nn)-(\nn,\tama \nn)\\
&
\geq
1-{\frac{1}{\k^2}
C\|\JI/\omega^{5/2}\|}
\mv -
\frac{\CCA}{\delta -\frac{3}{2}\CCA}.
\end{align*}
Note that $\sup_\ak\|\JI/\omega^{5/2}\|<\infty$ and
$\d \lim_{\k\rightarrow\infty} \frac{\CCA}{\delta-\frac{3}{2}\CCA} =0$ uniformly with respect to $\ak$.
Hence for sufficiently large $\k$,
$(\nn, \XX\otimes P_\Omega\nn)>\eta$
follows uniformly in $\amass'$ and $\ak$ with some $\eta>0$.
Take $\amass'\rightarrow 0 $ on
both sides above.
Since $\XX\otimes P_\Omega$ is a finite rank operator,
we see that
$\XX\otimes P_\Omega \nn\to \XX\otimes P_\Omega \gr$ strongly and
$(\gr, \XX\otimes P_\Omega \gr)>\eta$.
In particular $\gr\not=0$. Then $\gr$ is a ground state of $H(\k)$.
\qed

\begin{figure}[t]
\centering
\includegraphics[width=200pt]{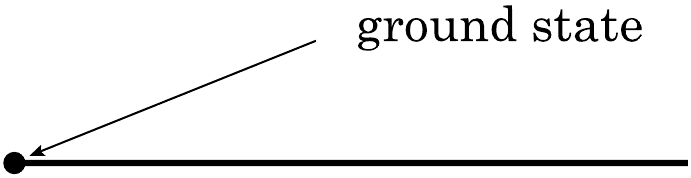}
\caption{Spectrum of massless Hamiltonian}
\label{picmassless}
\end{figure}

We can also show the existence of ground state for the Pauli-Fierz Hamiltonian without  scaling parameter.
\bt{noscaling}
\TTT{Enhanced binding, no scaling}
\index{enhanced binding!Pauli-Fierz Hamiltonian!without scaling}
{\rm \cite{hs01}}
Let $\k=1$, i.e.,  the Hamiltonian is not scaled.
Suppose Assumption~\ref{1v1},  and
(1) and (2) of Assumption   \ref{vs},
and
 that
\eq{alphabound}
\lim_{\ak\to\infty}
\|\JI/\omega^{n/2}\|=0,\quad n=3,4,5.
\en
Then
there exists $\ak_\ast>\ak_\eps$ such that
for all $\ak$ with $|\ak|>\ak_\ast$, $H$ has a ground state.
\et
\proof
By \kak{alphabound} we can see that
$\theta_\k\to 0$ and $\|\JI/\omega^{5/2}\|\to 0$ as $\ak\to\infty$.
Then
Lemma \ref{l8} holds for sufficiently large $\alpha$ with $\k=1$.
Then the massive ground state $\grn$ exists.
Furthermore we have
\begin{align*}
(\nn, \XX\otimes P_\Omega\nn)
\geq
1-{C\|\JI/\omega^{5/2}\|} \mv -
\frac{\theta_1}{\delta -\frac{3}{2}\theta_1},
\end{align*}
where $\theta_1$ is $\CCA$ with $\k=1$.
Since $\lim_{|\ak|\to\infty}\|\JI/\omega^{5/2}\|=0$ and
$\lim_{|\ak|\to\infty}
\frac{\theta_1}{\delta -\frac{3}{2}\theta_1}=0$,
we can conclude that
$\gr\not=0$ for sufficiently large $|\ak|$.
Then the corollary follows.
\qed
An example of \kak{alphabound} is given in Example \ref{goodexample2}.
See Figure \ref{picmassless} for the spectrum of massless Pauli-Fierz Hamiltonian.

\subsection{Transition from unbinding to binding}
In the previous sections we show the absence and the existence of ground state.
Combining these results we can construct examples of the Pauli-Fierz Hamiltonian
having
transition  from unbinding to binding according to the value of coupling constant  $\ak$.
\bl{abstra}
Suppose
Assumptions~\ref{1v1} and
  \ref{cc}.
  Then  $H(\k)$ has no ground state
for all $\k  >0$ and
all $\ak $ such that $|\ak|  <\ak _0$.
\el
\proof
Define the unitary operator $u_\k $ by $(u_\k  f)(x) = k^{d/2} f(x/\k )$.
Then we infer $V_\k  = \k ^{-2}u_\k  V u_\k ^{-1}$,
$-\Delta = \k ^{-2}u_\k  (-\Delta) u_\k ^{-1}$ and
\begin{align*}
\| |V_\k |^{1/2}(-\Delta)^{-1}|V_\k |^{1/2}\|
 = \k ^{-2} \| u_\k  |V|^{1/2} u_\k ^{-1}(-\Delta)^{-1} u_\k  |V|^{1/2} u_\k ^{-1} \|
 = \| K_0\|.
\end{align*}
Then the lemma follows from  Theorem \ref{absence}.
\qed

\bt{volume}
\TTT{Transition from unbinding to binding}
{\rm \cite{hss11}}
Suppose
Assumptions~\ref{1v1} , \ref{cc} and
  \ref{vs}.
Let arbitrary $\delta>0$ be given.
Then there exists
an external potential $\tilde V$ and constants
$\ak _+>\ak _-$
such that
\bi
\item[(1)]
$0<\ak _+-\ak _-<\delta$;
\item[(2)]
$H$
has a ground state for $|\ak| >\ak _+$
but  no ground state for $|\ak| <\ak _-$.
\ei
\et
\proof
For $\delta>0$  we take $\epsilon>0$ such that $\ak _\epsilon -\ak _0
<\delta$.
Take a sufficiently large $\k $,
 and set $\tilde V(x)=V(x/\k )/\k ^2$.
Define $H$ by the Pauli-Fierz Hamiltonian with potential $\tilde V$.
Then $H$ has a ground state for $|\ak|>\ak_\eps$ by Theorem \ref{maintheorem}, and
$H$ has no ground state for $|\ak|<\ak_0$ by Lemma \ref{abstra}.
Set $\alpha_\eps=\alpha_+$ and $\alpha_0=\alpha_-$. Then the proof is completed.
\qed


\subsection{Enhanced binding by UV cutoff}
We can also consider the enhanced binding by UV cutoff.
In Example \ref{goodexample} we give the example of UV cutoff function:
\ko{UV cutoff!sharp}
\eq{UVs}
\vp(k)=\one_{[\lambda,\Lambda]}(k).
\en
In this section we suppose that  $\vp$ is  \kak{UVs} and the dimension  $d=3$.
Thus  $\mass=m+\frac{8}{3}\pi \at (\Lambda-\lambda)$ and
 we have the corollary below.
\bc{absUV}
\TTT{Absence of ground state}
\index{absence of ground state!Pauli-Fierz Hamiltonian}
Suppose Assumptions \ref{1v1} and \ref{cc} and
\eq{uv}\Lambda< \frac{8}{3\pi} \ak^{-2}(m_c-m)+\lambda.
\en
Then
$H$ has no ground state.
\ec
\proof
\kak{uv} implies that $\mass<m_c$. Then the corollary follows from Theorem \ref{absence}.
\qed
We can also show the existence of ground state for sufficiently large $\Lambda$.
\bc{ex UV}
\TTT{Enhanced binding}
\index{enhanced binding!Pauli-Fierz model}
{\rm \cite{hir03}}
Suppose Assumption~\ref{1v1},  and
(1) and (2) of Assumption   \ref{vs}.
Then there exists $\Lambda_\ast$ such that $H$ has a ground state for $\Lambda>\Lambda_\ast$.
\ec
\proof
We notice that
\begin{align*}
\mass(k)&=
m+\at \frac{8\pi}{3}
 (\Lambda-\lambda)\non \\
&\label{tobira}-
\frac{\at}{2}
\frac{8\pi}{3}
\lk
|k|\log
\left|
\frac{(|k|+\Lambda)(|k|-\lambda)}
{(|k|+\lambda)(|k|-\Lambda)}
\right|
-i\pi
\one_{[\lambda,\Lambda]}(|k|)\sqrt{|k|}\rk.
\end{align*}
Then
we have
\begin{align*}
\|Q/\omega^{n/2}\|^2=\at \int_{\lambda\leq|k|\leq \Lambda}
\frac{1}{\mass(k)^2\omega(k)^n}dk
\end{align*}
and
$$\one_{[\lambda,\Lambda]}(k)\frac{1}{\mass(k)^2\omega(k)^n}\leq
\lk\frac{3}{4\pi^2 \at}\rk^2\frac{1}{\omega(k)^{n+1}}.$$
Since the right and side above is integrable  for $n=3,4$ and $5$.
The  the Lebesgue dominated convergence theorem yields that
$\lim_{\Lambda\to\infty} \|Q/\omega^{n/2}\|=0$.
Hence in a similar way to Theorem \ref{noscaling} we can prove the corollary.
\qed

\cleardoublepage
\part{The Nelson model}
\section{The Nelson Hamiltonian}
\subsection{The Nelson Hamiltonian}
We begin with giving the definition of the Nelson Hamiltonian.
Let $\fff$ be the Boson Fock space over $\LR$.
The creation operator and the annihilation operator  satisfy canonical commutation relations\ko{canonical commutation relation} on $\ffff$:
\eq{nel1}
[a(f), \add(g)]=(\bar f, g),\ \ \ [a(f), a(g)]=0=[\add(f),\add(g)].
\en
 For $h\in\LR$, the field operator is defined by
 \eq{field}
 \phi(h)=\frac{1}{\sqrt 2} \int
  \lk
  \add(k) \hat h(-k)+a(k) \hat
 h(k)\rk  dk.
 \en
 Let
$\hf$ be   the free field Hamiltonian
\ko{free field Hamiltonian!Nelson Hamiltonian} with
 the dispersion relation $\omega(k)=|k|$.

  In order to study the binding of $N$ particle system by
   the linear coupling with the scalar quantum field,
  $N$ particles assumed to be independent of each other, and then
  there is no external potential linking two particles.
  Thus the  $N$ particle Hamiltonian $\hp$
  is defined by
the self-adjoint operator  on $\LRNd$  by
 \eq{hp}
 \hp=\sum_{j=1}^N  \left( -\frac{1}{2m_j}\Delta_j+V_j \right),
 \en
 where $m_j$  is the mass of the $j$-th particle and $V_j=V_j(x_j)$
 external potential depending only on $x_j$.
 Hamiltonian  $\hp$  does not necessarily have ground states.
 For  example,  with  sufficiently shallow potential $V_j$'s,
 $\hp$ has no ground states.

 The state space of the $N$ particle coupled to
 the scalar quantum  field
 is
 \eq{state}
 \hhh=L^2(\RR^{dN})\otimes\fff.
 \en
Let
 \eq{dec}
 H_0=\hp\otimes
\one +\one \otimes \hf \en be
 the non-interacting Hamiltonian.
\begin{definition}
\TTT{Nelson Hamiltonian}\ko{Nelson Hamiltonian}
The Nelson Hamiltonian $H$ on
$\hhh$
  is defined by \eq{nelson} H=H_0+\hi, \en
where
 $\hi$ is given by
$$
\hi=\sum_{j=1}^N \ak _j\int^\oplus_{\RR^{dN}} \phi_j(x_j) dx.
$$
 Here $\ak _j$'s  are real coupling constants,
   we identify $\hhh$ as
 $\hhh \cong \int^\oplus_{\RR^{dN}} \fff dx$
 and
 $\phi_j(x)$, $x\in\BR$,  is given by
$$\phi_j(x)=
 \frac{1}{\sqrt 2}\int
\lk \add(k) \lambda _j(-k) e^{-ikx}+a(k) \lambda _j(k)
e^{ikx} \rk  dk.$$
\end{definition}
We will give assumptions on $\lambda_ j$   later.
 Since  the semigroup  $e^{-tH}$ is ergodic,
  the uniqueness of the ground state of $H$ follows.

The existence  of ground states of the Nelson
Hamiltonian  has been investigated in the last decade.
 \cite{bfs98-a,bfs98-b}  proved the existence of ground
 states under some conditions.
 \cite{ger00,spo99} remove the weak
 coupling condition, namely they show the existence of ground states
 of the Nelson Hamiltonian for arbitrary values of
  a coupling constant.
 \cite{sas05} shows the existence
 of  a ground state
  with general external potentials including the
 Coulomb potential, and \cite{hhs05} shows the existence of
 a ground state without cutoffs.

\subsection{Enhanced  binding}
 As is mentioned in the previous section,
  the existence of the ground state of the Nelson Hamiltonian
  has been proven under some general conditions.
  One of fundamental assumption among them is that $\hp$ has a ground state.
  In this note  we remove this condition.

\begin{figure}[t]
\centering
\includegraphics[width=200pt]{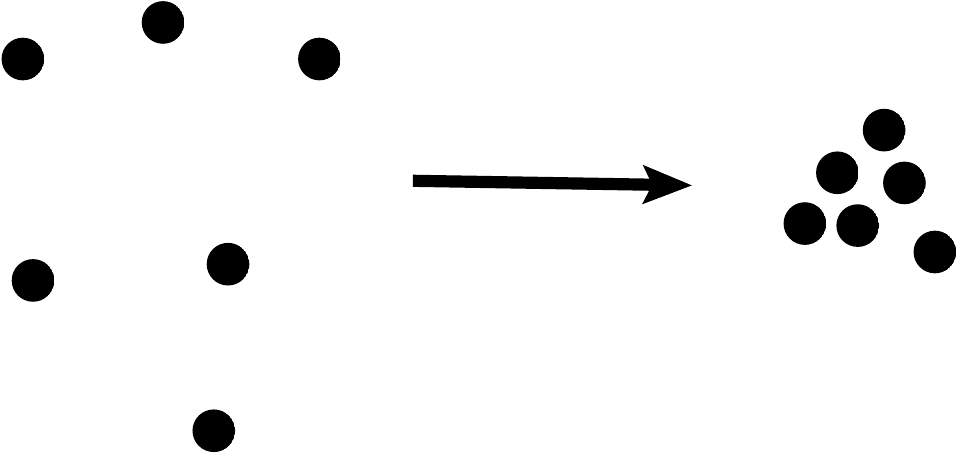}
\caption{Enhanced binding by effective potential}
\label{picpot}
\end{figure}

   If  there is no interaction between
particles, the $j$-th particle is governed only 
 by the potential $V_j$.
In this case, if $V_j$'s are sufficiently shallow,  external
potential $\sum_{j=1}^N V_j$ can not trap these
 particles.
But if these particles attractively interact with each other by an
effective potential derived from  the  scalar quantum  field,
 particles close up with each other and behave just like as one particle
 with { heavy}  mass $\sum_{j=1}^N m_j$.
 We will see that effective potential is of the form
\eq{vvv} \veff(x)=-\frac{1}{4}\sum_{i\not=j}^N \ak _i\ak _j
\int_\BR \frac{{\lambda_ i(-k)}\lambda_ j(k)}{\omega(k)} e^{-ik\cdot
(x_i-x_j)}dk.
\en
    Effective potential
$\veff$ depends on the choice of cutoff function $\lambda_ j$'s.
 A typical
example of $\veff$ is a three dimensional $N$-body smeared Coulomb
potential:
$$\veff(x_1,...,x_N)=-\frac{1}{8\pi}\sum_{i\not=j}^N
 \frac{\ak _i\ak _j}{|x_i-x_j|}  \varpi (|x_i-x_j|),$$
where $\varpi (|x|)>0$ holds for a sufficiently small $|x|$.
 For this case it is determined by signs of
$\ak _1,...,\ak _N$ whether  ${V_{\rm eff}}$ is attractive or repulsive
for sufficiently
 small $|x_i-x_j|$.
We can see from \kak{vv} that  an identical sign of coupling
constants and
 $${\rm supp}\lambda_ i\cap{\rm supp}\lambda_ j\not=\emptyset,\quad i\not =j,$$
 derive attractive effective potentials, and  which enhances binding
of the system.
 If $N$ is large enough,
 this \textit{one particle} has sufficiently
heavy mass and is bounded  by external potential $\sum_{j=1}^N V_j$,
and finally
  it is trapped.
  See Figure \ref{picpot}.
Heuristically we see that
\eq{fight}
H\sim
  -\frac{1}{2\sum_{j=1}^N m_j}\Delta+\sum_{j=1}^N V_j=
\frac{1}{\sum_{j=1}^N m_j}\lk -\frac{1}{2}\Delta+
 (\sum_{j=1}^N m_j)\sum_{j=1}^N V_j\rk.
 \en

\subsection{Weak coupling limit}
\ko{weak coupling limit}
In this section we see the
relationship between an enhanced binding and
a weak coupling limit, which has been seen in the case of the Pauli-Fierz model in Section \ref{pfweak}.
  In our model under consideration, it
is seen that the enhanced binding is derived from the effective
potential $\veff$ which is the sum of potentials between two
particles.
 Alternatively the  effective potential can be
derived from a  weak coupling limit \cite{dav77,dav79,hir98,hir99},
 which is
one of a key ingredient of this paper.
 We  outline a weak
coupling limit by path measures.
  Let us introduce a scaling in the Nelson model as
\eq{wn}
H(\k )=\hp +\k ^2 \hf+\k  \hi,
\en
where $\k >0$ is a scaling parameter.
  Let
  $${\ms X}_+=C([0,\infty);\RR^{dN})$$ be the set of continuous
   paths valued in $\RR^{dN}$.
 Then $e^{-tH(\k )}$ can be expressed by  a path measure as
  \eq{br2} (f\otimes \Omega, e^{-tH(\k )} g\otimes \Omega)_\hhh
  =
\int_{{\ms X}_+\times\RR^{dN}} \ov{f(X_0)} g(X_t) e^{-\int_0^T V(X_s)
ds} e^{W_\k }d{\cal W}^x dx, \en
 where
 $X_t=(X_t^1,...,X_t^N)$, $X_t^j(w)=w^j(t)\in\BR$, $w=(w^1,...,w^N)
 \in{\ms X}_+$,
  denotes the point evaluation of $w\in{\ms X}_+$,
 $d{\cal W}^x $  the Wiener measure on ${\ms X}$
starting from $x$ at $t=0$  and
 \begin{align}
 &V(X_s)=
 \sum_{j=1}^N V_j(X_s^j),\\
&
\label{br8} W_\k = \frac{1}{4}\sum_{i,j=1}^N
\ak _i\ak _j \int_0^T ds\int_0^T dt \int_\BR
{{\lambda_ i(-k)}\lambda_ j(k)} \k ^2 e^{-\k ^2 |s-t|\omega(k)}
e^{-ik\cdot (X_s^i -X_t^j)} dk.
 \end{align}
 Informally taking $\k \rightarrow \infty$ in \kak{br8}, we see
that only the diagonal part of  $\int_0^T ds \int_0^T dt$  survives
and the off diagonal part is dumped by
 the factor
 $$\k ^2 e^{-\k ^2 |s-t|\omega(k)}
 \rightarrow
 \delta(s-t)\frac{2}{\omega(k)}$$
as $\k \rightarrow \infty $.
Thus we have \eq{br3}
 W_\k \rightarrow
\frac{1}{2}\sum_{i,j=1}^N \ak _i\ak _j \int_0^T ds  \int_\BR
\frac{{\lambda_ i(-k)}\lambda_ j(k)} {\omega(k)}
 e^{-ik\cdot (X_s^i-X_s^j)}
dk \en
 as  $\k \rightarrow \infty$. Combining the right-hand side
of \kak{br3} with $\int_0^T V(X_s) ds$ in \kak{br2}, we can derive
the Feynman-Kac type formula:\ko{Feynman-Kac formula}
 \eq{br9}
 \lim_{\k \rightarrow\infty} \kak{br2}
  = \int_{{\ms X}_+\times\RR^{dN}} \ov{f(X_0)} g(X_t)
e^{-\int_0^T \lk
 V(X_s) +{V_{\rm eff}}(X_s)+G \rk ds}d{\cal W}^x  dx, \en
 where
 \eq{vv}
\veff(x_1,...,x_n)=-\frac{1}{4}\sum_{i\not=j}^N
\ak _i\ak _j \int_\BR \frac{{\lambda_ i(-k)}\lambda_ j(k)}{\omega(k)}
e^{-ik\cdot (x_i-x_j)}dk \en
 and $G$ is the constant derived from the diagonal part of
 \kak{br3},  which is
 given by
\eq{ll2}
G=-\frac{1}{4}\sum_{j=1}^N \ak _j^2 \int_\BR \frac{\lambda _j(-k)\lambda _j(k)}{\omega(k)}dk.
\en
Note that when ${\rm supp}\lambda_ i\cap{\rm supp}\lambda_ j=\emptyset$,
$i\not =j$, the effective potential ${V_{\rm eff}}$\ko{effective potential}  vanishes and only the
constant $G$ remains.
 Let
\eq{ll3}
\heff=\sum_{j=1}^N\left(-\frac{1}{2m_j} \Delta_j+V_j \right) +{V_{\rm eff}}.
\en
  Actually \kak{br9} can be shown rigorously.
\begin{proposition}
\TTT{Weak coupling limit}
{\rm \cite{dav77,dav79,hir98,hir99}}
\ko{weak coupling limit}
\label{br6}
 Let $t>0$. Then
 $$\slim_{\k \rightarrow\infty}
e^{-t H(\k )} =e^{-t(\heff+G)} \otimes P_\Omega,$$ where
$P_\Omega$ denotes the projection onto
$\{z\Omega|z\in\CC\}\subset\fff$.
  In particular
for $f,g\in L^2(\RR^{dN})$,
 $$ \lim_{\k \rightarrow\infty}
  (f\otimes \Omega, e^{-tH(\k )}g\otimes\Omega)_\hhh
  =(f,
 e^{-t(\heff+G)}g)_{\LRNd}.$$
 \end{proposition}

Proposition \ref{br6} is interesting in both the stochastic analysis
and the operator theory.
 Probabilistically, through a weak coupling
limit as is seen in Proposition \ref{br6}, one can derive a Markov
process from a non
 Markov process.
 We will see it below. The family of measures
 $\{\mu^\xx _\k \}_{\k >0}$
 on the path space ${\ms X}_+$ is given by
 \eq{br1}
\mu^x _\k (dX)=e^{-\int_0^t V(X_s) ds} e^{W_\k }  d{\cal W}^x . \en
 The double integral $W_\k $ in  \kak{br1} is independent of $x$
 and
  breaks a Markov property of the stochastic process $(X_s)_{s>0}$,
   and
$$T_{\k ,s}:f\longmapsto
\int _{{\ms X}_+} f(X_s)\mu_\k (dX)$$ does not define a semigroup
on $L^2(\RR^{dN})$.
 By Proposition \ref{br6}, however,
 the Markov property revives as
$\k \rightarrow \infty$,
 and we have $T_{\infty,
s}=e^{-s(\heff+G)}$ .

 Furthermore Proposition \ref{br6} also suggests
that $H(\k )\sim \heff+G$ for a sufficiently large $\k $.
Actually we can  show that  $H(\k )$ is isomorphic
  to
a self-adjoint operator of the form
 \eq{U} \heff+ \k ^2
\hf +\frac{1}{\k } H_1+\frac{1}{\k  ^2} H_2+{\rm
constant}\en
  with some operators $H_1$ and $H_2$.
 It is checked
that under some condition $\heff$ has a ground state for
$\ak _j$'s such that  $0<\ak _c<|\ak _j|$, $j=1,\ldots,N$, for
some $\ak _c$, which suggests by \kak{U} that for a sufficiently
large $\k $, $H(\k )$ also has a ground state for $\ak _j$
with $\ak _c<|\ak _j|<\ak _c(\k )$, $j=1,...,N$, for some
$\ak _c(\k )$.
  This is actually proved by checking stability  conditions for
  \kak{U} under some assumptions.
 This is an idea to show the enhanced binding for the Nelson model.
 Note that we do {\it not} need to assume the existence of ground
state of $\hp$,
namely $H(\k )$ with $\ak _1=\cdots=\ak _N=0$
may have no ground state.



\cleardoublepage
\section{Binding}
\subsection{Existence of ground states}
In order to show the enhanced binding  we check the so-called stability condition.
The stability condition implies that
 the lowest two cluster  threshold of $H$ is strictly larger than the ground state energy of $H$.
Then intuitively atom can not be ionized and thus the ground state is stable.
We introduce assumptions:
\begin{assumption}
\label{2v2}
For all $j=1,...,N$, (i),(ii),(iii) and (iv) are
fulfilled.
\bi
\item[ (i) ]
$\lambda_ j(-k)=\ov{\lambda_ j(k)}$ and $\lambda_ j,
\lambda_ j/\sqrt\omega\in\LR$.
\item[(ii)]
 There exists an open set $S\subset \BR$ such that
$\bar{S}=\supp \lambda_ j $ and
$\lambda_ j\in C^1(S)$.
\item[(iii)]
 For all $R>0$, $S_R=\{k\in S| |k|<R\}$ has a cone property.
\item[(iv)]
 For all $p\in [1,2)$ and all $R>0$, $|\nabla_k \lambda_j|\in L^p(S_R)$.
\ei
\end{assumption}
Condition (i)  guarantees that $\hi$ is a symmetric operator.
  In order
to show the existence of a ground state, we applied a method
invented in  \cite{gll01}. Precisely, we used the photon derivative
bound and the Rellich-Kondrachov theorem. The conditions (ii)-(iv)
are required to verify these procedures
  in the
proof of Proposition \ref{L2} below.
It is easily proven   that  $H$  is self-adjoint on
$D(H)=D(\hp)\cap D(\hf)$ and bounded from below
for an arbitrary $\ak _j\in\RR$.

Assumptions (V1) and (V2) are also introduced:

\bi
\item[ {(V1)} ] {\it There exists $\ak _c>0$ such that
$\inf\s(\heff)\in \s_{\rm disc}(\heff)$
for $\ak _j$  with $|\ak _j|>\ak _c$, $j=1,...,N$.}
\item[{(V2)} ] {\it  $V_j(-\Delta+1)^{-1}$, $j=1,\ldots,N$,
are compact.
}
\ei
The main theorem in Section 6 is stated below.
 \bt{main}
 \TTT{Enhanced binding}
\ko{enhanced binding!Nelson model}
 {\rm \cite{hs08}}
 Let
$\lambda_ j/\omega\in\LR$, $j=1,...,N$,  and assume Assumption \ref{2v2},(V1) and (V2).
Fix a sufficiently large $\k >0$. Then there exists $\alpha_c(\k)$ such that
 for $\ak _j$ with
$\ak _c<|\ak _j|<\ak _c(\k )$, $j=1,...,N$, $H(\k )$ has
a ground state, where $\ak _c(\k )$ is  possibly
infinity.
 \et
 \proof
 We give a proof in Section \ref{naga}
 \qed
  The scaling parameter $\k $ in Theorem \ref{main}
can be regarded as a dummy and absorbed into  $m_j$, $V_j$ and
 $\lambda_ j$, $j=1,...,N$.
  Let $\k $ be sufficiently large. Define
$$\hat H=\sum_{j=1}^N \lk
-\frac{1}{2\hat m_j}\Delta_j+ \hat V_j\rk
+ \sum_{j=1}^N
\ak _j \hat \phi_j+ \hf,$$ where $\hat m_j=m_j\k ^2$,
$\hat V_j=V_j/\k ^2$ and $\hat \phi_j$ is defined by $\phi_j$
with $\lambda_ j$ replaced by $\lambda_ j/\k $. \bc{main2}Let
$\lambda_ j/\omega\in\LR$, $j=1,...,N$,  and assume Assumption \ref{2v2},(V1) and (V2).
 Then
$\hat H$ has a ground state for
$\ak _c<|\ak _j|<\ak _c(\k )$, $j=1,...,N$, where
 $\ak _c(\k )$ is introduced in Theorem \ref{main}.
  \ec
\proof We have $\k ^{-2} H(\k )=\hat H$. Then by Theorem
\ref{main}, $\hat H$ has a ground state. \qed

\subsection{Stability conditions}
\label{naga}
Let $\lambda_ j/\omega\in\LR$, $j=1,...,N$,  and define the unitary operator $T$ on $\hhh$ by
$$\d T=\exp\lk -i\frac{1}{\k} \sum_{j=1}^N\ak_j \pi_j\rk,$$
where
$$\d \pi_j=\int_{\RR^{dN}}^\oplus \pi_j(x_j) dx$$
with
$$\pi_j(x)=\frac{i}{\sqrt2}
\int\lk \add(k) e^{-ik\cdot
x}\frac{\lambda_ j(-k)}{\omega(k)}-a(k)e^{ik \cdot
x}\frac{\lambda_ j(k)}{\omega(k)} \rk dk.$$
Then we can show that  $T$ maps $D(H)$ onto
itself and
\begin{align*}
 T^{-1} H(\k ) T
 & =\sum_{j=1}^N \lkk
\frac{1}{2m_j}\lk -i\nabla_j
 -\frac{\ak _j}{\k }A_j\rk^2+V_j
-\frac{\ak _j^2}{2}
\|\lambda_ j/\sqrt\omega\|^2\rkk+ \k  ^2 \hf+{V_{\rm eff}}\\
&=\heff+\k ^2 \hf+H'(\k ),
\end{align*}
where
$\d A_j=\int_{\RR^{dN}}^\oplus A_j(x_j) dx$ with
$$A_j(x)= \frac{1}{\sqrt 2}\int  k \lk \add(k) e^{-ikx}
\frac{\lambda_ j(-k)}{\omega(k)}+
a(k) e^{ikx} \frac{\lambda_ j(k)}{\omega(k)}\rk dk $$
and
\begin{align*}
H'(\k )
=\sum_{j=1}^N\lkk
\frac{1}{\k }\frac{\ak _j}{2m_j}
 \lk
 (-i\nabla_j)\!\cdot\!
  A_j+
  A_j
  \!\cdot\! (-i\nabla_j)
  \rk +
\frac{1}{\k ^2} \frac{\ak _j^2}{2m_j}
A_j^2 - \frac{\ak ^2_j}{2}\|\lambda_ j/\sqrt\omega\|^2\rkk.
\end{align*}
 Let us
set $\N=\{1,...,N\}$. For $\beta\subset \N$,
we define
\begin{align*}
& H_0(\beta)=H_0(\beta,\k ) = \sum_{j\in \beta}
           \frac{1}{2m_j} \left(-i\nabla_j -\frac{\ak _j}{\k }A_j\right)^2
           + \k ^2 H_f
        + V_\mathrm{eff}(\beta), \\
& V_\mathrm{eff}(\beta) =\lkk
\begin{array}{ll}
\displaystyle -\frac{1}{4}\sum_{i,j\in \beta, i\neq j}\ak _i\ak _j
 \int_{\BR}\frac{{\hat{\lambda}_i(-k)}\hat{\lambda}_j(k)}{\omega(k)}
 e^{-ik\cdot (x_i-x_j)}dk,&  |\beta|\geq 2,\\
0,&  |\beta | =0,1,\end{array}\right.
\\
&
H_V(\beta)=H_V(\beta,\k ) = H_0(\beta) + \sum_{j\in \beta} V_j,
\end{align*}
where
$|\beta|=\#\beta$.
 Simply we set
$ H_V=H_V(\N )$.
 \eq{hv}
 H_V=H(\k )-\frac{1}{4} \sum_{j=1}^N \ak _j^2\|\lambda_ j\|^2
 \en
  has ground states
if and only if $H(\k )$ does, since $\sum_{j=1}^N
\ak _j^2\|\lambda_ j\|^2/4$ is a fixed number.
The operators $H_0(\beta)$ and $H_V(\beta)$ are
self-adjoint operators acting on
$L^2({\RR}^{d|\beta|})\otimes\fff $. We set
 \begin{align*}
 & E_V(\k )=\is(H_V),\\
 &  E_V(\k ,\beta) = \inf \s(H_V(\beta)),\\
 &  E_0(\k ,\beta) = \inf \s(H_0(\beta)), \\
 &  E_V(\k ,\emptyset) = 0.
 \end{align*}
 The lowest two cluster threshold\ko{lowest two cluster threshold}
 $\Sigma_V(\k )$ is defined by
 \begin{align}
 \Sigma_V(\k )= \min\{E_V(\k ,\beta) + E_0(\k ,\beta^{\mathrm{c}})|
                            \beta\subsetneqq \N \}.
\end{align}
To establish the existence of ground state of $H(\k )$, we use
the next proposition:
\begin{proposition}
{\rm \cite{gll01}}
{\label{L2}}
Let
$  \Sigma_V(\k ) - E_V(\k ) >0.
$
Then
$H(\k )$ has a ground state.
\end{proposition}
For $\beta\subset\N $, we set the Schr\"odinger operators in
$L^2(\RR^{d|\beta|})$ by
\begin{align*}
h_0(\beta) &= -\sum_{j\in \beta} \frac{1}{2m_j}\Delta_j
          +V_\mathrm{eff}(\beta), \\
          h_V(\beta) &= h_0(\beta) + \sum_{j\in \beta}V_j, \\
\cE_0(\beta) &= \inf\s(h_0(\beta)),\\
\cE_V(\beta) &= \inf\s(h_V(\beta)),
\end{align*}
where
$ h_0(\emptyset) =0$ and $ h_V(\emptyset) = 0$.
Furthermore
we simply put
\eq{z123}
h_V=h_V(\N)=\heff,\quad
\cE _V=\inf\s(h_V).
\en
\begin{figure}[t]
\centering
\includegraphics[width=150pt]{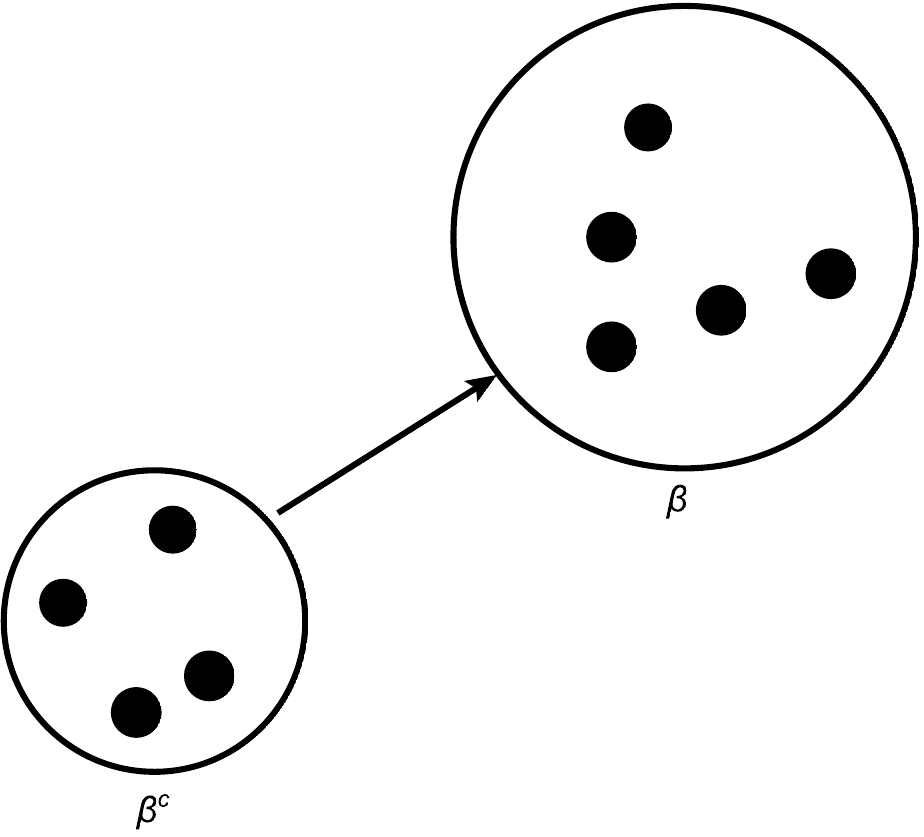}
\caption{Ionization ${\mathcal E}_V(\beta^c)+{\mathcal E}(\beta)$}
\label{picthr}
\end{figure}
We define the lowest two cluster threshold \ko{lowest two cluster threshold}for $h_V$ by (Figure  \ref{picthr})
\eq{th}
\Xi_V = \min\{\cE_V(\beta)+ \cE_0(\beta^\mathrm{c})|
\beta\subsetneqq \N \}
\en
and we set
$$\veff_{ij}(x)=
-\frac{1}{4}\ak _i\ak _j\int_{\RR^d} \frac{\lambda_ i(-k)\lambda_ j(k)}{\omega(k)}
e^{-ik\cdot x}
dk,\ \ \ i\not =j.$$
\bl{v-vv}
Effective potentials $\veff_{ij}$, $i,j=1,...,N$, are relatively compact with respect to
$-\Delta$.
\el
\proof
Since $\lambda_ i\lambda_ j/\omega\in L^1(\BR)$, $i,j=1,...,N$,
 we can see that
$\veff_{ij}(x)$ is continuous in $x$ and
$\lim_{|x|\rightarrow \infty}{V_{\rm eff}}_{ij}(x)=0$
 by the Riemann-Lebesgue
theorem.
In particular ${V_{\rm eff}}_{ij}$ is relatively compact with respect to the $d$-dimensional Laplacian.
\qed

We want to estimate $\is_{\rm ess}(\heff)$. For Hamiltonians with
the center of mass motion removed, the bottom of the essential
spectrum is estimated by HVZ theorem.
\begin{lemma}\label{hvz}
Assume (V2).
Then
$ \s_\mathrm{ess}(H_\mathrm{eff})~=~[\Xi_V,\infty) $.
\end{lemma}
\proof
We may assume that $V_i, \veff_{ij} \in C_0^\infty(\BR)$
by Proposition \ref{A2} below.
Then
there exists a normalized sequence  $\{g_n\}_n\subset
 C_0^\infty(\RR^{dN})$
such that
 $${\rm supp} g_n\subset
\left
 \{x\in \RR^{dN} \left|
 V_i(x)=0, {V_{\rm eff}}_{ij}(x_i-x_j)=0,
i,j=1,...,N \right. \right\}$$
 and
 $(g_n, h_V(\beta)g_n)=\sum_{j\in\beta}\lk g_n, -\frac{1}{2m_j} \Delta
g_n \rk \rightarrow 0$ as $n\rightarrow \infty$. Then we have
\eq{B7} {\cal E}_V(\beta)+{\cal E}_0(\beta^c)\leq 0. \en
Let
$\tilde{j}_\beta\in C^\infty(\mathbb{R}^d)$, $\beta\in \N$, be a
Ruelle-Simon partition of unity\ko{partition of unity}, which
satisfy
(i)-(v)  below:
\bi
\item[(i)] $
    \d \sum_{\beta\subseteq \N } \tilde{j}_\beta(x)^2 =1,
       $
\item[(ii)]
  $\d       \tilde{j}_\beta(Cx) = \tilde{j}_\beta(x)$
for $|x|=1$, $C\geq 1$ and $\beta\neq \N ,
           $
\item[(iii)]
$\d
  \mathrm{supp}\> \tilde{j}_\beta
  \subset  \left\{x\in\mathbb{R}^d
  \left|
  \min_{i\in\beta,j\in\beta^\mathrm{c}} \{|x_i-x_j|,|x_j|\}\geq c|x|\right.\right\}
$ for some $c>0$,
\item[(iv)] $\d
              \tilde{j}_\beta(x) = 0$ for $\d |x|<\frac{1}{2}$
 and $\beta\neq \N $,
\item[(v)]
 $\tilde{j}_{\N }$ has a compact
support.
\ei
For a constant $R>0$ we put $j_\beta(x) = \tilde{j}_\beta(x/R)$.
Note that for each $\beta\subset \N $,
$$H_\mathrm{eff}=h_V(\beta)\otimes \one +\one \otimes h_0(\beta^c)+I_\beta,$$
where
$$
I_\beta=
\sum_{i\in\beta^c}\one \otimes V_i(x_i)+
\sum_{\substack{i\in\beta,j\in\beta^\mathrm{c}
\\ i\in\beta^\mathrm{c},j\in\beta}} {V_{\rm eff}}_{ij}(x_i-x_j).$$
 Here
we identify as
$
  L^2(\RR^{dN})\cong L^2(\RR^{d|\beta|})\otimes
L^2(\RR^{d|\beta^c|})$.
  By the IMS localization formula\ko{IMS localization formula}
\cite[Theorem 3.2 and p. 34]{cfks87}, we have
$$
 H_\mathrm{eff}
=
 j_{\N } H_\mathrm{eff} j_{\N } +
      \sum_{\beta\subsetneqq\N } j_\beta \lk
      h_V(\beta)\otimes \one +\one \otimes h_0(\beta^\mathrm{c})
       \rk j_\beta
+\sum_{\beta\subsetneqq\N } j_\beta^2 I_\beta
  - \frac{1}{2}\sum_{\beta\subseteqq\N }|\nabla j_\beta|^2.
   \label{small}
$$
Since $j_{\N }^2  (\sum_{j=1}^N V_j+{V_{\rm eff}})  $ and
$\sum_{\beta\subsetneqq\N } j_\beta^2 I_\beta$ are relatively
compact with respect to the $dN$-dimensional Laplacian by the
property (iii) and (v), it is seen that
the essential spectrum of
$H_{\rm eff}$ coincides with
that of
$$
j_\N\lk -\half\sum_{j=1}^N\Delta_j \rk  j_\N+
\sum_{\beta\subsetneqq\N }
j_\beta
\lk
      h_V(\beta)\otimes \one +\one \otimes h_0(\beta^\mathrm{c})
\rk  j_\beta
     - \frac{1}{2}\sum_{\beta\subseteqq\N }|\nabla j_\beta|^2.
     $$
We have
$$
\sum_{\beta\subsetneqq\N } j_\beta \lk
 h_V(\beta)\otimes \one +\one \otimes
h_0(\beta^\mathrm{c})
       \rk j_\beta
 \geq  \sum_{\beta\subsetneqq\N }
   (\cE_V(\beta)+\cE_0(\beta^\mathrm{c}))j_\beta^2.
$$
By (ii) and (v),
$$
\Big\|
\frac{1}{2} \sum_{\beta\subseteqq\N }|\nabla j_\beta|^2\Big\|
 \leq \frac{C}{R^2}
$$ with some constant $C$ independent of $R$.
Hence we obtain
that $$
 \inf\s_\mathrm{ess}(H_\mathrm{eff})
 \geq
 \min_{x\in\mathbb{R}^d} \sum_{\beta\subsetneqq \N }
 (\cE_V(\beta)+ \cE_0(\beta^\mathrm{c}))j_\beta^2
 - \frac{C}{R^2}
 \geq
 \Xi_V -\frac{C}{R^2}
$$
for all $R>0$. Here we used (i) and \kak{B7}.
Thus
$
\s_\mathrm{ess}(H_\mathrm{eff}) \subset  [\Xi_V,\infty)$ follows.
Next we shall prove the reverse inclusion
$\s_\mathrm{ess}(H_\mathrm{eff}) \supset  [\Xi_V,\infty)$.
Fix $\beta\subsetneqq  \N $.
Let $\{\psi_n^V \}_{n=1}^\infty\subset C_0^\infty(\mathbb{R}^{d|\beta|})$
 be a minimizing sequence of $h_V(\beta)$ so that
\begin{align*}
 \lim_{n\to\infty}
 \|(h_V(\beta)-\cE_V(\beta))\psi_n^V\|
 =0, \quad \|\psi_n^V\|=1
\end{align*}
and $\{\psi_n^0\}_{n=1}^\infty\subset
C_0^\infty(\mathbb{R}^{d|\beta^\mathrm{c}|})$  a normalized
sequence such that
\begin{align}
 \lim_{n\to\infty}
 \| (h_0(\beta^\mathrm{c})-\cE_0(\beta^\mathrm{c})-K)\psi_n^0\|=0,
 \label{minim2}
\end{align}
where $K\geq 0$ is a constant. Note that since
$\s(h_0(\beta^c))=[{\cal E}_0(\beta^c),\infty)$, $\psi_n^0$ such as
\kak{minim2} exists. By the translation invariance of
$h_0(\beta^\mathrm{c})$, for any function $\tau_\cdot :\NN \to
\mathbb{R}^d$,
  the shifted  sequence
$\psi_n^0(x_{j_1}-\tau_n,\ldots ,x_{j_{|\beta^\mathrm{c}|}}-\tau_n)$
 also satisfies
\kak{minim2}.
Let $R_n>0$ be a constant satisfying
\begin{align*}
 \mathrm{supp}\> \psi_n^V \subset
 \left\{
 x=(x_{j_1},\cdots,x_{j_{|\beta|}})\in \mathbb{R}^{d|\beta|}
  \left |
  |x_{j_i}|<R_n,
 \, j_i\in\beta, i=1,...,|\beta| \right. \right \}.
\end{align*}
We take $\tau$  such that
\begin{align*}
& \mathrm{supp}\>\psi_n^0(\cdot-\tau_n,\cdots,\cdot-\tau_n) \\
&\subset \left \{x=(x_{k_1},\cdots,x_{k_{|\beta^c|}})\in
\mathbb{R}^{d|\beta^c|}
 \left |
 |x_{k_i}|\geq R_n+n,
 \, k_i\in\beta^c,i=1,...,|\beta^c| \right. \right \}.
\end{align*}
We set
$
 \Psi_n(x_1\cdots x_N) = \psi_n^V(x_{j_1}\cdots x_{j_{|\beta|}})
\otimes \psi_n^0(x_{k_1}-\tau_n\cdots x_{k_{|\beta^c|}}-\tau_n) \in
L^2(\RR^{dN})$.
Then,
for all $i,j$ with $i\in\beta, \, j\in\beta^\mathrm{c}$, we have
\begin{align*}
 \| \veff_{ij}(x_i-x_j)\Psi_n\| &\leq \sup_{x\in\mathbb{R}^d,|x|>n}
|\veff_{ij}(x)|
 \to 0, \quad (n\to\infty), \\
 \|V_j(x_j)\Psi_n\| &\leq \sup_{x\in\mathbb{R}^d, |x|\geq R_n+n}|V_j(x)|
 \to 0, \quad (n\to\infty).
\end{align*}
Hence, by a triangle inequality, we have that
\begin{align*}
 \|(H_{\mathrm{eff}}-\cE_V(\beta)-\cE_0(\beta^{\mathrm{c}})-K)\Psi_n\| \to 0, \quad
(n\to\infty).
\end{align*}
Therefore $[{\cal E}_V(\beta)+{\cal E}_0(\beta^c)+K,\infty)
\subset \s(H_{\rm eff})$.
Since $\beta\subsetneqq \N$ and $K>0$ are arbitrary,
 $ [\Xi_V,\infty) \subset \s_\mathrm{ess}(H_\mathrm{eff})$ follows.
Thus the proof is complete.
\qed
We define
$
  \Delta_\mathrm{p}(\ak _1,...,\ak _N ) =\Xi_V - \cE_V$.

\begin{corollary}{\label{L3}}
 Assume (V1) and (V.2).
 Then $\Delta_\mathrm{p}(\ak _1,...,\ak _N )>0$ follows
for $\ak _j$  with
$|\ak _j|>\ak _c$, $j=1,...,N$.
\end{corollary}
\proof
Since $\is_{\rm ess}(\heff)=\Xi_V$  by Lemma \ref{hvz} and
$\is(\heff)\in \s_{\rm disc}(\heff)$ by (V1),
the corollary follows from $\Delta_P(\ak _1,...,\ak _N )
=\is_{\rm ess}(\heff)-\is(\heff)>0$.
\qed

\begin{lemma}{\label{L4}} For an arbitrary $\k >0$,  it follows that
$\Sigma_V(\k )\geq \Xi_V$.
\end{lemma}
\proof It is well known that $H_V(\beta)$ can be realized as a
self-adjoint operator on a Hilbert space
$\hhh_Q=L^2(\RR^{|\beta|d})\otimes L^2(Q,d\mu)$
 with some probability
space $(Q, \mu)$, which is called a Schr\"odinger representation. It
is established in e.g., \cite{lhb11} that
$$(\Psi, e^{-tH_V(\beta)}\Phi)_{\hhh_Q}\leq (|\Psi|,
e^{-t(h_V(\beta)+\k ^2   \hf)}|\Phi|)_{\hhh_Q}.$$
Hence for any $\beta\subset\N $, it follows that
 $$\is(h_V(\beta)+\k ^2  \hf )\leq \is(H_V(\beta)).$$
 Since $\is(\hf)=0$ and $\is(h_V(\beta)\otimes \one +\k ^2 \one \otimes \hf )=\is(h_V(\beta))$,
  we can obtain
  $$ \is (h_V(\beta))\leq \is (H_V(\beta))$$
 for arbitrary $\beta\in C_N$.  Then the lemma
follows from  the definition of lowest two cluster thresholds.
\qed

\begin{lemma}{\label{L5}}
 Assume (V1). Then
$$  E(\k ) \leq \cE_V
  + \k ^{-2}\sum_{j=1}^N \ak _j^2\|\lambda_ j\|^2/(4m_j)
  $$
for $\ak _j$ with
$|\ak _j|>\ak _c$, $j=1,...,N$.
\end{lemma}
\proof
By (V1), $H_\mathrm{eff}$ has a normalized ground state $u$ for $\ak _j$ with
$|\ak _j|>\ak _c$, $j=1,...,N$.
Set $\Psi= u\otimes \Omega$.
Then
\begin{align*}
 E(\k )\leq(\Psi, H(\k )\Psi)
&\leq  (u, H_\mathrm{eff}u )
     + \sum_{j=1}^N \frac{\ak _j}{2m_j\k }
       2\Re ( i\nabla_j\Psi, A_j\Psi)
     + \sum_{j=1}^N \frac{\ak _j^2}{2m_j\k ^2}
       \| A_j\Psi\|^2 \\
   &=
      \cE_V +
      \sum_{j=1}^N \frac{\ak _j^2}{4m_j\k ^2} \|\lambda_j\|^2.
\end{align*}
 Here we used that
 $\d  ( \nabla_j\Psi, A_j\Psi)
  =\frac{1}{\sqrt2}\sum_{\mu=1}^ d
    (\nabla_{x_\mu^j}
u\otimes \Omega, u\otimes \add(k_\mu e^{-ik\cdot x} \lambda_ j /\omega
)\Omega)=0$.  Then the lemma follows.
\qed

\textit{Proof of Theorem \ref{main}}\\
By Lemmas \ref{L4} and \ref{L5}, we have
$$
 \Sigma_V(\k ) -E(\k )
 \geq \Xi_V -\cE_V- \sum_{j=1}^N \frac{\ak _j^2}{4m_j\k ^2} \|\lambda_j\|^2 \\
 = \Delta_\mathrm{p}(\ak _1,...,\ak _N )
      - \sum_{j=1}^N \frac{\ak _j^2}{4m_j\k ^2} \|\lambda_j\|^2.
$$
Note that $\Delta_p(\ak _1,...,\ak _N )>0$ is continuous in $\ak _1,...,\ak _N$.
Then for a sufficiently large $\k $,
there exists $\ak _c(\k )>\ak _c$ such that
for $\ak _c<|\ak _j|<\ak _c(\k )$, $j=1,...,N$,
$\Sigma_V(\k )-E(\k )>0$.
Thus  $H(\k )$ has
a ground state for such $\ak _j$'s by Proposition~\ref{L2}.
\qed

\subsection{Examples}
\subsubsection{Example of effective potential}
\label{ex}
 We show a typical example of
  cutoff function and effective potentials.
  We introduce the assumption below.
Let
 $   \lambda_ j=\rh_j/\sqrt\omega$, $
 j=1,...,N$,
   with rotation invariant
 nonnegative functions $\rh_j$.
In this case, by \kak{vv},
 effective potential\ko{effective potential}
${V_{\rm eff}}$ is explicitly computed as
\begin{align}
& \veff(x_1,\cdots,x_N)\non \\
&\label{gyate}
= -\frac{1}{4}\sum_{i\not= j}^N \ak _i\ak _j
\frac{\sqrt{(2\pi)^d}
}{|x_i-x_j|^{(d-1)/2}}\int_0^\infty\frac{r^{(d-1)/2}}{r^2}\rh_i(r)\rh_j(r)\sqrt{r|x_i-x_j|}
J_{\fdd }(r|x|)
dr.
\end{align}
Here $J_\nu$ is the Bessel function:
$J_\nu(x)=(\frac{x}{2})^\nu
 \sum_{n=0}^\infty\frac{(-1)^n}{n!\Gamma(n+\nu+1)}
  (\frac{x}{2})^{2n}$.
We can see that $\veff$ satisfies that
 \bi
  \item[(1)] $\veff_{ij}$ is continuous,
 \item[(2)] $\lim_{|x|\rightarrow\infty}V_{ij}(x)=0$,
 \item[(3)]
 $\veff_{ij}(0)<\veff_{ij}(x)$ for all $x\in \BR$ but
$x\not=0$.
\ei

In particular, when  $d=3$ and $\rh_j$ is the indicator function
such as
 \eq{rho}
  \rh_j(k)=\lkk\begin{array} {ll}
0&|k|<\k ,\\
1/\sqrt{(2\pi)^3}& \k <|k|<\Lambda,\\
0&|k|\geq \Lambda,
 \end{array}\right.
  \en
   we see that
 \eq{F}
\veff(x_1,\cdots,x_N)= -\frac{1}{8\pi^2} \sum_{i\not= j}^N
\frac{\ak _i\ak _j}{|x_i-x_j|}\int_{\k |x_i-x_j|}^{\Lambda|x_i-x_j|}
\frac{\sin r}{r}dr. \en
   For sufficiently small $|x_i-x_j|$, $i,j=1,...,N$,
 and $\ak _j$ with  an identical sign,
  the effective potential
   \kak{F} is attractive.

\subsubsection{Example of external potential}
We give an example of $V_1,\cdots,V_N$ satisfying assumption (V1).
Assume simply that $V_1=\cdots=V_N=V$,
 $\ak _1=\cdots=\ak _N=\ak $, $\lambda_ 1=\cdots=\lambda_ N=\la$ and
$m_1=\cdots=m_N=m$. Then
 $$\veff_{ij}(x)=
  W(x)=
   -\frac{\ak ^2 }{4}\int _\BR \frac{|\la(k)|^2}{\omega(k)} e^{-ik\cdot x} dk$$
  for all $i\not= j$. Let
\begin{equation*}
 h_{V}(\ak ) =
 \sum_{j=1}^N \left(-\frac{1}{2m}\Delta_j +V(x_j) \right)
 + \ak ^2 \sum_{j\neq l}^N
 \WWWWW (x_j-x_l),
\end{equation*}
which acts on $L^2(\mathbb{R}^{dN})$.
We assume (W1)-(W3) below:
\bi
\item[(W1)]
$V$ is  relatively compact with respect to
the $d$-dimensional Laplacian $\Delta$, and
$
 \s(-(\Delta /2m)+V) = [0,\infty)$.
\item[(W2)]
$\WWWWW $ satisfies that
$\d
 -\infty < \WWWWW (0)= \essinf _{|x|<\epsilon}\WWWWW (x) <
 \essinf_{|x|>\epsilon} \WWWWW (x) $ for all $\epsilon>0$.
\item[(W3)]
$\is( -(\Delta/(2Nm)
 + N V )\in \s_{\rm disc}
( -(\Delta/(2Nm)
 + N V )$.
\ei
\begin{remark}\label{remenber}
{\rm
 Note that examples of $\veff$ given in \kak{gyate}
satisfies (W2), and remember that $\lim_{|x|\rightarrow
\infty}W(x)=0$ and $W(x)$
 is relatively compact  with respect to the
$d$-dimensional Laplacian. See Lemma \ref{v-vv}.
 The condition (W1) means that the
external potential $V$ is shallow and the non-interacting
Hamiltonian $h^{V}(0)$ has no negative energy bound state.
}\end{remark}
\begin{theorem}{\label{Tex}}
 Assume (W1)-(W3).
Then, there exists $\ak _c>0$ such that for all $\ak $ with
$|\ak |>\ak _c$, $\is(h^{V}(\ak ))\in \s_{\rm
disc}(h_V(\ak ))$. Namely $h_V(\ak )$ for $|\ak |>\ak _c$
has a ground state.
\end{theorem}
To prove Theorem \ref{Tex} we need  several lemmas.
For  $\beta\subset  \N$,
we define
\begin{align*}
  h_0(\ak ,\beta) &= -\frac{1}{2m}\sum_{j\in\beta}\Delta_{j}
  +\ak ^2 \sum_{\substack{j,l\in\beta\\ j\neq l}}\WWWWW (x_j-x_l),\\
  h_V(\ak ,\beta) &= h_0(\ak ,\beta) +\sum_{j\in\beta}V(x_j), \\
  \cE_0(\ak ,\beta) &= \inf\s(h_0(\ak ,\beta)), \\
  \cE_V(\ak ,\beta) &=
\inf\s(h_V(\ak ,\beta)),
\end{align*}
where
$\cE_V(\ak ,\emptyset)=0$ and
$\cE_0(\ak ,\emptyset)=0$.
Simply we set
${\cal E}_V(\ak ,\N)={\cal E}_V(\ak )$
and ${\cal E}_0(\ak ,\N)={\cal E}_0(\ak )$.
Let $\Xi_V(\ak ) $ denote the lowest two cluster threshold of $h_V(\ak )$ defined by \kak{th}.
Then by  (W1) and Lemma \ref{hvz}, we have
\eq{sd}
 \s_\mathrm{ess}(h_V(\ak ))
 = [\Xi_V(\ak ),\infty).
\en
\begin{lemma}\label{ene}
 Let $\beta\subsetneqq \N $ but $\beta\neq\emptyset$.
Then there exists  $a_1>0$ such that,
for all $\ak $ with $|\ak |>a_1$,
\eq{sas}
   \cE_0(\ak ) < \cE_V(\ak ,\beta)+ \cE_0(\ak ,\beta^c).
  \en
\end{lemma}
\proof
 Since $h_0(\ak ,\beta)/\ak ^2 $ and
$h_V(\ak ,\beta)/\ak ^2$ converge to $\sum_{\substack{j,l\in\beta\\ j\neq l}}\WWWWW (x_j-x_l)$
 in the uniform resolvent sense,
by (W2), one can show that
  \begin{equation*}
    \lim_{\ak \to\infty}\frac{\cE_V(\ak , \beta)}{\ak ^2}
   = \lim_{\ak \to\infty}\frac{\cE_0(\ak , \beta)}{\ak ^2}
   = |\beta|(|\beta|-1) \WWWWW (0).
  \end{equation*}
\qed
Hence
\begin{align*}
\lim_{\ak \to\infty}\frac{\cE_0(\ak )}{\ak ^2}
  = N(N-1) \WWWWW (0)
  \end{align*}
  and
  \begin{align*}
\lim_{\ak \to\infty}
  \frac{1}{\at}
  \lk
{\cE_V(\ak ,\beta)+\cE_0(\ak ,\beta^c)}
\rk
  &= \big\{(|\beta|(|\beta|-1) + |\beta^c|(|\beta^c|-1)\big\}
     \WWWWW (0) \\
  &= \big\{N(N-1) +2|\beta|(|\beta|-N) \big\} \WWWWW (0).
\end{align*}
Since
$|\beta|(|\beta|-N)\leq -1$ and
$\WWWWW (0)<0$ by (W2),
we see that there exists  $a_1>0$ such that \kak{sas} holds for all $\ak $ with
$|\ak |>a_1$.
\qed

Let $X=(x_1,...,x_N)^t\in \RR^{dN}$ and $Y=(x_c,y_1,\ldots,y_{N-1})^t$ be
its  Jacobi coordinates\ko{Jacobi coordinates}:
\begin{align*}
  x_c = \frac{1}{N} \sum_{j=1}^N x_j, \quad
 y_j = x_{j+1} - \frac{1}{j}\sum_{i=1}^j x_i, \quad j=1,...,N-1.
\end{align*}
Let $T\in {\rm GL}(N,\RR)$ be  such that
$Y = TX$.
Note that
\begin{align*}
T&=\begin{bmatrix}
 \frac{1}{N}   &  \frac{1}{N}  & \frac{1}{N}   & \cdots&\cdots &\cdots & \frac{1}{N} \\
 -1            &  1            &  0            & \cdots & \cdots & \cdots & 0   \\
 -\frac{1}{2}  & -\frac{1}{2}  &  1            & 0 & 0  & \cdots & 0  \\
 -\frac{1}{3}  &  -\frac{1}{3} & -\frac{1}{3}  & 1      & 0 &\cdots & 0 \\
 \vdots        &    \vdots     & \vdots        & \cdots & \ddots& \cdots    & \cdots \\
 \vdots        &    \vdots     & \vdots        & \cdots & \cdots& \ddots    & \cdots \\
 -\frac{1}{N-1}&-\frac{1}{N-1} & -\frac{1}{N-1}& \cdots & \cdots& -\frac{1}{N-1} & 1
    \end{bmatrix}
    \end{align*}
 and
 \begin{align*}
 T^{-1}&=
\begin{bmatrix}
1 &-\frac{1}{2}&-\frac{1}{3}&-\frac{1}{4}&-\frac{1}{5}& \cdots&\cdots& -\frac{1}{N} \\
1 &\frac{1}{2}&-\frac{1}{3}&-\frac{1}{4}&-\frac{1}{5}&\cdots &\cdots& -\frac{1}{N} \\
1 &  0 &\frac{2}{3} & -\frac{1}{4} & -\frac{1}{5}&\cdots&\cdots& -\frac{1}{N}\\
1 & 0 & 0 & \frac{3}{4} & -\frac{1}{5} & -\frac{1}{6} & \cdots & -\frac{1}{N}\\
\vdots& \vdots&\cdots& \cdots&\ddots& \cdots&\cdots& \vdots \\
\vdots&\vdots&\cdots& \cdots&\cdots& \ddots&\cdots& \vdots \\
1 & 0 &\cdots& \cdots&\cdots&  0& \frac{N-2}{N-1}&  -\frac{1}{N} \\
1 & 0 & \cdots&\cdots&\cdots& \cdots&  0& \frac{N-1}{N}
\end{bmatrix}.
\end{align*}
Matrix $T$  induces
the  unitary operator
 $U:L^2(\mathbb{R}^{dN}_X)\to L^2(\mathbb{R}^{dN}_Y)$ defined
by
$$ (U\psi)(Y) = \psi\circ T^{-1}(Y).$$
We have
\begin{align*}
 & U h_0(\ak ) U^{-1}
 = -\frac{1}{2Nm} \Delta_{x_c}
   - \sum_{j=1}^N \frac{1}{2\mu_j}\Delta_{y_j}
      + \ak ^2\sum_{\substack{j\neq l}}^N \WWWWW (x_j(Y)-x_l(Y)), \\
 & U h_V(\ak ) U^{-1}
   = U h_0(\ak ) U^{-1}  + \sum_{j=1}^N V(x_j(Y)),
\end{align*}
where $\mu_j= jm/(j+1)$ is a
reduced mass\ko{reduced mass} and $x_j(Y)=(T^{-1}Y)_j$.
Let $k(\ak )$ be $h_0(\ak )$ with the center of mass motion removed:
\begin{align*}
  k(\ak ) = - \sum_{j=1}^N \frac{1}{2\mu_j}\Delta_{y_j}
      + \ak ^2\sum_{\substack{j\neq l}}^N \WWWWW (x_j(Y)-x_l(Y)).
\end{align*}
Set $\RR^{dN}=\RR^d_{x_c}\oplus\RR^{d(N-1)}_{y_1,...,y_{N-1}}=
\chi_c\oplus\chi_c^\perp$. Since   $x_j(Y)-x_i(Y)$, $i,j=1,...,N-1$,
depend only on $y_1,\ldots, y_{N-1}\in \chi_c^\perp$, $k(\ak )$
 is a
 self-adjoint operator acting in $L^2(\chi_c^\perp)$.
\begin{lemma}
  There exists  $a_2>0$ such that   $\is(k(\ak ))\in \s_{\rm disc}(k(\ak ))$ for all $\ak $ with $|\ak |>a_2$.
\end{lemma}
\proof
Note that $\lim_{|x|\to\infty}\WWWWW (x)=0$.
Let $\chi, \bar\chi \in C^\infty(\mathbb{R})$ be such that
$
   \chi(x)^2 + \bar\chi(x)^2 =1$ with
$ \chi(x) =
   \begin{cases}
     1, \quad |x|<1, \\
     0, \quad |x|>2.
   \end{cases}$
For a parameter $R$, we set
\begin{align*}
\chi_R(y_1)&=\chi(|y_1|/R), \quad \bar\chi_R(y_1)=\bar\chi(|y_1|/R),
 \quad  y_1\in \mathbb{R}^d, \\
\theta_R(Y_1)&= \chi(|Y_1|/2R), \quad \bar\theta_R(Y_1)= \bar\chi(|Y_1|/2R),
 \quad Y_1=(y_2,\ldots,y_{N-1}) \in \mathbb{R}^{d(N-2)}.
\end{align*}
By the IMS localization formula\ko{IMS localization formula}, we have
\begin{eqnarray}\label{kk2}
  k(\ak ) &=& \chi_R \theta_R k(\ak ) \theta_R \chi_R
             +\chi_R \bar\theta_R k(\ak ) \bar\theta_R \chi_R
             +\bar\chi_R k(\ak ) \bar\chi_R +B(R),
             \end{eqnarray}
             where
             $$B(R)=
{-\frac{1}{2}\chi_R^2 |\nabla\theta_R|^2
               -\frac{1}{2}\chi_R^2 |\nabla\bar\theta_R|^2
              -\frac{1}{2}|\nabla\chi_R|^2
             -\frac{1}{2}|\nabla\bar\chi_R|^2}.
$$
Here $B(R):L^2(\chi_c^\perp) \rightarrow L^2(\chi_c^\perp)$
 is a
bounded operator with the bound
$$\| B(R)\|
\leq \frac{C}{R^2},$$ where $C$ is  a constant independent of $R$.
Let us define $k'(\ak )$ by $k(\ak )$ with
 the firs term $\chi_R\theta_R k(\ak )\theta_R\chi_R$ in \kak{kk2}
 replaced by $\chi_R\theta_R (-\sum_{j=1}^N(1/2\mu_j)\Delta_{y_j})
\theta_R\chi_R$;
 $$k'(\ak )= \chi_R \theta_R \lk
 -\sum_{j=1}^N \frac{1}{2\mu_j}\Delta_{y_j} \rk
\theta_R \chi_R
             +\chi_R \bar\theta_R k(\ak ) \bar\theta_R \chi_R
             +\bar\chi_R k(\ak ) \bar\chi_R  +B(R).$$
Since the difference between $k(\ak )$ and $k'(\ak )$ is
$\chi_R^2\theta_R^2 \ak ^2 \sum_{j\not= l}^N W(x_j(Y)-x_l(Y))$,
and which  is relatively compact with respect to
 the kinetic term $-\sum_{j=1}^N (2\mu_j)^{-1}\Delta_{y_j}$
  by Remark \ref{remenber},
  we have $\s_{\rm ess}(k(\ak )) =\s_{\rm ess}(k'(\ak ))$.
Moreover $k'(\ak )$ can be estimated from below as
\begin{align}
 k'(\ak ) \geq
 & \chi_R^2\bar\theta_R^2E
\lk \frac{}{}k(\ak )-\ak ^2 \WWWWW (x_2(Y)-x_3(Y))
 -\ak ^2 \WWWWW (x_3(Y)-x_2(Y))\rk  \label{K44}\\
 & + \chi_R^2\bar\theta_R^2\ak ^2
  \lk \frac{}{}
  \WWWWW (x_2(Y)-x_3(Y))
   +  \WWWWW (x_3(Y)-x_2(Y))\rk  \label{K45}\\
 & + \bar\chi_R^2
E \lk\frac{}{}
 k(\ak ) -\ak ^2 \WWWWW (x_1(Y)-x_2(Y))
     - \ak ^2 \WWWWW (x_2(Y)-x_1(Y))\rk  \label{K46}\\
 & +  \bar\chi_R^2 \ak ^2
  \lk \frac{}{}
  \WWWWW (x_1(Y)-x_2(Y))
   +  \WWWWW (x_2(Y)-x_1(Y))\rk   \label{K47}\\
&-C/R^2\label{K43}.
\end{align}
Note that $y_1=x_2(Y)-x_1(Y)$ and $x_3(Y)-x_2(Y) = y_2-y_1/2$.
We have
\begin{align*}
  |(\ref{K45})| &\leq 2\sup_{\substack{y_1,y_2\\ |y_1|<2R, \, |y_2|>4R}}
                     \ak ^2 |\WWWWW (y_2-y_1/2)|
 \leq  2\ak ^2\sup_{|y|>3R} |\WWWWW (y)|, \\
  |(\ref{K47})| &\leq 2 \sup_{|y_1|>2R} \ak ^2 |\WWWWW (y_1)|.
\end{align*}
Since we assume that $\lim_{|x|\rightarrow \infty}\WWWWW (x)=0$, we obtain
that
$ \lim_{R\to\infty} \|(\ref{K45})\| =0$ and
$\lim_{R\to\infty} \|(\ref{K47})\| =0$.
Thus, for all $R>0$ we have
\begin{eqnarray}
&&
 \inf\s_{\mathrm{ess}}(k(\ak ))
 =\inf\s_{\mathrm{ess}}(k'(\ak ))\non\\
 && \geq
 \inf_{Y\in\mathbb{R}^{d(N-1)}}[(\ref{K44})+(\ref{K46})]
-\|\kak{K45}\|-\|\kak{K47}\|-C/R^2 \non \\
 &&
  \geq  \min\{ E(k(\ak )-\ak ^2 \WWWWW (x_1-x_2)
                    -\ak ^2 \WWWWW (x_2-x_1)), \non \\
 & &\quad  \label{y1}
                 E(k(\ak )-\ak ^2 \WWWWW (x_2-x_3)
                    -\ak ^2 \WWWWW (x_3-x_2))\} +o(R),
\end{eqnarray}
where $\lim_{R\rightarrow \infty}o(R)/R=0$.
It is seen  that
\begin{align}
& \lim_{\ak \to\infty}
\frac{1}{\at}
E \lk
k(\ak )-\ak ^2 \WWWWW (x_1-x_2)
-\ak ^2 \WWWWW (x_2-x_1) \rk
=[N(N-1)-2] \WWWWW (0), \\
 &\lim_{\ak \to\infty}
\frac{1}{\at}
 {E\lk
 \frac{}{}
 k(\ak )-\ak ^2 \WWWWW (x_2-x_3)
                    -\ak ^2 \WWWWW (x_3-x_2)\rk}
 =[N(N-1)-2] \WWWWW (0), \\
 &\lim_{\ak \to\infty}\frac{E(k(\ak ))}{\ak ^2} =N(N-1) \WWWWW (0).
\label{y2}
\end{align}
Therefore combining \kak{y1}-\kak{y2}
 we obtain that
 \eq{alpha}
  \lim_{\ak \rightarrow \infty}
 \frac{1}{\ak ^2}\lk\!\!\frac{}{}
  \inf\s_{\rm ess}(k(\ak ))-\is(k(\ak ))\rk \geq
 -2W(0).
  \en
   Since  $\WWWWW (0)<0$ by (W2),
     there exists  $a_2>0$ such that
$ \inf\s_{\mathrm{ess}}(k(\ak )) -\is (k(\ak )) > 0$ for
$|\ak |
>a_2$.
This implies the desired result.
\qed

\begin{lemma}{\label{delta}}
Let $u_\ak $ be a normalized ground state of $k(\ak )$, where $|\ak |>a_2$.
Then
$  |u_\ak (y_1,\ldots,y_{N-1})|^2 \to \delta(y_1)\cdots\delta(y_{N-1})$ as $\ak \to\infty$ in the
sense of distributions.
\end{lemma}
\proof
It suffices to show that for all $\epsilon>0$,
\eq{z6}
  \lim_{\ak \to\infty} \int_{|Y_0|>\epsilon} |u_\ak (Y_0)|^2 d Y_0 =0, \quad
 Y_0=(y_1,\ldots,y_{N-1}).
\en
We prove \kak{z6} by a reductive absurdity. Assume that
$$\d  \liminf_{\ell\to\infty} \int_{|Y_0|>\epsilon} |u_{\ak _\ell}(Y_0)|^2 d Y_0 >0 $$
for some constant $\epsilon>0$ and some sequence $\{\ak _\ell\}_{\ell=1}^\infty \subset \mathbb{R}$
such that $\ak _\ell\to\infty (\ell\to\infty)$.
We can take a subsequence $\{\hat\ak _\ell\}_{\ell=1}^\infty \subset \{\ak _\ell\}_{\ell=1}^\infty$ so that
$$ \gamma =\lim_{\ell\to\infty} \int_{|Y_0|>\epsilon} |u_{\hat\ak _\ell}(Y_0)|^2 d Y_0 >0. $$
Since
$k(\ak )/\ak ^2  \geq N(N-1)\WWWWW (0)$ and
$ \lim_{\ak \to\infty} E(k(\ak )/\ak ^2) = N(N-1)\WWWWW (0)$,
we have
\begin{align*}
  N(N-1)\WWWWW (0)
  &= \lim_{\ell\to\infty}\frac{1}{\hat\ak _\ell^2}(u_{\hat\ak _\ell}, k(\hat\ak _\ell) u_{\hat\ak _\ell})=
\lim_{\ell\to\infty} \lk
 u_{\hat\ak _\ell}, \sum_{j\neq l}^N  \WWWWW (x_j(Y_0)-x_l(Y_0)) u_{\hat\ak _\ell}
  \rk   \\
  &  \geq (1-\gamma) N(N-1)W(0) +
     \gamma \inf_{|Y_0|>\epsilon}\sum_{j\neq l}^N   \WWWWW (x_j(Y_0)-x_l(Y_0)) \\
  &  \geq N(N-1) \WWWWW (0).
\end{align*}
Thus we have
\eq{z7}
\inf_{|Y_0|>\epsilon}\sum_{j\neq l}^N   \WWWWW (x_j(Y_0)-x_l(Y_0)) = N(N-1)W(0).
\en
By (W2) and \kak{z7} there exists
a sequence $Z_n=(z_{1,n},\ldots,z_{(N-1),n})\in\mathbb{R}^{d(N-1)}$ such
that $|Z_n|>\epsilon$ and
$
 \lim_{n\to\infty}(  x_j(Z_n)-x_l(Z_n)) \to 0$
for $j\not =l$.
By the definition of $x_j(Y)$, we have
\begin{align*}
  &\lim_{n\to\infty}(x_2(Z_n)-x_1(Z_n)) = \lim_{n\to\infty}z_{1,n} =0,  \\
  &\lim_{n\to\infty}(x_3(Z_n)-x_2(Z_n)) = \lim_{n\to\infty}(z_{2,n}-\frac{1}{2}z_{1,n})
                                     = \lim_{n\to\infty}z_{2,n} =0, \\
  & \qquad    \cdots \\
  & \lim_{n\to\infty}(x_N(Z_n)-x_{N-1}(Z_n)) = \lim_{n\to\infty}z_{N-1,n}=0.
\end{align*}
This is a contradiction to $|Z_n|>\epsilon>0$ for all $n$. \qed

{\it Proof of Theorem \ref{Tex}}\\
Let $u_\ak $ be a ground state of $k(\ak )=Uh_0(\ak ) U^{-1}$.
By Proposition \ref{A2},
we may assume that $V \in C_0^\infty(\BR)$. Let $|\ak |>a_2$.
Let $v\in C_0^\infty(\mathbb{R}^d)$
 be a normalized vector such that
\eq{y3}
 \lk  v, \lk -\frac{1}{2Nm}\Delta_{x_c} + NV(x_c)\rk v \rk  < 0.
\en
Such a vector exists by $(W3)$.
We set
$ \Psi(Y)=\Psi(x_c,Y_0) = v(x_c)u_\ak (Y_0)$ for
$Y=(x_c,Y_0)=(x_c,y_1,\ldots,y_{N-1}) \in \RR^{dN}$.
Then
\eq{y4}
  (\Psi, Uh_V(\ak )U^{-1}\Psi)
  = - \frac{1}{2mN}(  v, \Delta_{x_c} v)
  + {\cal E}_0(\ak )
  + \lk  \Psi, \sum_{j=1}^N V(x_j(Y)) \Psi\rk.
\en
We define
\begin{align*}
  V_{j,\ast}^\ak (x_c) =
  \int_{\mathbb{R}^{d(N-1)}} d y_1 \cdots d y_{N-1}
  V(x_j(Y)) |u_\ak (y_1,\ldots,y_{N-1})|^2, \quad j=1,\ldots,N.
\end{align*}
By Lemma \ref{delta}, we have
\begin{align*}
  \lim_{\ak \to\infty} \lk
   \Psi, \sum_{j=1}^N V(x_j(Y)) \Psi\rk
 =\lim_{\ak \to\infty}\sum_{j=1}^N ( v, V_{j,\ast}^\ak  v
)
  = ( v, NV(x_c)v ).
\end{align*}
Therefore, by \kak{y3} and \kak{y4},
$(\Psi, h_V(\ak )\Psi) < {\cal E}_0(\ak )$
for $|\ak |>a_3$ with some $a_3>0$.
By this inequality, Lemma \ref{ene} and \kak{sd}, we conclude that for $\ak $ with
$|\ak |>\ak _c=\max\{a_1,a_3\}$,
 $$\Xi_V(\ak )-{\cal E}_V(\ak )\geq {\cal E}_0(\ak )-{\cal E}_V
  (\ak )>0.$$
  Then the theorem follows.
\qed
We give a general lemma.
\bl{B1}
Let $K_\epsilon$, $\epsilon>0$,  and $K$
 be  self-adjoint operators on a Hilbert space ${\cal K}$
and
$\s_{\rm ess}(K_\epsilon)=[\xi_\epsilon,\infty)$.
Suppose that $\lim_{\epsilon\rightarrow 0}
K_\epsilon=K$ in the uniform
resolvent sense,
and $\lim_{\epsilon\rightarrow 0} \xi_\epsilon=\xi$.
Then $\s_{\rm ess}(K)=[\xi,\infty)$. In particular $\lim_{\e\rightarrow 0}\is_{\rm ess}(K_\e)=\is_{\rm ess}(K)$.
\el
\proof
Let $a>\xi$. Then there
exists $\e_0$ such that
for all $\e$ with  $\e<\e_0$,
$\xi_\e<a$,
from which we have
$a\in \s(K_\e)$ for all $\e<\e_0$.
Since $K_\e$ uniformly converges to $K$ in the resolvent sense,
$a\in \s(K)$ follows from \cite[Theorem VIII.23 and p.291]{rs80}.
Since $a$ is arbitrary,
$(\xi, \infty)\subset \s(K)$ follows and then
$$[\xi,\infty)\subset \s_{\rm ess}(K).$$
 It is enough to show $\is_{\rm ess}(K)=\xi$. Let $\lambda \in
[\is_{\rm ess}(K),\xi)$ but $\lambda \not\in\s(K)$. Note that for
all sufficiently small $\e$, $\lambda \not\in\s(K_\e)$ by
\cite[Theorem VIII.24]{rs80}. Since $\RR\setminus\s(K)$ is an open
set, there exists $\delta>0$ such that $(\lambda -\delta, \lambda
+\delta)\not\subset \s(K)$. Let $P_A(T)$ denote the spectral
projection of a self-adjoint operator $T$ on a Borel set
$A\subset\RR$. We have $\lim_{\e\rightarrow 0} P_{(\is_{\rm
ess}(K)-\delta', \lambda )}(K_\e) = P_{(\is_{\rm ess}(K)-\delta',
\lambda )}(K)$ uniformly by \cite[Theorem VIII.23 (b)]{rs80}. In
particular, for some $\delta'>0$,
$$\| P_{(\is_{\rm ess}(K)-\delta', \lambda )}(K_\e)-
P_{(\is_{\rm ess}(K)-\delta', \lambda )}(K)\|<1.$$
Then
$P_{(\is_{\rm ess}(K)-\delta', \lambda )}(K_\e){\cal K}$ is isomorphic to
$P_{(\is_{\rm ess}(K)-\delta', \lambda )}(K){\cal K}$.
Hence
the dimension of $P_{(\is_{\rm ess}(K)-\delta', \lambda )}(K){\cal K}$ is
finite, since that of
$P_{(\is_{\rm ess}(K_\e)-\delta', \lambda )}(K){\cal K}$ is
finite.
Thus $ (\is_{\rm ess}(K)-\delta', \lambda )\cap \s(K)\subset \s_{\rm disc}(K)$.
This is a contradiction. Hence we have $[ \is_{\rm ess}(K),\xi)   \subset \s(K)$.
Suppose that $ \is_{\rm ess}(K) <\xi$.
Let $\tau>0$ be  sufficiently small.
Note that $(\is_{\rm ess}(K)-\tau, \is_{\rm ess}(K)+\tau)\subset \s_{\rm disc}(K_\e)$
for all sufficiently small $\e$.
Let $\theta\in C_0^\infty(\RR)$ satisfy that
$$\theta(z)=\lkk
\begin{array}{ll}1,& |z-\is_{\rm ess}(K)|<\tau,\\
0,&|z-\is_{\rm ess}(K)|>2\tau.
\end{array}\right.$$
Then we have
$\lim_{\e\rightarrow 0}\theta(K_\e)=\theta(K)$ uniformly by
\cite[Theorem VIII.20]{rs80}.
Since $ \theta(K_\e)$ is
a finite rank operator for all sufficiently small $\e$,
$\theta(K)$ has to be a compact operator.
It contradicts with the fact, however,
that  the spectrum of $\theta(K)$ is continuous.
Then we can conclude that $\is_{\rm ess}(K)=\xi$ and
 the proof is complete.
\qed

 Let $V: \BR\to \RR$ be a real-valued  measurable function.
 \bl{A1}
  Let
$\Delta$ be the $d$-dimensional Laplacian. Assume that
$V(-\Delta+1)^{-1}$ is  a compact operator. Then there exists a
sequence $\{V^\epsilon\}_{\epsilon>0}$ such that
$V^\epsilon\in
C_0^\infty(\BR)$ and $\lim_{\e\rightarrow 0} V^\epsilon(-\Delta
+1)^{-1}=V(-\Delta +1)\f $ uniformly. \el
\proof
Generally, let $A$ be a compact operator and $\{B_n\}_n$  bounded operators such that
 $\slim_{n\to\infty} B_n  = 0$, then $B_n A \to 0$ as $n\to\infty$  in the operator norm.
Since $V(-\Delta+1)^{-1}$ is a compact operator, we obtain that for a sufficiently large $R>0$,
\eq{z2}
 \| (1-\chi_R) V (-\Delta +1)^{-1} \| < \epsilon/3, \quad
\en
where $\chi_R$ characteristic function of $\{x\in\BR| |x|<R\}$.
Let $\chi^{(n)}$ denote the characteristic function of
$\{ x\in\BR| |V(x)|<n \}$.
Since
$(1-\chi^{(n)})\to 0$ strongly as $n\to\infty$,
\eq{z3}
\| (1-\chi^{(n)})\chi_R V (-\Delta +1)^{-1} \| < \epsilon/3
\en
for a sufficiently large $n$.
Since $C_0^\infty(\supp(\chi_R\chi^{(n)}))$ is dense in $L^2(\supp(\chi_R\chi^{(n)}))$,
there exists a sequence
$\{V_m\}_{m}\subset C_0^\infty(\supp(\chi_R\chi^{(n)}))$
$$
 \| V_m - \chi_R\chi^{(n)}V \|_{L^2(\BR)} \to 0$$ as $m\to\infty$.
Since $\chi_R\chi^{(n)}V$ has a compact support and is bounded,
we obtain that
$\slim_{m\to\infty}V_m= \chi_R\chi^{(n)}V$ as an operator.
Thus for a sufficiently large $m$,
\eq{z4}
  \| (V_m -\chi_R\chi^{(n)}V) (-\Delta +1)^{-1} \| < \epsilon/3.
\en By \kak{z2}-\kak{z4}  we can obtain that for an arbitrary
$\epsilon>0$, $ \| (V- V_m) (-\Delta + 1)^{-1} \| < \epsilon$ for
a sufficiently large $m$. Thus the lemma follows by setting
$V_m=V^\epsilon$. \qed
Let $\beta\subset\N$.
Set
$$k_0(\beta)=-\sum_{j\in \beta}\frac{1}{2m_j} \Delta_j
+\sum_{i,j\in\beta} V_{ij},\quad k_V(\beta)=h_0(\beta)+\sum_{j\in
\beta}V_j$$ with $V_i\in L_\mathrm{loc}^2(\BR)$ and $V_{ij} \in
L_\mathrm{loc}^2(\BR)$ such that $V_i (-\Delta +1)^{-1}$ and
$V_{ij}(-\Delta +1)^{-1}$ are compact operators. We define
$K=k_V(\N)$. Let \eq{BB3} \Xi_V=\min_{\beta\subsetneqq C_N}
 \left \{\is(k_0(\beta))+ \is(k_V(\beta))\right \} \en be the lowest two cluster
threshold of $K$.
\begin{proposition}{\label{A2}}
There exist sequences
$\{V_i^\e\}_\epsilon  , \{V_{ij}^\e\}_\epsilon
\subset  C_0^\infty(\BR)$, $i,j=1,...,N$,
such that
$$ (1)\ \lim_{\epsilon\to0} \Xi_V(\epsilon)=\Xi_V,\quad
(2) \ \lim_{\epsilon\to0} \is_{\rm ess}(K(\e))=\is_{\rm ess}(K),$$
where $\Xi_V(\epsilon)$ (resp. $K(\e)$ ) is
$\Xi_V$ (resp. $K$)   with $V_i$ and $ V_{ij}$ replaced by
$V_i^\e$ and $V_{ij}^\e$, respectively.
\end{proposition}
\proof
By Lemma \ref{A1},  there exist  sequences
$\{V_i^\e\}_{\epsilon>0}  , \{V_{ij}^\e\}_{\epsilon>0}  \subset  C_0^\infty(\BR)$,
such that
$$V_i^\e(x_i)(-\Delta_i+1)^{-1}\rightarrow V_i(x_i)(-\Delta_i+1)^{-1}$$ and
$$V_{ij}^\e(x_i-x_j)(-\Delta_i-\Delta_j+1)^{-1}\rightarrow V_{ij}(x_i-x_j)(-\Delta_i-\Delta_j+1)^{-1}$$
uniformly as $\epsilon\rightarrow 0$ for
$i,j=1,...,N$.
Hence
$\is(k_V(\epsilon))$ and $\is(k_0(\epsilon))$
converge to $\is(k_V)$ and $\is(k_0)$
as $\epsilon\rightarrow 0$, respectively.
Then (1)
follows from the definition \kak{BB3}.
By this and the uniform convergence of $K(\e)$ to $K$ in the resolvent sense,
Lemma \ref{B1} yields (2).
\qed

\cleardoublepage
\section{Absence of ground state}
\subsection{Introduction}

\subsubsection{Stability and the decay of variable mass}
In this  section we  study a general version of the Nelson model, i.e.,
the Nelson model with a variable coefficients.
This model is an extension of the standard Nelson model
 when the Minkowskian space-time   is replaced by a static pseudo Riemannian manifold. It is studied in the series of papers \cite{ghps09,ghps11,ghps12,ghps11preprint}.
In this section the absence of ground state of the  Nelson model with variable coefficients is discussed under infrared singularity condition.
Throughout this section we assume that
$$d=3.$$
The Hamiltonian of the Nelson Hamiltonian with a variable coefficients\ko{Nelson Hamiltonian!variable coefficients} is defined
formally by
\begin{eqnarray}
H&=&
\half
\sum_{\mu,\nu=1}^3 D_\mu A_{\mu\nu}(\xx ) D_\nu +W(\xx ) +
\int \omega(D,x)
\add (x) a(x) dx\non\\
&&
\label{g123}
+\frac{1}{\sqrt2}
\int \omega^{-\han}(D,x)
 \rho(x-\xx )(
 \add(x) +a(x)) dx,
\end{eqnarray}
where     $a(x)$ and $\add(x)$ the annihilation operator and the creation operator in the position representation, respectively,
$\rho $ a nonnegative cutoff function and $\omega=h^\han$
a dispersion relation\ko{dispersion relation!variable mass}
with a position dependent variable mass\index{variable mass}
$m(x)$:
\eq{g234}
h=h(D,x)=
\sum_{\mu,\nu=1}^3
c(x)^{-1}
D_\mu a_{\mu\nu}(x)
D_\nu
c(x)^{-1}
+m^2(x).
\en
We give examples of \kak{g234} in the next section.
In  \cite{ghps11} the existence of ground states of $H$
is shown when
\eq{g4}
m(x)\geq a \langle x\rangle\f,
\en
where $\langle x\rangle=(1+|x|^2)^\han$.
Then  we study the case of
$$m(x)\leq a\langle x\rangle^{-\beta/2},\quad \beta<2
$$ in this lecture  note. See Figure \ref{picexab}. 

The standard Nelson model is defined by $H$  with $A_{\mu\nu}(\xx)$ replaced by $\delta_{\mu\nu}$, $a_{\mu\nu}(x)$ by $\delta_{\mu\nu}$ and
$m(x)$ by a constant $m\geq0$.
By the condition $\rho\geq 0$,
$\hat \rho(0)>0$ follows, and the integral $\int|\hat\rho(k)|^2/\omega(k)^3 dk$ is finite
if and only if $m>0$ since $d=3$.
Thus
$m>0$  corresponds to the infrared regular condition and $m=0$ to
the infrared singular condition.

\begin{figure}[t]
\begin{center}
\arrayrulewidth=1pt
\def\arraystretch{2.0}
\begin{tabular}{|c|c|c|}
\hline
\       & $m(x)\geq a\ab{x}^{-1}$  &
$m(x)\leq a\ab{x}^{-\beta}$, $\beta>1$ \\
\hline
ground state
&
exist&
not exist
 \\
\hline
\end{tabular}
\end{center}
\caption{Existence and absence of ground state}
\label{picexab}
\end{figure}%
\subsubsection{Klein-Gordon equation on  pseudo Riemannian manifold}
In quantum field theory
the dispersion relation $\omega=\sqrt{-\Delta+m^2}$ can be derived from the Klein-Gordon equation\index{Klein-Gordon equation}:
\eq{wave}
\frac{\partial^2 \phi}{\partial t^2}=-\omega^2 \phi.
\en
On the other hand the dispersion relation with variable coefficients can be derived from
the Klein-Gordon equation\index{Klein-Gordon equation!pseudo Riemannian manifld} on a pseudo Riemannian manifold\index{pseudo Riemannian manifold}.
We here give an example
of a Klein-Gordon equation defined on a static pseudo Riemannian manifold $\ms M$ such that a short range potential $v(x)=\mathcal O(\ab{x}^{-\beta-2})$ appears.

Let $\underline x=(t,x)=(x_0,x)\in\RR\times\RR^3$
and
$\ms M$
the  $4$ dimensional pseudo Riemannian manifold equipped with
the metric tensor:
\eq{ex10}
g(\underline x)=g(x)=\lk
\begin{array}{cccc}
 {e^{-\theta(x)}} & 0&  0&   0\\
 0& {-e^{-\theta(x)}} & 0& 0 \\
 0& 0& {-e^{-\theta(x)}} & 0  \\
 0& 0& 0& {-e^{-\theta(x)}}
 \end{array}
 \rk.
 \en
Note that $g$ depends on $x$ but independent of $t$.
The line element associated with $g$ is given by
\eq{line}
ds^2=\eb dt\otimes dt -\eb \sum_{j=1}^3 dx^j\otimes dx^j .
\en
The Klein-Gordon equation on $\ms M$ is
\eq{ex11}
\square _g\phi+m^2\phi=0,
\en
where
the d'Alembertian operator\index{d'Alembertian operator} is defined by
\eq{dar1}
\square _g=e^{\theta(x)}\partial_t^2 -e^{2\theta(x)}
\sum_j \partial_j e^{-\theta(x)}\partial_j.
\en
Thus the
Klein-Gordon equation \kak{ex11} is
reduced to the equation
\eq{ex12}
\frac{\partial^2\phi}{\partial t^2}=K_0 \phi ,\en
where
\eq{dar2}
K_0= e^{\theta(x)}
\sum_j
\partial_j
\eb \partial_j -\eb m^2.
\en
The operator $K_0$ is symmetric on the weighted $L^2$ space
$L^2(\BR;\eb dx)$.
Now we transform the operator $K_0$
to the one on $\LRT$.
This is done by   the unitary map $U_0:L^2(\BR;\eb dx)
\rightarrow \LR$,
$f\mapsto e^{-(\han)\theta}f$.
\bl{ex16}
There exist functions $\theta$ and
$v$ such that
$U_0K_0U_0^{-1}=\Delta-v$,
   $v(x)=\mathcal O(\ab{x}^{-\beta-2})$ for $\beta\geq 0$, and
   $-\Delta +v$ has no non-positive eigenvalues.
\el
Hence the Klein-Gordon equation
\kak{ex12} is transformed
to the equation
\eq{ex12-1}
\frac{\partial^2\phi}{\partial t^2}
-\Delta\phi+v\phi=0
\en
on $\LRT$, and the dispersion relation is given by
$\sqrt{-\Delta+v}$.
Although the proof of Lemma \ref{ex16} is straightforward, we shall show this statement through a more general scheme in what follows.

Suppose that
 $g=(g_{\mu\nu})$, $\mu,\nu=0,1,2,3$,
is a metric tensor on $\RR^4$ such that
\bi
\item[(1)]
$g_{\mu\nu}(\underline x)=g_{\mu\nu}(x)$, i.e., it is independent of time $t$,
\item[(2)] $g_{0j}(\underline x)=g_{j0}(\underline x)=0$, $j=1,2,3$,
    \item[(3)]
    $g_{ij}(\underline{x})=-\gamma_{ij}(x)$, where $\gamma=(\gamma_{ij})$ denotes a
    $3$-dimensional Riemannian metric.
\ei
Namely
\eq{me1}
g=\MMM {g_{00}} 0 0 {-\gamma}.
\en
Let  $\ms M$ be a pseudo Riemannian manifold equipped with the metric tensor   $g$ satisfying (1)-(3) above.
Then
the line element on $\ms M$ is given by
$$ds^2=g_{00}(x)dt\otimes dt -\sum_{i,j=1}^3\gamma_{ij}(x) dx^i \otimes dx^j.$$
Let $g^{-1}=(g^{\mu\nu})$ denote the inverse of $g$. In particular $1/g_{00}=g^{00}$. We also denote the inverse of $\gamma$ by $\gamma^{-1}=(\gamma^{ij})$.
The Klein-Gordon equation on the static pseudo Riemannian manifold $\ms M$ is generally given by
\eq{ex1}
\square _g \phi+
(m^2+\rieman {\mathcal R})\phi=0,
\en
where $\rieman$ is a constant,
  ${\mathcal R}$ the scalar curvature of $\ms M$,
  and $\square _g$ is
 the d'Alembertian operator\index{d'Alembertian operator} on $\ms M$, which  is   given by
\eq{ex23}
\square _g=\sum_{\mu,\nu=0}^3
    \frac{1}{\sg}\partial_\mu g^{\mu\nu} \sg \partial_\nu .
    \en
      Let us assume that $g_{00}(x)>0$.
Then
\kak{ex1} is rewritten as
\eq{ex22}
\frac{\partial^2\phi}{\partial t^2}=K\phi,
\en
where
\eq{dar4}
K=g_{00}\lk
\frac{1}{\sg}\sum_{i,j=1}^3
\partial_j
\sg \gamma^{ji}\partial_i-m^2-\rieman {\mathcal R}\rk.\en
The operator
$K$ is symmetric on $L^2(\RR^3;\rho(x) dx)$, where
\eq{ex20}
\rho=\frac{\sg}{g_{00}}=g_{00}^{-1/2}\sqrt{|{\rm det} \gamma|}.
\en
Now let us transform the operator $K$ on
$L^2(\RR^3;\rho(x) dx)$ to the one on $\LRT$.
Define the unitary operator
$U:L^2(\RR^3;\rho(x) dx)\rightarrow \LRT$ by
\eq{dar5}
Uf=\rho^\han f.
\en
Let $\rho_i=\partial_i \rho$ and $\partial_i\partial_j\rho=\rho_{ij}$ for notational simplicity. Furthermore we set $\alpha^{ij}=g_{00}\gamma^{ij}$ and $\partial_k\alpha^{ij}=\alpha^{ij}_k$.
Since $U^{-1}\partial_j U=\partial_j +\frac{\rho_j}{2\rho}$,
we see that
as an operator identity
\eq{ex3}
U^{-1}\lk
\sum_{i,j=1}^3 \partial_i g_{00}\gamma^{ij}
 \partial_j \rk
 U=g_{00}\sum_{i,j=1}^3 \gamma^{ij} \partial_i\partial_j +V_1+V_2,
 \en
 where
\begin{eqnarray*}
V_1&=&
\sum_{i,j=1}^3 \lk
\alpha^{ij}_i +\alpha^{ij}
\frac{\rho_i }{\rho}\rk\partial_j,\\
V_2&=& \frac{1}{4}
\sum_{i,j=1}^3
\lk
2
\alpha^{ij}_{i}
\frac{ \rho_j}{\rho}
 +2\alpha^{ij}
\frac{\rho_{ij}}{\rho}-
\alpha^{ij}
\frac{\rho_i}{\rho}
\frac{\rho_j}{\rho}
\rk.
\end{eqnarray*}
Set
$|{\rm det} g|=G$ and $\partial_i G=G_i$.
Hence
we have
$$V_1=g_{00}\sum_{i,j=1}^3 \lk \gamma^{ij}_i+\frac{G_i}{2G}\rk\partial _j,$$where $\gamma^{ij}_i=\partial_i\gamma^{ij}$,
and directly we can see that
\eq{ex4}
g_{00}\frac{1}{\sg}
\sum_{i,j=1}^3 \partial_i
\sg \gamma^{ij}\partial_j=
V_1+g_{00}\sum_{i,j=1}^3
\gamma^{ij}\partial_i\partial_j.
\en
Comparing \kak{ex3} with \kak{ex4} we obtain that
\eq{ex5}
U^{-1}\lk
 \sum_{i,j=1}^3 \partial_i g_{00}\gamma^{ij}\partial_j-V_2
 \rk U=
 g_{00}\frac{1}{\sg}\sum_{i,j=1}^3 \partial_i
\sg \gamma^{ij}\partial_j.
\en
Then we proved the lemma below.
\bl{ex14}
It follows that
\eq{ex6}
UKU^{-1}=
 \sum_{i,j=1}^3 \partial_i g_{00}\gamma^{ij}\partial_j-v,
 \en where
 $v=g_{00}(m^2+\rieman {\mathcal R})+V_2$.
 \el
By
Lemma \ref{ex14},
\kak{ex22} is transformed to
the equation:
 \eq{23p}
\frac{\partial^2\phi}{\partial t^2}=
\lk
 \sum_{i,j=1}^3 \partial_i g_{00}\gamma^{ij}\partial_j-v\rk
 \phi
 \en
on $\LR$.

{\it Proof of Lemma \ref{ex16}}:
Now we come back to the proof of Lemma \ref{ex16}.
Set
$$g_{\mu\nu}(x)=\lkk
\begin{array}{ll}
e^{-\theta(x)},&\mu=\nu=0,\\
-e^{-\theta(x)},& \mu=\nu=1,2,3,\\
0,&\mu\not=\nu.
\end{array}
\right.
$$
Then
\eq{ex7}
\rho=\frac{\sg}{g_{00}}=e^{-\theta}, \quad \alpha^{ij}=g_{00}\gamma^{ij}=\delta_{ij},
\en
and $UKU^{-1}=\Delta -v$ follows by \kak{ex6},
where,
inserting \kak{ex7} to $v$,
we have
\eq{ex8}
v=e^{-\theta}(m^2+\rieman {\mathcal R})
-\frac{\Delta \theta}{2}
+\frac{|\nabla\theta|^2}{4}.
\en
Taking $\rieman=0$, $m=0$, and $\theta(x)=2a \ab{x}^{-\beta}$,
we obtain
\eq{ex21}
v(x)=-a\ab{x}^{-\beta-4}(\beta(\beta-1)|x|^2-3\beta) +a^2\beta^2 \ab{x}^{-2\beta-4}|x|^2.
\en
In the case of $0\leq \beta \leq 1$ and $a>0$, we see
that $v\geq 0$ and $v=\mathcal O(\ab{x}^{-\beta-2})$.
Furthermore $-\Delta+v$ has no non-positive eigenvalues.
In the case of $\beta >1$ and $a<0$, we see that however
 $v\not\geq0$.
We can estimate the number of non-positive eigenvalues of  $-\Delta +v$
by the Lieb-Thirring inequality\index{Lieb-Thirring inequality} \cite{lie76}:
\eq{lt}
\# \{\mbox{ eigenvalues of }-\Delta+v\leq  0\}\leq a_3
\int |v_-(x)|^{3/2} dx,\en
 where $v_-$ denotes the negative part of $v$ and $a_3$ is a constant independent of $v$.
This yields that
  $-\Delta+v$ has no non-positive eigenvalues for sufficiently small $a$.
Thus the lemma  holds.
\qed

\subsection{The Nelson model with
a variable mass}

\subsubsection{Dirichlet forms and symmetric semigroups}
Before going to study the Nelson Hamiltonian with variable coefficients we review
fundamental properties of
Dirichlet forms and symmetric semigroups, which will be  used in the next sections.
The general reference in this section is \cite{dav89}.

We assume that the dimension of the configuration space is $d$.
\label{a1}
Let $({\ms E}, D)$ be a symmetric quadratic form on
$\LR$ with a form domain $D$. $({\ms E}, D)$
is  Markovian
if and only if
 for arbitrary $\epsilon>0$, there
exists $\rho_\epsilon$ such that
\bi
\item
[(1)]
$\rho_\epsilon(t)=t$ for
$t\in [0,1]$,
 $-\epsilon\leq \rho_\epsilon(t)\leq 1+\epsilon$ for all $t\in\RR$,
 $0\leq \rho_\epsilon(t)-\rho_\epsilon(s)\leq t-s$ for $s<t$,
 \item[(2)]
$\rho_\epsilon\circ f\in D$ and
$
\ms E(\rho_\epsilon\circ f,\rho_\epsilon\circ f)\leq \ms E (f,f)
$ holds for $f\in D$.
\ei
A Markovian closed symmetric form $({\ms E},D)$ is called the Dirichlet form\ko{Dirichlet form}. When $C_0^\infty(\BR)$ is a form core of the Dirichlet form $({\ms E},D)$,
it is called a {\it regular} Dirichlet form\ko{Dirichlet form!regular}.
When $f,g\in D$ satisfies ${\rm supp} f\cap {\rm supp}g=\emptyset$,
$\ms E(f,g)=0$.
Then $({\ms E},D)$ is called a {\it local} Dirichlet form\ko{Dirichlet form!local}.

Let $g_{\mu\nu}\in L_{\rm loc}^1(\RR^d)$ and
$(g_{\mu\nu}(x))_{1\leq\mu,\nu\leq d}=g(x)$
satisfy
\eq{51}
\la_1(x)\one
\leq g(x)\leq \la_2(x)\one
\en
with strictly positive
continuous
functions $\la_j$.
Define
\eq{oota3}
{\ms E}_g(f,g)=\sum_{\mu,\nu=1}^d
\int_{\RR^d}
 g_{\mu\nu}(x)\ov{
 \partial_\nu f(x)} \partial_\nu g(x) dx
 \en
 for $f,g\in C_0^\infty(\RR^d)$.
\bp{el0}
${\ms E}_g$ is closable quadratic
 form on $C_0^\infty(\RR^d)$.
\ep
\proof
See \cite[Theorem 1.2.6]{dav89}.
\qed
We denote the closure by $\bar {\ms E}_g$.
\bp{el}
Let $L$ be the self-adjoint operator
 associated with
 $\bar {\ms E}_g$.
 Then
\bi
\item
[(i)]
 $e^{-tL}$, $t\geq0$,
  is contraction from
  $L^p(\RR^d)$ to itself for
 all $1\leq p\leq\infty$,
\item[(ii)]
 $e^{-tL}$, $t\geq0$, is positivity
 preserving.
\ei
\ep
\proof
See
\cite[Theorem 1.3.5]{dav89}.
\qed
\bp{el2}
Suppose that
$K>0$ be a self-adjoint operator
such that $e^{-tK}$ is positivity
preserving and $e^{-tK}$
is bounded on  $L^\infty(\RR^d)$.
Let ${\ms E}$ denote
the quadratic form associated with
$K$.   Then
\bi
\item[(1)]
A bound of the form
$$\|e^{-tK} f\|_\infty\leq C_1 t^{-\alpha/4}\|f\|_2$$
with $\alpha>2$ for all $t\geq 0$
and all $f\in L^2(\RR^d)$
is equivalent to
$$\|f\|^2_{2\alpha/(\alpha-2)}
\leq C_2 {\ms E}(f,f).
$$
\item[(2)]
Suppose a bound
$\|e^{-tK} f\|_\infty\leq
C_t\|f\|_2$
follows for all $t\geq 0$ and all
$f\in L^2(\RR^d)$.
Then $e^{-tK} f$ has
an integral
kernel
$e^{-tK}(x,y)$ for all $t\geq0$
which satisfies that
$0\leq  e^{-tK}(x,y)\leq C_{t/2}^2$
almost everywhere.
\ei
\ep
\proof
See
\cite[Theorem 2.4.2]{dav89} for (1), and
\cite[Lemma 2.1.2]{dav89} for (2).
\qed
\begin{remark}
{\rm Let $K>0$ be a self-adjoint operator
in $\LR$ such that $e^{-tK}$ is positivity
preserving and $e^{-tK}$
is bounded on
$L^\infty(\RR^d)$.
Then $e^{-tK}$ is bounded on $L^p(\BR)$ for
$1\leq p\leq \infty$.
}\end{remark}
Suppose that $L$ is the self-adjoint
operator associated with
the quadratic form $\bar{\ms E}_g$ defined by \kak{oota3}
but
$\la_1(x)$ and $\la_2(x)$ in \kak{51}
are replaced by
positive constants $\la_1$ and
$\la_2$, respectively.
Then $L$ is called a strictly elliptic operator.

\bp{el3}
Let $L$ be a strictly elliptic operator.
Then $e^{-tL}$ has an integral kernel
$e^{-tL}(x,y)$
and has Gaussian bounds:
$$
C_1
e^{C_2 t\Delta}(x,y)
\leq
e^{-tL}(x,y)\leq
C_3
e^{C_4 t\Delta}(x,y).$$
\ep
\proof
The upper and lower Gaussian bounds are
proven in
\cite[Corollary 3.2.8]{dav89}
 and \cite[Theorem 3.3.4]{dav89},
respectively.
\qed

\subsubsection{Schr\"odinger operators with divergence form}
We define the Schr\"odinger operator $K$ on $\LRT$ with variable coefficients.
Let $K_0$ be defined formally
by
\eq{1}
K_0=\half \sum_{\mu,\nu=1}^3
D_\mu A_{\mu\nu}(\xx)D_\nu,
\en
where
$D_\mu=-i\nabla_\mu$ with the domain
$
\dom(D_\mu)=H^1(\BT)$
describes the momentum operator and
 $A=A(\xx )=(A_{\mu\nu}(\xx ))_{1\leq\mu,\nu\leq 3}$ is a $3\times 3$
 symmetric matrix
for each $\xx \in\BT$.
We give the rigorous definition of $K_0$ through
a quadratic form.
We introduce an assumption on $A(\xx )$.
\begin{assumption}
\label{ass}
{\bf (Uniform elliptic condition)}
Suppose that
each $A_{\mu\nu}$, $1\leq \mu,\nu\leq 3$,
 is a measurable function,
 and
 $A$ is uniformly elliptic, i.e.,
 there exist  constants $C_0>0$ and $C_1>0$
 such that
 \begin{equation}
 \label{y222}
C_0\one \leq A(\xx )\leq C_1\one.
 \end{equation}
\end{assumption}

Let  ${\ms E}_\one (f,g)$ and
${\ms E}_A(f,g)$ be the quadratic forms defined  by
\eq{y1as}
{\ms E}_A(f,g)=\half \sum_{\mu,\nu=1}^3
\int A_{\mu\nu}(\xx )
\ov{\partial_\mu f(\xx )}
\cdot  \partial_\nu g(\xx )
d\xx
\en
and
\eq{y5}
{\ms E}_\one (f,g)=
\half \sum_{\mu=1}^3
\int
\ov{\partial_\mu f(\xx )}
\cdot  \partial_\mu g(\xx )
d\xx
\en
with the form domain
$H^1(\BT)$.
Under Assumption \ref{ass},
we have
\eq{y4ask}
C_0
{\ms E}_\one (f,f)
\leq
{\ms E}_A (f,f)
\leq
C_1 {\ms E}_\one (f,f),
\quad f\in H^1(\BT).\en
From this inequality we can see that
$({\ms E}_A, H^1(\BT))$ is a closed semibounded form.
We define $K_0$ by the unique self-adjoint operator associated with $\ms E_A$:
there exists a nonnegative self-adjoint operator
$K_0 $ such that
\eq{ko5}
\ms E_A(f,g)=(K_0 ^\han f, K_0 ^\han g)
\en
 with $
 H^1(\BT)=\dom(K_0 ^\han)$.
In general it is not easy to specify the operator
domain of $K_0$.
We can however specify it under some regularity conditions on $A_{\mu\nu}(\xx )$.
Let
$$W^{n,\infty}=\lkk
f\in L^\infty(\BT)|
\partial ^z f\in L^\infty(\BT) \mbox{ for } |z|\leq n\rkk,$$
where $\partial $
denotes
the distributional
differential operator on $L_{\rm loc}^1(\BT)$.
It is fundamental  that
for $f\in W^{1,\infty}(\BT)$ and $u\in H^1(\BT)$, we have $fu\in H^1(\BT)$ and
$\partial_\mu(fu)=(\partial_\mu f) u+f\partial_\mu u$
 for $\mu=1,2,3$.
\bl{domain}
Suppose that each $A_{\mu\nu}$, $1\leq \mu,\nu\leq 3$,
satisfies  $A_{\mu\nu}\in W^{1,\infty}(\BT)$,
and
Assumptions \ref{ass}. Then
$\dom(K_0)=H^2(\BT)$ and
$$K_0 f=\sum_{\mu,\nu=1}^3 D_\mu(A_{\mu\nu}(\xx ) D_\nu f).$$
\el
\proof
Since  $H^2(\BT)\subset \dom(K_0)$ is trivial, it is enough to see $H^2(\BT)\supset \dom(K_0)$.
Let $f\in K_0$ and
 $T_t^\mu=e^{itD_\mu}$.
Note that
$T_t^\mu f(\xx )=f(\xx +te_\mu)$, where $e_\mu$ is the unit vector in $\BT$ to the $\mu$th direction, and
$D_\nu T_t^\mu  f=T_t^\mu  D_\nu f$ follows for $f\in H^1(\BT)$ with  $\mu,\nu=1,...,d$.
It is a fundamental fact that
$f\in H^1(\BT)$ if and only if
\eq{ko2}
\sup_{0\in (0,1]}\left\|
\frac{1}{t}(T_t^\mu-1)
f\right\|_{L^2}<\infty,\quad \mu=1,...,d.
\en
Furthermore
\eq{ko21}
\sup_{0\in (0,1]}\left\|\frac{1}{t}(T_t^\mu-1)
f\right\|\leq\|f\|_{H^1(\BT)},\quad \mu=1,...,d
\en
holds for $f\in H^1(\BT)$.
Then
if $f\in H^1(\BT)$ satisfies that
\eq{ko13}
\sup_{0\in (0,1]}
\left\|
\frac{1}{t}(T_t^\mu-1) f\right\|_{H^1(\BT)}
<\infty, \quad \mu=1,...,d,
\en
then $f\in H^2(\BT)$.
Let
$\|f\|_{\ms E_A}^2=\|f\|^2+\ms E_A(f,f)$. We fix $\alpha=1,...,d$,
and
set
$$\Delta_tf(\xx )
=
\frac{1}{t}(T_t^\alpha-1)f(\xx )=
\frac{1}{t}(f(\xx +te_\alpha)-f(\xx )).$$
Let $f\in \dom(K_0)(\subset H^1(\BT))$ and set $f_t=\Delta _t f$.
We will show that
\eq{ko133}
\sup_{t\in (0,1]}
\|f_t\|_{H^1(\BT)}<\infty.
\en
We have
$
\|f_t\|_{\ms E_A}^2
=
(\Delta_t f, f_t)_{\ms E_A}
=P_t+Q_t
$,
where
\begin{align*}
P_t&=
-(f, \Delta_{-t}f_t)_{L^2}
-(K_0 f, \Delta_{-t}f_t)_{L^2},\\
Q_t
&=
(f, \Delta_{-t}f_t)_{\ms E_A}
+
(\Delta_tf, f_t)_{\ms E_A}.
\end{align*}
We have
$$|P_t|\leq \|f_t\|_{H^1(\BT)}
(\|f\|+\|K_0 f\|)\leq
\|f_t\|_{\ms E_A}
(\|f\|+\|K_0 f\|),$$
while
\begin{align*}
Q_t
&=&
(f, \Delta_{-t}f_t)
+
(\Delta_tf, f_t)
+\sum_{\mu,\nu=1}^3 \lk
(A_{\nu\mu}D_\mu f, D_\nu \Delta_{-t}f_t)+
(A_{\nu\mu}D_\mu \Delta_t f, D_\nu f_t)\rk
.\end{align*}
We have
\begin{align*}
(A_{\nu\mu}D_\mu f, D_\nu \Delta_{-t}f_t)+
(A_{\nu\mu}D_\mu \Delta_t f, D_\nu f_t)
&=
(A_{\nu\mu}D_\mu \Delta_tf-\Delta_t(A_{\nu\mu}D_\nu f),
D_\nu f_t)\\
&=
(-\Delta_t A_{\nu\mu}
\cdot
T_t(D_\mu f),
D_\nu f_t).
\end{align*}
Then
$$
|(A_{\nu\mu}D_\mu f, D_\nu \Delta_{-t}f_t)+
(A_{\nu\mu}D_\mu \Delta_t f, D_\nu f_t)|\leq
\|\Delta_t A_{\nu\mu}\|_\infty
\|f\|_{H^1(\BT)}\|f_t\|_{H^1(\BT)}
$$
and
$$|Q_t|\leq C\|f\|_{H^1(\BT)}
\|f_t\|_{\ms E_A}$$ follows
with some constant $C$ independent of $t$.
Then we see that
$$
\|f_t\|_{\ms E_A}^2
\leq
\|f_t\|_{
H^1(\BT)}(\|f\|+\|K_0 f\|)+
C\|f\|_{H^1(\BT)}
\|f_t\|_{\ms E_A}$$
and
$$
\sup_{t\in (0,1]}
\|f_t\|_{H^1(\BT)}
\leq
\sup_{t\in (0,1]}
\|f_t\|_{\ms E_A}
\leq
\|f\|+\|K_0 f\|
+
C\|f\|_{H^1(\BT)}<\infty.
$$
Then \kak{ko133}
follows and the lemma  is proven.
\qed
We furthermore introduce the assumption on external potentials $W$.
\begin{assumption}
\label{v}
{\bf (Confining potential)}
$\VV\in L_{\rm loc}^1(\BT)$ and there exist $\delta >0$ and
$C>0$  such that
\eq{cv}
W(\xx )\geq C\ab {\xx }^{2\delta}.
\en
\end{assumption}
 The Schr\"odinger operator
 $K$ on $\LRT$
 with kinetic term $K_0$
 and an external potential
 $W$ satisfying
Assumption \ref{v}
is defined by the quadratic form sum.
Let
\eq{yoko11}
\ms E(f,g)=\ms E_A(f,g)+(W^\han f, W^\han g)
\en
with the form domain
$C_0^\infty(\BT)$.
The quadratic form
$\ms E$ is semibounded and then closable.
We denote the closure of $\ms E$ by the same symbol.
We define $K$ as the unique self-adjoint operator associated with the quadratic form $\ms E$:
\eq{yoko30}
\ms E(f,g)=
(K^\han f, K^\han g)
\en
 for $f, g$ in the quadratic form domain of
 $\ms E$.
We describe it as
\eq{y1-1}
K=K_0 \,\,  \dot +\,\,  W .
\en
\bl{ground}
{\bf (Compact resolvent)}
Suppose Assumptions \ref{ass} and \ref{v}. Then
$K$ has a compact resolvent and in particular it has a ground state.
\el
\proof
In general a nonnegative self-adjoint operator $T$ has a compact resolvent if and only if
$$D_T(b)=\{f\in \dom(T^\han)\ |\  \|f\|<1, \|T^\han f\|\leq b\}$$
is a compact set for all $b>0$.
See e.g., \cite[Theorem XIII.64]{rs4}.
Let $L=-\half \Delta\, \dot +\,  W$.
Then $L$ is  essentially self-adjoint on $C_0^\infty(\BT)$ by Kato's inequality,
and since
$D_L(b)$ is compact for all $b$,
$L$ has a compact resolvent.
By Assumption \ref{ass}
we see that
\eq{yoko12}
\|L^\han f\|
\leq
C_0\f \|K^\han f\|
\en
 for $f\in C_0^\infty(\BT)$, where
 constant $C_0$ is given by \kak{y222}.
 By a limiting argument
\kak{yoko12} holds true for $f\in \dom(K^\han)$,
and
$D_{K}(b)\subset D_L(b/C_0)$
follows.
Then $D_{K}(b)$ is compact for all $b>0$, thus $K$ has a compact resolvent.
\qed
In addition to Assumptions \ref{ass} and
\ref {v},
suppose Assumption \ref{ass1} (Lipshitz condition)
mentioned later.
It will be  proven
 in Corollary \ref{pii} that
the normalized ground state
$\varphi_{\rm p}$
of $K$ is strictly positive and unique.
Define the probability measure on $\BT$ by
\eq{vo1}
\mup=\varphi_{\rm p}^2(\xx) d\xx
\en
and we set
\eq{vo2}
\HP=L^2(\BT;d \mu_{\rm p}).\en
We transform $K$ by the ground state transformation for later use.
Let
$$ U_{\rm p}:\HP\to \LRT,\quad f\mapsto \varphi_{\rm p} f.$$
Let  $\lp$ be
the unitarily transformed operator of $K-\is({K})$ defined by
\eq{y1-2}
\lp= U_{\rm p} \f (K-\is({K}))  U_{\rm p}
\en
with the domain $\dom(\lp)= U_{\rm p}\f \dom(K)$. We note that
$$(f, \lp g)_{\HP}=(\varphi_{\rm p} f, K \varphi_{\rm p} g)_{L^2}-
\is(K)(\varphi_{\rm p} f, \varphi_{\rm p} g)_{L^2}
.$$

\subsubsection{Scalar quantum fields}
In the previous section we discuss the particle part.
In the present section we introduce a scalar quantum field.
Let us begin with defining
a scalar field
in the Sch\"odinger representation.
We use the notation $\mathbb E_P$ for the expectation with respect to
a probability measure $P$, i.e.,
$$\int \cdots dP=\mathbb E_P[\cdots].$$

Let $\ms Q=\ms S_{\RR}(\BT)$ be the set of real-valued rapidly decreasing and infinite-times differentiable  functions
on $\BT$.
There exist a  $\sigma$-field $\Sigma$,
a probability measure
$\mu$ on
$(\ms Q, \Sigma)$
and a Gaussian random variable $\phi(f)$ indexed by $f\in
 L^2_{\RR}(\BT)$
such that
\eq{21}
\mathbb E_\mu[\phi(f)]=0
\en
and the covariance
given by
\eq{22--1}
\mathbb E_\mu[\phi(f)\phi(g)]=\half
(f,g)_{L^2},
\en
and henceforth
\eq{yoko2}
\mathbb E_\mu\lkkk
 e^{z\phi(f)}
 \rkkk=e^{(z^2/4)\|f\|^2},\quad z\in\CC.
\en
For general $f\in
  \LRT$, $\phi(f)$ is defined  by $\phi(f)=\phi(\Re f)+i\phi(\Im f)$.
 Thus $\phi(f)$ is  linear in $f$ over $\CC$.
 The boson Fock space\ko{boson Fock space!function space} is defined by $L^2(\ms Q,d\mu)=L^2({\ms Q})$.
 The identity function  $\one\in L^2({\ms Q})$ is called the Fock vacuum\ko{Fock vacuum!function space}.
 It is know that the linear hull of
\eq{23}
\{\one\}\cup
\{:\phi(f_1)\cdots\phi(f_n):|f_j\in \LRT,j=1,,.,n,n\geq 1\}
\en
is dense in $L^2({\ms Q})$, where $:\phi(f_1)\cdots\phi(f_n):$
denotes the Wick product\ko{Wick product!function space}  inductively defined by
\begin{align*}
&:\phi(f):=\phi(f),\\
&:\phi(f)\prod_{j=1}^ n\phi(f_j):
=\phi(f):\prod_{j=1}^ n\phi(f_j):
-\half\sum_{k=1}^n (f,f_k):\prod_{j\not=k}^ n\phi(f_j):.
\end{align*}
For a contraction operator $T$ on $ \LRT$,
define the second quantization\ko{second quantization!function space}  $\Gamma(T):L^2({\ms Q})\to L^2({\ms Q})$ by
$\Gamma(T)\one =\one$ and
\eq{25}
\Gamma(T)
: \phi(f_1)\cdots\phi(f_n):=
:\phi( T f_1)\cdots\phi(T f_n):.
\en
Then $\Gamma(T)$ is also contraction on \kak{23} and can be uniquely extended to the contraction operator on the hole space $L^2({\ms Q})$, which is denoted by the same symbol $\Gamma (T)$.
We can check that $\Gamma(T)\Gamma(S)=\Gamma(TS)$.
Then $\{\Gamma(e^{-ith})\}_{t\in\RR}$ for a self-adjoint operator $h$ defines the strongly continuous one-parameter unitary group on $L^2({\ms Q})$. The unique self-adjoint generator\ko{differential second quantization!function space} of $\{\Gamma(e^{-ith})\}_{t\in\RR}$ is denoted by $d\Gamma(h)$, i.e.,
\eq{ka6}
\Gamma(e^{-ith})=e^{-itd\Gamma(h)},\quad t\in\RR.
\en

\subsubsection{The Nelson model with a variable mass}
For the standard Nelson model
the dispersion relation
is given by
$\lk
-\Delta+m^2\rk^\han$ with
a constant  $m\geq0$.
In this note  $m$ is replaced by a  positive function
$m(x)$
and $-\Delta$ by the divergence form\index{divergence form}
$\sum_{\mu,\nu=1}^3
c(x)\f
D_\mu a_{\mu\nu}(x)D_\nu
c(x)\f $.
Let
\eq{yoko13}
h=h(D,x)=\sum_{\mu,\nu=1}^3
c(x)^{-1}
D_\mu a_{\mu\nu}(x)D_\nu
c(x)^{-1}
+m^2(x).
\en
In the same way as $K_0$ we define $h$ by the quadratic form, then
   the following assumption is introduced.
\begin{assumption}\label{w}
{\bf (Condition on $\omega$)}
Let $a=a(x)=(a_{\mu\nu}(x))_{1\leq\mu,\nu\leq 3}$.
\bi
\item[(1)]
{\bf (Uniform elliptic condition)}
$a_{\mu\nu}\in W^{1,\infty}$
and
there exist  constants $C_0>0$ and $C_1>0$ such that \eq{yoko17}
C_0\one
\leq a(x)\leq C_1\one.
\en
\item[(2)]
{\bf (Uniform bound)}
There exist $0<C_0$ and $0<C_1$ such
that $C_0\leq c(x)\leq C_1$ and $c\in W^{2,\infty}$.
\item[(3)]{\bf (Decay of the variable mass)}
There exists $\beta>2$ such that
\eq{variable mass}
m(x)\leq
\langle x\rangle^{-\beta/2}.
\en
\ei
\end{assumption}
Under Assumption \ref{w}
let us define the semibounded quadratic form
\begin{align*}
(f,g)&\mapsto
{\ms E}_a(f, g)\\
&=
\sum_{\mu,\nu=1}^3
\int a_{\mu\nu}(x)
{\partial_\mu
\lk
\frac{1}{c(x)} \ov{f(x)}\rk
}
\cdot \partial_\nu
\lk
\frac{1}{c(x)} g(x)\rk dx+\int m^2(x)\ov{f(x)}g(x) dx
\end{align*}
for $f,g\in H^1(\BT)$,
which is closable.
Notice that $c\f f \in H^1(\BT)$ if $f\in H^1(\BT)$,
 since $c\f \in W^{2,\infty}$, and
 $\partial_\mu (c\f f)=\partial c\f\cdot f+c\f\cdot \partial _\mu f$.
\begin{definition}\TTT{Dispersion relation with a variable mass}
\ko{dispersion relation!variable mass}
Operator $h$ is defined by  the  nonnegative self-adjoint operator associated with
the closure of ${\ms E}_a$, and
the self-adjoint operator $\omega$ on $\LRT$ is defined by
\eq{1100}
\omega=h^\han.
\en
\end{definition}
\bl{self-1}
Suppose Assumption \ref{w}.
Then
 $h$ is self-adjoint on $H^2(\BT)$,
and $\is (h)=0$ but
$0$ is not an eigenvalue of $h$.
In particular ${\rm Ker}\  \omega =0$.
\el
\proof
 Directly we can
 see that $c\f f\in H^2(\BT)$ if $f\in
 H^2(\BT)$ and
 \eq{hero2}
 hf=
h_0f
+vf,
\en
where
\eq{hero}
h_0f=\sum_{\mu,\nu=1}^3
D_\mu (c\f a_{\mu\nu}c\f D_\nu f),
\en
and, by assumptions $c\in W^{2,\infty}$ and $a_{\mu\nu}\in W^{1,\infty}$,  $v$ is the bounded multiplication
operator
given by
$$v=m^2+
\sum_{\mu,\nu=1}^3
\lk c\f (\partial_\nu
a_{\mu\nu})
(\partial_\mu c\f)+c\f
a_{\mu\nu}
(\partial_\mu\partial_\nu c\f)
\rk.$$
Since
$D(h_0)
=H^2(\BT)$ by $c\f
a_{\mu\nu} c\f\in W^{1,\infty}$,
$h$ is self-adjoint on
$H^2(\BT)$ by the Kato-Rellich theorem.
By \kak{hero2}
we have
\begin{align}
\label{z}
&{\ms E}_a(f,f)
\leq
D_1
{\ms
  E}_\one  (c\f f,c\f f)+
  (mf, mf),\\
&
D_2{\ms E}_\one  (c\f f,c\f f)+(mf, mf)\leq
{\ms E}_a(f,f)
\end{align}
with some constants $D_1$ and $D_2$.
Notice that
$-D_j\Delta +m^2$ has no zero eigenvector
and $\s(-D_j\Delta +m^2)=[0,\infty)$,
since $m^2$
is a compact perturbation of $-C\Delta$.
By \kak{z},
$h$ has also no zero eigenvector
and $\is(h)=0$.
\qed

\begin{definition}
\TTT{The Nelson Hamiltonian with a variable mass}
\ko{Nelson Hamiltonian!variable mass}

The Nelson Hamiltonian  with a variable mass
$m(x)$ and a cutoff function $\rho$ is defined by
\eq{123}
H=\lp\otimes\one+\one\otimes \hf+\phi_\rho\en
on the tensor product Hilbert space
$
\hhh=\HP\otimes L^2({\ms Q})$,
where
we set the coupling constant
$\alpha$ as $\alpha=1$,
$\lp$ is defined by \kak{y1-2},
the free field Hamiltonian $\hf$
\ko{free field Hamiltonian!Nelson Hamiltonian!variable mass}by
$
\hf=
d\Gamma(\omega)
$
and
the field operator $\phi_\rho$ is given by
\eq{haha}
\phi_\rho=\int_\BT^\oplus
\phi_\rho(\xx) \mup
\en
with
$
\phi_\rho(\xx)=\phi (\omega^{-\han} \rho(\cdot-\xx))$.
Here we used the identification
$
\hhh\cong \int_\BT^\oplus L^2({\ms Q}) \mup$.
\end{definition}
Thus the Nelson Hamiltonian  is a linear operator defined on the
$L^2$-space over the probability space
$(\BT\times \ms Q, d\mu_{\rm p}
\otimes d \mu)$.
\begin{assumption}
\label{chi}
{\bf (Condition on $\rho $)}
The ultraviolet cutoff function $\rho $ satisfies that
\eq{chi1}
(1)\  \rho \geq0,\quad
(2)\
\hat \rho /\sqrt{|k|}\in\LRT,
\quad
(3)\  \hat \rho /|k|\in \LRT.
\en
\end{assumption}
We will use (1) of Assumption \ref{chi} in the proof of Lemma \ref{yui1}.
\bp{self22}
Suppose
Assumptions \ref{ass}, \ref{v},
\ref{w}
and \ref{chi}. Then the  Nelson Hamiltonian $H$ is self-adjoint on
$\dom(\lp )\cap \dom( \hf)$, and bounded from below. Furthermore it is essentially self-adjoint on any core of $\lp+\hf$.
\ep
\proof
By Assumption \ref{w}
it follows that
\eq{this}
\sup_\xx \|\omega^{-n/2}
\rho (\cdot-\xx)\|\leq C
\|\hat \rho /|k|^{n/2}\|
\en
for $n=1,2$ with some $C$.
\kak{this}
is shown in Corollary \ref{suzuki1} later.
Then $\phi_\rho (\xx)$ is
infinitesimally small with respect to $\hf$ for each $\xx\in\BT$.
Then $\phi_\rho $
is infinitesimally small with respect to
$L_{\rm p}+\hf$,
and the proposition follows by the Kato-Rellich theorem. \qed

\subsection{Feynman-Kac formula and diffusions}
In this section we construct a functional integral representation of the one-parameter heat semigroup $e^{-tH}$.

\subsubsection{Super-exponential decay}
The following assumption ensures the existence and uniqueness
of a stochastic differential equation.
Let $C_{\rm b}^n(\BT)=\{f\in C^n(\BT)|f^m\in L^\infty(\BT), |m|\leq n\}$.
\begin{assumption}
\label{ass1}
{\bf (Lipshitz condition)}
Suppose that
$A_{\mu\nu}\in C_{\rm b}^1(\BT)$, $\mu,\nu=1,2,3$, and
 $\d b_\nu(\xx )=\half \sum_{\mu=1}^3
 {\partial_\mu} A_{\mu\nu}(\xx )$
and the $3\times 3$ matrix  $(\s_{\mu\nu}(\xx ))_{1\leq\mu, \nu\leq 3}=\s(\xx )=\sqrt{A(\xx )}$ satisfy
the Lipshitz condition:
\eq{n2}
|b(\xx )-b(\yyy )|+|\s(\xx )-\s(\yyy )|\leq D|\xx -\yyy  |
\en
for arbitrary $\xx ,\yyy  \in\BT$ with some constant $D$
independent of $\xx $ and $\yyy $,
where $|\s(\xx )|=\sqrt{\sum_{\mu,\nu=1}^3|\s_{\mu\nu}(\xx )|^2}$.
\end{assumption}
\bl{domain2}
Suppose Assumptions \ref{ass} and \ref{ass1}.
Then
$\dom(K_0 )=H^2(\BT)$ and
$K_0 f=\sum_{\mu,\nu=1}^3 D_\mu(A_{\mu\nu}D_\nu f)$ for $f\in H^2(\BT)$.
\el
\proof
We see that
$A_{\mu\nu}\in W^{1,\infty}(\BT)$. Then the lemma immediately follows from
Lemma~\ref{domain}.
\qed
Let us consider the stochastic differential equation:
\eq{n1}
\lkk
\begin{array}{lcl}
dX_t^\nu&=&\s_\nu(X_t)\cdot dB_t+
b_\nu(X_t) dt,\\
 X_0^\nu&=&
 \xx _\nu,
 \end{array}\right.
\en
on the probability space
$({\ms X}_+ , \ms B({\ms X}_+ ), {\mathcal W} )$,
where
we recall
$
{\ms X}_+ =C([0,\infty);\BT)$,
$\ms B({\ms X}_+ )$ is the $\s$-field generated by cylinder sets and
${\mathcal W} $ the Wiener measure starting at $0$.
We denote $\Ebb_{{\cal W}}$ by $\Ebb$ unless confusions may arise.
$\pro B$ denotes
the $3$-dimensional
Brownian motion on
$({\ms X}_+ , \ms B({\ms X}_+ ), {\mathcal W})$.
The drift term  $b_\nu$ and
the diffusion term
$\s_\nu=(\s_{\nu1},
\s_{\nu 2},
\s_{\nu 3}
)$ are defined in Assumption \ref{ass1}.
Note that $b_\nu$ and $\s_{\mu\nu}$ are bounded; $\|b_\nu\|_\infty<\infty$ and $\|\s_{\mu\nu}\|_\infty<\infty$.
 Let $\pro \FFF$ be the natural
 filtration of the Brownian motion:  $\FFF_t=\s(B_s, 0\leq s\leq t)$.
\bp{funda}
Suppose Assumption \ref{ass1}.
Then \kak{n1} has the unique solution $X^\xx=\pro{X^\xx}$ which is a diffusion process with respect to the filtration $\pro {{\mathcal F}}$.
Namely
$X^\xx$ has
continuous sample paths and Markov property:
\eq{ina}
\Ew{f(X_{s+t}^\xx )|{\cal F}_s}=
\Ew{f(X_t^{X_s^\xx})},
\en
where $\Ew{f(X_t^{X_s^\xx})}$ is
$\Ew{f(X_t^y)}$ evaluated at $y=X_s^\xx$.
\ep
From \kak{ina}
we can show that
\eq{I11}
T_tf(\xx)=\Ew{f(X_t^\xx)}
\en
satisfies the semigroup property $T_sT_tf=T_{s+t}f$ on $L^\infty(\BT)$. In the next proposition we show indeed that $T_tf\in \LRT$ for $f\in\LRT$. Namely $T_t$ defines a semigroup not only on $L^\infty(\BT)$ but also on $\LRT$.

In order to show that $T_t:L^\infty\to L^\infty$ can
be extended to a semigroup
on $\LRT$, we introduce a Dirichlet form.
Suppose Assumption \ref{ass}.
We see that
$({\ms E}_A,H^1(\BT))$ is
a  local and regular Dirichlet form
\ko{Dirichlet form!local}
\ko{Dirichlet form!regular}.
It is a fundamental fact
that there exist 
a probability measure $\nu^\xx $ on
$({\ms X}_+ , \ms B({\ms X}_+ ))$
and
a  coordinate  process\index{coordinate process}
$Z=\pro Z$
such
that
(1)
$\nu^\xx(Z_0=\xx)=1$,
(2)
$Z$ is
a symmetric diffusion process
with respect to the natural filtration
$\ms M_t=\s(Z_s,0\leq s\leq t)$,
(3)
\eq{n10}
S_t f(\xx)=\mathbb E_{\nu^\xx}[f(Z_t)]
\en
defines the semigroup, and
(4) for each $t\geq0$,
\eq{y7}
\lk
e^{-t K_0 }f\rk
(x)=
\lk
S_t f\rk
(\xx),\quad a.e. \quad  \xx\in\BT.
\en
See e.g.,\cite[Lemma 4.3.1]{fuk80}.

\bp{I10}
\TTT{$L^2$ extension}
Let $f\in \LRT\cap L^\infty(\BT)$.
Suppose Assumptions \ref{ass} and
\ref{ass1}.
Then
\eq{yoko10}
T_t f =e^{-tK_0 }f,\quad t\geq0,\quad a.e.
\en
In particular
$\ov{T_t\lceil_{L^2\cap
L^\infty}}=e^{-tK_0}$, where $\ov{\{\cdots\}}$ denotes
the closure in $\LRT$.
\ep
\proof
Let $f\in C_0^\infty(\BT)$. We set
\begin{align*}
&
M_t=f(Z_t)-\int_0^t(K_0 f)(Z_s) ds,\\
&
N_t=
f(X_t^\xx )-
\int_0^t(K_0 f)(X_s^\xx ) ds.
\end{align*}
By the It\^o formula we have $$
f(X_t^\xx )-f(\xx )
=\int_0^t (K_0 f)(X_s^\xx ) ds+\sum_{\mu=1}^3
\int_0^t (\partial_\mu f)(X_s^\xx )\s_\mu(X_s^\xx )\cdot dB_s.$$
Hence
$$N_t=f(\xx )+\sum_{\mu=1}^3
\int_0^t
(\partial_\mu f)(X_s^\xx )\s_\mu(X_s^\xx )\cdot dB_s$$
and
then
$\pro N$ is martingale on $({\ms X}_+ , \ms B({\ms X}_+ ), {\mathcal W})$
with respect to $\pro
{\mathcal F}$,
while
we can see that
\begin{align*}
\mathbb E_{\nu^\xx }[M_{t+s}|\ms Z_s]
=
\mathbb E[f(Z_{t+s})|\ms Z_s]
-\int_0^s K_0 f(Z_r)dr-
\mathbb E\lkkk \int_s^{t+s}K_0 f(Z_r)dr\left|
\ms Z_s\right.\rkkk.
\end{align*}
Here $\ms Z_t=\s(Z_s,0\leq s\leq t)$.
Let $p(t, y, A)$ be the
probability transition kernel of $Z_t$ under $\nu^\xx $.
Then by the Markov
property of $Z$ we have
$$
 \mathbb E[f(Z_{t+s})|
 \ms Z_s]
=
\int f(\yyy) p(t, Z_s, d\yyy)=
(e^{-tK_0 }f)(Z_s)
$$
and
\begin{align*}
&
\mathbb E\lkkk
\int_s^{t+s}(K_0 f)(Z_r)dr\left|
\ms Z_s\right.\rkkk
=
\int_s^{t+s}dr
\int (K_0 f)(\yyy)p(r-s,Z_s,d\yyy)\\
&=
\int_0^t dr
\int (K_0 f)(\yyy)p(r,Z_s,d\yyy)
=
\int_0^t(e^{-rK_0 }K_0 f)(Z_s) dr
=
(e^{-tK_0 }f)(Z_s)-f(Z_s).
\end{align*}
Then we see that
\begin{align*}
\mathbb E[M_{t+s}|\ms Z_s]=
f(Z_s)-
\int_0^s K_0 f(Z_r) dr=M_s
\end{align*}
and we conclude that
$\pro M$ is also martingale
with respect to $\pro {\ms Z}$.
 By the uniqueness of martingale problem \ko{uniqueness of martingale problem} (e.g., \cite[Theorem (24.1)]{rw}, \cite[Chapter 8]{SV},\cite[Section 5.4.E]{ks}),
it follows that $\nu^\xx ={\mathcal W} \circ (X^\xx )\f $.
  In particular
  $$\mathbb E_{\nu^\xx }[f(Z_t)]=\mathbb E_{\mathcal W}[f(X_t^\xx )],$$
  which is equivalent to $T_tf=S_t f$. Then the proposition follows from \kak{y7}.
  \qed

In order to see some
properties of
$e^{-tK_0}$, we give
a Gaussian bound of
the integral kernel
of $e^{-tK_0}$.
When
$\|e^{-tL} f\|_\infty
\leq C_t \| f\|_2$
is satisfied
 for all $t>0$ and all $f\in\LRT$,
$e^{-tL}$ is called
{\it ultracontractivity}.
\ko{ultracontractivity}.
\bp{ultra1}
\TTT{Kernels}
Suppose Assumption \ref{ass}.
Then $e^{-tK_0}$ is
ultracontractive,
has an integral kernel,
 and
the kernel satisfies that
\eq{da1}
C_1 e^{tC_2\Delta}(\xx,\yyy)
\leq e^{-tK_0}(\xx,\yyy)\leq
C_3 e^{tC_4\Delta}(\xx,\yyy)
\en
with some constants $C_j$, $j=1,2,3,4$,
 where
$$e^{T\Delta}(\xx,\yyy)=(2\pi T)^{-3/2}\exp(-|\xx-\yyy|^2/(2T))$$
is the 3-dimensional heat kernel.
\ep
\proof
See Propositions \ref{el},
\ref{el2} and \ref{el3}.
 \qed

We prove the Feynman-Kac formula\ko{Feynman-Kac formula} of $e^{-t(K_0+W)}$   for  general $W$.
Let $h_0 =(-\han)\Delta$.
Suppose that $W$ is form bounded with respect to $h_0 $ with a relative bound $b$, i.e.,
$$\lim _{E\to\infty}
\||W|^\han (h_0 +E)^{-\han}f\|/\|f\|=b.$$
By Proposition \ref{ultra1}
we notice that
\eq{uchi2}
|(f, e^{-tK_0 }g)|\leq
(|f|, C'e^{-t C h_0 }|g|),
\en
where
$C'$ and $C$ are nonnegative constants.
Let $T$ be nonnegative self-adjoint operator.
Then
$(T+E)^{-\han}=\pi^{-\han}
\int_0^\infty t^{-\han}e^{-(T+E)}dt$ for $E>0$.
From this formula and \kak{uchi2}
 it follows that
\eq{uchi3}
|(K_0 +E)^{-\han}f|(x)\leq
C'(Ch_0 +E)^{-\han}|f|(x).
\en
Hence
\eq{uchi4}
{
\|
|W|^\han (K_0 +E)^{-\han}f\|}/{\|f\|}
\leq
C'
{
\||W|^\han (Ch_0 +E)^{-\han}|f|\|}/{\|f\|}
\en
and we have
$$\lim_{E\to\infty}
{
\|
|W|^\han (K_0 +E)^{-\han}f\|}/{\|f\|}
=C'C^{-\han}b.$$
Then $W$ is also relatively form bounded with respect to $K_0 $ with a relative bound $<C'C^{-\han}b$.
We introduce an assumption on $\VV$.
\begin{assumption}
\label{W}
  Let $W=W_+-W_-$, where $W_\pm=\max\{\pm W,0\}$.
Suppose $W_+\in L_{\rm loc}^1(\BR)$ and $W_-$ is relatively form bounded with respect to $h_0 $ with
a relative bound $b$ such that \eq{naka2}
C'C^{-\han}b<1,
\en
where constants $C,C'$ are in \kak{uchi2}.
\end{assumption}
Suppose
 Assumptions \ref{ass} and \ref{W}.
Then by the KLMN theorem
\eq{nakajima}
K=K_0 \,\,\dot +\,\,W_+\,\,\dot-\,\,W_-
\en
can be defined as
a self-adjoint operator.
Here $\dot\pm$ denotes the quadratic form sum.

\bp{6}
\TTT{Feynman-Kac formula}
Suppose
 Assumptions \ref{ass}, \ref{ass1} and \ref{W}.
Let $K$ be given by \kak{nakajima}.
 Then
\eq{71-1}
\lk g,  e^{-tK}f\rk=
\int \mup
{\mathbb E}\lkkk
\ov{g(\xx)} f(X_t^\xx)e^{-\int_0^t W(X_s^\xx) ds}\rkkk.
\en
In particular
\eq{71}
\lk e^{-tK}f\rk (\xx)=
{\mathbb E}\lkkk
f(X_t^\xx)e^{-\int_0^t W(X_s^\xx) ds}\rkkk.
\en
\ep
\proof
Suppose first that $W$ is bounded and continuous.
By the Trotter-Kato product formula  we have
\eq{I14}
(f, e^{-tK}g)=
\lim_{n\to \infty}
(f, (e^{-(t/n)K_0}e^{-(t/n)W})^n g).
\en
By Proposition \ref{I10}
we have
for $t_0\leq t_1\leq \cdots\leq t_n$,
\begin{align*}
&
 \lk
 e^{-(t_1-t_0)K_0}f_1\cdots e^{-(t_n-t_{n-1})K_0}f_n
 \rk(\xx )\\
 &=
\Ew{f_1(X_{t_1-t_0}^\xx )
\lk
e^{-(t_2-t_1)K_0}f_1\cdots e^{-(t_n-t_{n-1})K_0}f_n\rk(X_{t_1-t_0}^\xx )}.
\end{align*}
By the Markov property \kak{ina} we also have
\begin{align*}
&
=
\Ew
{
f_1(X_{t_1-t_0}^\xx )
\Ew
{
f_2(X_{t_2-t_1}^{X_{t_1-t_0}^\xx })
\lk
e^{-(t_3-t_2)K_0}f_1\cdots e^{-(t_n-t_{n-1})K_0}f_n\rk(X_{t_2-t_1}^{X_{t_1-t_0}^\xx })
}}\\
&
=
\Ew{
f_1(X_{t_1-t_0}^\xx )
\Ew{
f_2(X_{t_2-t_0}^\xx )
\lk
e^{-(t_3-t_2)K_0}f_1\cdots e^{-(t_n-t_{n-1})K_0}f_n\rk(X_{t_2-t_0}^\xx )\left|
\mathcal F_{t_1-t_0}\right.
}}\\
&
=\Ew{
f_1(X_{t_1-t_0}^\xx )
f_2(X_{t_2-t_0}^\xx )
\lk
e^{-(t_3-t_2)K_0}f_1\cdots e^{-(t_n-t_{n-1})K_0}f_n\rk(X_{t_2-t_0}^\xx )
}.
\end{align*}
Inductively we obtain that
\eq{I13}
 \lk
 e^{-(t_1-t_0)K_0}f_1\cdots e^{-(t_n-t_{n-1})K_0}f_n
 \rk(\xx )
=\Ew{
\prod_{j=1}^n f_j(X_{t_j-t_{j-1}}^\xx )}.
\en
Combining the Trotter product formula \kak{I14} and \kak{I13} with $t_j=tj/n$,
we have
\eq{I15}
(f, e^{-tK}g)=
\lim_{n\to \infty}
\int d\xx  \bar f(\xx )
\Ew{
e^{-(t/n)\sum_{j=1}^ nW(X_{tj/n}^\xx )}
g(X_t^\xx)
}.
\en
 Since $s\mapsto X_s^\xx(\omega)$ has continuous paths,
$\VV(X_s^\xx(\omega))$ is continuous in $s\in [0,t]$ for each $\omega$.
Therefore
$\sum_{j=1}^n \frac{t}{n} \VV(X_{{tj/n}}^\xx)
\rightarrow \int_0^t \VV(X_{s}^\xx)ds$
as $n\to \infty$ for each $\omega$ and exists as a Riemann integral.

In order to extend $\VV$ to more general class, we use a standard limiting argument. To do that, suppose that $\VV\in L^\infty$ and $\VV_n(\xx )=\phi(\xx /n)(\VV\ast j_n)(\xx )$, where $j_n=n^3\phi(\xx n)$
with $\phi\in C_0^\infty(\BT)$ such that $0\leq \phi\leq 1$, $\int\phi(\xx )d\xx =1$ and $\phi(0)=1$.
Then $\VV_n$ is bounded and continuous, moreover
$\VV_n(\yyy )
\rightarrow \VV(\yyy )$ as $n\rightarrow \infty$ for
$\yyy \not\in \ms N$
with some null set $\ms N$.
Notice that
$\mathbb E[\one_{X_t^\xx\in\ms N}]=\int \one_{\yyy \in\ms N}e^{-tK_0 }(\xx ,\yyy ) d\yyy =0$ and
thus
$\int_0^t ds \mathbb E[\one_{X_s^\xx\in\ms N}]=
\mathbb E\lkkk
\int_0^t ds \one_{X_s^\xx\in\ms N}\rkkk=0$ by  Fubini's lemma.
Thus for each $\xx \in\BT$  the measure of
$\{t\in [0,\infty)\,|\,X_t^\xx(\omega)\in \ms N\}$ is zero
for almost every $\omega$.
Hence $\int_0^t \VV_n(X_s^\xx)ds
\rightarrow \int_0^t \VV(X_s^\xx)ds$ as $n\rightarrow \infty$ almost surely,
and
$$
\int\!\! \mup
 \Ew
{ \ov{f(\xx )}g(X_t^\xx)
e^{-\int_0^t W_n(X_s^\xx)ds} }
\rightarrow
\int\!\! \mup
\Ew
{
\ov{f(\xx )}g(X_t^\xx)
e^{-\int_0^t W(X_s^\xx)ds} }
$$
as $n\rightarrow\infty$.
On the other hand, $e^{-t(K_0 +W_n)}\rightarrow e^{-t(K_0 + W)}$
strongly as $n\rightarrow \infty$,
since $K_0 +W_n$ converges to $K_0 +W$ on the common
domain $H^2(\BT)$.
Next define
\begin{eqnarray*}
W_{+,n}(\xx ) =
\left\{
\begin{array}{ll}
W_+(\xx ),&W_+(\xx )<n,\\
n,& W_+(\xx ) \geq n,
\end{array}
\right.\quad
W_{-,m}(\xx )=
\left\{
\begin{array}{ll}
W_-(\xx ),&W_-(\xx )<m,\\
m,& W_-(\xx ) \geq m.
\end{array}
\right.
\end{eqnarray*}
Note that $\domf(K_0 )=H^1(\BT)$,
where $\domf(T)$
denotes the form domain of $T$, i.e., $\domf (T)=\dom(|T|^\han)$.
Define the closed quadratic forms
\begin{eqnarray*}
{\rm q}_{n,m}(f,f)
&=&
(K_0  ^\han f, K_0  ^\han f) + (W_{+,n}^{\han}f,W_{+,n}^{\han}f)-(W_{-,m}^{\han}f,W_{-,m}^{\han}f), \\
{\rm q}_{n,\infty}(f,f)
&=&
(K_0  ^\han f,K_0  ^\han f)+(W_{+,n}^{\han}f,W_{+,n}^{\han}f)-(W_{-}^{\han}f,W_{-}^{\han}f), \\
{\rm q}_{\infty, \infty}(f,f)
&=&
(K_0  ^\han f,K_0  ^\han f)+(W_{+}^{\han}f,W_{+}^{\han}f)-(W_{-}^{\han}f,W_{-}^{\han}f),
\end{eqnarray*}
where the form domains are given by
$$
\domf({\rm q}_{n,m})=H^1(\BT),\quad
\domf({\rm q}_{n,\infty})=H^1(\BT),
\quad
\domf({\rm q}_{\infty, \infty})=
H^1(\BT)\cap \domf(W_+).
$$
Note that
$${\rm q}_{n,m}\geq {\rm q}_{n,m+1}\geq {\rm q}_{n,m+2}\geq...\geq {\rm q}_{n,\infty}$$ and
${\rm q}_{n,m}\rightarrow {\rm q}_{n,\infty}$
in the sense of quadratic forms on $\cup_m \domf({\rm q}_{n,m})=
H^1(\BT)$. Since ${\rm q}_{n,\infty}$ is closed on
$H^1(\BT)$, by the monotone convergence theorem
for a non-increasing sequence of forms (see
\cite[VIII. Theorem 3.11]{kat76})
the associated positive self-adjoint operators satisfy $K_0  \, \, \dot{+}\, \, W_{+,n}\, \, \dot{-}\, \,
W_{-,m} \rightarrow K_0   \, \, \dot{+}\, \, W_{+,n}\, \, \dot{-}\, \, W_{-}$ in strong resolvent sense,
which implies that
\eq{mono1}
e^{-t\left({K_0   }
\, \, \dot{+}\, \, W_{+,n}\, \, \dot{-}\, \, W_{-,m}\right)}
\rightarrow
e^{-t \left({K_0   }
\, \, \dot{+}\, \, W_{+,n}\, \, \dot{-}\, \, W_{-}\right)}
\en
strongly as
$m\rightarrow \infty$
for all $t\geq0$. Similarly, we have
$${\rm q}_{n,\infty}\leq {\rm q}_{n+1,\infty}
\leq {\rm q}_{n+2,\infty}\leq ...\leq {\rm q}_{\infty,\infty}$$ and ${\rm q}_{n,\infty}\rightarrow
{\rm q}_{\infty,\infty}$ in quadratic form sense on
$$\{f\in\cap_n
\domf({\rm q}_{n,\infty}) \,|\,\sup_n
{\rm q}_{n,\infty}(f,f)<\infty \} =
H^1(\BT)\cap \domf(W_+).$$
Hence by the monotone convergence
theorem for a non-decreasing sequence of forms (see 
\cite[VIII. Theorem 3.13 and p.575]{kat76})
we obtain
\eq{mono2}
e^{-t \left({K_0   }\,\, \dot{+}\,\, W_{+,n}\,\,\dot{-}\,\, W_-\right)} \rightarrow
e^{-t \left({K_0   }\, \, \dot{+}\, \, W_{+}\, \, \dot{-}\, \, W_{-}\right)},
\en
strongly as $n\rightarrow\infty$.
On the other hand, we can see that
\begin{eqnarray*}
\int \!d\xx {\mathbb E} \lkkk  e^{-\int_0^t(W_{+,n}-W_{-,m})(
X_s^\xx ) ds}  \rkkk
\longrightarrow
\int \!d\xx{\mathbb E} \lkkk   e^{-\int_0^t(W_{+,n}-W_{-})(X_s^\xx ) ds}  \rkkk
\end{eqnarray*}
as $m\rightarrow\infty$.
Moreover,
$$
\int \!d\xx {\mathbb E} \lkkk   e^{-\int_0^t(W_{+,n}-W_{-})(X_s^\xx ) ds} \rkkk
\longrightarrow \int \!d\xx {\mathbb E} \lkkk   e^{-\int_0^t(W_{+}-W_{-})(X_s^\xx ) ds}
\rkkk
$$
as $n\rightarrow\infty$, by \kak{mono2} and the dominated convergence theorem. Thus the proof is
complete.
\qed

\bc{pii}\TTT{Positivity improving}
\ko{positivity improving}
Suppose Assumptions
\ref{ass}, \ref{v}
 and \ref{ass1}.
Then $e^{-tK}$ is positivity improving.
In particular the ground state of $K$ is strictly positive and unique.
\ec
\proof
Let $f\geq 0$ and $g\geq 0$ but $f\not\equiv 0$ and $g\not\equiv 0$.
It is enough to show that $(f, e^{-tK} g)>0$.
Let ${\rm supp} f=D_f$ and ${\rm supp}g=D_g$.
We first show that
for each $\xx \in\BT$,
\eq{yume}
{\mathcal W}\lk
\int_0^t \VV(X_s^\xx) ds=\infty\rk
=0.
\en
Let us recall that $\pro B$ is the Brownian
motion on $({\ms X}_+ , \ms B({\ms X}_+ ), {\mathcal W})$.
Let $N\in\mathbb N$.
Since $\VV\in
L_{\rm loc}^1(\BT)$,
 $\one_N\VV\in L^1(\BT)$
and then by
Proposition \ref{ultra1}
\begin{align*}
&
\int d\xx
\Ew{\int_0^t \one_N\VV(X_s^\xx) ds}
=
\int d\xx
{\int_0^t \Ew{
\one_N\VV(X_s^\xx)} ds}\\
&\leq
\int d\xx
C_3 {\int_0^t \Ew{
\one_N\VV(B_{2C_4 t}+\xx )} ds}
\leq
\|\one_N\VV\|_{L^1}C_3 t<\infty.
\end{align*}
Thus
${\mathcal W}
(\int_0^t \one_N\VV(X_s^\xx) ds<\infty)=1$
and there exists $\ms N_N$ such that
${\mathcal W}(\ms N_N)=0$ and
$\int_0^t \one_N
\VV(X_s^\xx(\omega))ds<\infty$ for  arbitrary $\omega\in
{\ms X}_+\setminus\ms N_N$.
Let $\ms N=\cup_N\ms N_N$.
Thus
$$\int_0^t
\one_N
\VV(X_s^\xx(\omega))ds<\infty$$
for arbitrary $N\in\mathbb N$ and $\omega\in
{\ms X}_+\setminus\ms N$.
Since $X_s^\xx$ is continuous in $s$,
for each $\omega\in {\ms X}_+\setminus \ms N$, there exists $N=N(\omega)$ such that $\sup_{0\leq s\leq t}X_s^\xx(\omega)<N$.
 Then $\VV(X_s^\xx(\omega))=
 \one_N\VV(X_s^\xx(\omega))$
 for $0\leq s\leq t$, and
$$\int_0^t \VV(X_s^\xx(\omega)) ds=
\int_0^t \one _N \VV
(X_s^\xx(\omega))
 ds<\infty.$$
This implies \kak{yume} and
$e^{-\int_0^t W(X_s^\xx(\omega)) ds}>0$
for a.e. $\omega\in{\ms X}_+ $.
By
the Feynman-Kac formula,
it is sufficient to see that
$\int d\xx
\Ew{
f(X_0^\xx )g(X_t^\xx)}>0$.
Let
$$D_g^\xx =
\{\omega\in{\ms X}_+|
X_t^\xx(\omega)\in D_g\}.$$
Thus
$$\int_{D_f} d\xx \Ew{\one_{D_g^\xx }}=
(\one_{D_f}, e^{-tK_0 }\one_{D_g})
\geq C_1(\one_{D_f}, e^{C_2t\Delta}\one_{D_g})>0
.$$
Then the measure of $\cup_{\xx \in D_f}  D_g^\xx (\subset \BT\times {\ms X}_+ )$ is strictly positive with respect to $d\xx \otimes d{\mathcal W}$
and $f(X_0^\xx )g(X_t^\xx)>0$ on $\cup_{\xx \in D_f} D_g^\xx $.
Then
$$
\int d\xx
\Ew{
f(X_0^\xx )g(X_t^\xx)}\geq
\int_{D_f}
 d\xx
\int_{D_g^\xx }
f(X_0^\xx )g(X_t^\xx)d{\mathcal W}>0$$
and the corollary follows.
\qed
\bc{ultra}\TTT{Ultracontractivity}
\ko{ultracontractivity}Suppose Assumptions
\ref{ass}, \ref{v}
 and \ref{ass1}.
Then $e^{-tK}$ is ultracontractive.
\ec
\proof
Note that
$e^{-\int_0^t W(X_s^\xx) ds}\leq 1$.
By the Feynman-Kac  formula,
we have
$$\left|\lk e^{-tK}f\rk (\xx )\right|
\leq
\lk
\Ew{|f(X_t^\xx)|^2
}\rk^\han.
$$
By Proposition
\ref{ultra1}
we have
$$
\Ew{|f(X_t^\xx)|^2
}
=
\lk
e^{-tK_0}
|f|^2\rk(\xx)
\leq
C_3
\lk
e^{C_4 t\Delta}|f|^2\rk(\xx)
\leq
Ct^{-3/2}
\|f\|_{L^2}^2.
$$
Then
$\|e^{-tK}f\|_\infty\leq
Ct^{-3/4}\|f\|_{L^2}$ and the corollary follows.
\qed
We can
also prove the theorem below.
\bt{expdecay}\TTT{Super-exponential decay}
\ko{super-exponential decay}
Suppose Assumptions
\ref{ass}, \ref{v} and
\ref{ass1}.
Then there exists a constant
 $\gamma >0$ such that
\eq{car1}
e^{\gamma |\xx|^{\delta+1}}
\varphi_{\rm p}\in H^1(\BT).
\en
\et
\proof
Let $F\in C^\infty(\BT)$ be real, bounded with all derivatives. Then for $u\in D(K)$ we have the Agmon identity\ko{Agmon identity}:
\begin{align*}
&\int \half \nabla(e^F \bar u)\cdot A\nabla(e^F u) d\xx+\int e^{2F}(\VV-\half \nabla F\cdot A\nabla F) |u|^2 d\xx\\
&=
\int e^{2F}\bar u K u d\xx+2i {\rm Im} \int e^{2F}\nabla\bar u\cdot A\nabla F d\xx.
\end{align*}
Applying this identity to the ground state
$\varphi_{\rm p}$, we obtain that $e^{\gamma\ab{\xx}^{\delta+1}} \varphi_{\rm p}\in \LRT$ and
$\nabla(e^{\gamma\ab{\xx}^{\delta+1}} \varphi_{\rm p})\in \LRT$.
\qed

\subsubsection{Diffusion processes}
We can also construct a
Markov process
$X=(X_t)_{t\in\RR}$
on the hole real line $\RR$
associated with the semigroup $e^{-tL_{\rm p}}$ by a stochastic differential equation.
Let
$$\ms X=C(\RR;\BT).$$
\bp{main1}
\TTT{Diffusion process associated with $e^{-tL_{\rm p}}$}
\ko{diffusion process}
Let
$X_t(\omega)=\omega(t)$ be the coordinate  process\index{coordinate process}
on $(\ms X, \ms B(\ms X))$.
Suppose Assumptions
\ref{ass}, \ref{v} and \ref{ass1}.
Then there  exists
a probability measure $P^\xx $ on $(\ms X, \ms B(\ms X))$
satisfying (1)-(4) below:
\bi
\item[(1)]
{\bf (Initial distribution)}
 $P^\xx (X_0=\xx )=1$.
\item[(2)]
{\bf (Reflection symmetry)}
\ko{reflection symmetry}
Two processes $\pro X$ and $(X_s)_{s\leq0}$
are independent
 and $X_{-t}\stackrel{\rm d}{=}
 X_t$.\footnote{
   $X\stackrel{\rm d}{=}
Y$ means that $X$ and $Y$ has the same distribution.
}
\item
[(3)]
{\bf (Diffusion property)}
Let
$\pro
{\mathcal F^+}=\s(X_s, 0\leq s\leq t)$
and
$(\mathcal F_t^-)_{t\leq 0}=\s(X_s, -t\leq s\leq 0)$ be filtrations.
Then
$\pro X$ and $(X_s)_{s\leq 0}$
are diffusion processes with respect to $\pro {\mathcal F^+}$ and $(\mathcal F_t^-)_{t\leq 0}$, respectively,
 i.e.,
\begin{align*}
&
\EE_{P^\xx }[X_{t+s}|\mathcal F_s^+]=
\EE_{P^\xx }[X_{t+s}|\s(X_s)]
=
\EE_{P^{X_s}}[X_{t}],\\
&
\EE_{P^\xx }[X_{-t-s}|
\mathcal F_{-s}^-]=
\EE_{P^\xx }[X_{-t-s}|\s(X_{-s})]
=\EE_{P^{X_{-s}}}[X_{-t}]
\end{align*}
for $s,t\geq0$, and $X_t$
is continuous in $t\in\RR$, where
$\EE_{P^{X_s}}$ means $\EE_{P^y}$
evaluated at $y=X_s$.
\item[(4)]{\bf (Shift invariance)}
\ko{shift invariance}
It follows that
\eq{I4-1ko}
\int\mup
\EE_{P^\xx }
\lkkk
 {f_0(X_{t_0})\cdots f_n(X_{t_n})}\rkkk
=(f_0, e^{-(t_1-t_0)L_{\rm p}}f_1\cdots e^{-(t_n-t_{n-1})L_{\rm p}} f_n)_{\HP}
\en
for $f_j\in L^\infty(\BT)$, $j=1,...,n$, and then the finite dimensional distribution of $X$
is shift-invariant, i.e.,
$$
\int\mup
\EE_{P^\xx }
\lkkk
\prod_{j=1}^n f_j(X_{t_j})\rkkk
=
\int\mup
\EE_{P^\xx }
\lkkk
\prod_{j=1}^n f_j(X_{t_j+s})\rkkk,
\quad
s\in\RR,
$$ for any bounded Borel measurable functions $f_j$, $j=1,...,n$.
\ei
\ep
\ko{diffusion process}
In order to prove Proposition\ref{main1} we need several steps.
An outline of constructing a diffusion process $(X_t)_{t\in\RR}$ is as follows.

For $0\leq t_0\leq t_1\leq \cdots\leq t_n$
let the set function
$\nu_{t_0,...,t_n}:\prod_{j=0}^n \ms B(\BT)\to \RR$
be given by
\eq{uchii}
\nu_{t_0,...,t_n}
\lk
\prod_{i=0}^n  A_i
\rk
=
(\one_{A_0}, e^{-(t_1-t_0)L_{\rm p} }\one_{A_1}
\cdots e^{-(t_n-t_{n-1}) L_{\rm p} }\one_{A_n})_{\HP}
\en
and
for $0\leq t$, $\nu_t:\ms B(\BT)\to \RR$ by
\eq{uchi}
\nu_t\lk
A\rk=
(\one, e^{-tL_{\rm p}}\one_A)_{\HP}
=(\one, \one_A)_{\HP},
\en
where $\ms B(\BT)$ denotes the Borel $\s$-field of $\BT$.
We show an outline of steps of the proof.
\begin{itemize}
\item[(Step 1)]
By the Kolmogorov extension theorem
we can construct a probability measure $\nu_\infty$
on $(\BT)^{[0,\infty)}$
from the family of
probability measures given by \kak{uchii} and \kak{uchi},
and
we define
a stochastic process $Y=\pro Y$ on a
probability space $((\BT)^{[0,\infty)}, \ms B((\BT)^{[0,\infty)}), \nu_\infty)$ such that finite dimensional distributions of $Y$ is given by the right-hand side of \kak{uchii} and \kak{uchi}.
 We also show the existence of the continuous version $\pro {\bar Y}$ on the same probability space.
 \item[(Step 2)]
Let  $Q=
\nu_\infty\circ \bar Y\f $ be the image measure of $\nu_\infty$
 on $({\ms X}_+ , \ms B({\ms X}_+ ))$, where ${\ms X}_+ =C([0,\infty);\BT)$.
Let
$\pro {\wY }$ be the
coordinate process on
the probability space $({\ms X}_+ , \ms B({\ms X}_+ ), Q)$, i.e.,
$\wY _t(\omega)=\omega(t)$ for $\omega\in {\ms X}_+ $.
Notice that
$\wY \stackrel{\rm d}{=}\bar Y$.
\item[(Step 3)]
Define a regular conditional probability measure by $Q^\xx (\cdot)=\domf(\cdot|\wY _0=\xx )$. Then
the stochastic process $\pro {\wY }$ on a probability space
$({\ms X}_+ , \ms B({\ms X}_+ ), Q^\xx )$
satisfies
\eq{ptkk}
(f_0, e^{-(t_1-t_0)L_{\rm p} }f_1
\cdots e^{-(t_n-t_{n-1}) L_{\rm p} }f_n)_{\HP}
=\int \mup   \EE_{Q^\xx }
\left[\prod_{j=0}^n f_j(\wY _{t_j})\right]
\en
for $0\leq t_0\leq t_1\leq\cdots\leq t_n$
and
we can show that
$\wY $
is a diffusion process
with respect to the natural filtration
$\s(\wY _s,0\leq s\leq t)$.
\item[(Step 4)]
We extend $\wY $ to a process of the hole real line.
Define a stochastic process
$
\tilde X_t(\omega)=\lkk
\begin{array}{ll}
\wY_t(\omega_1),& t\geq 0,\\
\wY_{-t}(\omega_2),& t<0
\end{array}
\right.
$
on the product probability space
$(\tilde {{\ms X}_+ }, \tilde
{\ms M}, \tilde {Q}^\xx )=
({{\ms X}_+ }\times{\ms X}_+ ,
{\ms B}({\ms X}_+ )\times\ms B({\ms X}_+ ),
{Q}^\xx \times Q^\xx )$.
This is a continuous process.
\item[(Step 5)]
We will prove Proposition \ref{main1} in this step.
Let $P^\xx $ be the image measure
given by
$P^\xx =
   \tilde Q_x\circ \tilde X\f$
   on $(\ms X, \ms B(\ms X))$,
   where
   $\ms X=C(\RR;\BT)$.
   Then the coordinate process  $(X_t)_{t\in\RR}$
 on the probability space $X=(\ms X, \ms B (\ms X), P^\xx )$
 satisfies that
 \eq{I4-1}
\ixp {f_0(X_{t_0})\cdots f_n(X_{t_n})}
=(f_0, e^{-(t_1-t_0)L_{\rm p}}f_1\cdots e^{-(t_n-t_{n-1})L_{\rm p}} f_n)_{\HP}
\en
for
$-\infty <t_0\leq t_1\leq \cdots\leq t_n$.
By  this we can see that
$X=(X_t)_{t\in\RR}$
satisfies (1)-(4) of Proposition \ref{main1}.
\end{itemize}

{\bf (Step 1)}
The family of set functions
$\{\nu_\xi\}_{\xi\subset\RR,\#\xi<\infty}$
given by \kak{uchii} and \kak{uchi} satisfies
the consistency condition:
$$
\nu_{t_0,...,t_{n+m}}\lk
\prod_{i=0}^n  A_i \times \prod_{i=n+1}^{n+m}  \BT
\rk
=\nu_{t_0,...,t_n}\lk
\prod_{i=0}^n  A_i\rk
$$ and
by the Kolmogorov extension  theorem \cite[Theorem 2.2]{ks}
there exists a
a probability measure $\nu_\infty$
on
$((\BT)^{[0, \infty)},\ms B((\BT)^{[0,\infty)}))$ such that
\begin{eqnarray}
&&
\label{na3}
\nu_t\lk
A\rk
=
\mathbb E_{\nu_\infty}\left[
 \one_{A}(Y_t)\right],\\
&&
\nu_{t_0,...,t_n}\lk
\prod_{i=0}^n  A_i\rk
=
\mathbb E_{\nu_\infty}\left[
\prod_{j=0}^n  \one_{A_j}(Y_{t_j})\right], \quad n\geq 1,
\end{eqnarray}
where
$
Y_t(\omega)=\omega(t)$,
$
\omega\in
(\BT)^{[0, \infty)}$,
 is
the coordinate  process.
Then
the process $Y=\pro Y$ on
the probability space
$((\BT)^{[0, \infty)}, \ms B((\BT)^{[0,\infty)}), \nu_\infty)$
satisfies that
\begin{eqnarray}
&&
\label{t1}
(f_0, e^{-(t_1-t_0)L_{\rm p} }f_1
\cdots e^{-(t_n-t_{n-1}) L_{\rm p} }f_n)_{\HP}
=\mathbb E_{\nu_\infty}\lkkk
\prod_{j=0}^n f_j(Y_{t_j})\rkkk,\\
&&
\label{t2}
(\one,  f)_{\HP}=
(\one,  e^{-tL_{\rm p}} f)_{\HP}
=
\mathbb E_{\nu_\infty}\lkkk
f(Y_t)\rkkk
=
\mathbb E_{\nu_\infty}\lkkk
f(Y_0)\rkkk
\end{eqnarray}
for $f_j\in L^\infty(\BT)$, $j=0,1,...,n$.

{\bf(Step 2)}
We now see that the process $Y$ has a continuous version.
\bl{n14}
The process $Y$
on
$((\BT)^{[0, \infty)}, \ms B((\BT)^{[0,\infty)}), \nu_\infty)$ has a continuous version.
\el
\proof
We note that by \kak{t1},
  \kak{t2}
and
Proposition \ref{6},
$E_{\nu_\infty}[|Y_t-Y_s|^{2n}]$ can be expressed in terms of the diffusion process $X=\pro{X^\xx}$.
Since
$$E_{\nu_\infty}[|Y_t-Y_s|^{2n}]=\sum_{k=0}^{2n}
\lkkk\!
\begin{array}{c} 2n\\ k
\end{array}\rkkk
(-1)^k\mathbb E_{\nu_\infty}\lkkk Y_t^{2n-k} Y_s^k\rkkk,$$
the left hand side above can be expressed in terms of $e^{-tL_{\rm p}}$ as
\begin{align*}
E_{\nu_\infty}[|Y_t-Y_s|^{2n}]
&=
\sum_{k=0}^{2n}\lkkk\!
\begin{array}{c} 2n\\ k
\end{array}
\!\rkkk (-1)^k
\lk
\xx^{2n-k}, e^{-(t-s)L_{\rm p}}
\xx^k\rk_{\HP}\\
&=
\sum_{k=0}^{2n}\lkkk\!
\begin{array}{c} 2n\\ k
\end{array}
\!\rkkk (-1)^k
\lk
\xx^{2n-k}
\varphi_{\rm p},
e^{-(t-s){K}}
\xx^k\varphi_{\rm p} \rk_{L^2}
e^{(t-s)\is(\lp)}.
\end{align*}
Furthermore
by Feynman-Kac  formula, i.e., Proposition \ref{6},
the right-hand side above can be expressed
in terms of $X^\xx=\pro{X^\xx}$ as
\begin{align*}
&
\mathbb E_{\nu_\infty}
  [|Y_t-Y_s|^{2n}]\\
  &
  =
\int \mup
\Ew
{
 {
 |X_{t-s}^\xx -X_0^\xx |^{2n}
 \varphi_{\rm p}(X_0^\xx )
 \varphi_{\rm p}(X_{t-s}^\xx )
 e^{-\int_0^{t-s}W(X_r^\xx )dr}
 }} e^{(t-s)\is(\lp)}.
 \end{align*}
Since $\VV\geq 0$,
$$
\mathbb E_{\nu_\infty}
  [|Y_t-Y_s|^{2n}]\leq
  \|\varphi_{\rm p}\|_\infty^2
e^{(t-s)\is(\lp)}\int \mup
\Ew
{
 {
 |X_{t-s}^\xx -X_0^\xx |^{2n}
 }} .
$$
We next estimate
$\Ew
{|X_t^{\xx}-X_s^{\xx}|^{2n}}$.
Since $X_t^\xx$ is the solution to the stochastic
differential equation:
$$X_t^{\xx,\mu}-X_s^{\xx,\mu}=
\int_s^t b_\mu(X_r^\xx ) dr+\int _s^t \s_\mu(X_r^\xx )\cdot dB_r,$$
we have
$$\Ew{|X_t^{\xx,\mu}-X_s^{\xx,\mu}|^{2n}}
\leq 2^{2n-1}
\Ew{\frac{|t-s|^{2n}}{2^{2n}}
\|b_\mu\|_\infty^{2n}
+\sum_{\nu=1}^3 \left|
\int_s^t \s_{\mu\nu}
(X_r^\xx )dB_r^\nu \right|^{2n}
}.$$
By the Burkholder-Davies-Gundy inequality \cite[Theorem 3.28]{ks}, we have
\begin{align*}
\Ew{\left|\int_s^t \s_{\mu\nu}(X_r^\xx )dB_r^\nu \right|^{2n}}
&\leq
 (n(2n-1))^n |t-s|^{n-1}
\Ew{\int_s^t |\s_{\mu\nu}(X_r^\xx )|^{2n} dr}\\
&\leq
 (n(2n-1))^n |t-s|^n
\|\s_{\mu\nu}\|_\infty^{2n}.
\end{align*}
Then
$\Ew{
|X_t^\xx-X_s^\xx|^{2n}}
\leq C |t-s|^n$ with some constant $C$ independent of $s$ and $t$, and
\eq{20}
\mathbb E_{\nu_\infty}\lkkk
|Y_t-Y_s|^{2n}
\rkkk
\leq C |t-s|^n
\en
follows.
Thus   $Y=\pro Y$
has a continuous version by Kolmogorov-\v{C}entov continuity  theorem \cite[Theorem 2.8]{ks}.
\qed
Let
$\ov Y=\pro {\ov Y }$ be the continuous version of $Y$ on
$((\BT)^{[0, \infty)}, \ms B((\BT)^{[0,\infty)}), \nu_\infty)$.
The image measure of $\nu_\infty$
on
$({\ms X}_+ , \ms B({\ms X}_+ ))$  with
respect to $\ov Y $
is denoted by
$Q$, i.e.,
$Q=\nu_\infty\circ \ov Y\f$,
and
$\wY _t(\omega)=\omega(t)$
for $\omega\in {\ms X}_+ $ is the coordinate process.
Then
we constructed a stochastic process $\wY=\pro{\wY} $ on $({\ms X}_+ , \ms B({\ms X}_+ ), Q)$
such that
$\bar Y\stackrel{\rm d}{=}\wY$.
Then
\kak{t1} and \kak{t2}
  can be expressed in terms of $\wY $ as
  \begin{align*}
&
(f_0, e^{-(t_1-t_0)L_{\rm p} }f_1
\cdots e^{-(t_n-t_{n-1}) L_{\rm p} }f_n)_{\HP}
=\mathbb E_Q\lkkk
\prod_{j=0}^n f_j(\wY _{t_j})\rkkk,\\
&
(\one,  f)_{\HP}=
(\one,  e^{-tL_{\rm p}} f)_{\HP}
=
\mathbb E_Q\lkkk
f(\wY _t)\rkkk
=
\mathbb E_{Q}\lkkk
f(\wY _0)\rkkk
\end{align*}
for $0\leq t$ and $0\leq t_0\leq t_1\leq\cdots
\leq t_n$.

{\bf(Step 3)}
Define
the regular conditional probability measure on ${\ms X}_+ $ by
\eq{n15}
Q^\xx (\cdot)=Q(\cdot|\wY  _0=\xx )
\en
for
each $\xx \in\BT$.
It is well defined,
since ${\ms X}_+ $ is a Polish space (completely separable metrizable  space).  See e.g., \cite[Theorems 3.18. and 3.19]{ks}.
Since the
 distribution of $\wY_0$ equals to $\mup $,
 note
 that
$Q(A)
=\int \mup  \mathbb E_{Q^\xx }[\one_A]$.
Then the stochastic
process $\wY =\pro{\wY }$
on $({\ms X}_+ , \ms B({\ms X}_+ ), Q^\xx )$
satisfies
\begin{align}
&
\label
{ptk}
(f_0, e^{-(t_1-t_0)L_{\rm p} }f_1
\cdots e^{-(t_n-t_{n-1}) L_{\rm p} }f_n)_{\HP}
=\int \mup   \EE_{Q^\xx }
\left[\prod_{j=0}^n f_j(\wY _{t_j})\right],\\
&
(\one, e^{-tL_{\rm p}}f)_{\HP}=(\one, f)_{\HP}
=\int d\xx \varphi_{\rm p}^2(\xx ) \EE_{Q^\xx }\lkkk f(\wY_0)\rkkk
=\int d\xx \varphi_{\rm p}^2(\xx ) f(\xx ).
\end{align}
\bl
 {I2}
 $\wY$ is a Markov process on $({\ms X}_+ , \ms B({\ms X}_+ ), Q^\xx )$
 with respect to the natural filtration
 $\pro {\ms M}$, where
 $\ms M_s=\s(\wY_r,0\leq r\leq s)$.
\el
\proof
Let \eq{ptk2}
p_t(\xx , A)=\lk e^{-tL_{\rm p} }\one_A\rk(\xx ),\quad A\in {\ms B} (\BT),\quad t\geq0.
\en
Notice that
$p_t(\xx ,A)=\Ew{\one_A(X_t^\xx)}$.
Then the finite dimensional
distribution of $\wY $
is
$$
\EE_{Q^\xx }
\left[\prod_{j=1}^n \one_{A_j}(\wY _{t_j})\right]
=\int \prod_{j=1}^n \one_{A_j}(\xx _j) \prod_{j=1}^n p_{t_j-t_{j-1}}(\xx _{j-1}, d\xx _j)$$
with $t_0=0$ and $\xx _0=\xx $ by
\kak{ptk}.
We show that
$p_t(\xx ,A)$ is a probability transition kernel, i.e.,
(1)
$p_t(\xx , \cdot)$ is a probability measure on $\ms B(\BT)$,
(2) $p_t(\xx,A)$ is Borel measurable with respect to $\xx $,
(3) the Chapman-Kolmogorov equality\index{Chapman-Kolmogorov equality}
\eq{chap}
\int p(s,\yyy,A)p(t,\xx ,d\yyy)=p(s+t,\xx ,A)
\en
is satisfied.
Note that
$e^{-tL_{\rm p} }$
is positivity improving.
Then $0\leq e^{-tL_{\rm p} } f\leq \one$
for all
function $f$ such that
$0\leq f\leq \one$,
and $e^{-tL_{\rm p} }\one=\one$ follows.
Then
 $p_t(\xx ,\cdot)$ is the probability measure on $\BT$ with $p_t(\xx , \BT)=1$, and (1) follows.
(2) is trivial.
From the semigroup property $e^{-sL_{\rm p}}e^{-tL_{\rm p}}\one_{A}=
e^{-(s+t)L_{\rm p}}\one_A$,
the Chapman-Kolmogorov equality \kak{chap} follows.
Hence $p_t(\xx ,A)$ is a probability transition kernel.
We write $\EE$ for $\EE_{Q^\xx }$ for notational simplicity.
From the identity
$\EE[\one_A(\wY_t)
\EE[f(\wY_r)|\s(\wY_t)]]=\EE[\one_A(\wY_t)f(\wY_r)]$ for $r>t$,
 it follows that
$$\int \one_A(\yyy )\EE[f(\wY_r)|\wY_t=\yyy ]
P_t(d\yyy )=\int P_t(d\yyy )
\one_A(\yyy )
\int
f(\yyy ')p(r-t, \yyy , d\yyy '),$$ where $P_t(d\yyy )$ denotes the distribution of $\wY_t$ on $\BT$. Thus
$$\EE[f(\wY_r)|\wY_t=\yyy ]
=\int f(\yyy ')p(r-t, \yyy ,d\yyy ')$$
follows almost everywhere $\yyy $ with respect to $P_t(d\yyy )$.
Then
$$\EE[f(\wY_r)|\s(\wY_t)]=\int f(\yyy ) p(r-t, \wY_t,d\yyy )$$
and
\eq{t3}
\EE[\one_A(\wY_r)|\s(\wY_t)]=p(r-t,\wY_t,A)
\en
follow.
By using \kak{t3} and the Chapman-Kolmogorov equality
we can show that
$$
\EE\lkkk \one_{A}(\wY_{t+s})
\prod_{j=0}^n \one_{A_j}(\wY_{t_j})\rkkk
=
\EE\lkkk \EE\lkkk
\one_{A}(\wY_{t})|\s(\wY_s)\rkkk \prod_{j=0}^n \one_{A_j}(\wY_{t_j})\rkkk$$
for $t_0\leq\cdots\leq t_n\leq s$. This implies that
$
\EE[\one_A(\wY_{t+s})|\ms M_t]=
\EE[\one_A(\wY_{t})|\s(\wY_s)]$.
Then $\wY $ is Markov with respect to the natural filtration under the measure $Q^\xx $.
\qed

{\bf(Step 4)}
We extend $\wY =\pro{\wY }$
to a process on the hole real line $\RR$.
Set
$\tilde {{\ms X}_+ }={\ms X}_+ \times {\ms X}_+ $,
$\tilde{\ms M}=\ms B({\ms X}_+ )\times\ms B({\ms X}_+ )$  and
${\tilde Q}^\xx =Q^\xx \times Q^\xx $.
Let $(\tilde X_t)_{t\in\RR}$
be the stochastic process on
the product space
$(\tilde {{\ms X}_+ }, \tilde {\ms M}, \tilde Q^\xx )$,
defined by
for $\omega=(\omega_1,\omega_2)\in\tilde
{{\ms X}_+ }$,
\eq{I3}
\tilde X_t(\omega)=\lkk
\begin{array}{ll}
\wY_t(\omega_1),& t\geq 0,\\
\wY_{-t}(\omega_2),& t<0.
\end{array}
\right.
\en
Note that
 $\tilde X_0=\xx $ almost surely with respect to $\tilde Q^\xx $ and
$\tilde X_t$ is continuous in $t$ almost surely.
It is trivial to see that
$\tilde X_t$, $t\geq0$,
and $\tilde X_s$, $s\leq 0$,
 are independent,
 and $\tilde X_t\stackrel{\rm d}{=}
 \tilde X_{-t}$.

{\bf(Step 5)}
{\it Proof of Proposition \ref{main1}:}\\
The image measure of $\tilde Q^\xx $ on
$(\ms X, \ms B(\ms X))$ with respect to $\tilde X$ is denoted by
$P^\xx $, i.e.,
$P^\xx =\tilde Q^\xx \circ\tilde X\f$.
Let
$
X_t(\omega)=\omega(t)$,
$t\in\RR$,
$\omega\in{\ms X}$, be the coordinate
 process.
Then we can  see that
\eq{cent}
X_t\stackrel{\rm d}{=}\wY _t\quad
(t\geq 0),
\quad \quad
X_t\stackrel{\rm d}{=}\wY _{-t}\quad
(t\leq 0).
\en
Since by (Step 3), $\pro {\wY }$ and
$(\wY _{-t})_{t\leq 0}$ are
Markov processes  with respect to
the
natural filtration
$\s(\wY _s, 0\leq s\leq t)$ and
$\s(\wY _s, -t\leq s\leq 0)$, respectively,
 $\pro X$
 and
$(X_t)_{t\leq 0}$
are also Markov processes with respect to
$(\ms F_t^+)_{t\geq 0}$ and
$(\ms F_t^-)_{t\leq 0}$,
respectively, where
$$
\ms F_t^+=\s(X_s,0\leq s\leq t),  \quad
\ms F_{t}^-=\s(X_s, -t\leq s\leq 0).$$
Thus the diffusion property (3) follows.
We also see that
$(X_{s})_{s\leq 0}$
and $\pro X$  are
independent and
$X_{-t}\stackrel{\rm d}{=}
X_t$
by \kak{cent} and (Step 4).
Thus reflection symmetry (2) follows.
\bl{final}
Let $-\infty <t_0\leq t_1\leq \cdots\leq
t_n$. Then
\eq{I4}
\ixp {f_0(X_{t_0})\cdots f_n(X_{t_n})}
=(f_0, e^{-(t_1-t_0)L_{\rm p}}f_1\cdots e^{-(t_n-t_{n-1})L_{\rm p}} f_n)_{\HP}.
\en
\el
\proof
Let $t_0\leq\cdots \leq t_n\leq 0\leq t_{n+1}\leq\cdots t_{n+m}$.
Then we have by the independence of
$(X_{s})_{s\leq 0}$
and $\pro X$,
\begin{align*}
&
\ixp {f_0(X_{t_0})\cdots f_{n+m}(X_{t_{n+m}})}\\
&=
\ixp
{f_0(X_{t_0})\cdots f_{n}(X_{t_{n}})}
\mathbb E_{P^\xx} \lkkk
f_{n+1}(X_{t_{n+1}})\cdots f_{n+m}(X_{t_{n+m}})\rkkk.
\end{align*}
Since we have
\begin{eqnarray}
&&\mathbb E_{P^\xx} \lkkk
f_{n+1}(X_{t_{n+1}})\cdots f_{n+m}(X_{t_{n+m}})\rkkk\non \\
&&\label{I5}
=
\lk
e^{-t_{n+1}L_{\rm p}}
f_{n+1}
e^{-(t_{n+2}-t_{n+1})L_{\rm p}}
f_{n+2}\cdots e^{-(t_{n+m}-t_{n+m-1})L_{\rm p}}f_{n+m}\rk(\xx )
\end{eqnarray}
and
\begin{eqnarray}
\mathbb E_{P^\xx }\lkkk
f_{0}(X_{t_0})\cdots f_{n}(X_{t_{n}})\rkkk
&=&
\mathbb E_{P^\xx }\lkkk
f_{0}(\wY_{-t_0})\cdots f_{n}(\wY_{-t_{n}})\rkkk\non \\
&=&
\label{I6}
\lk
e^{+t_{n}L_{\rm p}}
f_{n}
e^{-(t_{n}-t_{n-1})L_{\rm p}}
f_{n-1}\cdots e^{-(t_{1}-t_{0})L_{\rm p}}f_1
\rk(\xx ). \ \ \ \
\end{eqnarray}
By \kak{I5} and \kak{I6} we have
\begin{align*}
&
\ixp {f_0(X_{t_0})\cdots f_{n+m}(X_{t_{n+m}})}\\
&=
(e^{+t_{n}L_{\rm p}}
f_{n}
\cdots e^{-(t_{1}-t_{0})L_{\rm p}}f_1,
e^{-t_{n+1}L_{\rm p}}
f_{n+1}
\cdots e^{-(t_{n+m}-t_{n+m-1})L_{\rm p}}f_{n+m})_{\HP}\\
&=
(f_1,
e^{-(t_1-t_0)L_{\rm p}}
f_2
\cdots
e^{-(t_{n+m}-t_{n+m-1})L_{\rm p}}f_{n+m})_{\HP}.
\end{align*}
Hence \kak{I4} follows.
\qed
From Lemma \ref{final}
it follows that
for any $s\in\RR$,
$$\ixp{\prod_{j=0}^ nf_j(X_{t_j})}
=
\ixp{\prod_{j=0}^ nf_j(X_{t_j+s})}.$$
Hence shift invariance (4) is obtained.
\qed
We denote $\EE^\xx $ for $\EE_{P^\xx }$ in what follows.

\subsection{Absence of ground state}
\subsubsection{The Nelson model by path measures}
Now we construct a Feynman-Kac formula for $e^{-tH}$
by using the diffusion process $X$.
Let $\phi_{\rm E}(f)$ be
the Euclidean scalar field\ko{Euclidean scalar field} on a
probability space
$(\ms Q_{\rm E}, \Sigma_{\rm E}, \mu_{\rm E})$,
which is the Gaussian random variable indexed by $f\in L^2(\RR^4)$
with
$\mathbb E_{\mu_{\rm E}}[\phi_{\rm E}(f)]=0$ and
the covariance given by
$$\mathbb E_{\mu_{\rm E}}\lkkk
  \phi_{\rm E}(f)\phi_{\rm E}(g)\rkkk
  =\half (\hat f, \hat g).$$
Euclidean scalar field
$L^2({\ms Q}_{\rm E})$ and $L^2({\ms Q})$ are connected through some isometry $\jj_t$.
Let $\jj_t:\LRT\to L^2(\RR^4)$ be given by
\eq{16}
 \jj_t f(x_0,x)=\frac{1}{2\pi}
 \int
  dk_0
  {e^{-i(t-x_0)k_0}}
\lk
\omega^\han (\omega^2+|k_0|^2)^{-\han}f\rk (x).
\en
Then we can have the formula
$(\jj_s f,\jj_t g)=(f, e^{-|t-s|\omega}g)$.
In particular
\eq{17-1}
\jj_t^\ast \jj_s=e^{-|t-s|\omega}.
\en
Let
$\JJ_t=\Gamma(\jj_t):L^2({\ms Q})\to L^2({\ms Q}_{\rm E})$
be the isometry defined by $\JJ_t \one =\one_{\rm E}$ and
\begin{align*}
\JJ_t\wick{\prod_{j=1}^n \phi(f_j)}=
\wick{\prod_{j=1}^n \phi_{\rm E}(\jj_t f_j)}.
\end{align*}
From the definition of $\JJ_t$,
the identity
\eq{1888}
\JJ_t ^\ast \JJ_s=e^{-|t-s|\hf}
\en follows.
Thus the semigroup
$e^{-t\hf}$ can be
factorized by $\JJ_t$ and it can be expressed
as
\eq{19}
(\Phi, e^{-t\hf}\Psi)_{L^2({\ms Q})}
=
 (\JJ_0\Phi,
  \JJ_t\Psi)_{L^2({\ms Q}_{\rm E})}.
\en
\bt{14}
\TTT{Feynman-Kac formula}
\ko{Feynman-Kac formula}

Suppose Assumptions
\ref{ass}, \ref{v}, \ref{w}, \ref{chi}
 and \ref{ass1}.
Then
we have
\eq{15}
(F, e^{-TH}G)_\hhh
=\ix {
(\JJ_0 F(X_0), e^{
 \phi_{\rm E}(K_T)}\JJ_T G(X_T))_{L^2({\ms Q}_{\rm E})}},
\en
where $K_T=\int_0^T \jj_s\omega^{-\han}
\rho(\cdot-X_s) ds$ is an $L^2(\RR^4)$-valued integral.
In particular it follows that
\eq{I20}
(\one, e^{-TH}\one)_\hhh=
\ix {
e^{(1/2) \int_0^T dt \int_0^T ds
\double( X_t, X_s, |t-s|)}},
\en
where $\one\in L^2(\BT\times \ms Q)$ and
\eq{num}
\double(X,Y, |t|)=\half
(\rho(\cdot-X),
\omega\f
e^{-|t|\omega}
\rho(\cdot-Y)).
\en
\et
\proof
By the Trotter-Kato
product formula:
$$e^{-tH}=
\slim_{n\rightarrow\infty}
\lk e^{-(t/n)L_{\rm p} }e^{-(t/n)
 \phi_\rho }
e^{-(t/n)\hf}\rk^n,$$
the factorization formula
\kak{1888},
and
Markov property of $E_t=\JJ_t \JJ_t^\ast$,
 we have
\eq{yoko3}
(F,e^{-tH}G)
=
\lim_{n\rightarrow \infty}
\int \mup \EE^\xx \!\!\left[
\lk
\JJ_0F(X_0),
e^{-
\sum_{j=0}^n
\frac{t}{n}
\phi_{\rm E}
(\jj_{\frac{tj}{n}}
\rho (\cdot-X_{tj/n}))}
\JJ_tG(X_t)
\rk
\right].
\en
Note that
the map $\RR\rightarrow \LRT$,
$s\mapsto \omega^{-\han}\rho (\cdot-X_s)$,
is strongly continuous
 almost surely.
Hence
the map
$\RR\rightarrow L^2({\ms Q}_{\rm E})$,
$s\mapsto \phi_{\rm  E}(\jj_s\rho (\cdot-X_{s}))$,
is also strongly continuous.
By a simple limiting argument
\kak{15} follows.
Let $F=G=\one$.
Since  $\phi_{\rm E}$ is a Gaussian random variable,
we have
$$
(\one, e^{-tH}\one)
=
\int\mup\mathbb E^\xx \lkkk
(\one, e^{\phi_{\rm E}(K_T)}\one)\rkkk
=
\int\mup\mathbb E^\xx  \lkkk
e^{(1/4) \|K_T\|_{L^2(\RR^4)}^2}
\rkkk.
$$
Hence
$\|K_T\|_{L^2(\RR^4)}^2=\int_0^T dt \int_0^T ds
\double (X_t,X_s,|t-s|)$ and
\kak{I20} follows.
\qed

\subsubsection{Absence of ground states}
From Theorem \ref{14}
we can obtain
a useful lemma to show
the absence of ground states.

\begin{figure}[t]
\centering
\includegraphics[width=300pt]{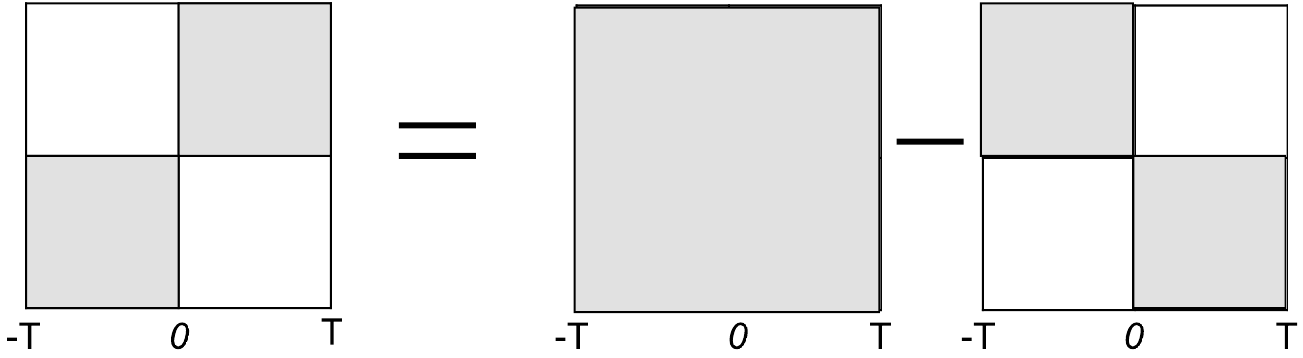}
\caption{$\d \int_{-T}^0
\int_{-T}^0
+\int_0^T
\int_0^T =
\int_{-T}^T
\int_{-T}^T
-2\int_{-T}^0
\int_0^T $}
\label{picbox}
\end{figure}

\bt
{pi2}
\TTT{Positivity improving}
\ko{positivity improving}
Suppose Assumptions
\ref{ass}, \ref{v}, \ref{w}, \ref{chi}
 and \ref{ass1}.
Then
 $e^{-tH}$ is positivity improving for
 all $t>0$.
\et
\proof
Let $F,G\in L^2(\BT\times \ms Q)$ be such that $F\geq 0$ and $G\geq 0$ but $F\not\equiv 0$ and $G\not\equiv 0$.
Define
$\ms D_F=\{x\in\BT| F(x,\cdot)\not\equiv 0\}$
and
$\ms D_G=\{x\in\BT| G(x,\cdot)\not\equiv 0\}$.
Note that $\int_{\ms D_F} dx>0$ and
$\int_{\ms D_G}dx>0$.
Let $\ms K^\xx  =
\{\omega\in\ms X|X_0(\omega)=x, X_t(\omega)\in \ms D_F\}$.
It follows that
$$\int_{\ms D_F}
\mup \int_{\ms K^\xx }  dP^\xx =
(\one_{\ms D_G}, e^{-tL_{\rm p}}
\one_{\ms D_F})_{\HP}=
(\varphi_{\rm p}\one_{\ms D_G},
e^{-tK}\varphi_{\rm p} \one_{\ms D_F})
_{L^2} e^{t\is(K)}>0$$
by Lemma \ref{pi}.
Thus $\BT\times \ms Q\supset \cup_{\xx\in \ms D_F}\ms K^\xx $ has a positive measure with respect to $\mup  dP^\xx $ and
\begin{align*}
(F, e^{-tH}G)\geq
\int_{\ms D_F}
\mup
\int_{\ms K^\xx }  dP^\xx
(\JJ_0 F, e^{\phi(K_t)}\JJ_t G(X_t))>0,
\end{align*}
since
$(\JJ_0 F(X_0(\omega)),
e^{\phi(K_t(\omega))}\JJ_t G(X_t(\omega)))>0$ for $\omega\in\cup_{\xx\in\ms D_F}
\ms K^\xx  $. Then
the theorem  follows.
\qed
By Theorem  \ref{pi2}
and Perron-Frobenius arguments, we immediately have the corollary.
 \bc{uni}
\TTT{Uniqueness of ground state}
 Suppose Assumptions
 \ref{ass}, \ref{v}, \ref{w}, \ref{chi}
 and \ref{ass1}.
Then the ground state of $\gr$ is unique and $\gr>0$ if it exists.
In particular $(\one, \gr)>0$.
 \ec
From Corollary \ref{uni}
  we  can see  that
$e^{-TH}\one/\|e^{-TH}\one\|$ converges to the ground state as $T\to \infty$ if the ground state exists.
We then define
$\gamma(T)$ by
\eq{I22}
\gamma(T)=\frac{(\one, e^{-TH}\one)^2}{(\one, e^{-2TH}\one)},\quad T>0.
\en
\bl{I23}
Suppose Assumptions
\ref{ass}, \ref{v}, \ref{w}, \ref{chi}
 and \ref{ass1}.
Let $P_{\Delta}$,
$\Delta\subset \RR$,  denote the spectral projection of $H$ associated with
$\Delta\cap \s(H)$.
Let $E=\is (H)$. Then
it follows that
$\lim_{T\to \infty}\gamma(T)=\|P_{\{E\}}\one\|^2$.
In particular $H$ has a ground state if and only if $\lim_{T\to \infty }\gamma(T)\not=0$.
\el
\proof
Assume that $E=0$.
Thus $\lim_{T\to\infty }e^{-TH}=P_{\{E\}}$.
If $0$ is an eigenvalue,
then by Corollary
\ref{uni} and Perron-Frobenius arguments,
$P_{\{E\}}=(u,\cdot)u$ for some $u>0$.
It follows that
$\lim_{T\to\infty }\gamma(T)=(u,\one)^2>0$.
Next we prove
the sufficient part.
Assume now that there exists a sequence $T_n\to +\infty$ such that
$\delta(T_n)\geq \epsilon ^2>0$.
This implies that
$(\one, e^{-T_n H}\one)\geq
\epsilon
(\one, e^{-2T_nH}\one)^\han$.
Letting $n\to \infty$,
we obtain that $\|P_{\{E\}}\one\|\geq\epsilon$.
Then $H$ has a ground state.
\qed
The denominator of $\gamma(T)$ is
computed as
 \begin{eqnarray*}
\|e^{-TH}\one \|^2
=(\one, e^{-2TH}\one)
=\ix{e^{(1/2)\int_{-T}^T
ds\int_{-T}^T dt
\double(X_s,X_t,|s-t|)}}
\end{eqnarray*}
by the shift invariance (Proposition  \ref{main1}) of $X_t$.
Then $\gamma(T)$ can be expressed as
\eq{17}
\gamma(T)=
\frac{\lk
\ix{e^{(1/2)\int_0^Tds\int_0^Tdt
\double( X_s,X_t,|s-t|)}}\rk^2}
{\ix{e^{(1/2)\WTT}}}
.
\en
Let $\mu_T$ be
the probability measure on
$(\BT\times \ms X ,
\ms B(\BT)\times
\ms B(\ms X))$ defined by for
$\mathcal O\in \ms B(\BT)\times\ms B(\ms X)$,
\eq{17-2}
\mu_T(\mathcal O)
=\frac{1}{Z_T}\ix
{\one_{\mathcal O} e^{(1/2) \WTT}},
\en
where $Z_T$ denotes the normalizing constant
such that $\mu_T$ becomes a probability measure.
\bl{upper}
Suppose Assumptions
\ref{ass}, \ref{v}, \ref{w}, \ref{chi}
 and \ref{ass1}.
Then it follows that
\eq{18}
\gamma(T)\leq \EE_{\mu_T}
\left[
e^{-\WT}
\right].
\en
\el
\proof
The numerator of \kak{17}
can be estimated by
the Schwartz inequality with respect to $\mup$
and
the reflection symmetry  of $X$, and then
\begin{align*}
&\lk
\ix{e^{(1/2)
\int_0^Tds\int_0^Tdt
\double }}\rk^2\\
&\leq
\int \mup
\lk
\EE ^\xx \!\!\left[
e^{(1/2)\int_0^T ds \int_0^T dt \double
}\right]\rk
\lk\EE ^\xx \!\!\left[
e^{(1/2)\int_{-T}^0
ds
\int_{-T}^0 dt \double }
\right]\rk.
\end{align*}
Since $X_t$ and $X_s$ for $s\leq 0\leq t$ are independent,
thus we have
$$
\lk
\ix{e^{(1/2)\int_0^Tds\int_0^Tdt
\double }}\rk^2\leq
\int \mup
\EE ^\xx \!\!\left[
e^{(1/2)
\lk
\int_0^Tds\int_0^T dt \double
+\int_{-T}^0
ds\int_{-T}^0dt \double \rk}
\right].
$$
Moreover
$\int_{-T}^0
\int_{-T}^0
+\int_0^T
\int_0^T =
\int_{-T}^T
\int_{-T}^T
-2\int_{-T}^0
\int_0^T $ yields
that (Figure \ref{picbox})
$$
\lk
\ix{e^{(1/2)\int_0^Tds\int_0^Tdt
\double }}\rk^2\leq
\ix{e^{- \wt+(1/2)\wtt}}.$$
Then the lemma follows.
\qed

In order to show  that the right-hand side of
\kak{18} converges to zero as $T\to\infty$, we estimate its upper bound.
Let
\eq{ko1}
\double_\infty (X,Y,|t|)=\half \int dk\frac{\hat\rho (k)^2
e^{-ik\cdot(X-Y)}}{|k|}e^{-|t||k|}
\en
or it is expressed in the position representation as
\eq{ko4}
\double_\infty (X,Y,|t|)=\half
(\rho (\cdot-X), \omega_\infty\f e^{-|t|\omega_\infty}  \rho (\cdot-Y))
\en
with
\eq{infty}
\omega_\infty=\sqrt{-\Delta}.
\en
The next proposition on
the  upper and lower Gaussian
bound of
 the integral kernel
 $e^{-t\omega^2}(x,y)$
is the key ingredient of
the proof of the
absence of ground states of $H$.
\bp{gerard}
Suppose Assumption \ref{w}.
Then the semigroup $e^{-t\omega^2}$
has an integral kernel
$e^{-t\omega^2}(x,y)$,
 and
 there exist constants
$C_1,\cdots, C_4$ such that
\eq{g1}
C_1 e^{-C_2t\omega_\infty^2}(x,y)
\leq e^{-t\omega^2}(x,y)
\leq
C_3 e^{-C_4t\omega_\infty^2}(x,y)
\en
for $t\geq0$ and a.e. $x,y\in\BT$.
\ep
\proof
Conjugating
by the unitary $U:\LRT\to L^2(\BT, c^2(x) dx)$, $f\mapsto c\f f$,
we obtain
$$\tilde \omega^2=U \omega^2 U\f =
\tilde h_0+ m^2(x),$$
where
$$\tilde h_0=\tilde h_0(D,x)
=c^{-2}(x)\sum_{\mu\nu=1}^3
D_\mu a_{\mu\nu}(x) D_\nu.$$
Throughout
$(f,g)_2$ denotes
$\int \bar f(x) g(x) c^2(x) dx$
and $\|f\|_p^p=\int|f(x)|^p c^2(x) dx$.
Note that
$C_0\int|f(x)|^p dx\leq
\|f\|_p^p
\leq
C_1
\int|f(x)|^p dx$.
Let
$f\in C_0^\infty(\BT)$. Then
$$
C_0
(f, \tilde h_0 f)_2
\leq
{\ms E}_\one (f,f)
\leq
C_1
(f, \tilde h_0 f)_2.
$$
From this $D(\tilde h_0^\han)=H^1(\BT)$ follows and
\eq{oota}
C_0
(\tilde h_0^\han f, \tilde h_0^\han f)_2
\leq
{\ms E}_\one (f,f)
\leq
C_1
(\tilde h_0^\han f, \tilde h_0^\han
f)_2
\en
for $f\in H^1(\BT)$.
Notice that
$e^{-t\tilde\omega}(x,y)$
denotes an integral kernel
of
$e^{-t\tilde\omega}$ with respect to
the measure $c^2(x) dx$, while
$e^{-t\omega}(x,y)$ is that of
$e^{-t\omega}$ with respect to
$dx$.
Since
\begin{align*}
(f, e^{-t\omega^2}g)_{\LRT}
&=
\int dx \int dy
 \ov{f(x)}
 e^{-t\omega^2}(x,y)g(y)=
(Uf, e^{-t\tilde \omega^2} U g)_2
\\
&=
 \int c^2(x)dx
 \int c^2(y)dy
{ \frac{1}{c(x)} \ov{f(x)}}
e^{-t\tilde \omega^2}(x,y)
\frac{1}{c(y)} g(y),
\end{align*}
we note that
$$
e^{-t \omega^2}(x,y)
=c(x)
e^{-t\tilde \omega^2}(x,y)c(y)$$
almost everywhere.
So it suffices to prove proposition
for $e^{-t\tilde \omega^2}$.
We know from \cite[Theorems 3.4 and 3.6]{PE}
that $e^{-t\tilde h_0}
$
has an
integral kernel
with
\eq{yu1}
C_1 e^{-C_2t\omega_\infty}(x,y)
\leq
e^{-t\tilde h_0}(x,y)
\leq
C_3 e^{-C_4t\omega_\infty}(x,y)
\en
for a.e. $x,y\in\BT$ and all $t>0$.
Notice that
$e^{-t(\tilde h_0+V)}$, $\sup_x|V(x)|<\infty$,
 is positivity preserving
and
bounded on $L^\infty(\BT)$.
This can be proven by the Trotter product formula.
Then by \kak{yu1}
we see that
$\|e^{-t\tilde h_0}f\|_\infty
\leq Ct^{-3/4}
\|f\|_2$,
which is equivalent to
\eq{tabi2}
\|f\|_6^2\leq C(\tilde h_0^\han f, \tilde h_0^\han f)_2
\en
by Proposition \ref{el2}.
 Since $m^2(x)\geq 0$, the upper bound follows from the Trotter product formula.
Let us now prove the lower bound, following \cite[Theorem 6.1]{se}.
Since $m^2(x)\leq \langle x\rangle^{-\beta}$ with $\beta>2$,
we see that
$m^2\in L^{3/2}(\BT)$.
Then we have
$(m^2 f, f)\leq \|m^2\|_{3/2}
\|f\|^2_6$.
By the Sobolev inequality
we see that
$\|f\|_6^2\leq C_1
{
{\ms E}_\one (f,f)}$,
and then together with \kak{oota},
$$
\gamma
(m^2 f, f)_2\leq
(\tilde h_0^\han f,
\tilde h_0^\han f)_2$$
for some  $\gamma >0$.
Set now $w(x)=-\gamma m^2(x)/4$.
Then
$\tilde h_0+2w\geq \half \tilde h_0$
in the sense of
form.
We see, together with
\kak{tabi2}, that
\eq{oota2}
\|f\|_6^2\leq
C_2((\tilde h_0+2w)^\han f,
(\tilde h_0+2w)^\han f)_2.
\en
By
Proposition \ref{el2}
 and
the fact that $e^{-t(\tilde h_0+w)}$
is positivity preserving and bounded on
$L^\infty(\BT)$,
\kak{oota2}
is equivalent to
$$\|e^{-t(\tilde h_0+2w)} f\|_\infty
\leq
C_3 t^{-3/4}\|f\|_2$$
and
$e^{-t(\tilde h_0+2w)}$
is ultracontractive.
Then $e^{-t(\tilde h_0+2w)}$
  has an integral kernel,
furthermore
it can be
estimated as
$$e^{-t(\tilde h_0+2w)}(x,y)
\leq C_4t^{-3/2}$$
almost everywhere.
We prove
in Lemma \ref{narita}
that
$\la\mapsto e^{-t(\tilde h_0+\la w)}(x,y)$
is logarithmically convex
for all $t>0$
and a.e. $x,y\in\BT$.
Then
we have
\eq{a19}
e^{-t(\tilde h_0+w)}(x,y)
=
e^{-t(\tilde h_0+\half 0w +\half 2w)}
(x,y)
\leq t^{-3/4}
e^{-t\tilde h_0}(x,y)^\han.
\en
Applying again the log-convexity we get that
$$
e^{-t\tilde h_0}(x,y)
=
e^{-t(\tilde h_0+sm^2+(1-s)w)}(x,y)
\leq
e^{-t(\tilde h_0+m^2)}(x,y)^s
e^{-t(\tilde h_0+w)}(x,y)^{1-s}$$
with $s=\gamma /(4+\gamma )$.
Hence
using \kak{a19}
we obtain
$$e^{-t\tilde h_0}(x,y)^{(1+s)/2} t^{(1-s)3/4}\leq
e^{-t(\tilde h_0+m^2)}(x,y)^s,$$
which, together with
\kak{yu1},
 implies the proposition.
\qed
It remains to show the log-convexity
of $e^{-t(\tilde h_0+w)}(x,y)$.
\bl{narita}
$\RR\ni\la\mapsto
e^{-t(\tilde h_0+\la w)}(x,y)\in\RR
$ is logarithmically convex for
all $t>0$ and a.e. $x,y\in\BT$,
i.e.,
for $0\leq s\leq 1$,
$$e^{-t(\tilde h_0+(s\la+(1-s)\la') w)}
(x,y)\leq
e^{-t(\tilde h_0+\la w)}(x,y)^s
e^{-t(\tilde h_0+\la' w)}(x,y)^{1-s}.
$$
\el
\proof
Set $t=1$.
By the Trotter product formula we have
\eq{tro}
e^{-(\tilde h_0+\la w)}(x,y)=
\lk
s-\limn
\lk
e^{-\tilde h_0/n}
e^{-\la w/n}\rk^n \rk
(x,y).
\en
Let $A_\la(x,y)$ and $B_\la(x,y)$
be the kernels of two operators
$A_\la$ and $B_\la$ assumed to be log-convex in $\la$.
Then the kernel
of $A_\la B_\la$:
$$A_\la B_\la(x,y)=
\int _\BT A_\la(x,z) B_\la(z,y) dz$$
is also log-convex in $\la$.
Then the kernel of
$e^{-\tilde h_0/n}e^{-\la w/n}(x,y)
= e^{-\tilde h_0/n}(x,y)
e^{-\la w(y)/n}$ is log-convex in $\la$.
Then the lemma follows from
the Trotter product formula \kak{tro}.
\qed
\bc{h}
\TTT{Positivity improving}
Suppose Assumption \ref{w}. Then
$e^{-t\omega^2}$ is positivity improving.
\ko{positivity improving}\ec
\proof
This immediately follows
from the Gaussian bound \kak{g1}.
\qed

\bc{suzuki1}
Suppose Assumption \ref{w}.
Then it follows that
$$\|\omega^{-n/2} f
\|\leq C \|\omega_\infty^{-n/2} f\|.$$
\ec
\proof
Since $\omega^{-n/2}=
C_n
\int_0^\infty e^{-t\omega^2}
t^{(n+4)/4}dt$
with
$C_n=(\int_0^\infty
e^{-s}s^{(n+4)/4}ds)\f$.
Hence the corollary
follows from Proposition \ref{gerard}.
\qed

\bl{suzuki}
Suppose Assumptions \ref{w} and \ref{chi}.
Then
$$\double(X,Y,|t|)\geq 0,\quad
\double_\infty (X,Y,|t|)\geq 0$$
 and there  exist constants
$C_j>0$, $j=1,2,3,4$,
 such that
\eq{ko2ko}
C_1 \double_\infty (X,Y,C_2|t|)
\leq \double(X,Y,|t|)
\leq
C_3 \double_\infty (X,Y,C_4|t|)
\en
for all $X,Y\in \BT$ and $t\in\RR$.
In particular it follows that
\eq{yoko5}
\gamma(T)\leq \EE_{\mu_T}
\lkkk
e^{- \WTZ}
\rkkk.
\en
\el
\proof
Set $\rho_X(x)=\rho (x-X)$.
We note that the function
$f(x)=e^{-\sqrt x}$ on $[0,\infty)$ is completely monotone, i.e.,
$(-1)^n  df(x)/dx^n\geq0$ and that $f(+0)=0$. Then
 there exists a Borel probability measure $m$ on $[0,\infty)$ such that
$$e^{-\sqrt x}=
\int_0^\infty e^{-sx} dm(s)$$
and
it is indeed exactly given by
$$dm(s)=\frac{1}{2\sqrt\pi}\frac{e^{-1/(4s)}}{s^{3/2}}ds.$$
Hence
$$e^{-t\omega}=\int_0^\infty e^{-st^2\omega^2}dm(s)=
\frac{1}{2\sqrt\pi}
\int_0^\infty
\frac{t e^{-t^2/(4s)}}
{ s^{3/2}}
e^{-s \omega^2} ds.$$
It follows that
$$\double( X,Y,|t|)=
\half
\int_t^\infty dr
(\rho_X, e^{-r\omega} \rho_Y)
=
\frac{1}{4\sqrt \pi}
\int_t^\infty dr
\int_0^\infty
\frac{r e^{-r^2/(4p)}}
{p^{3/2}}
(\rho_X, e^{-p\omega^2}\rho_Y)
 dp.$$
Hence
$\double( X,Y,|t|)>0$ follows,
since $e^{-p\omega^2}$
is
positivity improving for  $p>0$.
$\double_\infty  (X,Y,|t|)>0$ also follows in the same way as above.
Since $\rho_X$ and $\rho_Y$ are nonnegative,
by the Gaussian bound
$
c_1 e^{-c_2t\omega_\infty}(x,y)
\leq e^{-t\omega^2}(x,y)\leq
c_3 e^{-c_4t\omega_\infty}(x,y)
$,
 we can see that by changing a variable,
$$
c_1c_2 \double_\infty (X,Y, \sqrt{c_2}|t|)
\leq \double( X,Y,|t|)
\leq
c_3c_4 \double_\infty
(X,Y, \sqrt{c_4}|t|).
$$
Then the lemma follows.
\qed
Let us take $\la$ such that
\eq{assalp}
\frac{1}{\delta  +1}<\la<1,
\en
where $\delta  $ is the positive constant given in Assumption \ref{v}.
Let
\eq{ko11}
A_T=\BT\times
\lkk
\sup_{|s|\leq T}|X_s|\leq T^\lambda\rkk
\subset\BT\times\ms X.
\en
We divide the right-hand side of \kak{yoko5} into
$
\mathbb E_{\mu_T}[\one_{A_T}\cdots]
+
\mathbb E_{\mu_T}[\one_{A_T^c}\cdots]$.
Then in order to prove the absence of ground state it is enough to show that
$
\lim_{T\rightarrow \infty}
\EE_{\mu_T}\left[
\one_{A_T}\cdots\rkkk=0$
and
$\lim_{T\rightarrow \infty}
\EE_{\mu_T}\left[
\one_{A_T^c}
\cdots\rkkk=0$.
\bl{yui1}
{\rm  \cite{lms02}}
Suppose Assumptions \ref{ass}, \ref{v},
\ref{w}, \ref{chi} and \ref{ass1}.
Then it follows that
\eq{yui3}
\d \lim_{T\rightarrow \infty}
\EE_{\mu_T}\left[
\one_{A_T}e^{- \WTZ}\right]=0.
\en
\el
\proof
Since
the integral kernel of
$e^{-|t|\omega_\infty}$ is
$$
e^{-|t|\omega_\infty}(x,y)=
\frac{1}{\pi^2}
\frac{|t|}{(|x-y|^2+|t|^2)^2},$$
we have
\eq{81}
\double_\infty (X,Y,|t|)=
\frac{1}{4\pi^2}
 \int dx\int dy
\frac{{\rho  (x)}\rho  (y)}
{|(x-X)-(y-Y)|^2+|t|^2}.
\en
On $A_T$ we know that
$|(X_s-x)-(X_t-y)|^2+|t-s|^2\leq
8T^{2\la}+2|x-y|^2+|t-s|^2$.
Let \begin{align*}
\Delta_T&=\{(s,t)|0\leq s\leq T, 0\leq t\leq T, 0\leq s+t\leq T/\sqrt 2\},\\
\Delta_T'&=\{(s,t)|0\leq s\leq T/\sqrt 2, -s\leq t\leq s\}.
\end{align*}
Since
\begin{align*}
\int_{-T}^0 dt \int_0^T dt \frac{1}{a^2+|t-s|^2}
&\geq \int\int_{\Delta_T}ds dt \frac{1}{a^2+|s+t|^2}\\
&=
\int\int_{\Delta'_T}dsdt\frac{1}{a^2+s^2}=\log\lk
\frac{a^2+T^2/2}{a^2}\rk,
\end{align*}
we have
\begin{align*}
&\one_{A_T}
\int_{-T}^0 ds \int_0^T dt
\double_\infty (X_s,X_t,C_2|s-t|)\non \\
&
\geq
\frac{1}{4\pi^2}
\one_{A_T}
\int_{-T}^0 ds \int_0^T dt\int dx dy
\frac{\rho (x)\rho (y)}{
8T^{2\la}+2|x-y|^2+C_2|t-s|^2}\non
\\
&
\geq
\frac{1}{4C_2\pi^2}
\one_{A_T}
\int dxdy\rho  (x)\rho  (y)
\log\lk
\frac
{8T^{2\la}+2|x-y|^2+C_2T^2/2}
{8T^{2\la}+2|x-y|^2}
\rk.
\end{align*}
Note that $\rho \geq0$ and $\la<1$.
Since the right-hand side above goes to $+\infty$  as $T\to \infty$, \kak{yui3}
follows.
\bl{yui2}
Suppose Assumptions \ref{ass}, \ref{v},
\ref{w}, \ref{chi} and \ref{ass1}.
Then it follows that
\eq{yui4}
\lim_{T\rightarrow \infty}
\EE_{\mu_T}\left[
\one_{A_T^c}
e^{- \WTZ}\right]=0.
\en
\el
\proof
Note that
\eq{sa3}
\WTI
\leq \frac{T}{2}
 \|\omegai\f\rho\|^2
 \en
  and
\eq{sa4}
\WTTI
\leq 4T
 \|\omegai\f\rho\|^2.
 \en
 Then
$$
\EE_{\mu_T}
\left[
\one_{A_T^c}
e^{- \wti}
\right]
\leq
e^{ (T/2)\|\omegai\f\rho\|^2}
\EE_{\mu_T}
\left[
\one_{A_T^c}
\right].$$
By the Schwartz inequality we have
\begin{align}
&
e^{ (T/2)\|\omegai\f\rho\|^2}
\EE_{\mu_T}
\left[
\one_{A_T^c}
\right]=
e^{ (T/2)\|\omegai\f\rho\|^2}
\frac{
\ix{\one_{A_T^c}e^{(1/2)\wtt}}}
{\ix{e^{(1/2)\wtt}}}\non
\\
&
\label{67}
\leq
e^{ (T/2)\|\omegai\f\rho\|^2}
\frac{
\lk
\ix{e^{ \wtt}}\rk^\han}
{\ix{e^{(1/2)\wtt}}}
\ix{\one_{A_T^c}}.
\end{align}
By Lemma \ref{suzuki}
bounds
$$
C_1
\int_{-T}^Tds\int_{-T}^T dt \double_\infty(X_s,X_t, C_2|s-t|)
\leq
\WTT
$$
and
$$
\WTT\leq
C_3
\int_{-T}^Tds\int_{-T}^T dt
\double_\infty(X_s,X_t, C_4|s-t|)
$$
are derived.
Then
we obtain
$$
\frac{
\lk
\ix{e^{ \wtt}}\rk^\han}
{\ix{e^{(1/2)\wtt}}}
\leq
\frac{
\lk
\ix{e^{C_3\int_{-T}^T \!ds\int_{-T}^T\!dt \double_\infty(X_s,X_t,C_4|s-t|)
}}\rk^\han}
{\ix{e^{(C_1/2)\int_{-T}^T \!ds
\int_{-T}^T\!dt \double_\infty
(X_s,X_t,C_2|s-t|)
}}}
$$
and by \kak{sa4}
there exists $\epsilon>0$ such that
\eq{26}
\frac{
\lk
\ix{e^{ \wtt}}\rk^\han}
{\ix{e^{(1/2)\wtt}}}\leq
e^{\epsilon T\|\omegai\f\rho\|^2}.
\en
It remains to  estimate
$\ix{\one_{A_T^c}}$ in \kak{67}.
There exists
an at most polynomially growth
function $\xi(T)$ such that
\eq{27}
\ix{\one_{A_T^c}}\leq \xi(T) \exp\lk
-cT^{\lambda(\delta  +1)}\rk
\en
with some constant $c>0$. This is proven in
Lemma \ref{lms2} below.
By  \kak{67}, \kak{26} and
\kak{27}
we have
\eq{28}
\lim_{T\rightarrow\infty}
\EE_{\mu_T}[\one_{A_T^c}]
\leq
\lim_{T\rightarrow\infty}
\xi(T) e^{-cT^{\lambda(\delta  +1)}}
e^{ (\epsilon+\han)
 T\|\omegai\f\rho\|^2}=0,
\en
since $\frac{1}{\delta  +1}<
\lambda<1$.
Then \kak{yui4} follows.
\qed
Let us now consider some path properties of $X$ to show \kak{27}.
\bp{kv}
Let $P(B)=\int \one_B \mup dP^\xx $ be
the probability measure on $\BT\times \ms X$ and $\La>0$.
Suppose Assumptions \ref{ass},
\ref{v},
and \ref{ass1}.
Suppose that $f\in C(\BT)\cap \dom(L_{\rm p}^\han)$.
Then
it  follows that
\eq{d1}
P\lk
\sup_{0\leq s\leq T}|f(X_s)|
\geq \La
\rk
\leq \frac{e}{\La}
\sqrtt
{(f, f)_{\HP}+T(L_{\rm p}^\han f, L_{\rm p}^\han f)_{\HP}}.
\en
\ep
\proof
The proof is a  modification of
that of \cite[Lemma 1.4 and Theorem 1.12]{var}.
Set $T_j=Tj/2^n$, $j=0,1,...,2^n$ and we fix $T$ and $n$.
Let $G=\{x\in\BT| |f(x)|\geq \La\}$, then the stopping time $\tau$ is defined by
$$\tau=
\inf\{T_j \geq 0|X_{T_j}\in G\}.$$
Then it follows that
$$P\lk\sup_{j=0,...,2^n}
|f(X_{T_j})|\geq \La\rk
=P(\tau\leq T).$$
 We estimate the right-hand side above.
 Let $0<\chi<1$ be fixed and we choose
 a suitable $\chi$ later.
We see that
\begin{eqnarray}
P(\tau\leq T)
&=&
\ix{\one_{\tau\leq T}}
\leq
\ix {\chi ^{\tau-T}}\non \\
&\leq& \label{d9}
\chi ^{-T}\ix {\chi ^{\tau}}
\leq
\chi ^{-T} \lk
\int \mup (\EE^\xx [\chi ^\tau])^2\rk^\han.
\end{eqnarray}
Let $0\leq \psi$ be any function such that
$\psi(\xx)\geq 1$ on $G$.
Then
the Dirichlet principle
\ko{Dirichlet principle}\eq{d5}
\int \mup (\EE^\xx [\chi ^\tau])^2
\leq
(\psi, \psi)_{\HP}+\frac{\chi ^{T/2^n}}{1-\chi ^{T/2^n}}(\psi, (\one-\eee)\psi)_{\HP}
\en
follows.
We prove this in the next lemma.
Inserting
$$|f(x)|/\La=\lkk
\begin{array}{ll}\geq 1,& x\in G,\\
|f(x)|/\La,&x\in G^c,
\end{array}
\right.$$ into $\psi$ in \kak{d5},
we have
\eq{d88}
\int \mup (\EE^\xx [\chi ^\tau])^2
\leq
\frac{1}{\La^2}
(f,f)_{\HP}+\frac{\chi ^{T/2^n}}{1-\chi ^{T/2^n}}\frac{1}{\La^2}
(|f|, (\one-\eee)|f|)_{\HP}.
\en
Since $\eee$ is positivity improving and then
$$(|f|, (\one-\eee)|f|)_{\HP}
\leq
(f, (\one-\eee)f)_{\HP},$$
we have by \kak{d9},
$$
P\lk\sup_{j=0,...,2^n}
|f(X_{T_j})|\geq \La\rk
\leq
\frac{\chi ^{-T}}{\La}
\sqrtt{(f,f)_{\HP}+
\frac{\chi ^{T/2^n}}{1-\chi ^{T/2^n}}
(f, (\one-\eee)f)_{\HP}
}.$$
Set $\chi =e^{-1/T}$.
Then by $\frac{\chi ^{T/2^n}}{1-\chi ^{T/2^n}}
\leq 2^n$,
we have
\eq{d6}
P\lk\sup_{j=0,...,2^n}
|f(X_{T_j})|\geq \La\rk
\leq
\frac{e}{\La}
\sqrtt{(f,f)_{\HP}+
2^n
(f, (\one-\eee)f)_{\HP}
}.
\en
Since
$(f, (\one-\eee)f)_{\HP}\leq (T/2^n)(L_{\rm p}^\han f , L_{\rm p}^\han f)$, we obtain
that
\eq{d666}
P\lk\sup_{j=0,...,2^n}
|f(X_{T_j})|\geq \La\rk
\leq
\frac{e}{\La}
\sqrtt{(f,f)_{\HP}+
T
(L_{\rm p}^\han f, L_{\rm p}^\han f)_{\HP}
}.
\en
Take $n\to \infty$ on both sides of \kak{d666}.
By the Lebesgue dominated convergence theorem,
$$\lim_{n\to\infty}
P\lk\sup_{j=0,...,2^n}
|f(X_{T_j})|\geq \La\rk
=
P\lk\lim_{n\to\infty}\sup_{j=0,...,2^n}
|f(X_{T_j})|\geq \La\rk.$$
Since
$f(X_t)$ is continuous in $t$,
$\lim_{n\to\infty}\sup_{j=0,...,2^n}
|f(X_{T_j})|=\sup_{0\leq s\leq T}|f(X_s)|$ follows.
Then
we complete the proposition.
\qed
It remains to show the Dirichlet principle \kak{d5}.
\bl{d7}
\TTT{Dirichlet principle}
\ko{Dirichlet principle}
Suppose Assumptions \ref{ass}, \ref{v}
and \ref{ass1}.
Then  it follows that
\eq{dpr}
\int \mup (\EE^\xx [\chi ^\tau])^2
\leq
(\psi, \psi)_{\HP}+\frac{\chi ^{T/2^n}}{1-\chi ^{T/2^n}}(\psi, (\one-\eee)\psi)_{\HP}
\en
for  any function $\psi\geq 0$ such that
$\psi(\xx)\geq 1$ on $G$.
\el
\proof
Define the function $\psi_\chi$ by $\psi_\chi(\xx)=\EE^\xx [\chi ^\tau]$.
By the definition of $\tau$ we can see that
\eq{d3}
\psi_\chi(\xx)=1,\quad x\in G,
\en
 since $\tau=0$ when
 $X_s$ stars from the inside of $G$.
Let $\mathcal F_t=\s(X_s,0\leq s\leq t)$ be the natural filtration of $\pro X$.
By the Markov property of $X$,
we can directly see that
\eq{429}
\eee \psi_\chi(\xx)=\EE^\xx [\EE^{X_{T/2^n}}
[\chi ^\tau]]=
\EE^\xx [\EE^\xx [\chi ^
{\tau\circ \theta_{T/2^n}}|\mathcal F_{T/2^n}]]
=
\EE^\xx [\chi ^
{\tau\circ \theta_{T/2^n}}],
\en
where $\theta_t$ is the shift on $\ms X$,
which is  defined by  $(\theta_t\omega)(s)=\omega(s+t)$ for $\omega\in \ms X$.
Note that
\eq{ast}
(\tau\circ \theta_{T/2^n})(\omega)=\tau(\omega)-T/2^n
\geq 0,
\en
when $\xx=X_0(\omega) \in G^c$. Hence
by \kak{429} and \kak{ast} we have the identity:
\eq{d4}
\chi ^{T/2^n}\eee \psi_\chi(\xx)=\psi_\chi(\xx),\quad
\xx\in G^c.
\en
It is trivial to see that
$$
\int \mup (\EE^\xx [\chi ^\tau])^2
=(\psi_\chi, \psi_\chi)_{\HP}\leq
(\psi_\chi, \psi_\chi)_{\HP}+\frac{\chi ^{T/2^n}}{1-\chi ^{T/2^n}}(\psi_\chi, (\one-\eee)\psi_\chi)_{\HP}.$$
Let us define
$(f, g)_G=\int_G\mup \bar f(\xx)g(\xx)$.
By the relation \kak{d4} we can compute the right-hand side above as
\eq{d10}
(\psi_\chi, \psi_\chi)_G+\frac{\chi ^{T/2^n}}{1-\chi ^{T/2^n}}
(\psi_\chi, (\one-\eee)\psi_\chi)_G.
\en
Since
\begin{align*}
&
(\psi_\chi, (\one-\eee)\psi_\chi)_G\\
&=
(\psi_\chi\one_G, (\one-\eee)\psi_\chi\one_G)_\HP+
(\psi_\chi\one_G, (\one-\eee)\psi_\chi\one_{G^c})_\HP\\
&=
(\psi_\chi\one_G, (\one-\eee)\psi_\chi\one_G)_\HP-
(\psi_\chi\one_G, \eee \psi_\chi\one_{G^c})_\HP\\
&\leq
(\psi_\chi\one_G, (\one-\eee)\psi_\chi\one_G)_\HP.
\end{align*}
Hence
\eq{d12}
\int \mup (\EE^\xx [\chi ^\tau])^2
\leq
(\psi_\chi\one_G, \psi_\chi\one_G)_{\HP}
+
\frac{\chi ^{T/2^n}}{1-\chi ^{T/2^n}}
(\psi_\chi\one_G, (\one-\eee)\psi_\chi\one_G)_{\HP}
.
\en
Note that $\psi_\chi \one_G(\xx)\leq \psi(\xx)$ for all $\xx\in\BT$.
Then
\begin{eqnarray}
&&
(\psi_\chi\one_G, \psi_\chi\one_G)_{\HP}
+
\frac{\chi ^{T/2^n}}{1-\chi ^{T/2^n}}
(\psi_\chi\one_G, (\one-\eee)\psi_\chi\one_G)_{\HP}\non\\
&&\label{d13}
\leq
(\psi, \psi)_{\HP}+
\frac{\chi ^{T/2^n}}{1-\chi ^{T/2^n}}
(\psi,  (\one-\eee)\psi)_{\HP}.
\end{eqnarray}
Combining \kak{d12} with  \kak{d13},
we prove the lemma.
\qed
\bl
{lms2}
\kak{27} holds.
\el
\proof
Suppose that
$f \in C^\infty(\BT)$, $f(-x)=f(x)$
and
$$f (x)=\lkk
\begin{array}{ll}
|x|,& |x|\geq T^\la,\\
\leq |x|,& T^\la-1<|x|<T^\la,\\
0,& |x|\leq T^\la-1.
\end{array}\right.
$$
Then
we see that
\eq{va123}
\ix{\one_{A_T^c}}=
\ix {\one_{\sup_{|s|<T}|X_s|>T^\la}}
=\ix {\one_{
\sup_{|s|<T}|f(X_s)|>T^\la}}.
\en
By the reflection symmetry of $(X_t)_{t\in\RR}$ we have
$$\ix {\one_{
\sup_{|s|<T}|f(X_s)|>T^\la}}
=2\ix {\one_{
\sup_{0\leq s\leq T}|f(X_s)|>T^\la}}
$$
and by Proposition \ref{kv}
we have
\eq{va1}
\ix {\sup_{|s|<T}|f(X_s)|>T^\la}
\leq
 \frac{2 e}{T^\la}\sqrtt{(f,f)_{\HP}+T (L_{\rm p}^\han f, L_{\rm p}^\han  f)_{\HP}}.
\en
We estimate the right-hand side of \kak{va1}.
First we show $f\gr\in \dom(K)$.
Let $C_0^\infty(\BT)\ni f_R(\xx)=
{\mathcal X} (\xx/R)f(\xx)$,
where ${\mathcal X} \in C_0^\infty(\BT)$ and $${\mathcal X} (\xx)=
\lkk\begin{array}{ll}1,&|\xx|<1,\\
<1,&1\leq |\xx|\leq 2,\\
0,&|\xx|> 2.
\end{array}
\right.
$$
Since
$A_{\mu\nu}$ satisfies the Lipshitz condition (Assumption \ref{ass1}),
$A_{\mu\nu}\in W^{1,\infty}(\BT)$. Then by Lemma \ref{domain}, we see that $\varphi_{\rm p}\in H^2(\BT)$,
$f_R\varphi_{\rm p}\in \dom(K)$, $(K-E) \varphi_{\rm p}=0$, and
$$K f_R\varphi_{\rm p}=\sum_{\mu,\nu=1}^3
\lk D_\mu A_{\mu\nu}
(D_\nu f_R) \varphi_{\rm p}+
(D_\mu f) A_{\mu\nu} D_\nu \varphi_{\rm p}\rk+Ef_R\varphi_{\rm p}.$$
We see that
\begin{align*}
&f_R\varphi_{\rm p}\to  f\varphi_{\rm p},\\
&(D_\mu f_R) A_{\mu\nu}
D_\nu \varphi_{\rm p}
\to
(D_\mu f) A_{\mu\nu} D_\nu\varphi_{\rm p},\\
&
D_\mu A_{\mu\nu}(D_\nu f_R) \varphi_{\rm p}
=
D_\mu A_{\mu\nu}\cdot (D_\nu f_R) \varphi_{\rm p}
 +A_{\mu\nu}\cdot D_\mu (D_\nu f_R)\cdot  \varphi_{\rm p}
 +A_{\mu\nu}(D_\nu f_R) \cdot D_\mu\varphi_{\rm p}\\
&\hspace{3cm}
\to
\lk
(D_\mu A_{\mu\nu}) (D_\nu f)+
A_{\mu\nu} (D_\mu D_\nu f)
\rk \varphi_{\rm p}
\end{align*}
as $R\to\infty$ in $\LRT$.
Since $K$ is closed, $f\varphi_{\rm p}\in \dom(K)$ follows.
By the estimate above we also see that
\begin{align*}
(L_{\rm p}^\han f, L_{\rm p}^\han f)_{\HP}
=\lk
f\varphi_{\rm p},
(D_\mu f) A_{\mu\nu}D_\nu\varphi_{\rm p}+
(D_\nu A_{\mu\nu}) (D_\nu f) \varphi_{\rm p}+
A_{\mu\nu}f^{\nu\mu}\varphi_{\rm p}\rk.
\end{align*}
By the spatial super-exponential decay
$\varphi_{\rm p}(\xx)\leq Ce^{-\gamma |\xx|^{\delta+1}}$
derived in \kak{car1}
we have
\eq{va3}
\|f\varphi_{\rm p} \|^2=
  \int f (\xx)^2\varphi_{\rm p}^2(\xx)d\xx\leq
C^2 e^{-2\gamma  T^{\la(\delta  +1)}}
\int |\xx|^2
e^{-2\gamma  |\xx|^{\delta  +1}}
d\xx.
\en
Note that $D_\mu f,D_\mu D_\nu f\in L^\infty(\BT)$.
Then
\begin{align*}
(L_{\rm p}^\han f, L_{\rm p}^\han f)_{\HP}
&\leq
\|f\varphi_{\rm p}\|
\|
A_{\mu\nu}D_\nu\varphi_{\rm p}+A_{\mu\nu}^\nu \varphi_{\rm p}+A_{\mu\nu}\varphi_{\rm p}\|\\
&\leq
C'
 e^{-\gamma   T^{\la(\delta  +1)}}
\|
A_{\mu\nu}D_\nu\varphi_{\rm p}+A_{\mu\nu}^\nu \varphi_{\rm p}+A_{\mu\nu}\varphi_{\rm p}\|
\end{align*}
follows. Similarly
$$(f,f)_\HP\leq
C^2 e^{-2\gamma  T^{\la(\delta  +1)}}
\int |\xx|^2
e^{-2\gamma  |\xx|^{\delta  +1}}
d\xx$$
is also derived.
Hence
we have
$$ \mathbb E_{\mu_T}\left[
 \one_{A_T^c}
 \right]
\leq
{T^{-\la}}
 \sqrt{a+Tb}
 e^{-(\gamma /2) T^{\la(\delta  +1)}}$$
with some constant $a$ and $b$.
This completes the proof.
 \qed
Now we are in the position to
state the main theorem.
\bt{hiroshima4}
\TTT{Absence of ground state}
\ko{absence of ground state!Nelson Hamiltonian}
Suppose Assumptions \ref{ass}, \ref{v}, \ref{w}, \ref{chi} and \ref{ass1}.
Then there  is no ground states of $H$.
\et
\proof
Since
$
\gamma(T)\leq \EE_{\mu_T}
\lkkk
e^{- \WTZ}
\rkkk$
and
$$\lim_{T\to\infty}
 \EE_{\mu_T}
\lkkk
e^{- \WTZ}
\rkkk=0$$
by Lemmas \ref{yui1} and \ref{yui2},
we obtain $\lim_{T\to\infty}\gamma(T)=0$. Then the theorem follows.
\qed

\section*{Acknowledgments}
FH acknowledges support of Grant-in-Aid for Science Research (B) 20340032 from JSPS and
Grant-in-Aid for Challenging Exploratory Research 22654018 from JSPS.
HS is thankful to the hospitality of Mathematics-for-Industry of Kyushu University from October 22 of 2009 to January 7 of 2010,
where part of this work has been done.

  \printindex
\end{document}